\documentclass[10pt,psfig,a4paper]{dmjthesis}

\newcommand{\etalcite}[2]
{#1 \emph{et al.}~\cite{#2}}

\newcommand{\subsize}
{\scriptsize}

\newcommand{\subsubsize}
{\tiny}

\newcommand{\sub}[1]
{_\textrm{\subsize #1}}

\newcommand{\subsub}[1]
{_\textrm{\subsubsize #1}}

\newcommand{\super}[1]
{^\textrm{\subsize #1}}

\newcommand{\gammafkt}[1]
{\Gamma\big(\frac{#1}{2}\big)}

\newcommand{\gammafktB}[2]
{\frac{\Gamma\big(\frac{#1}{2}\big)}{\Gamma\big(\frac{#2}{2}\big)}}

\newcommand{\angleint}[2]
{\int_0^{\pi/2}d#1\sin^2#1\cos^{#2}#1}

\newcommand{\BE}[1]
{Bose-Einstein}

\newcommand{\hatR}{\hat{\mathrm R}}
\newcommand{\JacP}{{\mathcal P}}
\newcommand{\JacQ}{{\mathcal Q}}
\newcommand{\permO}{{\mathrm P}}
\newcommand{\hypgeoF}{{\mathcal F}}
\newcommand{\cmplxI}{\mathrm i}

\newcommand{\Smartfig}[1]{\Reffig.~\ref{#1}}
\newcommand{\smartfig}[1]{\reffig.~\ref{#1}}
\newcommand{\reffig}[1] {figure}
\newcommand{\Reffig}[1] {Figure}
\newcommand{\reffigs}[1] {figures}

\newcommand{\Smarteq}[1] {\Refeq.~(\ref{#1})}
\newcommand{\smarteq}[1] {\refeq.~(\ref{#1})}
\newcommand{\smarteqs}[1] {\refeqs.~(\ref{#1})}

\newcommand{\refeq}[1] {equation}
\newcommand{\Refeq}[1] {Equation}
\newcommand{\refeqs}[1] {equations}
\newcommand{\Refeqs}[1] {Equations}

\newcommand{\refsec}[1] {section}
\newcommand{\refchap}[1] {chapter}

\newcommand{\publikation}[5] {\begin{itemize}\item[#1.]
    ``\textbf{#2}'';\\ #3;\\#4;\\ \emph{Internet}: #5.\end{itemize}}

\newcommand{\Index}[1]{#1\index{#1}}

\newcommand{\angval}{\index{angular eigenvalue}}
\newcommand{\angvar}{\index{angular variational equation}}
\newcommand{\bec}{\index{Bose-Einstein condensation}}
\newcommand{\born}{\index{scattering length!Born approximation}}
\newcommand{\couplingterms}{\index{coupling terms}}
\newcommand{\efithree}{\index{Efimov effect!for three particles}}
\newcommand{\efiN}{\index{Efimov effect!for $N$ particles}}
\newcommand{\faddec}{\index{Faddeev decomposition}}
\newcommand{\fadeqthree}{\index{Faddeev equations!for three particles}}
\newcommand{\fadeqN}{\index{Faddeev equations!for $N$ particles}}
\newcommand{\fadyak}{\index{Faddeev-Yakubovski{\u\i}}}
\newcommand{\fermion}{\index{fermions}}

\newcommand{\gaussian}{\index{Gaussian two-body interaction}}
\newcommand{\gpe}{\index{Gross-Pitaevskii equation}}
\newcommand{\hartree}{\index{Hartree}}
\newcommand{\hypcor}{\index{hyperspherical coordinates}}
\newcommand{\jaccor}{\index{Jacobi coordinates}}
\newcommand{\jacpol}{\index{Jacobi function}}
\newcommand{\jastrow}{\index{Jastrow}}
\newcommand{\lambdadelta}{\index{angular eigenvalue!lambdadelta@$\lambda_\delta$}}
\newcommand{\lambdainfty}{\index{angular eigenvalue!lambdainfty@$\lambda_\infty$}}
\newcommand{\lambdaschematic}{\index{angular eigenvalue!parametrization}}
\newcommand{\maccol}{\index{macroscopic collapse}}
\newcommand{\macstab}{\index{macroscopic stability}}
\newcommand{\mactun}{\index{macroscopic tunneling}}
\newcommand{\radpot}{\index{radial potential}}
\newcommand{\recthree}{\index{three-body recombination}}
\newcommand{\scatlen}{\index{scattering length}}
\newcommand{\threecor}{\index{three-body correlations}}
\newcommand{\traplength}{\index{trap length}}
\newcommand{\twobound}{\index{two-body bound state}}
\newcommand{\twothreshold}{\index{two-body threshold}}
\newcommand{\zerorangepot}{\index{zero-range interaction}}
\newcommand{\zerorangeappr}{\index{zero-range approximation}}

\usepackage{makeidx}
\makeindex

\usepackage{longtable}

\usepackage{t1enc}      
\usepackage[english]{babel}
\usepackage{layout}     
\usepackage{epsf}
\usepackage{psfrag}     
\usepackage[dvips]{graphicx}   
\usepackage{makeidx}    
\usepackage{amssymb}    
\usepackage{amsbsy}     
\usepackage{amstext}    
\usepackage{rotating}   
\usepackage{epsfig}
\usepackage{afterpage}
\usepackage{float}
\makeatletter
   \newcommand{\chapapp}{\MakeLowercase\@chapapp}
   \newcommand{\Chapapp}{\@chapapp}
\makeatother
\begin{document}

\pagenumbering{roman}                                         
\newfont{\bigtitlefont}{cmss17 scaled 1500}
\newfont{\smalltitlefont}{cmss12 scaled 1000}
\title{\vspace{-2cm}
       \rule{\linewidth}{1mm}\\[10pt]
       \vspace{0.3cm}
       {\bigtitlefont Brief Encounters}\\
       {\smalltitlefont Binary Correlations among Bosons}
       \rule{\linewidth}{1mm}\\[1.5cm]
       Ph.D. Thesis\vspace{1.5cm}
        } 
\author{
        {Ole S\o rensen}\\[1cm]
        \includegraphics[width=4cm]{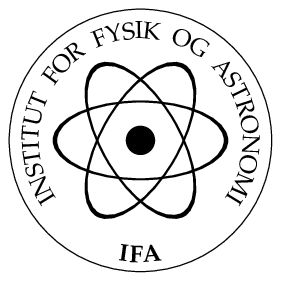}\\[0.5cm]
        \smalltitlefont Department of Physics and Astronomy\\
        \smalltitlefont University of Aarhus\\[0.3cm]}
\date{\smalltitlefont January 2004}
\maketitle

\cleardoublepage

\chapter*{Preface}
\addcontentsline{toc}{chapter}{Preface}

This thesis is presented for the Faculty of Science at the University
of Aarhus, Denmark, in order to fulfil the requirements for the Ph.D.
degree in physics.

The thesis gives an account of the work done during my Ph.D.  studies
at the Department of Physics and Astronomy at the University of
Aarhus.  First of all, I thank my supervisors Aksel Jensen and Dmitri
Fedorov.  I am grateful to their encouragement and our many
discussions through the four years.  Also a thank you to Esben
Nielsen, Han Guangze, and Takaaki Sogo, and thanks to the people at
the institute, at conferences, and at other physics meetings.  Thanks
to Karen Marie Hilligsøe for proof-reading.

Thanks to fellow students and friends.

\vspace{.5cm}

\textsl{Tak til familie og venner.}

\vspace{\stretch{1}}

\hspace{\stretch{1}}Ole Sørensen

\hspace{\stretch{1}}Århus, January 2004

\vspace{\stretch{9}}

\chapter*{List of publications and preprints}
\addcontentsline{toc}{chapter}{List of publications and preprints}

Most subjects in this thesis are also discussed in the following
publications and preprints:

\publikation{1}{Two-body correlations in Bose-Einstein condensates}
{Ole S\o rensen, Dmitri V.~Fedorov, Aksel S.~Jensen, and Esben
  Nielsen} {Phys.~Rev.~A \textbf{65}, 051601(R) (2002)}
{http://arxiv.org/abs/cond-mat/0110069}

\publikation{2}{Towards Treating Correlations in Bose Condensates}
{Aksel S.~Jensen, Ole S\o rensen, Dmitri V.~Fedorov, and Esben
  Nielsen} { Few-Body Systems \textbf{31}, 261 (2002)}
  {http://link.springer.de/link/service/journals/00601/bibs\\
  \phantom{Internet:x}/2031002/20310261.htm}

\publikation{3}{Correlated Trapped Bosons and the Many-Body {E}fimov
  Effect}{Ole Sørensen, Dmitri V.~Fedorov, and Aksel
  S.~Jensen}{Phys.~Rev.~Lett.~\textbf{89}, 173002 (2002)}
{http://arxiv.org/abs/cond-mat/0203400}

\publikation{4}{Two-body correlations in {$N$}-body boson systems}
{Ole Sørensen, Dmitri V.~Fedorov, and Aksel S.~Jensen} {Phys.~Rev.~A
  \textbf{66}, 032507 (2002)} {http://arxiv.org/abs/cond-mat/0210095}

\publikation{5}{Decay of boson systems with large scattering length}
{Ole Sørensen, Dmitri V.~Fedorov, and Aksel S.~Jensen} {J.~Opt.~B:
  Quantum Semiclass.~Opt.~\textbf{5}, S388 (2003)}
{http://www.iop.org/EJ/abstract/1464-4266/5/3/374}

\publikation{6}{Two-Body Correlations and the Structure of
  Bose-Einstein Condensates} {Ole Sørensen, Dmitri V.~Fedorov, and
  Aksel S.~Jensen} {Few-Body Systems Suppl.~\textbf{14}, 373 (2003);\\
  Proceedings of the XVIIIth European Conference on Few-Body Problems
  in Physics, Bled, Slovenia, 2002}
  {http://www.springer.at/main\\
  \phantom{Internet:x}/book.jsp?bookID=3-211-83900-3\&categoryID=13}

\newpage

\publikation{7}{Correlated {$N$}-boson systems for arbitrary
  scattering length} {Ole Sørensen, Dmitri V.~Fedorov, and Aksel
  S.~Jensen} {Phys.~Rev.~A \textbf{68}, 063618 (2003)}
{http://arxiv.org/abs/cond-mat/0305040}

\publikation{8}{Structure of boson systems beyond the mean-field} {Ole
  Sørensen, Dmitri V.~Fedorov, and Aksel S.~Jensen} {J.~Phys.~B:
  At.~Mol.~Opt.~Phys.~\textbf{37}, 93 (2004)}
{http://arxiv.org/abs/cond-mat/0306564}

\publikation{9}{Condensates and Correlated Boson Systems} {Ole
  Sørensen, Dmitri V.~Fedorov, and Aksel S.~Jensen} {to be published
  in Few-Body Systems;\\ Conference on Critical Stability III, Trento,
  Italy, 2003} {http://arxiv.org/abs/cond-mat/0310304}

\publikation{10}{Stability, effective dimensions, and interactions for
  bosons in deformed fields} {Ole Sørensen, Dmitri V.~Fedorov, and
  Aksel S.~Jensen} {revised version will be submitted to Phys.~Rev.~A}
{http://arxiv.org/abs/cond-mat/0310462}


\tableofcontents                                              

\cleardoublepage
\pagenumbering{arabic}

\chapter{Introduction}                                        
\label{kap:introduction}
A description of a large system of particles is often sought in a
derivation from the detailed behaviour of just a few of the particles.
The present thesis deals with the connection between such microscopic
features and the nature of a collection of many particles.  A study of
identical bosons is an obvious first investigation of this link, but
the ideas might also be applied to fermion systems or to systems with
mixed symmetry.  The relatively well-known, flexible properties of
cold alkali gases have presented questions which might be addressed by
a study of few-body correlations.  In this chapter we comment on basic
features of bosonic systems and motivate a description of few-body
correlations within a many-particle system.\footnote{The use of ``we''
  refers to the author with the reader's participation.}

\fermion

\section{Bosons}

All particles can be classified as either bosons or fermions.  The
distinction is important when identical particles approach each other.
Electrons, nucleons, and quarks are fermions and obey the Fermi-Dirac
statistics, while force carriers like photons and gluons are bosons
and obey the Bose-Einstein statistics.  Atomic nuclei, atoms, and
molecules obey one of the statistics depending on the number of
contained fermions.  In this thesis we as far as possible consider
bosons generally, but often relate to bosonic neutral atoms with an
even number of neutrons, in particular alkali atoms like $^{87}$Rb or
$^{23}$Na.  Although bosons are the main objects, we will a couple of
times consider extensions of the methods to deal with fermions.

\hartree\fermion

At most one fermion can occupy the same quantum state, whereas bosons
are not restricted.  An example is Bose-Einstein condensation of a
vast number of bosons in the same single-particle state, which was
experimentally achieved in 1995 by cooling dilute alkali gases
\cite{and95,bra95,dav95}.

\gpe

A mean field is the basis for the description of dilute alkali gases
by the Gross-Pitaevskii equation \cite{edw95,bay96}.  In mean-field
self-consistent theories \cite{bra83} the influence from interactions
is included as an average, hence the name mean field.  Such a
description is reasonable when the interaction between particles is so
weak that each particle only feels the other particles as an average
background cloud in which they move.  The rigorous criterion is that
the mean free path is long compared to the spatial extension of the
system.  Reviews of theoretical developments before and after the
experimental realization of Bose-Einstein condensation are given in
the references \cite{dal99,pet01,pit03}.

At strong interaction or large densities each particle might interact
strongly with one or a few of the other particles.  The particles then
adapt to the local environment, but still feel the (weaker) mean-field
influence from the remaining particles.  This competition between the
background and the local surroundings is important when we formulate
the theory in chapter~\ref{kap:hyperspherical_method}.

\gpe

When the attraction is too large, for example when condensates
collapse \cite{sac98,sac99,don01,rob01}, correlations beyond the mean
field are crucial.  Descriptions based on the mean-field
Gross-Pitaevskii equation \cite{adh02b,ued03} can account for this
collapse, but avoid direct relations to underlying microscopic
processes.

\hartree

Some methods go beyond the mean field, but still avoid the explicit
inclusion of correlations.  An example is the Skyrme-Hartree-Fock
method with density-dependent interactions \cite{sie87}.  Related are
low-density expansions of the total energy for a many-boson system
\cite{lee57,bra02b}.

\zerorangepot

For large densities particle encounters are more frequent and at least
two-body correlations need to be explicitly included.  This cannot be
done directly on top of the usual mean field with a two-body contact
interaction, i.e.~of zero range, which would lead to a wave function
with zero separation and diverging energy \cite{fed01}.  On the other
hand, the use of realistic potentials in self-consistent mean-field
calculations leads to disastrous results because the Hilbert space
does not include correlations as needed to describe both the short-
and long-range asymptotic behaviour \cite{esr99b}.

\jastrow

An explicit inclusion of correlations is done by the Jastrow method
\cite{jas55} where the many-body wave function is written as a product
of two-body amplitudes instead of one-body amplitudes.  With a few
assumptions about the asymptotic behaviour of the amplitudes, this
results in variational numerical procedures that can be carried
through for many-boson systems also for large densities
\cite{mou01,cow01}.

\section{Two-body properties}

\scatlen

A study of the properties of a many-particle system requires
understanding of the two-body problem.  The interaction between
neutral atoms is repulsive at short distances and attractive at large
distances.  There may for alkali atoms be a large number of bound
two-body states that are sensitive to the details of the interaction.
However, when two atoms approach each other slowly from afar, their
encounter can be described in a universal way irrespective of the
short-distance details of the interaction.  The determining parameter
is then the scattering length.  This is infinitely large in the
presence of a two-body bound state with vanishing energy.  Then the
two particles correlate in all space, which is the opposite limit than
assumed by a mean field where no correlations are allowed.

At a \Index{Feshbach resonance}, when the energy in a scattering
channel coincides with the energy of a bound state in another channel
\cite{pet01}, the scattering length diverges in the same way as when a
two-body bound state occurs.  In recent years such a \Index{Feshbach
  resonance} has been investigated for both sodium \cite{ino98,ste99}
and rubidium gases \cite{rob98,rob01b}.  For $^{85}$Rb atoms an
external magnetic field can slightly change the effective two-body
potential curves, which have resulted in a tool for tuning the
two-body scattering length \cite{cor00}.

Mean-field studies of dilute boson systems often include a two-body
contact interaction with a coupling strength given by the scattering
length.  At large densities this becomes a problem since the
interaction energy for the bosons then diverges.  It is also difficult
to handle large scattering lengths close to a Feshbach resonance.  The
alternative use of a boundary condition given by the scattering length
at zero separation between the bosons allows larger density and
scattering length \cite{fed01b}.

Ultimately, the best description of the physics properties could be
obtained by using realistic potentials which besides the correct
large-distance two-body behaviour also incorporate high-energy
features and the correct nature of bound states.  However, this is
especially difficult when the two-body system contains innumerable
bound states, as is the case for the alkali atoms in experiments.
Furthermore, if the goal is a description of the low-energy two-body
properties within the many-body system, it would be an investment of
too much effort in the wrong place.  A more rewarding method is to use
a simpler finite-range potential with the correct scattering length,
for instance a linear combination of Gaussian potentials \cite{blu01}.
Considerations about the two-body interaction return in chapters
\ref{kap:angular} and \ref{kap:meanfield_validity}.

\section{Few-body physics}

\jastrow\fadyak

A system with only a few particles can be described accurately without
crude assumptions.  The related methods can provide insight into
complicated problems, for instance how two or three particles approach
each other within a many-body system.  If the spatial extension of the
system is large, an encounter of two particles can be considered as a
pure two-body process with an average background influence from the
other particles.  Alternatively, it might be considered as a
three-body process, the third body being the collection of the other
particles.  In smaller systems three particles approach each other
more frequently, which then demands a description of a true three-body
process.  Faddeev \cite{fad61simple} wrote a wave function as a sum of
terms that account for pairwise encounters, which by Yakubovski{\u\i}
\cite{yak67simple} was extended to account for the right behaviour of
three-body clusters, four-body clusters, and so on.  Within each
cluster it is then possible to treat the degrees of freedom rather
accurately, while keeping in mind that the present collection of
particles moves relative to the particles outside the cluster.  The
art of these Faddeev-Yakubovski{\u\i} techniques is to make the proper
assumption about the dominating structure of the many-body system;
otherwise little is accomplished and the calculations turn out as
complicated as for some methods based on a Jastrow ansatz.

\fadeqN\fadyak

Specifically for bosons, but implemented in nuclear physics, an
approach with inclusion of two-body correlations was worked out by
\etalcite{de la Ripelle}{rip84,rip88}.  This approach is equivalent to
the Faddeev-like equations to be described in
\refchap.~\ref{kap:hyperspherical_method}.  The result is an
eigenvalue equation in only one variable.  Barnea \cite{bar99a}
proposed the inclusion of higher-order correlations in a method which
reminded of the Yakubovski{\u\i} technique.

The hyperspherical adiabatic method, which was formulated for a study
of the helium atom by Macek \cite{mac68}, separates the description of
the three-body system into a common length scale, i.e.~the
hyperradius, and an additional hyperangle.  The extension of this
method is now frequently used in atomic physics for descriptions of
many-electron systems \cite{lin95}.

\gpe

After the experimental realization of Bose-Einstein condensation, most
descriptions of this phenomenon were based on the mean-field
Gross-Pitaevskii equation.  However, \etalcite{Bohn}{boh98} introduced
a hyperspherical method to study the effect of interactions within the
many-boson system.  This hyperspherical method was simplified by the
inclusion of a contact interaction and the assumption that the only
dependence is on the average distance from the centre of mass, i.e.~no
correlations were allowed, which reminds of a mean field.  An
advantage of the hyperspherical method is that it provides an
effective potential in a linear eigenvalue equation, in contrast to
the non-linear nature of the Gross-Pitaevskii equation.  In a related
study Blume and Greene \cite{blu02c} calculated the properties of
three bosons in an external trap with no assumptions about the
structure of the wave function, and thus confirmed some of the
behaviours of the effective potentials in the hyperspherical model for
the many-boson system.

\scatlen\fadeqthree

The detailed study of three particles provides an important first step
in the formulation of a theory for clusterizations within a many-body
system.  The Faddeev-formalism is often applied within the
hyperspherical adiabatic approximation when the emphasis is on the
asymptotic two-body properties \cite{jen97,nie01}.  The threshold
phenomenon of infinitely many bound three-body states in the case of a
two-body bound state with zero energy \cite{efi70} can be described by
just the inclusion of two-body correlations \cite{fed93}.  This
anticipates that a generalization of the method to a many-boson system
might describe the case of large scattering length where
non-correlated models become inadequate.

\section{The thesis work}

The work for this thesis started from a few-body description of
two-body correlations within a many-body system, and it has been
centred on solving the many-boson problem in a hyperspherical frame.
Central questions are formulated as follows.

\subsection*{What is the effect of two-body correlations?}

The outset for the thesis work was to understand how two-body
correlations influence the properties of a many-boson system.  This
work is closely related to studies of the three-body system, see e.g.
\etalcite{Nielsen}{nie01}, and to the hyperspherical investigation of
the average properties of the many-boson system by
\etalcite{Bohn}{boh98}.  The formulation of the main techniques behind
the inclusion of such correlations was given in the publications
\cite{sor01,sor02b}.  This is collected in chapter
\ref{kap:hyperspherical_method} and appendix
\ref{sec:hyper-matr-elem}.  The effects of two-body correlations were
mainly discussed in the references \cite{sor02b,sor03a} and are here
collected in chapters \ref{kap:angular} and \ref{kap:radial}.

\subsection*{What happens close to a resonance?}

\scatlen

The case of large scattering length provides for the three-body case
the Efimov phenomenon of many bound three-body states
\cite{efi70,fed93}.  The possibility of similar threshold effects for
the many-boson system was investigated during the thesis work and the
results were published in \cite{sor02}.  This is discussed mainly in
chapters \ref{kap:angular} and \ref{kap:radial}.

\subsection*{Are the deviations from the mean field trustworthy?}

The relations to the mean field and the deviations of the results were
published in \cite{sor03b}, where also ranges of validity were
considered.  This is here included in chapter
\ref{kap:hyperspherical_method} and in chapter
\ref{kap:meanfield_validity}.

\subsection*{Do two-body correlations influence stability?}

\recthree \maccol \mactun

The macroscopic stability problems for a many-boson system were
briefly discussed in the previous publications, but were further
addressed in relation to the observed phenomenon of macroscopic
collapse \cite{rob01,don01} in the reference \cite{sor03a}, which also
included a discussion of the competition between three-body
recombinations, macroscopic tunneling, and macroscopic collapse.
Chapter \ref{kap:stability_validity} collects these considerations.

\subsection*{How does the geometry influence the system?}

Deformation effects were discussed in a preprint \cite{sor03d}, which
gave stability criteria, effective dimensions, and effective
interactions for bosons in a deformed external field.  This discussion
is continued in chapter \ref{kap:deformed}.  Since an inclusion of
correlations in the deformed case is not presently implemented, this
treatment is somehow similar to a mean-field treatment.

\section{Thesis outline}

\gpe

The structure of the thesis can be summarized as follows.
Chapter~\ref{kap:hyperspherical_method} presents the hyperspherical
frame for studying few-body correlations within the many-boson system.
This includes a discussion of the structure of the wave function and
some comments on the mean-field wave function.
We present the necessary assumptions when restricting to two-body
correlations and the main features of the resulting equations of
motion.
Chapter~\ref{kap:angular} contains a discussion of analytical and
numerical solutions to the hyperangular part of the Hamiltonian.
Chapter~\ref{kap:radial} contains a discussion of the hyperradial part
of the description of a many-boson system, as well as a discussion of
Bose-Einstein condensation in the hyperspherical model.
In chapter \ref{kap:meanfield_validity} we compare results from the
previous chapters to the results from the mean-field Gross-Pitaevskii
equation and ranges of validity are estimated.
In chapter~\ref{kap:stability_validity} we discuss stability criteria,
approaches to the dynamical evolution of the many-boson system, and
relevant time scales.
Chapter~\ref{kap:deformed} deals with the effects of a deformed
external trap.
Finally, chapter~\ref{kap:conclusion} contains conclusions and
discussions of the results.

\chapter{Hyperspherical description of correlations}          
\label{kap:hyperspherical_method}
\label{sec:theory}

Hyperspherical methods are used in studies of both few-body and
many-body problems, for example in atomic physics \cite{mac68},
nuclear physics \cite{jen97}, and within atomic physics especially for
many-electron systems \cite{lin95}.  Especially relevant in the
present context is a study of a many-boson system in a hyperspherical
frame performed by \etalcite{Bohn}{boh98}.  A large number of degrees
of freedom are described in terms of one length, the hyperradius, and
some angles called hyperangles.  This reminds of the characterization
of a two-body system by the relative two-body distance and the angular
degrees of freedom.  In the hyperspherical description, the angular
degrees of freedom are usually integrated or averaged such that the
many-body system is described by only the hyperradius.  An effective
potential depending on the hyperradius then carries the information
about the average angular properties.  This is once more analogous to
the description of the two-body problem by only the radial distance
with inclusion of the angular momentum in an effective centrifugal
potential.

\faddec \fadeqN

In section~\ref{sec:n-body-problem} we first define a set of
coordinates appropriate for a study of correlations in the many-body
system.  The steps are similar to the ones written by Barnea
\cite{bar99a,bar99b}.  For clarity we formulate this in three spatial
dimensions, but it can as well be written in lower dimensions.  We
then rewrite the Hamiltonian and the Schr\"odinger equation according
to this choice of coordinate system in section~\ref{sec:schr-equat-n}.
Appendix~\ref{sec:app.coord-transf} contains further details.  As
basic input we need an anzatz for the many-body wave function.  We
discuss the basis for assuming a Faddeev-like decomposition of the
wave function and relate this to other common approaches in section
\ref{sec:wave-function}.  We also discuss possible extensions to
fermion symmetry and three-body correlations.  We are then equipped to
solve the angular equation in section \ref{sec:angular-equation},
which is done with the inclusion of two-body correlations both by a
Faddeev-like equation and by a variational equation.  Some details of
the derivations are given in appendix~\ref{sec:hyper-matr-elem}.  We
end the chapter by considering the complications involved in an
angular equation with the inclusion of three-body correlations.

\section{Jacobi vectors and hyperspherical coordinates}

\label{sec:n-body-problem}

\jaccor

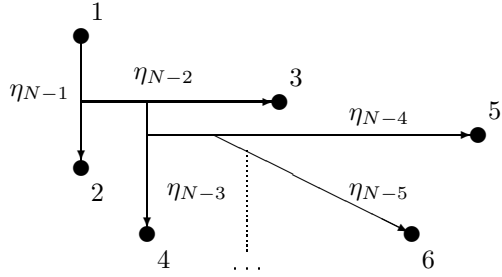
\begin{figure}[htb]
  \setlength{\unitlength}{2.2mm}
  \centering
  \begin{picture}(30,16.5)(-4.5,-2.5)
\linethickness{1pt}
\thinlines
\put(0,12){\vector(0,-1){7.6}}
\put(0,8){\vector(1,0){11.6}}
\put(4,8){\vector(0,-1){7.6}}
\put(4,6){\vector(1,0){19.6}}
\put(8,6){\vector(2,-1){11.8}}
\put(1,14){\makebox(0,0)[t]{$1$}}
\put(1,3){\makebox(0,0)[t]{$2$}}
\put(13,10){\makebox(0,0)[t]{$3$}}
\put(5,-1){\makebox(0,0)[t]{$4$}}
\put(25,8){\makebox(0,0)[t]{$5$}}
\put(21,-1){\makebox(0,0)[t]{$6$}}
\put(0,4){\circle*{1}}
\put(0,12){\circle*{1}}
\put(12,8){\circle*{1}}
\put(4,0){\circle*{1}}
\put(24,6){\circle*{1}}
\put(20,0){\circle*{1}}
\put(-2.5,8){\makebox(0,0)[b]{$\eta_{N-1}$}}
\put(5,9){\makebox(0,0)[b]{$\eta_{N-2}$}}
\put(7,2){\makebox(0,0)[b]{$\eta_{N-3}$}}
\put(18,6.5){\makebox(0,0)[b]{$\eta_{N-4}$}}
\put(18,2){\makebox(0,0)[b]{$\eta_{N-5}$}}
\put(10.2,-2){\makebox(0,0)[t]{\ldots}}
\qbezier[20](10,5)(10,2)(10,-1)
\end{picture}
  \caption[Jacobi vectors for six particles]
  {Jacobi vectors connecting the first six particles.}
  \label{fig:jacobi.tree}
\end{figure}

The system of $N$ identical particles may be described by $N$
coordinate vectors $\boldsymbol{r}_{i}$ and momenta
$\boldsymbol{p}_{i}$, labeling the particles by the index
$i=1,\ldots,N$.  Here a more suitable choice of coordinates is the
centre-of-mass coordinates $\boldsymbol R = \sum_{i=1}^N\boldsymbol
r_i/N$, the $N-1$ relative Jacobi vectors $\boldsymbol{\eta}_k$ with
\begin{eqnarray}
  \label{eq:jacobi_vectors}
  \boldsymbol{\eta}_k
  =
  \sqrt{\frac{N-k}{N-k+1}}
  \biggl(
  \boldsymbol{r}_{N-k+1}-
  \frac1{N-k}\sum_{j=1}^{N-k}\boldsymbol{r}_j
  \biggr)
\end{eqnarray}
and $k=1,2,\ldots,N-1$, and their associated momenta.  These Jacobi
coordinates are illustrated for the first six particles in
\reffig.~\ref{fig:jacobi.tree}.
The notation is $\eta_k\equiv|\boldsymbol\eta_k|$, so $\eta_{N-1}$ is
proportional to the distance between particles $1$ and $2$,
$\eta_{N-2}$ is proportional to the distance between particle $3$ and
the centre of mass of $1$ and $2$, $\eta_{N-3}$ is proportional to the
distance between particle $4$ and the centre of mass of the first
three particles, and so on.\footnote{Throughout the thesis normal font
  for a corresponding vector denotes the length of that vector,
  i.e.~$\eta_k=|\boldsymbol\eta_k|$, $r_i=|\boldsymbol r_i|$, etc.}

\hypcor

Hyperspherical coordinates are now defined in relation to the Jacobi
vectors.  One length, the hyperradius $\rho$, is defined by
\begin{eqnarray}
  \label{eq:hyperradius}
  \rho_l^2
  \equiv
  \sum_{k=1}^l\eta_k^2
  \;,\qquad
  \rho^2
  \equiv
  \rho_{N-1}^2
  =
  \frac{1}{N} \sum_{i<j}^N r_{ij}^2
  =\sum_{i=1}^N(\boldsymbol r_i-\boldsymbol R)^2
  \;,
\end{eqnarray}
where $r_{ij} \equiv |\boldsymbol{r}_{i}-\boldsymbol{r}_{j}|$.  The
last two equalities show that the hyperradius can be interpreted
either as $\sqrt{(N-1)/2}$ times the root-mean-square (rms) distance
between particles or as $\sqrt N$ times the rms distance between
particles and the centre of mass.

\hypcor

In three spatial dimensions, the $N-2$ hyperangles $\alpha_k \in
[0,\pi/2]$ for $k=2,3,\ldots,N-1$ relate the length of the Jacobi
vectors to the hyperradius via the definition
\begin{eqnarray}
  \label{eq:hyperangles}
  \sin\alpha_k
  \equiv
  \frac{\eta_k}{\rho_k}
  \;.
\end{eqnarray}
Since $\rho_1=\eta_1$, the fixed angle $\alpha_1=\pi/2$ is
superfluous, but is for convenience often included in the notation.
Remaining are the $2(N-1)$ angles $\Omega_\eta^{(k)}= (\vartheta_k,
\varphi_k)$, for $k=1,2,\ldots,N-1$, that define the directions of the
$N-1$ vectors $\boldsymbol\eta_k$, that is $\vartheta_k\in[0,\pi]$ and
$\varphi_k \in [0,2\pi]$.  All angles are collectively denoted by
$\Omega \equiv \{\alpha_k,\vartheta_k,\varphi_k\}$ with
$k=1,2,\ldots,N-1$.  In total the hyperangles $\Omega$ and the
hyperradius $\rho$ amount to $3(N-1)$ degrees of freedom and the
centre-of-mass coordinates $\boldsymbol{R}$ amount to three.  These
coordinates are also related by
\begin{eqnarray}
  \sum_{i=1}^Nr_i^2
  =\frac1N\sum_{i<j}^N r_{ij}^2
  +\frac1N\bigg(\sum_{i=1}^N\boldsymbol{r}_i\bigg)^2
  =\rho^2
  +NR^2
  \label{eq:r_rho_R}
  \;.
\end{eqnarray}

\hypcor

The total volume element is 
\begin{eqnarray}
  \prod_{i=1}^N d\boldsymbol r_i 
  =N^{3/2}d\boldsymbol R\prod_{k=1}^{N-1}d\boldsymbol\eta_k
  \;,
\end{eqnarray}
where the part depending on relative coordinates is
$\prod_{k=1}^{N-1}d\boldsymbol\eta_k$.\footnote{The notation is
$d\boldsymbol r_i=dr_{ix}dr_{iy}dr_{iz}$ with $\boldsymbol
r_i=(r_{ix},r_{iy},r_{iz})$.}  In hyperspherical coordinates this
relative part becomes
\begin{eqnarray}
  &&\prod_{k=1}^{N-1}d\boldsymbol\eta_k
  =d\rho \rho^{3N-4} 
  \; d\Omega_{N-1}
  \label{eq:vol_eta}
  \;,
  \\
  &&
  d\Omega_k
  =
  d\Omega_\alpha^{(k)}
  \;
  d\Omega_\eta^{(k)}
  \;
  d\Omega_{k-1}
  \;,
  \qquad
  d\Omega_1=d\Omega_\eta^{(1)}
  \label{eq:vol_omegak}
  \;,
  \\
  \label{eq:vol_alphak}
  &&d\Omega_\alpha^{(k)}
  =d\alpha_k\sin^2\alpha_k\cos^{3k-4}\alpha_k
  \;,\qquad
  d\Omega_\eta^{(k)} = d\vartheta_k\sin\vartheta_kd\varphi_k
  \;,
\end{eqnarray}
where $d\Omega_\eta^{(k)}$ is the familiar angular volume element in
spherical coordinates.  Since the angle $\alpha_{N-1}$ is related
directly to the two-body distance $r_{12}$ by $\sin\alpha_{N-1} \\=
\eta_{N-1}/\rho_{N-1}=r_{12}/(\sqrt2\rho)$, the volume element in
\refeq.~(\ref{eq:vol_alphak}) related to this angle is especially
important, that is $d\Omega_\alpha^{(N-1)} = d\alpha_{N-1}
\sin^2\alpha_{N-1} \cos^{3N-7}\alpha_{N-1}$.

\hypcor

The angular volume integrals can be computed to
\begin{eqnarray}
  \int
  d\Omega_\alpha^{(k)} 
  =
  \frac{\sqrt\pi\Gamma[3(k-1)/2]}{4\Gamma(3k/2)}
  \;,\qquad
  \int d\Omega_\eta^{(k)} = 4\pi
  \;,
\end{eqnarray}
where $\Gamma$ is the gamma function \cite{sch68}.  An angular matrix
element of an operator $\hat{O}$ with two arbitrary functions $\Psi$
and $\Phi$ is with \refeq.~(\ref{eq:vol_eta}) for fixed $\rho$ then
given by
\begin{eqnarray}
  \langle\Psi|\hat{O}|\Phi\rangle_\Omega
  =
  \int d\Omega_{N-1}\;\Psi^*(\rho,\Omega)\;\hat{O}\;\Phi(\rho,\Omega)
  \label{eq:angular_matrix_element}
  \;,
\end{eqnarray}
which in general is a function of $\rho$.

\hypcor 

In this section the many-body system is described by a straightforward
ordering of particles as $\{1,2,3,\ldots,N\}$.  Furthermore, we use
the configuration principle that a Jacobi vector connects one particle
with the centre of mass of some other particles.  Both the ordering of
particles and the recursive configurations have to be done differently
when we later evaluate matrix elements.  However, the present
formulation of the basic equations is independent of such
considerations and therefore the simplest structure is presented.

\jaccor

\section{Schr\"odinger equation for $N$ identical particles}

\label{sec:schr-equat-n}

With the preceding choice of coordinates we obtain the Hamiltonian and
next rewrite the Schr\"odinger equation by an adiabatic expansion,
which was first used in a study of a helium atom \cite{mac68}.

\subsection{Hamiltonian in hyperspherical coordinates}

\index{Hamiltonian!centre-of-mass}

We consider $N$ identical particles of mass $m$ interacting only
through two-body potentials $V_{ij}=V(\boldsymbol r_{ij})$.  We do not
consider effects due to spin, and thus omit explicit spin dependence
throughout this thesis.

An external trapping field $V\sub{trap}$ confines all particles to a
limited region of space.  This is written explicitly as an isotropic
harmonic-oscillator potential of angular frequency $\omega$, i.e.~for
particle $i$ it is given by $V\sub{trap}(\boldsymbol
r_i)=m\omega^2r_i^2/2$.  This confining field is relevant for studying
trapped atomic gases, but can later be omitted from the general
results by putting $\omega=0$.  The total Hamiltonian is here given by
\begin{eqnarray}
  \hat H\sub{total}
  =
  \sum_{i=1}^N
  \bigg(
  \frac{\hat {\boldsymbol p}_i^2}{2m}+\frac12m\omega^2r_i^2
  \bigg)
  +\sum_{i<j}^N V(\boldsymbol r_{ij})
  \;,
\end{eqnarray}
which with \refeq.~(\ref{eq:r_rho_R}) is separable into a part only
involving the centre-of-mass coordinates and a part only involving
relative coordinates.  The centre-of-mass Hamiltonian is
\begin{eqnarray}
  \hat H_R
  \equiv
  \frac{\hat{\boldsymbol P}_{R}^2}{2M}
  +\frac12M\omega^2R^2
  \label{eq:cm_hamiltonian}
  \;,
\end{eqnarray}
where $\boldsymbol P_{R} \equiv \sum_{i=1}^N \boldsymbol p_i$ is the
total momentum and $M=Nm$ is the total mass of the system.  We
subtract this from the total Hamiltonian and get
\begin{eqnarray}
  \hat H
  &\equiv&
  \hat H\sub{total}-\hat H_R
  \\ \nonumber
  &=&
  \sum_{i=1}^N\frac{\hat{\boldsymbol p}_i^2}{2m}
  -\frac{\hat{\boldsymbol P}_R^2}{2M}+
  \sum_{i=1}^N\frac12m\omega^2r_i^2
  -\frac12M\omega^2R^2
  +\sum_{i<j}^N V_{ij}
  \;.
\end{eqnarray}
Using \smarteq{eq:r_rho_R} we can write this as
\begin{eqnarray}
  \hat H
  =
  \hat T
  +\frac12m\omega^2\rho^2
  +\sum_{i<j}^N V_{ij}
  \;,\qquad
  \hat T 
  \equiv
  \sum_{i=1}^N
  \frac{\hat{\boldsymbol p}_i^2}{2m}
  -\frac{\hat{\boldsymbol P}_R^2}{2M}
  \;.
\end{eqnarray}
Here $\hat T$ is the intrinsic kinetic-energy operator which in
hyperspherical coordinates can be rewritten as
\begin{eqnarray}
  \hat T
  =
  -\frac{\hbar^2}{2m}
  \bigg(\frac{1}{\rho^{3N-4}}\frac{\partial}{\partial\rho}
  \rho^{3N-4} \frac{\partial}{\partial\rho}
  -\frac{\hat\Lambda_{N-1}^2}{\rho^2}
  \bigg)
  \label{eq:hyperspherical_kinetic_operator}
  \;.
\end{eqnarray}
The dimensionless angular kinetic-energy operator
$\hat\Lambda_{N-1}^2$ is recursively defined by
\begin{eqnarray}
  &&
  \hat\Lambda_k^2
  =\hat\Pi_k^2+
  \frac{\hat\Lambda_{k-1}^2}{\cos^2\alpha_k}
  +\frac{\hat{\boldsymbol l}_k^2}{\sin^2\alpha_k}
  \;,\qquad
  \hat\Lambda_1^2=\hat{\boldsymbol l}_1^2
  \;,
  \label{eq:angular_kinetic_operator}
  \\
  &&
  \hat\Pi_k^2
  =-\frac{\partial^2}{\partial\alpha_k^2}+
  \frac{3k-6-(3k-2)\cos 2\alpha_k}{\sin 2 \alpha_k}
  \frac{\partial}{\partial\alpha_k}
  \label{eq:alphak_kinetic_operator}
  \;,
\end{eqnarray}
where $\hbar\hat{\boldsymbol l}_k$ is the angular-momentum operator
associated with $\boldsymbol\eta_k$.  Thus, the angular kinetic-energy
operator is a sum of derivatives with respect to the various
hyperspherical angles.  Convenient transformations to avoid first
derivatives in \smarteqs{eq:hyperspherical_kinetic_operator} and
(\ref{eq:alphak_kinetic_operator}) are
\begin{eqnarray}
  -\frac{2m}{\hbar^2}\hat T_\rho
  &\equiv&
  \rho^{-(3N-4)}
  \frac{\partial}{\partial\rho} \rho^{3N-4} \frac{\partial}{\partial\rho} 
  \label{eq:hyperradial_kinetic_operator}
  \\
  \nonumber
  &=&
  \rho^{-(3N-4)/2}
  \bigg[
  \frac{\partial^2}{\partial\rho^2}-\frac{(3N-4)(3N-6)}{4\rho^2}
  \bigg]
  \rho^{(3N-4)/2}
  \;,
  \\
  \hat\Pi_k^2
  &=&\sin^{-1}\alpha_k\cos^{-(3k-4)/2}\alpha_k
  \bigg[-\frac{\partial^2}{\partial\alpha_k^2}-\frac{9k-10}2+
  \\ \nonumber
  &&
  \frac{(3k-4)(3k-6)}{4}\tan^2\alpha_k
  \bigg]
  \sin\alpha_k\cos^{(3k-4)/2}\alpha_k
  \label{eq:transformed_alphak_kinetic}
  \;.
\end{eqnarray}

The Hamiltonian $\hat H$ can now be collected as
\begin{eqnarray}
  \label{eq:simplified_hamiltonian}
  \hat H
  &=&
  \hat T_\rho
  +\frac12m\omega^2\rho^2
  +\frac{\hbar^2}{2m\rho^2}
  \hat h_\Omega
  \;,
  \\
  \label{eq:angular_hamiltonian}
  \hat h_\Omega
  &\equiv&
  \hat\Lambda^2_{N-1}
  +\sum_{i<j}^N v_{ij}
  \;,\qquad
  v_{ij}=\frac{2m\rho^2}{\hbar^2}V_{ij}
  \;,
\end{eqnarray}
where $\hat T_\rho$ is the radial kinetic-energy operator, $\hat
h_\Omega$ is a dimensionless angular Hamiltonian, and $v_{ij}$ is a
dimensionless potential.  Thus, the intrinsic Hamiltonian $\hat H$
contains a part which only depends on $\rho$ and a part $\hat
h_\Omega$ which depends on $\Omega$ and on $\rho$ through the two-body
potentials $v_{ij}$.

\subsection{Adiabatic expansion and equations of motion}

Since the total Hamiltonian is given as $\hat H\sub{total} = \hat
H_R+\hat H$, the total wave function for the $N$-particle system can
be written as a product of a function $\Upsilon$ depending only on
$\boldsymbol R$ and a function $\Psi$ depending on $\rho$ and the
$3N-4$ angular degrees of freedom collected in $\Omega$, i.e.
\begin{eqnarray}
  \Psi\sub{total}
  =
  \Upsilon(\boldsymbol R)
  \Psi(\rho,\Omega)
  \label{eq:totalwavefunction}
  \;.
\end{eqnarray}
The centre-of-mass motion for the total mass $M=Nm$ is determined by
\begin{eqnarray}
  \hat H_R\Upsilon(\boldsymbol R)
  =
  E_R\Upsilon(\boldsymbol R)
  \;.
\end{eqnarray}
From \refeq.~(\ref{eq:cm_hamiltonian}) the corresponding energy
spectrum is obtained as that of a harmonic oscillator, that is
$E_{R,n} = \hbar\omega(2n+3/2)$ with $n=0,1,2,\ldots$.

\angval

The relative wave function $\Psi(\rho,\Omega)$ obeys the stationary
Schr\"odinger equation
\begin{eqnarray}
  \hat H\Psi(\rho,\Omega)
  =
  E\Psi(\rho,\Omega)
  \label{eq:equations-motion}
  \;,
\end{eqnarray}
where $E$ is the energy in the centre-of-mass system.  This is solved
in two steps.  First, for a fixed value of the hyperradius $\rho$ we
solve the angular eigenvalue equation
\begin{eqnarray}
  (\hat h_\Omega-\lambda_\nu)\Phi_\nu(\rho,\Omega)
  =0
  \label{eq:angular_eigenvalue_eq}
  \;.
\end{eqnarray}
The angular eigenvalue $\lambda_\nu(\rho)$ depends on $\rho$.  Second,
the collection of angular eigenfunctions $\Phi_\nu(\rho,\Omega)$ is
used as a complete set of basis functions in an expansion of the
relative wave function.  This is for each value of the hyperradius
$\rho$ written as
\begin{eqnarray}
  \Psi(\rho,\Omega)
  =
  \sum_{\nu=0}^\infty
  F_\nu(\rho)\Phi_\nu(\rho,\Omega)
  \;,\qquad
  F_\nu(\rho)
  =
  \rho^{-(3N-4)/2}
  f_\nu(\rho)
  \;,
  \label{eq:hyperspherical_wave_function}
\end{eqnarray}
where the factor $\rho^{-(3N-4)/2}$ is introduced to eliminate first
derivatives in $\rho$, see \smarteq{eq:hyperradial_kinetic_operator}.
The expansion coefficients for fixed $\rho$, $f_\nu$ or $F_\nu$, are
then considered as hyperradial wave functions.

\couplingterms\radpot\traplength

In analogy to the technique for $N=3$ \cite{jen97},
\refeq.~(\ref{eq:hyperspherical_wave_function}) is inserted in
\refeq.~(\ref{eq:equations-motion}),
\refeqs.~(\ref{eq:simplified_hamiltonian}) and
(\ref{eq:angular_eigenvalue_eq}) are used, and the resulting equation
is projected onto an angular eigenfunction $\Phi_\nu$.  The result is
a set of radial equations
\begin{eqnarray}
  &&
  \bigg[
  -\frac{d^2}{d\rho^2}-\frac{2mE}{\hbar^2}+
  \frac{\lambda_\nu(\rho)}{\rho^2}+
  \frac{(3N-4)(3N-6)}{4\rho^2}+
  \frac{\rho^2}{b\sub t^4}
  -Q_{\nu\nu}^{(2)}(\rho)
  \bigg]
  f_\nu(\rho)
  \nonumber\\
  &&
  \qquad
  = 
  \sum_{\nu'\ne \nu}\bigg[
  2Q_{\nu\nu'}^{(1)}(\rho)\frac{d}{d\rho}+Q_{\nu\nu'}^{(2)}(\rho)
  \bigg]
  f_{\nu'}(\rho)
  \label{eq:radial.equation}
  \;,
\end{eqnarray}
that couple the different angular channels.  Here $b\sub t$ is the
trap length given by $b\sub t\equiv\sqrt{\hbar/(m\omega)}$, and the
coupling terms $Q_{\nu\nu'}^{(i)}$ are defined as
\begin{eqnarray}
  Q_{\nu\nu'}^{(i)}(\rho)
  \equiv
  \frac
  {\big\langle
    \Phi_\nu(\rho,\Omega)
    \big|
    \big(\frac{\partial}{\partial\rho}\big)^i
    \big|
    \Phi_{\nu'}(\rho,\Omega)
    \big\rangle_\Omega}
  {\big\langle
    \Phi_\nu(\rho,\Omega)
    \big|
    \Phi_\nu(\rho,\Omega)
    \big\rangle_\Omega}
  \label{eq:coupling.terms}
  \;.
\end{eqnarray}
A special result is $Q_{\nu\nu}^{(1)}=0$ \cite{nie01}.  The angular
eigenvalues $\lambda_\nu$ enter these coupled equations as a part of a
radial potential.  The total radial potential $U_\nu(\rho)$, entering
on the left hand side of \smarteq{eq:radial.equation}, is:
\begin{eqnarray}
  \frac{2mU_\nu(\rho)}{\hbar^2}
  \equiv
  \frac{\lambda_\nu(\rho)}{\rho^2}+
  \frac{(3N-4)(3N-6)}{4\rho^2}+
  \frac{\rho^2}{b\sub t^4}
  -Q_{\nu\nu}^{(2)}(\rho)
  \label{eq:radial.potential}
  \;.
\end{eqnarray}
This includes a $\rho^2$-term due to the external harmonic field, a
$\rho^{-2}$ centrifugal barrier-term due to the transformation of the
radial kinetic-energy operator, the angular potential $\lambda_\nu$,
and the diagonal term $Q_{\nu\nu}^{(2)}$.

\couplingterms

The expansion in \smarteq{eq:hyperspherical_wave_function} is called
the hyperspherical adiabatic expansion.  Its efficiency relies on
small coupling terms $Q_{\nu\nu'}^{(i)}$ which then requires inclusion
of fewer channels $\nu$.  In the following investigations of the
dilute boson system, the non-diagonal terms are often found to be
smaller than $1\%$ of the diagonal terms.  Without these couplings the
right-hand side of \smarteq{eq:radial.equation} vanishes, and the
equation simplifies significantly to
\begin{eqnarray}
  \bigg[
  -\frac{\hbar^2}{2m}
  \frac{d^2}{d\rho^2}
  +U_\nu(\rho)-E\bigg]
  f_\nu(\rho)=0
  \label{eq:noncoupled_radial_eq}
  \;.
\end{eqnarray}
In this thesis small couplings are generally assumed and only results
of this non-coupled treatment are shown.

\couplingterms

Thus, the centre-of-mass motion is separated out and the
hyperspherical adiabatic method turns out to be promising for a
sufficiently dilute system due to small coupling terms.  The remaining
problem is the determination of the angular potential $\lambda$ from
the angular eigenvalue equation.

\section{Wave function for identical particles}

\label{sec:wave-function}

So far no specific structures are assumed.  The allowed Hilbert space
for the many-body wave function in principle includes any structure of
the system.  However, at this point an ansatz for or approximation of
the angular wave function $\Phi_{\nu}(\rho,\Omega)$ is necessary.

\jastrow \fadyak \hartree

The Hartree wave function with a product of single-particle amplitudes
\cite{bra83} is the basis for mean-field treatments of many-particle
systems.  In contrast, Faddeev-Yakubovski{\u\i} formulations
\cite{fad61simple,yak67simple} contain additive decompositions of the
wave function which explicitly reflect the possible asymptotic
large-distance behaviours of cluster subsystems.  A different starting
point is the Jastrow factorization into products of two-body
amplitudes \cite{jas55}.  The Jastrow form is more efficient for large
densities, while Faddeev-Yakubovski{\u\i} methods are more successful
for smaller densities where the system separates into smaller
clusters.

\threecor

In the present hyperspherical formulation, the angular wave function
can be written as a general expansion in the full angular space and
subsequently be reduced to yield a practical wave function.  The
result of such considerations is an ansatz which is well suited for
the low densities encountered for Bose-Einstein condensates.  Two-body
correlations are expected to be most important.  Possible extensions
of the method to include three-body correlations or fermion
antisymmetry are briefly discussed.

\subsection{Hartree: single-particle product}

\label{sec:mean-field-descr}

\hartree

The Hartree ansatz with a product of single-particle amplitudes
\cite{bra83} is
\begin{eqnarray}
  \Psi\sub{H}(\boldsymbol r_1,\boldsymbol r_2,\ldots,\boldsymbol r_N)
  =
  \prod_{i=1}^N
  \psi\sub{H}(\boldsymbol r_i)
  \label{eq:hartree}
  \;.
\end{eqnarray}
For the ground state of non-interacting bosons trapped by the
spherically symmetric external field of trap length $b\sub t$, the
amplitudes are given by
\begin{eqnarray}
  \psi\sub{H}(\boldsymbol r_i)
  =
  C
  e^{-r_i^2/(2b\sub t^2)}
  \;,\qquad
  C^{-1}=\pi^{3/4}b\sub t^{3/2}
  \label{eq:hartree.ground_state}
  \;.
\end{eqnarray}
With the relation $\sum_{i=1}^N r_i^2=\rho^2+NR^2$ this is rewritten
as
\begin{eqnarray}
  &&
  \Psi\sub{H}(\boldsymbol r_1,\boldsymbol r_2,\ldots,\boldsymbol r_N)
  =
  C^N
  \exp\bigg( -\sum_{i=1}^N \frac{r_i^2}{2b\sub t^2} \bigg)
  \nonumber\\
  &&
  \qquad
  =
  C^N
  e^{-\rho^2/(2b\sub t^2)}e^{-NR^2/(2b\sub t^2)}
  =
  \Upsilon_0(\boldsymbol R)F_0(\rho)\Phi_0
  \label{eq:Gauss_product}
  \;,
\end{eqnarray}
which turns out as a product similar to
\refeqs.~(\ref{eq:totalwavefunction}) and
(\ref{eq:hyperspherical_wave_function}).  The separation of the
centre-of-mass motion assures that the ground-state centre-of-mass
function always is $\Upsilon_0(\boldsymbol R) = CN^{3/4}
\exp[-NR^2/(2b\sub t^2)]$.  Then \refeq.~(\ref{eq:Gauss_product}) is a
product of the ground-state wave function for the motion of the
centre-of-mass in a trap and the lowest hyperspherical wave function
$F_0 \Phi_0$ in \refeq.~(\ref{eq:hyperspherical_wave_function}), where
$F_0(\rho) \propto \exp[-\rho^2/(2b\sub t^2)]$ and the angular part
$\Phi_0$ is a constant.  This implies equivalence between a
Hartree-Gaussian wave function and lack of dependence on the
hyperangles $\Omega$.

\hartree

The interactions produce correlations in such a way that the
hyperspherical wave function $\Psi$ deviates from a hyperradial
Gaussian multiplied by a constant hyperangular part.  Therefore the
Hartree product wave function is strictly not exact.  However, a
measure can be obtained by calculating the single-particle density $n$
from the obtained function $\Psi$, that is
\begin{eqnarray}
  n(\boldsymbol r_1)
  =
  \int d\boldsymbol r_2 d\boldsymbol r_3\cdots d\boldsymbol r_N
  | \Upsilon_0(\boldsymbol R) \Psi(\rho,\Omega) | ^2
  \label{eq:singleparticledensity}
  \;.
\end{eqnarray}
This can then be compared with the Hartree analogue
$|\psi\sub{H}(\boldsymbol r_1)|^2$.  When the numerical hyperspherical
solution is inserted, the $3(N-1)$-dimensional integral in
\refeq.~(\ref{eq:singleparticledensity}) is rather complicated.  In
order to get an idea of the possible relations, we instead assume a
constant angular part $\Phi_0$.  Then the hyperradial density
distribution is expanded on Gaussian amplitudes with different length
parameters $a_j$
\begin{eqnarray}
  |F(\rho)|^2
  =
  \sum_j
  c_j\;
  \frac{2}{\gammafkt{3N-3}a_j^{3N-3}}
  e^{-\rho^2/a_j^2}
  \;,
\end{eqnarray}
where $\sum_j c_j=1$ assures that $F(\rho)$ is properly normalized as
$\int_0^\infty d\rho \rho^{3N-4} |F(\rho)|^2 \\ = 1$.  This yields the
single-particle density
\begin{eqnarray}
  n(\boldsymbol r_1)
  =
  \sum_j 
  c_j\;
  \frac{1}{\pi^{3/2}B_j^3}
  e^{-r_1^2/B_j^2}
  \;,\qquad
  B_j^2 \equiv \frac{(N-1)a_j^2+b\sub t^2}{N}
  \;,
\end{eqnarray}
which is equivalent to $\langle r_1^2\rangle = \int d\boldsymbol r_1\;
n(\boldsymbol r_1)r_1^2$ since
\begin{eqnarray}
  \langle r_1^2\rangle
  =
  \frac{1}{N}\langle\rho^2\rangle+\langle R^2\rangle
  =
  \frac{3}{2}\bigg(1-\frac{1}{N}\bigg)\sum_j c_j a_j^2
  +\frac{3}{2}\frac{1}{N}b\sub{t}^2
\end{eqnarray}
and
\begin{eqnarray}
  &&
  \int d\boldsymbol r_1\; n(\boldsymbol r_1)r_1^2
  =
  \frac{3}{2}\sum_j c_jB_j^2
  \nonumber\\
  &&
  \qquad
  =
  \frac{3}{2}\bigg(1-\frac{1}{N}\bigg)\sum_j c_j a_j^2
  +\frac{3}{2}\frac{1}{N}b\sub{t}^2\sum_jc_j
  =
  \langle r_1^2\rangle
  \;.
\end{eqnarray}
The mean-square distance between the particles is then obtained by the
relation
\begin{eqnarray}
  \langle r_{12}^2\rangle
  =
  \frac{2N}{N-1}
  \Big(\langle r_1^2\rangle-\langle R^2\rangle\Big)
  =
  \frac{2N}{N-1}
  \bigg(\langle r_1^2\rangle-\frac{1}{N}\frac{3}{2}b\sub t^2\bigg)
  \label{e21}
  \;.
\end{eqnarray}

\hartree

These relations are derived and valid only for Gaussian wave
functions.  However, the true Hartree solution is not strictly a
Gaussian although such an approximation rather efficiently describes
the dilute boson system \cite{pet01}.  The above results relate a
Hartree density distribution to a similar hyperradial distribution
provided that the angular wave function is assumed to be a constant
which corresponds to an uncorrelated structure.

\subsection{Faddeev-Yakubovski{\u\i}: cluster expansion}

\fadyak

The effect of correlations is beyond a mean field where particles only
feel each other on average and do not correlate.  With the
Faddeev-Yakubovski{\u\i} techniques the proper asymptotic behaviours
of the wave functions are directly taken into account
\cite{fad61simple,yak67simple}.  These formulations are well suited
when the large-distance asymptotics are crucial, as expected for
low-density systems.

Faddeev \cite{fad61simple} studied three-particle systems where one of
the two-body subsystems was bound and the other subsystems were
unbound.  The wave function was written as $\Phi =
\Phi_{12}+\Phi_{13} +\Phi_{23}$ with the three terms given by suitable
permutations of 
\begin{eqnarray} 
  \Phi_{23}
  =
  \phi_{23}(\boldsymbol r_{23})
  e^{\cmplxI \boldsymbol \kappa_1\boldsymbol r_1+
    \cmplxI \boldsymbol \kappa_{23}\boldsymbol R_{23}}
  \label{eq:faddeev_historical}
  \;,
\end{eqnarray}
where $\boldsymbol R_{23} = (m_2\boldsymbol r_2 + m_3\boldsymbol
r_3)/(m_2 + m_3)$ is the centre of mass of the bound subsystem and
$\hbar\boldsymbol \kappa_{1}$ and $\hbar\boldsymbol\kappa_{23}$ are
the momenta of particle 1 and of particle pair 2-3,
respectively.\footnote{The complex number $\sqrt{-1}$ is here denoted
  by $\cmplxI$.}  This form accounts for the details of the possibly
bound pair $2$-$3$ and considers other effects as low-energy plane
waves.  A generalization of this three-body wave function is
\begin{eqnarray}
  \Phi_{ij}=
  \phi_{ij}(\boldsymbol r_{ij})
  \exp\bigg({\cmplxI \sum_{\kappa\ne i,j}\boldsymbol \kappa_k\boldsymbol r_k
    +\cmplxI \boldsymbol \kappa_{ij} \boldsymbol R_{ij}} \bigg)
  \;,\qquad
  \Phi=\sum_{i<j}^N \Phi_{ij}
  \label{eq:fadN}
  \;.
\end{eqnarray}
When all relative energies are small, that is when
$\kappa_{ij}\simeq0$ and $\kappa_k\simeq0$, the result is
$\Phi_{ij}\simeq\phi_{ij}(\boldsymbol r_{ij})$.

A generalization to an $N$-particle system was formulated by
Yakubovski{\u\i} \cite{yak67simple}, who arranged the particles into
possible groups of subsystems and thereby included the correct
large-distance asymptotic behaviour for all cluster divisions.  The
decisive physical properties are related to the division into
clusters, which for $N=3$ amounts to three possibilities.  The three
Faddeev components are related to the number of divisions and not the
number of particles.  For $N > 3$ the number of cluster divisions is
much larger than $N$.  For $N$ particles the wave function is
therefore written as a sum over possible clusters
\begin{eqnarray}
  \Psi\sub{Y}
  =
  \sum\sub{clusters}\Phi\sub{Y}(\textrm{cluster})
  \label{eq:cluster}
  \;.
\end{eqnarray}
This method is often applied in nuclear physics \cite{cie98,fil02b}.
In a dilute system two close-lying particles are found more frequently
than other cluster configurations.  Then the dominating terms in the
cluster expression in \refeq.~(\ref{eq:cluster}) are due to the
two-body clusters, and the remaining particles are considered
uncorrelated and described by plane waves or as a mean-field
background.  The Yakubovski{\u\i} wave function then reduces to a
Faddeev-like form similar to \refeq.~(\ref{eq:fadN})
\begin{eqnarray} 
  \Psi\sub{Y}
  \to
  \Phi(\rho,\Omega)
  =
  \sum_{i<j}^N \Phi_{ij}(\rho,\Omega)
  \label{eq:faddeev_expansion}
  \;.
\end{eqnarray}

\fadyak

\subsection{Jastrow: two-body factorization}

\label{sec:jastrow} 

\jastrow

The Jastrow variational formulation \cite{bij40,din49,jas55} was
designed to account for correlations in a Bose system.  The Jastrow
ansatz
\begin{eqnarray}
  \Psi\sub J
  =
  \prod_{i<j}^N \psi\sub J(\boldsymbol r_{ij})
  \;,\qquad
  \boldsymbol r_{ij}\equiv \boldsymbol r_j-\boldsymbol r_i
  \;,
\end{eqnarray}
provides an argument for writing the wave function as a sum of
two-body terms in the dilute limit.  The two-body Jastrow component
can be written as a Gaussian term, which corresponds to mean-field
amplitudes, multiplied by a modification expected to be important only
at small separation, i.e.
\begin{eqnarray}
  \psi\sub J(\boldsymbol r_{ij})
  =
  e^{-r_{ij}^2/(2Nb\sub t^2)}
  [1+\phi\sub J(\boldsymbol r_{ij})]
  \;,\qquad
  \phi\sub J(\boldsymbol r_{ij})=0 \quad\textrm{for}\quad r_{ij}>r_0
  \;.
\end{eqnarray}
Beyond some length scale $r_0$, deviations due to correlations vanish.
With \refeq.~(\ref{eq:r_rho_R}) this leads to the relative wave
function
\begin{eqnarray}
  \Psi\sub J
  &=&
  e^{-\rho^2/(2b\sub t^2)}
  \prod_{i<j}^N
  \Big[1 + \phi\sub J(\boldsymbol r_{ij})\Big]
  \label{e29}
  \\ \nonumber
  &=&
  e^{-\rho^2/(2b\sub t^2)}
  \Bigg[ 1
  +\sum_{i<j}^N \phi\sub J(\boldsymbol r_{ij})+
  \sum_{i<j\ne k<l}^N
  \phi\sub J(\boldsymbol r_{ij})\phi\sub J(\boldsymbol r_{kl})
  +\ldots  \Bigg]
  \;.
\end{eqnarray}
For a non-interacting system the sums are zero and the Gaussian
mean-field Hartree ansatz from \refeq.~(\ref{eq:Gauss_product}) is
obtained.  For a sufficiently dilute system it is unlikely that more
than two particles simultaneously are close in space, that is when
both $r_{ij}<r_0$ and $r_{kl}<r_0$.  Therefore the expansion in
\refeq.~(\ref{e29}) can be truncated after the first two terms, i.e.
\begin{eqnarray}
  \prod_{i<j}^N 
  \Big[1+\phi\sub J(\boldsymbol r_{ij})\Big]
  \simeq
  1+\sum_{i<j}^N \phi\sub J(\boldsymbol r_{ij})
  =
  \sum_{i<j}^N \bigg[
  \frac{1}{N(N-1)/2}+\phi\sub J(\boldsymbol r_{ij})
  \bigg]
  \;.
\end{eqnarray}
A redefinition of the two-body amplitude results in a Faddeev-like sum
as in \refeq.~(\ref{eq:faddeev_expansion}).

\jastrow

\subsection{Hyperharmonic expansion of two-body components}

\label{sec:full-struct-fadd} 

\faddec \index{hyperspherical harmonics} \jacpol

The sum of two-body terms can be formally obtained as the $s$-wave
reduction of an expansion on a properly symmetrized complete set of
basis functions.  Appropriate are the hyperspherical harmonics
$\mathcal Y$ that are eigenfunctions of the grand angular
kinetic-energy operator $\hat\Lambda_{N-1}^2$,
\refeq.~(\ref{eq:angular_kinetic_operator}) \cite{smi60}.  These are
for collective angular momentum $L_k$ and projection $M_k$ given by
\cite{bar99b}
\begin{eqnarray}
  \hat\Lambda_k^2\mathcal Y_{[K_k,L_k,M_k]}^{\{q_k\}}
  =K_k(K_k+3k-2)\mathcal Y_{[K_k,L_k,M_k]} ^{\{q_k\}}
  \label{eq:ang_kin_eigenvalue_eq} 
  \;,
\end{eqnarray}
where $q_k$ denotes the set of quantum numbers $\{q_k\} =
\{l_1,\ldots,l_k,\nu_2,\ldots,\nu_k\}$, and the hyperangular momentum
$K_k$ is given by
\begin{eqnarray}
  K_k =
  2\nu_k+K_{k-1}+l_k
  \;,\qquad
  K_1 = l_1
  \label{eq:angularKnumber}
  \;.
\end{eqnarray}
The expression for $\mathcal Y$ is
\begin{eqnarray}
  &&
  \mathcal Y_{[K_k,L_k,M_k]}^{\{q_k\}}
  =
  \Big[
  Y_{l_1}(\vartheta_1,\varphi_1)\otimes
  Y_{l_2}(\vartheta_2,\varphi_2)\otimes
  \cdots\otimes
  Y_{l_k}(\vartheta_k,\varphi_k)
  \Big]_{L_k,M_k}
  \label{eq:hyperspherical_harmonics}
  \times
  \nonumber\\
  &&
  \qquad
  \Bigg\{\prod_{j=2}^k
  \;
  \sin^{l_j}\alpha_j\cos^{K_{j-1}}\alpha_j
  \;
  \JacP_{\nu_j}^{[l_j+1/2,K_{j-1}+(3j-5)/2]}(\cos2\alpha_j)\Bigg\}
  \;.
\end{eqnarray}
Here the $Y_{l,m}$'s are the usual spherical harmonics, $\JacP_\nu$ is the
Jacobi function, and the coupling of angular momenta is given by the
Clebsch-Gordan coefficients through
\begin{eqnarray}
  &&
  \Big[
  Y_{l_1}(\vartheta_1,\varphi_1)\otimes
  Y_{l_2}(\vartheta_2,\varphi_2)\otimes
  \cdots\otimes
  Y_{l_k}(\vartheta_k,\varphi_k)
  \Big]_{L_k,M_k}
  =
  \nonumber\\
  &&\qquad
  \bigg[\sum_{m_1,m_2,\ldots,m_k}
  \langle l_1m_1l_2m_2|L_2M_2 \rangle
  \;
  \langle L_2M_2l_3m_3|L_3M_3 \rangle
  \;
  \cdots
  \times
  \nonumber\\
  &&\qquad
  \cdots
  \;
  \langle L_{k-1}M_{k-1}l_km_k|L_kM_k \rangle
  \;
  \prod_{j=1}^k Y_{l_j,m_j}(\vartheta_j,\varphi_j)\bigg]
  \label{eq:ang_momentum_coupling}
  \;.
\end{eqnarray}
Omitting dependence on $\alpha_k$, we also define a reduced function
$\tilde{\mathcal Y}$ by
\begin{eqnarray}
  &&
  \tilde{\mathcal Y}_{[K_{k-1},L_k,M_k]}^{\{\tilde q_k\}}
  =
  \Big[
  Y_{l_1}(\vartheta_1,\varphi_1)\otimes
  Y_{l_2}(\vartheta_2,\varphi_2)\otimes
  \cdots\otimes
  Y_{l_k}(\vartheta_k,\varphi_k)
  \Big]_{L_k,M_k}
  \label{eq:reduced_hyperspherical_harmonics}
  \times
  \nonumber\\
  &&
  \qquad
  \Bigg\{\prod_{j=2}^{k-1}
  \;
  \sin^{l_j}\alpha_j\cos^{K_{j-1}}\alpha_j
  \;
  \JacP_{\nu_j}^{[l_j+1/2,K_{j-1}+(3j-5)/2]}(\cos2\alpha_j)\Bigg\}
  \;,
\end{eqnarray}
where $\{\tilde q_k\} = \{l_1,\ldots,l_k,\nu_2,\ldots,\nu_{k-1}\}$.

The wave function for each fixed value of $\rho$ is for fixed relative
angular momentum $\tilde L\equiv L_{N-1}$ and projection $\tilde
M\equiv M_{N-1}$ decomposed as
\begin{eqnarray}
  \Phi_{[\tilde L,\tilde M]}(\rho,\Omega)
  =\sum_{i<j}^N
  \Phi_{ij\;[\tilde L,\tilde M]}(\rho,\Omega)
  \label{eq:LM_expansion}
  \;,
\end{eqnarray}
where the component $\Phi_{ij}$ is focused on the particle pair
$i$-$j$.  Each of these components, for instance $\Phi_{12}$, can be
written as the complete expansion
\begin{eqnarray}
  \Phi_{12\;[\tilde L,\tilde M]}(\rho,\Omega)=
  \sum_{\{\tilde q_{N-1}\}}
  \phi_{12\;[\tilde L,\tilde M]}^{\{\tilde q_{N-1}\}}(\rho,r_{12})
  \tilde{\mathcal Y}_{[\tilde K,\tilde L,\tilde M]}^{\{\tilde q_{N-1}\}}
  \label{eq:12component_LM_expansion}
  \;,
\end{eqnarray}
where the sum runs over all possible quantum numbers $\tilde q_{N-1}$.
Analogies are given in \cite{nie01,bar99b}.\footnote{See page 384 in
  \cite{nie01} and page 1137 in \cite{bar99b}.}  The number $\tilde
K=K_{N-2}$ is used for reference to the kinetic-energy eigenvalue.  No
assumptions are made yet.

In order to obtain an explicitly symmetric boson wave function in
\smarteq{eq:LM_expansion}, we need to rewrite
\smarteq{eq:12component_LM_expansion} as
\begin{eqnarray}
  \Phi_{12\;[\tilde L,\tilde M]}(\rho,\Omega)
  =
  \sum_{\{\tilde q_{N-1}\}}
  \phi_{12\;[\tilde L,\tilde M]}^{\{\tilde q_{N-1}\}}(\rho,r_{12})
  \sum_{\permO_{12}}\hat\permO_{12}
  \tilde{\mathcal Y}^{\{\tilde q_{N-1}\}}_{[\tilde K,\tilde L,\tilde M]}
  \label{eq:sym12component_LM_expansion}
  \;,
\end{eqnarray}
where the second sum accounts for all possible permutations
$\hat\permO_{12}$ of particles apart from the pair $1$-$2$.

Zero angular momenta $\tilde L=\tilde M=0$ is reasonable in the
short-range limit.  Vanishing hyperangular quantum number $\tilde K=0$
for the remaining degrees of freedom yields only zero quantum numbers
$\{\tilde q_{N-1}\}=\{0\}$, and the sum is truncated to include only
the term
\begin{eqnarray}
  \Phi_{12\;[0,0]}(\rho,\Omega)
  =\phi_{12\;[0,0]}^{\{0\}}(\rho,r_{12})
  \tilde{\mathcal Y}_{[0,0,0]}^{\{0\}}
  \;.
  \label{eq:trunc_sym12component_LM_expansion}
\end{eqnarray}
The remaining terms $\Phi_{ij\ne12}$ are obtained in similar ways, so
the angular wave function is
\begin{eqnarray}
  \Phi(\rho,\Omega)=\sum_{i<j}^N \phi_{ij}(\rho,r_{ij})
  \label{eq:simplified_expansion}
  \;,
\end{eqnarray}
omitting the constant $\tilde{\mathcal Y}_{[0,0,0]}$ and superfluous
indices.

\subsection{Fermion antisymmetry}

\fermion

For identical fermions the total wave function has to be antisymmetric
under permutation of any two particles.  If the spin wave function is
symmetric, then the spatial wave function can be antisymmetrized by an
extension of the method written for bosons
\begin{eqnarray}
  &&\Phi_{[\tilde L,\tilde M]}(\rho,\Omega)
  =
  \sum_{i\ne j}^N \sum_{\permO_{ij}}
  (-1)^p\;\hat\permO_{ij}
  \Phi_{ij\;[\tilde L,\tilde M]}(\rho,\Omega)
  \;,\qquad
  \\
  &&\Phi_{12\;[\tilde L,\tilde M]}(\rho,\Omega)
  =
  \sum_{\{\tilde q_{N-1}\}}
  \phi_{12\;[\tilde L,\tilde M]}^{\{\tilde q_{N-1}\}}(\rho,r_{12})
  \tilde{\mathcal Y}_{[\tilde K,\tilde L,\tilde M]}^{\{\tilde q_{N-1}\}}
  \;.
\end{eqnarray}
The second sum of the first line accounts for all different
permutations $\hat\permO_{ij}$ of the $N-2$ particles apart from
$i$ and $j$.  We assume that a specific permutation resulted in the
particles ordered as $ijkl\ldots$.  The number $p$ is then the total
number of permutations of two particles needed to transform the
straightforward ordering $1234\ldots$ into the ordering $ijkl\ldots$.

If no term depends on more than $N-2$ particles' positions, it is not
possible to write a properly symmetrized wave function for a system of
identical fermions as a sum of terms.  Therefore, the dependences can
not be truncated as roughly as was done for bosons in the steps
leading to \refeq.~(\ref{eq:trunc_sym12component_LM_expansion}).
However, the antisymmetric function with lowest possible quantum
numbers might provide a useful fermion wave function which can be
implemented in calculations.  In this thesis we do not discuss
fermions further, but restrict ourselves to the case of bosons.

\fermion

\subsection{Two-boson direction-independent correlations}

\faddec

As seen in the preceding sections, a relative wave function of the
form
\begin{eqnarray}
  \Psi(\rho,\Omega)
  =
  F(\rho)\sum_{i<j}^N \Phi_{ij}(\rho,\Omega)
\end{eqnarray}
incorporates both mean-field properties through $F(\rho)$ and
correlations beyond the mean field through the Faddeev-components
$\Phi$.  We therefore decompose the angular wave function $\Phi$,
\refeq.~(\ref{eq:hyperspherical_wave_function}), into the symmetric
expression of Faddeev components $\Phi_{ij}$ for fixed values of the
hyperradius $\rho$
\begin{eqnarray}
  \label{eq:faddeev_decomposition}
  \Phi(\rho,\Omega)
  =
  \sum_{i<j}^N \Phi_{ij}(\rho,\Omega)
  \;,
\end{eqnarray}
where each term $\Phi_{ij}$ is a function of $\rho$ and all angular
coordinates $\Omega$.  Since each term in itself is sufficient when
all $\Omega$ degrees of freedom are allowed, this decomposition is
exact.  At first this ansatz seems clumsy by introducing an
overcomplete basis.  However, the indices $i$ and $j$ indicate a
special emphasis on the particle pair $i$-$j$.  The component
$\Phi_{ij}$ is expected to carry the information associated with
binary correlations of this particular pair.

\efithree

\scatlen

This wave function is a natural choice for the trivial case of $N=2$.
A wave function rewritten as a sum of terms has also been successful
in three-body computations.  The advantage is that the correct
boundary conditions are simpler to incorporate, as expressed in the
original formulation by Faddeev \cite{fad61simple} intended for
scattering.  Still, mathematically nothing is gained or lost in this
Faddeev-type of decomposition.  For weakly bound and spatially
extended three-body systems, $s$-waves in each of the Faddeev
components are sufficient to describe the system \cite{nie01}.  This
is exceedingly pronounced for large scattering lengths where the
Efimov states appear \cite{fed93,jen97,nie01}.

The present $N$-body problem is in general more complicated.  However,
for dilute systems essential similarities remain, i.e.~the relative
motion of two particles that on average are far from each other is
most likely dominated by $s$-wave contributions.  Each particle cannot
detect any directional preference arising from higher partial waves.
Implementation of these ideas in the present context implies that each
amplitude $\Phi_{ij}$ for a fixed $\rho$ only should depend on the
distance $r_{ij}$ between the two particles.  Thus, we assume
\begin{eqnarray}
  \Phi_{ij}(\rho,\Omega)
  \simeq
  \phi_{ij}(\rho,\alpha_{ij})
  \;,
\end{eqnarray}
where the two-index parameter $\alpha_{ij}$ is defined by
\begin{eqnarray}
  \sin\alpha_{ij}
  \equiv
  \frac{r_{ij}}{\sqrt2\rho}
  \label{eq:alphaij}
  \;.
\end{eqnarray}
These $\alpha_{ij}$'s are distinctively different from the
$\alpha_k$'s of \refeq.~(\ref{eq:hyperangles}).

The boson symmetry implies that the functions $\phi_{ij}$ are
non-distinguishable, so the indices are omitted.  The resulting
angular wave function is
\begin{eqnarray}
  \Phi(\rho,\Omega)
  =
  \sum_{i<j}^N \phi(\rho,\alpha_{ij}) = \sum_{i<j}^N \phi(\alpha_{ij}) 
  \label{eq:faddeev_decomposition_swaves}
  \;,
\end{eqnarray}
where $\phi_{ij}(\rho,\alpha_{ij}) = \phi(\rho,\alpha_{ij}) \equiv
\phi(\alpha_{ij})$ with the notational convenient omission of the
coordinate $\rho$.  The wave function in
\refeq.~(\ref{eq:faddeev_decomposition_swaves}) is symmetric with
respect to the interchange of two particles, $i\leftrightarrow j$,
since $\alpha_{ij}=\alpha_{ji}$ and since terms like
$\phi(\alpha_{ik})+\phi(\alpha_{jk})$ always appear symmetrically.

This ansatz of only $s$-waves dramatically simplifies the angular wave
function.  The original overcomplete Hilbert space is reduced such
that some angular wave functions can not be expressed in this
remaining basis.  Thus, the reduction resulted in an incomplete basis,
but the degrees of freedom remaining in
\smarteq{eq:faddeev_decomposition_swaves} are expected to be
those needed to describe the features of a dilute system.

In \refsec.~\ref{sec:full-struct-fadd} the Faddeev ansatz
\refeq.~(\ref{eq:faddeev_expansion}) was formally established as a
generalized partial wave expansion in terms of the hyperspherical
harmonic kinetic-energy eigenfunctions.  The two-body $s$-wave
simplification then appears as a truncation of this expansion, which
also leads to \refeq.~(\ref{eq:faddeev_decomposition_swaves}).

In conclusion, when the system is dilute, the Faddeev ansatz with
two-body amplitudes is expected to account sufficiently for the
correlations and at the same time keep the mean-field-like information
about motion relative to the remaining particles.

\subsection{Three-body correlations}

\recthree \threecor

An extension of the inclusion of pairwise correlations to study
three-body correlations in denser systems is possible and could yield
insight into the process of three-body recombination within $N$-boson
systems.  Since all degrees of freedom are kept in every term, the
Faddeev-like decomposition of the wave function can describe all kinds
of clusterizations in a particle system.  The indices $ij$ just refer
to the correct asymptotic behaviour of the two-body scattering
properties between particle $i$ and $j$ in a given amplitude
$\phi_{ij}$.  A higher-order correlated wave function is in that sense
included in the general expansion.  However, for actual application it
does not provide any solvable method.  Therefore, we proceed as
follows with what might be an applicable three-body expansion of the
many-body wave function.

A symmetric boson wave function with three-body amplitudes is
\begin{eqnarray}
  \Phi(\rho,\Omega)
  =
  \sum_{i<j}^N\sum_{k\ne i,j}
  \phi_{ij,k}
  \;,
\end{eqnarray}
where $\phi_{ij,k}$ depends on the distances between the three
particles $i$, $j$, and $k$.  In hyperspherical coordinates the
dependence, exemplified for the term $\phi_{12,3}$, can be reduced to
be on $\rho$, $\alpha_{N-1}$ ($\equiv\alpha_{12}$), $\alpha_{N-2}$
($\equiv\alpha_{12,3}$), and $\vartheta_{N-2}$
($\equiv\vartheta_{12,3}$), where $\vartheta_{N-2}$ is the angle
between $\boldsymbol\eta_{N-1}$ and $\boldsymbol\eta_{N-2}$.  A
general term $\phi_{ij,k}$ depends on $\rho$, $\alpha_{ij}$,
$\alpha_{ij,k}$, and $\vartheta_{ij,k}$.  It is written as
\begin{eqnarray}
  \phi_{ij,k}(\rho,\alpha_{ij},\alpha_{ij,k},\vartheta_{ij,k})
  =
  \phi_2(\rho,\alpha_{ij})
  +
  \phi_3(\rho,\alpha_{ij},\alpha_{ij,k},\vartheta_{ij,k})
  \label{eq:3bodycorr}
  \;,
\end{eqnarray}
where the term $\phi_n$ accounts for an $n$-body correlation.  This is
analogous to a proposal by Barnea \cite{bar99a}.  Since the functional
dependence is the same for all terms, the symmetry is explicitly
included.  The two-body correlated method reappears with $\phi_3=0$
since then $\phi_{ij,k} \to \phi_2 \to \phi(\rho,\alpha_{ij})$.
However, $\phi_3$ provides the three-body correlation on top of the
two-body correlation.  Terms with $k<j$ are responsible for an
overdetermined expansion since three terms, in principle, describe the
same three-body correlation.  In the case $i<j<k$ the three similar
terms are $\phi_{ij,k}$, $\phi_{ik,j}$, and $\phi_{jk,i}$.  With only
one term $\phi_{ij,k}$ we would have to impose other symmetry
restrictions on this single term such that $\phi_{ij,k} = \phi_{ik,j}
= \phi_{jk,i}$.  With the sum $\phi_{ij,k} + \phi_{ik,j} +
\phi_{jk,i}$ in the ansatz for the wave function, the symmetry is
explicitly built in and is independent of the amplitude's functional
form.  A simpler description is obtained when neglecting
$\vartheta_{ij,k}$ in $\phi_3$, and thus yielding a term that accounts
for one particle's relations to the pair of particles.

\section{Angular eigenvalue equation for two-boson correlations}
\label{sec:angular-equation}

\angval

Since the eigenvalue $\lambda$ from
\refeq.~(\ref{eq:angular_eigenvalue_eq}) carries information about the
two-body interactions and kinetic energy due to internal structure and
as well about possible correlations, the techniques and approximations
used to find $\lambda$ are especially important.

This section contains the essential rewritings of the angular equation
with the ansatz from \refeq.~(\ref{eq:faddeev_decomposition_swaves})
for two-body correlations.  We first present the Faddeev-like
equations and next construct a variational equation as an alternative
which is solvable under the additional assumption of short-range
interactions, i.e.~the system must be relatively dilute.  The
Faddeev-like equations were previously written in this form for a
boson system by \etalcite{de la Ripelle}{rip88}, whereas the angular
variational equation according to the author's knowledge is an
original contribution by the author and co-workers and first presented
in \cite{sor01}.  We end the section by briefly considering the
inclusion of higher-order correlations.

\subsection{Faddeev-like equation}

\label{sec:faddeev_eq} 

\fadeqN \fadeqthree \scatlen\angval

Insertion of the ansatz for the boson wave function in
\refeq.~(\ref{eq:faddeev_decomposition_swaves}) along with
\refeq.~(\ref{eq:angular_hamiltonian}) into
\refeq.~(\ref{eq:angular_eigenvalue_eq}) yields
\begin{eqnarray}
  \Big(
  \hat\Lambda_{N-1}^2+\sum_{k<l}^Nv_{kl}-\lambda
  \Big)
  \sum_{i<j}^N \phi_{ij}
  =0  
  \label{eq:angular_eq}
  \;,
\end{eqnarray}
with $\phi_{ij}=\phi(\alpha_{ij})$.  Rearrangement of summations leads
to
\begin{eqnarray}
  \sum_{k<l}^N \Bigg[
  \Big(\hat\Lambda_{N-1}^2-\lambda\Big)\phi_{kl}
  +v_{kl}\sum_{i<j}^N\phi_{ij}
  \Bigg]
  =0
  \label{eq:angular_eq2}
  \;.
\end{eqnarray}
For three particles and with the assumption that each term in the
square brackets separately is zero, the Faddeev equations are
obtained.  They have been applied for the three-body case to describe
particularly the regime of large scattering length \cite{jen97}.  The
same assumption for the $N$-particle system results in the $N(N-1)/2$
Faddeev-like equations
\begin{eqnarray}
  \Big(\hat\Lambda_{N-1}^2-\lambda\Big)
  \phi_{kl}
  +v_{kl}\sum_{i<j}^N\phi_{ij}
  =
  0
  \label{eq:angular-equation1}
  \;,
\end{eqnarray}
which are identical due to symmetry.  A description of the many-boson
system with such Faddeev-like equations was previously performed by
\etalcite{de la Ripelle}{rip88}, who concentrated on systems within
the realms of nuclear physics.

With $k=1$ and $l=2$ the kinetic-energy operator $\hat\Lambda^2_{N-1}$
from \refeq.~(\ref{eq:angular_kinetic_operator}) reduces to
$\hat\Pi^2_{N-1}$ because ${\hat\Lambda}_{N-2}^2 \phi_{12} = 0$ and
$\hat{\boldsymbol l}_{N-1}^2 \phi_{12} =0$.  Since $\boldsymbol
\eta_{N-1} = (\boldsymbol r_2 -\boldsymbol r_1)/\sqrt{2}$ and
$\rho_{N-1} = \rho$, then \smarteqs{eq:hyperangles} and
(\ref{eq:alphaij}) yield $\alpha_{N-1} = \alpha_{12}$.  Therefore,
only derivatives with respect to $\alpha_{12}$ remain, and it is
convenient to introduce the notation $\hat\Pi^2_{12} \equiv
\hat\Pi^2_{N-1}$.

In the sum over angular wave function components in
\refeq.~(\ref{eq:angular-equation1}), only three different types of
terms appear.  When $k=1$ and $l=2$, these types are classified by the
set $\{i,j\}$ either having two, one, or zero numbers coinciding with
the set $\{1,2\}$.  Then \refeq.~(\ref{eq:angular-equation1}) is
rewritten as
\begin{eqnarray}
  0
  &=&
  \Big[\hat\Pi^2_{12}+v(\alpha_{12})-\lambda\Big]\phi(\alpha_{12})+
  \label{eq:angular-equation2}
  \\
  &&
  v(\alpha_{12})\Bigg[
  \sum_{j=3}^N\phi(\alpha_{1j})+
  \sum_{j=3}^N\phi(\alpha_{2j})+
  \sum_{3\le i < j}^N\phi(\alpha_{ij})
  \Bigg]
  \nonumber
  \;,\\
  v(\alpha_{kl}) 
  &=&
  \frac{2m\rho^2}{\hbar^2}
  V \big( \sqrt2\rho\sin\alpha_{kl} \big)
  \;,
\end{eqnarray}
for a central potential $V(r)$.  Multiplication of
\smarteq{eq:angular-equation2} from the left by $\phi(\alpha_{12})$
followed by integration over all angular space except $\alpha_{12}$
results in an integro-differential equation in $\alpha \equiv
\alpha_{12}$ of the form\footnote{Throughout we interchange $\alpha$,
$\alpha_{N-1}$, and $\alpha_{12}$.}
\begin{eqnarray}
  \label{eq:angular-equation3}
  0
  &=&
  \big[\hat\Pi^2_{12}+v(\alpha)-\lambda\big]\phi(\alpha)+
  v(\alpha)
  2(N-2)\int d\tau\;\phi(\alpha_{13})
  \nonumber\\
  &&
  +v(\alpha)\frac12(N-2)(N-3)\int d\tau\;\phi(\alpha_{34})
  \;.
\end{eqnarray}
Here $d\tau\propto d\Omega_{N-2}$ is the angular volume element,
excluding the $\alpha$ dependence, with the normalization $\int d\tau
= 1$.  Due to symmetry between the first and second sums in
\refeq.~(\ref{eq:angular-equation2}), this projection leaves for every
value of $\alpha$ only two different integrals.  Both can analytically
be reduced to one-dimensional integrals.  The results are collected in
appendix \ref{sec:app.fadd-like-equat}.  For brevity here the terms
are denoted by
\begin{eqnarray}
  \int d\tau\; \phi(\alpha_{34})
  \equiv
  \hatR_{34}^{(N-2)} \phi (\alpha)
  \label{eq:intf34}
  \;,
  \\
  \int d\tau\;\phi(\alpha_{13})
  \equiv
  \hatR_{13}^{(N-2)} \phi (\alpha)
  \label{eq:intf13}
  \;,
\end{eqnarray}
where $\hatR_{ij}^{(N-2)}$ is an operator acting on the function
$\phi$ resulting in a function of $\alpha$.\footnote{Mathematically
  $\hatR$ resembles a rotation operator, hence the choice of
  notation.}  \Refeq.~(\ref{eq:angular-equation3}) can now be written
as
\begin{eqnarray}
  0
  &=&
  \Big[
  \hat\Pi^2_{12}+v(\alpha)-\lambda+
  2(N-2)v(\alpha)\hatR_{13}^{(N-2)}
  \label{eq:faddeev_eq}
  \nonumber\\
  &&+
  \frac12(N-2)(N-3)v(\alpha)\hatR_{34}^{(N-2)}
  \Big]
  \phi(\alpha)
  \;,
\end{eqnarray}
which is linear in the function $\phi$.  An advantage of this equation
is that it does not become more complicated as the number $N$ of
particles increases.

In this equation, the potential $v$ and the kinetic-energy operator
$\hat\Pi^2$ are diagonal in the sense that they only act in the space
of particle pair $1$-$2$, whereas the angular wave function is also
evaluated for other two-body pairs.  In the Faddeev-like
\smarteq{eq:faddeev_eq} for the many-body case, all wave-function
components are projected onto $s$ waves in the 1-2 system.  This means
that effective contributions from higher partial waves in the
hyperangular space are omitted.

Another problem is that the Faddeev approximation is not variational,
i.e.~the energy may be underestimated \cite{nie01}.  This is
explicitly obvious when we add and subtract a constant $v_0$ from the
interaction potential to rewrite \refeq.~(\ref{eq:angular_eq}) as
\begin{eqnarray}
  &&
  \Big(
  \hat\Lambda_{N-1}^2+\sum_{k<l}^Nv'_{kl}-\lambda'
  \Big)
  \sum_{i<j}^N \phi_{ij}
  =
  0  
  \label{eq:angular_eq_u0}
  \;,\\
  &&
  v_{kl}'=v_{kl}-v_0
  \;,\quad
  \lambda'=\lambda-\frac12N(N-1)v_0
  \;.
\end{eqnarray}
The Faddeev approximation then results in an angular eigenvalue
$\lambda$ which depends on the choice of $v_0$.  This shows that the
Faddeev-like equation has to be handled with care and at worst that it
is inconsistent with the present assumption of $s$ waves.

\subsection{Variational angular equation}

\label{sec:angul-vari-equat}

\angval\angvar

Proceeding with the Faddeev-like \refeq.~(\ref{eq:faddeev_eq}) is one
option, but as discussed this equation shows inadequacies under the
required assumptions about the many-body wave function.  In this
section we therefore rely on the full angular equation and discuss a
variational equation where the Faddeev approximation is not necessary.
Thus, we might effectively catch influences due to higher partial
waves from the different subsystems in the many-body system and
benefit from the maintained validity of the variational principle.
However, in some of the model calculations in the following chapters,
we study results obtained from the Faddeev-like equation and then
compare to results from the following variational equation.

First the optimal angular equation is derived within the Hilbert space
defined by the form of the angular wave function in
\refeq.~(\ref{eq:faddeev_decomposition_swaves}).  The very short range
of the two-body interaction compared with the size of the system
simplifies the problem as we shall see in the next section.

The angular Schr\"odinger equation for fixed $\rho$ in
\refeq.~(\ref{eq:angular_eigenvalue_eq}) and the ansatz for the wave
function in \refeq.~(\ref{eq:faddeev_decomposition_swaves}) allow the
eigenvalue expressed as an expectation value, i.e.
\begin{eqnarray}
  \lambda
  =
  \frac{\langle\Phi|\hat h_\Omega|\Phi\rangle_\Omega}
  {\langle\Phi|\Phi\rangle_\Omega}
  =
  \frac{ \big\langle \sum_{i'<j'}^N \phi_{i'j'}
    \big| \hat h_\Omega \big| \Phi \big\rangle_\Omega }
  { \big \langle\sum_{i'<j'}^N\phi_{i'j'} \big| \Phi \big\rangle_\Omega }
  \label{eq:lambda_expectation0}
  \;.
\end{eqnarray}
For an operator $\hat O$ which is invariant when interchanging any two
particles, the terms $\langle\phi_{i'j'}|\hat O|\Phi\rangle_\Omega =
\langle\phi_{i''j''}|\hat O|\Phi\rangle_\Omega$ are identical since
the possible differences vanish when averaging over all angles
$\Omega$.  Since $\hat h_\Omega$ is invariant with respect to
interchange of particles, this identity holds for both numerator and
denominator, so
\refeq.~(\ref{eq:lambda_expectation0}) simplifies to
\begin{eqnarray}
  \label{eq:lambda_expectation}
  \lambda=
  \frac
  {\big\langle\phi_{12}\big|\hat h_\Omega\big|\sum_{i<j}^N\phi_{ij}
    \big\rangle_\Omega}
  {\big\langle\phi_{12}\big|\sum_{i<j}^N\phi_{ij}\big\rangle_\Omega}
  \;.
\end{eqnarray}
The total angular volume element is $d\Omega_{N-1} =
d\Omega_\alpha^{(N-1)}d\Omega_\eta^{(N-1)}d\Omega_{N-2}$, see
\refeq.~(\ref{eq:vol_omegak}).  Since the integrands are independent
of $\Omega_\eta^{(N-1)}$, then $d\Omega_\eta^{(N-1)}$ can be omitted
from the integrations.  Using
\refeq.~(\ref{eq:angular_matrix_element}) we then obtain
\begin{eqnarray}
  \label{eq:angvar1}
  \int d\Omega_\alpha^{(N-1)}\;\phi_{12}^*\int d\Omega_{N-2}
  \big(\hat h_\Omega-\lambda\big)\sum_{i<j}^N\phi_{ij}
  =
  0
  \;.
\end{eqnarray}
The wave-function component $\phi^*_{12}$ is varied until the lowest
eigenvalue is obtained.  This gives the integro-differential equation
\begin{eqnarray}
  \label{eq:angvar2}
  \int d\Omega_{N-2}
  \sum_{k<l}^N
  \Bigg[
  (\hat\Lambda^2_{N-1}-\lambda)\phi_{kl}+
  v_{kl}\sum_{i<j}^N\phi_{ij}
  \Bigg]
  =0
  \;,\quad
\end{eqnarray}
where the unknown functions $\phi_{ij}=\phi(\alpha_{ij})$ all are the
same identical function of the different coordinates $\alpha_{ij}$.
Many terms are identical, e.g.~$\int d\Omega_{N-2}\;v_{12}\phi_{34} =
\int d\Omega_{N-2} \;v_{12}\phi_{56}$, since particles $1$ and $2$
cannot distinguish between other pairs of particles, see
appendix~\ref{sec:counting-terms} for the details.  Collecting all
terms yields
\begin{eqnarray}
  \label{eq:angvar3}
  \int d\Omega_{N-2}\Big[\big(\hat\Pi^2_{12}+v_{12}-\lambda\big)
  \phi_{12}+G(\tau,\alpha_{12})\Big]
  =
  0
  \;,\;
\end{eqnarray}
where $\tau$ denotes angular coordinates apart from $\alpha_{12}$.
The kernel $G$ contains all non-diagonal parts involving other
particles than $1$ and $2$.  This is given by
\begin{eqnarray}
  \label{eq:Gkernel}
  &&
  G(\tau,\alpha_{12})
  =
  \frac12 n_2 \Big[\hat\Pi^2_{34}
  +v(\alpha_{12})+v(\alpha_{34})-\lambda\Big]\phi(\alpha_{34})
  \nonumber\\
  &&
  \quad
  +\frac12 n_2 v(\alpha_{34})\phi(\alpha_{12})+2n_1v(\alpha_{13})
  \big[\phi(\alpha_{12})+\phi(\alpha_{23})\big]
  \nonumber\\
  &&
  \quad
  +2n_1\Big[\hat\Pi^2_{13}+v(\alpha_{12})+v(\alpha_{13})-\lambda\Big]
  \phi(\alpha_{13})
  \nonumber\\
  &&
  \quad
  +n_3\Big\{v(\alpha_{34})
  \big[\phi(\alpha_{35})+\phi(\alpha_{15})\big]+
  v(\alpha_{13})\phi(\alpha_{45})\Big\}
  \nonumber\\
  &&
  \quad
  +2n_2v(\alpha_{13})\big[\phi(\alpha_{14})+\phi(\alpha_{24})+
  \phi(\alpha_{34})\big]
  \nonumber\\
  &&
  \quad
  +2n_2v(\alpha_{34})\phi(\alpha_{13})
  +\frac{1}{4} n_4 v(\alpha_{34})\phi(\alpha_{56})
  \;,
\end{eqnarray}
where $n_i=\prod_{j=1}^i(N-j-1)$ and $\hat\Pi_{ij}^2$ is defined from
\refeq.~(\ref{eq:alphak_kinetic_operator}) with $k=N-1$ and with
$\alpha_k$ replaced by $\alpha_{ij}$.  In \refeq.~(\ref{eq:Gkernel})
all terms depend at most on coordinates of the six particles $1$-$6$.
The first three terms in \refeq.~(\ref{eq:angvar3}) do not depend on
the integration variables $\tau$ leaving only $G(\tau,\alpha_{12})$
for integration.

By appropriate choices of Jacobi systems \cite{smi77simple}, the
relevant degrees of freedom can be expressed in terms of the five
vectors $\boldsymbol\eta_{N-1},\ldots,\boldsymbol\eta_{N-5}$.  One is
the argument of the variational function and not an integration
variable.  The remaining twelve-dimensional integral is then evaluated
with the corresponding volume element $d\tau\propto \prod_{i=2}^{5}
d\Omega_\alpha^{(N-i)} d\Omega_\eta^{(N-i)}$ where the normalization
is $\int d\tau=1$.  Then \refeq.~(\ref{eq:angvar3}) becomes
\begin{eqnarray}
  \label{eq:simplified_var_eq}
  \Big[\hat\Pi^2_{12}+v(\alpha_{12})-\lambda\Big]
  \phi(\alpha_{12})+\int d\tau\; G(\tau,\alpha_{12})
  = 0
  \;,\quad
\end{eqnarray}
where the first terms are independent of the integration variables.
\Refeq.~(\ref{eq:simplified_var_eq}) is a linear integro-differential
equation in one variable containing up to five-dimensional integrals,
see appendix~\ref{sec:evaluation-terms}.  As is the case for the
Faddeev-like equation, this equation does not complicate further at
large $N$, i.e.~when $N$ increases beyond $N=6$, the structure does
not change.

\subsection{Short-range approximation}

\label{sec:shortrangeappr}

The two-body potentials $V(r_{ij})$ are assumed to be characterized by
a length scale $b$ beyond which the interaction vanishes, that is when
$r_{ij}\gg b$.
The angular eigenvalue \smarteq{eq:simplified_var_eq} simplifies in
the limit when this two-body interaction range $b$ is much smaller
than $\rho$.  Then the integrals are either analytical or reduce to
one-dimensional integrals.  This reduction could in principle be
accounted for by substituting the interaction potential with a
$\delta$ function, but is done generally for any finite-range
interaction as long as the range is small compared to the hyperradius.
However, this is only possible for the potentials appearing under the
integrals.  Thus, apart from the local terms containing $v(\alpha)$,
the results mainly depend on a parameter $a\sub B$ related to the
volume average of the potential by the definition
\begin{eqnarray}
  a\sub B 
  \equiv 
  \frac{m}{4\pi\hbar^2}
  \int d\boldsymbol r\; V(\boldsymbol r)
  \label{eq:Born_appr_copy1}
  \;.
\end{eqnarray}
A finite value of this volume integral is essential for the validity
of the method.  This is obeyed for short-range potentials that fall
off faster than $1/r^2$.

As an example of the reductions, when $\rho\cos\alpha\gg b$, the $\int
d\tau\;v(\alpha_{34})$-term reduces to
\begin{eqnarray}
  \int d\tau\; v(\alpha_{34})
  \equiv
  v_1(\alpha)
  \simeq
  2\sqrt{\frac{2}{\pi}}
  \;
  \gammafktB{3N-6}{3N-9}
  \frac{a\sub B}{\rho\cos^3\alpha}
  \;.\;
  \label{eq:v1-term}
\end{eqnarray}
Similarly the $\int d\tau\;v(\alpha_{13})$-term reduces to
\begin{eqnarray}
  \int d\tau\; v(\alpha_{13})
  \equiv
  v_{2}(\alpha)
  \simeq
  \frac{8}{3\sqrt3}\cos^{3N-11}\beta_0\;
  \Theta(\alpha<\pi/3)\;v_{1}(\alpha)
  \label{sec:short-range-appr-2}
\end{eqnarray}
in the limit when $\rho\cos\alpha\cos\beta_0\gg b$, where
$\sin\beta_0\equiv\tan\alpha/\sqrt3$.  Here $\Theta$ is the truth
function, i.e.~it equals unity when the argument is true and zero
otherwise.  The remaining terms can in this limit be expressed through
$v_1(\alpha)$, $v_2(\alpha)$, $\hatR_{ij}^{(k)}$ from
\refeqs.~(\ref{eq:intf34}) and (\ref{eq:intf13}), and other related
operators $\hatR_{ijkl}^{(n)}$.  Corresponding definitions are given
in appendix~\ref{sec:results-delta-limit}.

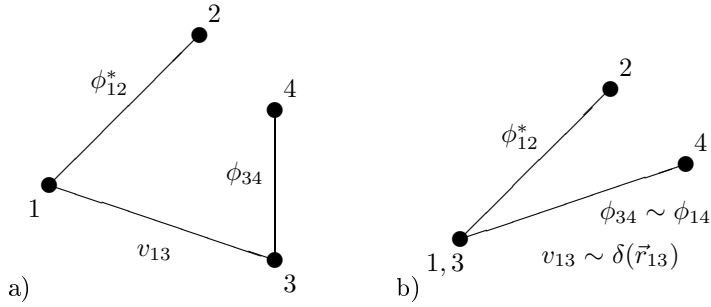
\begin{figure}[htbp]
  \centering
  \setlength{\unitlength}{2mm}
  a)
\begin{picture}(17.5,19.5)(0.8,-0.4)
\linethickness{1pt}
\thinlines
\put(2,7){\line(1,1){10}}
\put(2,7){\line(3,-1){15}}
\put(17,2){\line(0,1){10}}
\put(1,6){\makebox(0,0)[t]{$1$}}
\put(13,19){\makebox(0,0)[t]{$2$}}
\put(18,1){\makebox(0,0)[t]{$3$}}
\put(18,14){\makebox(0,0)[t]{$4$}}
\put(2,7){\circle*{1}}
\put(12,17){\circle*{1}}
\put(17,2){\circle*{1}}
\put(17,12){\circle*{1}}
\put(6,13){\makebox(0,0)[b]{$\phi_{12}^*$}}
\put(9,2){\makebox(0,0)[b]{$v_{13}$}}
\put(15,7){\makebox(0,0)[b]{$\phi_{34}$}}
\end{picture}
  \hspace{1cm}
  b)
\begin{picture}(18.9,16.3)(-0.6,3.2)
\linethickness{1pt}
\thinlines
\put(2,7){\line(1,1){10}}
\put(1,6){\makebox(0,0)[t]{$1,3$}}
\put(13,19){\makebox(0,0)[t]{$2$}}
\put(18,14){\makebox(0,0)[t]{$4$}}
\put(2,7){\circle*{1}}
\put(12,17){\circle*{1}}
\put(17,12){\circle*{1}}
\put(2,7){\line(3,1){15}}
\put(6,13){\makebox(0,0)[b]{$\phi_{12}^*$}}
\put(12,5){\makebox(0,0)[b]{$v_{13}\sim\delta(\vec r_{13})$}}
\put(15,8){\makebox(0,0)[b]{$\phi_{34}\sim\phi_{14}$}}
\end{picture}
  \caption  [Simplifications due to short-range potentials]
  {Simplifications due to short-range potentials.}
  \label{fig:variationsled.delta}
\end{figure}

The reductions can be understood qualitatively via 
\reffig.~\ref{fig:variationsled.delta} which shows the geometry when
the short-range interaction contributes to the integrals.  In the
integral $\int d\tau\; v(\alpha_{13}) \phi(\alpha_{34})$, see
\reffig.~\ref{fig:variationsled.delta}a, the dominant contributions
occur when particles $1$ and $3$ are close together as shown in
\reffig.~\ref{fig:variationsled.delta}b.  Then the distance between
particles $3$ and $4$ appearing in $\phi_{34}$ is approximately equal
to the distance between particles $1$ and $4$.  Therefore $\int
d\tau\;v(\alpha_{13})\phi(\alpha_{34}) \simeq \int
d\tau\;v(\alpha_{13})\phi(\alpha_{14})$.
  
So, for $b \ll \rho$ the exact short-range shapes of the potential are
not important and the integral in
\refeq.~(\ref{eq:simplified_var_eq}) of \refeq.~(\ref{eq:Gkernel}) can
be written as
\begin{eqnarray}
  &&
  \int d\tau\; G(\tau,\alpha)
  \simeq
  \Big[\frac{n_2}{2}v_{1}(\alpha)
  +4n_1v_{2}(\alpha)\Big]\phi(\alpha)
  \nonumber\\
  &&
  \quad
  + \frac{n_2}{2} \hatR_{34}^{(N-2)}v\phi(\alpha)
  + 2n_1 \hatR_{13}^{(N-2)}v\phi(\alpha)
  \nonumber\\
  &&
  \quad
  +\frac{n_2}{2}\Big\{\hatR_{34}^{(N-2)}\hat\Pi^2_{34}\phi(\alpha)+
  [v(\alpha)-\lambda]\hatR_{34}^{(N-2)}\phi(\alpha)\Big\}
  \nonumber\\
  &&
  \quad
  +2n_1\Big\{\hatR_{13}^{(N-2)}\hat\Pi^2_{13}\phi(\alpha)+
  [v(\alpha)-\lambda]\hatR_{13}^{(N-2)}\phi(\alpha)\Big\}
  \nonumber\\
  &&
  \quad
  +\frac{n_4}4v_1(\alpha)\hatR_{34}^{(N-3)}\phi(\alpha)
  +n_3v_1(\alpha)\hatR_{3435}^{(1)}\phi(\alpha)
  +n_3v_2(\alpha)\hatR_{1345}^{(1)}\phi(\alpha)
  \nonumber\\
  &&
  \quad
  +n_3v_1(\alpha)\hatR_{13}^{(N-3)}\phi(\alpha)
  +2n_2v_1(\alpha)\hatR_{3413}^{(2)}\phi(\alpha)
  \nonumber\\
  &&
  \quad
  +2n_2v_2(\alpha)\big[2\hatR_{1314}^{(2)}\phi(\alpha)+
  \hatR_{1324}^{(2)}\phi(\alpha)\big]
  \label{eq:Gkernel_0}
  \;.
\end{eqnarray}
The variational equation with these reductions is the basis for the
calculations in chapter~\ref{kap:angular}.

\subsection{Variational equation for three-body correlations}

\label{sec:inclusion-three-body}

\threecor

The ansatz for the angular wave function from
\refeq.~(\ref{eq:3bodycorr}) includes a general correlation within all
three-body subsystems.  We choose a trial wave function and write the
angular potential $\lambda$ as an expectation value analogous to
\refeq.~(\ref{eq:lambda_expectation0})
\begin{eqnarray}
  \int d\Omega_{N-1}
  \sum_{i'<j'}^N\sum_{k'\ne i',j'}\phi_{i'j',k'}^*
  \bigg(
  \hat\Lambda^2_{N-1}
  + \sum_{i''<j''}^Nv_{i''j''}
  -\lambda
  \bigg)
  \sum_{i<j}^N\sum_{k\ne i,j}\phi_{ij,k}
  =0
  \;.\quad
\end{eqnarray}
The calculation of these expectation values requires at most twelve
degrees of freedom which with a short-range potential for $b\ll\rho$
reduces to at most nine degrees of freedom.

Performing the variation $\phi_{12,3}^* \to \phi_{12,3}^* +
\delta\phi_{12,3}^*$ leads to the angular variational
integro-differential equation in $\alpha_{N-1}$, $\alpha_{N-2}$, and
$\vartheta_{N-2}$:
\begin{eqnarray}
  \int d\tilde\tau
  \bigg(
  \hat\Lambda^2_{N-1}
  + \sum_{i'<j'}^N v_{i'j'}
  -\lambda
  \bigg)
  \sum_{i<j}^N\sum_{k\ne i,j}\phi_{ij,k}
  =0
  \;,
\end{eqnarray}
where $d\tilde\tau$ denotes the angular volume element for all angles
apart from $\alpha_{N-1}$, $\alpha_{N-2}$, and $\vartheta_{N-2}$.
There are 126 different $V$-terms (38 for $N=4$), 12 different
$\hat\Lambda^2$-terms, and 12 different $\lambda$-terms.  In the
short-range limit many terms are identical and thus reduce the
complications.  The integrals in the integro-differential equation are
three dimensions lower than those in the expectation value since three
angles are fixed.  Thus, the short-range approximation results in an
integro-differential equation in three variables with up to
six-dimensional integrals.  This is beyond the scope of the present
work, but indicates the complications when including higher-order
correlations.

\chapter{Interactions and the hyperangular spectrum}          
\label{kap:angular}

\angval\radpot

In the hyperspherical formulation of the many-body problem in
chapter~\ref{kap:hyperspherical_method}, the tedious problems are
``hidden'' in the angular equation.  The angular solutions carry
essential information about interactions between the particles and
about internal kinetic energy.  The correlations were assumed to be
two-body for sufficiently dilute systems, and this was built into the
wave function.  The key quantity is then the function $\lambda$,
\smarteq{eq:angular_eigenvalue_eq}, which determines the properties of
the radial potential, \smarteq{eq:radial.potential}.  The angular wave
functions potentially carry information about couplings between the
different adiabatic channels.

\angvar

First, in section~\ref{sec:model-an-interacting} we discuss how to
model two-body interactions in the $N$-particle system.  Analytical
derivations of angular potentials in various regimes are given in
section~\ref{sec:analyt-angul-prop}.  Then we comment on the numerical
procedure before solving the angular variational equation.
Section~\ref{sec:numer-angul-solut} presents the attributes of the
found wave functions and angular potentials for various kinds of
interaction strengths.  Section~\ref{sec:summ-angul-prop} summarizes
the nature of the angular potentials, which can be parametrized by the
interaction parameters and the number of particles.  The details
behind this parametrization were previously published \cite{sor03a}
and hence collected in appendix~\ref{sec:scal-with-scatt}.

\section{Interactions between neutral bosons}

\label{sec:model-an-interacting}


The effective two-body interactions vary enormously for different
boson systems depending both on the nature of the bosons in question
and on the surroundings.  Here we consider bosons with short-range
interactions in the sense that the volume integral of the two-body
interaction potential is finite.  Neutral atoms, that are frequently
encountered in experiments with dilute boson systems, interact via a
potential of sufficiently short range and can be considered by this
method.

\begin{figure}[htb]
  \centering
  \input{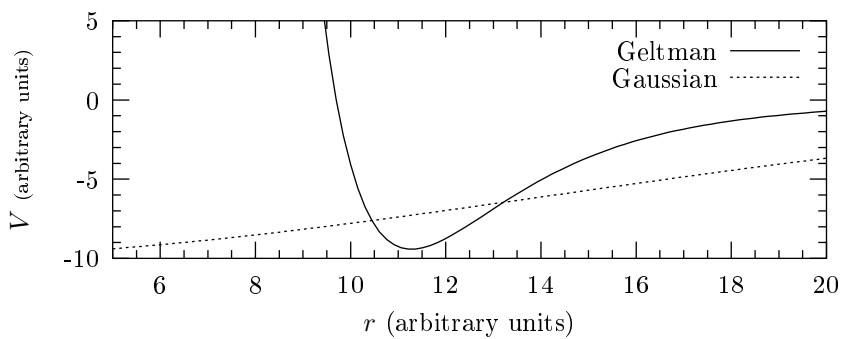}
  \caption
  [Two possible two-body potentials: realistic and simplified]
  {Two-body potentials.  The solid line is the potential in
    \refeq.~(\ref{eq:vpot_geltman}) from Geltman and Bambini
    \cite{gel01}, and the dotted line is the Gaussian potential
    from \refeq.~(\ref{eq:vpotGaussian}).}
  \label{fig:twobodypots}
\end{figure}

The interaction between atoms is repulsive at short distances due to
the Pauli exclusion principle which forbids overlapping centres.
Neutral atoms attract each other at longer distances due to mutual
polarization which induces a dipole moment.  The interaction between
two particles, e.g.~$1$ and $2$, can be modelled by the two-body
potential \cite{gel01}
\begin{eqnarray}
  V\sub{vdW}(r)
  =\frac{C_6}{r_0^6}
  \bigg[
  e^{-c(r-r_0)}-\bigg(\frac{r_0}{r}\bigg)^6
  \bigg]
  \;,\qquad
  \boldsymbol r
  \equiv
  \boldsymbol r_2
  -\boldsymbol r_1
  \;.
  \label{eq:vpot_geltman}
\end{eqnarray}
This potential has the van der Waals (vdW) tail $-C_6/r^6$ when $r\gg
r_0$, and is thus of short range in the sense that it decays faster
than $1/r^2$.  The important part of this potential is illustrated in
\reffig.~\ref{fig:twobodypots} for some choice of the parameters
$C_6$, $r_0$, and $c$.

\gaussian

\scatlen

At large particle separations, a direction-independent behaviour is
expected, which means that zero relative angular momentum is
preferred.  Then the asymptotic two-body wave function for particles
interacting via short-range potentials behaves as
\begin{eqnarray}
  u(r)=\sin[\kappa r+\chi(\kappa)]
  \label{eq:twobody_wavefunction}
  \;,
\end{eqnarray}
where $\chi$ is the phase shift and $\hbar\boldsymbol\kappa$ is the
relative momentum.  The phase shift depends on the relative energy
$\hbar^2\kappa^2/m$ and is at low energy given by the expansion
\begin{eqnarray}
  \kappa\cot\big[\chi(\kappa)\big]
  =-\frac{1}{a_s}+\frac{1}{2}\kappa^2R\sub{eff}
  +\mathcal{O}(\kappa^4)
  \label{eq:phaseshift}
  \;,
\end{eqnarray}
where $a_s$ is the $s$-wave scattering length and $R\sub{eff}$ is the
effective range.  The convention applied here is that for a purely
repulsive interaction the scattering length is positive, while for a
purely attractive interaction without any bound states the scattering
length is negative.  The effective range and higher-order terms can be
neglected at sufficiently low relative energy.  Thus, at low energy
the properties of the two-body system are basically determined by the
scattering length $a_s$.

\scatlen

The $s$-wave scattering length for a given two-body potential $V(r)$
can be obtained by solving the radial Schr\"odinger equation for two
identical particles of mass $m$ for zero angular momentum, zero energy
($\kappa=0$), and boundary condition $u(0)=0$:
\begin{eqnarray}
  \bigg[- \frac{\hbar^2}{m}\frac{d^2}{dr^2}
  +V(r)\bigg]u(r)=0
  \;.
\end{eqnarray}
Outside the two-body potential the solution is a straight line.
According to a Taylor expansion of \smarteq{eq:twobody_wavefunction},
the wave function for small $\kappa$ is
\begin{eqnarray}
  u(r) 
  \simeq
  \big\{1+\kappa r\cot\big[\chi(\kappa)\big]\big\}
  \sin\big[\chi(\kappa)\big]
  \propto
  1-\frac{r}{a_s}
  \label{eq:u_asymptote}
  \;.
\end{eqnarray}
Thus, the scattering length can be determined by the intersection of
the asymptotic wave function with zero, that is $u(a_s)=0$.

\scatlen\born

It is often convenient to also define the parameter $a\sub B$ by
\begin{eqnarray}
  a\sub B 
  \equiv 
  \frac{m}{4\pi\hbar^2}
  \int d\boldsymbol r\; V(\boldsymbol r)
  =
  \frac{m}{\hbar^2}
  \int_0^\infty dr \;r^2V(r)
  \label{eq:Born_appr}
  \;,
\end{eqnarray}
which is the Born approximation to the scattering length $a_s$.  The
last equality holds for a central potential.  The strength of the
interaction is then proportional to $a\sub B$.

\scatlen\gaussian

Since the finer details of the interaction potential are superfluous,
a finite-range Gaussian potential 
\begin{eqnarray}
  V\sub G(r)=V_0 e^{-r^2/b^2}
  \;,\qquad
  V_0=\frac{4\hbar^2a\sub B}{\sqrt\pi mb^3}
  \label{eq:vpotGaussian}
  \;,
\end{eqnarray}
see dotted line in \reffig.~\ref{fig:twobodypots}, is sufficient for a
study of the dependence on the scattering length and possibly a few
more of the low-energy parameters in the expansion of the phase shift.
The strength $V_0$ is then related to $a\sub B$ as indicated.

\twobound\scatlen\born\gaussian

\begin{figure}[htb]
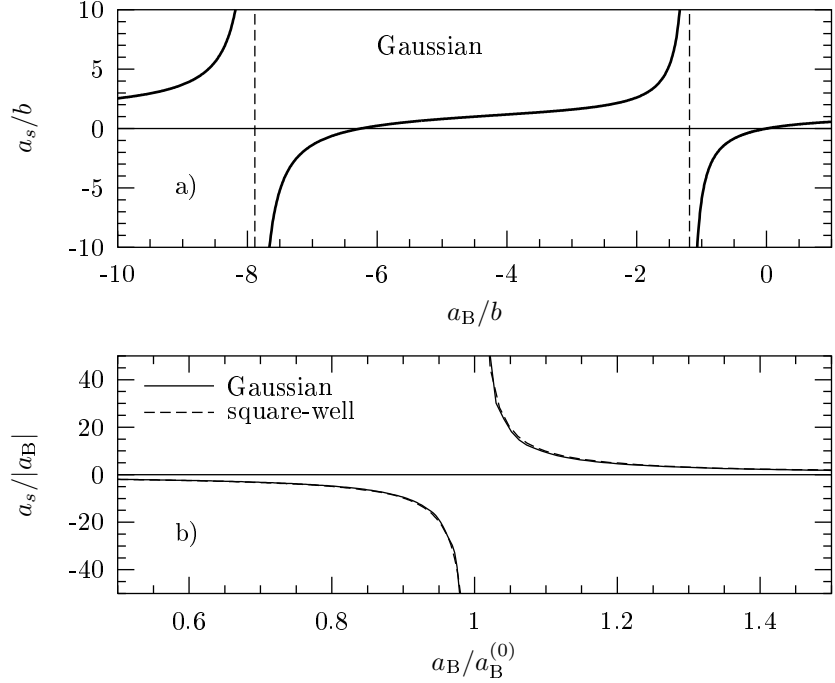

  \centering
  \input{as_schematic}
  \input{as_ap.scaled}
  \caption  [Scattering length for Gaussian and square-well potentials]
  {a) Scattering length $a_s$ divided by the potential range $b$ as a
    function of $a\sub B$ divided by $b$ for the Gaussian potential
    from \refeq.~(\ref{eq:vpotGaussian}).  b) Scattering length $a_s$
    divided by $a\sub B$ as a function of $a\sub B$ divided by $a\sub
    B^{(0)}$, defined as the value of $a\sub B$ where the first bound
    state occurs.  Results are shown for the Gaussian potential with
    $a\sub B^{(0)}/b=-1.1893$ and for the square-well potential
    $V\sub{sw}(r) = V\sub{sw,0}\Theta(r<b)$ with $a\sub B^{(0)}/b =
    -0.8225$.}
  \label{fig:scatlen}
\end{figure}

\Reffig.~\ref{fig:scatlen}a shows $a_s$ as a function of the strength
parameter $a\sub B$ for the Gaussian potential.  When the parameter
$a\sub B$ decreases from zero to negative values, the scattering
length varies slowly and roughly linearly with $a\sub B$ for small
$a\sub{B}$, until a value $a\sub{B}^{(0)}$ where $a_s$ diverges as a
signal of the appearance of the first two-body bound state.  For
increasing attraction $a_s$ turns positive when this state is slightly
bound.  Then the scattering length decreases and turns negative again.
This pattern repeats itself as the second bound state appears, and so
on at each subsequent threshold.

\scatlen\gaussian

\scatlen\gaussian

For a square-well potential $V\sub{sw}(r)=V\sub{sw,0}\Theta(r<b)$ the
threshold value of $a\sub B^{(0)}$ differs from the value for the
Gaussian potential, but $a_s/a\sub B$ as a function of $a\sub B/ a\sub
B^{(0)}$ results in virtually the same curves, see
\reffig.~\ref{fig:scatlen}b.  This indicates that for simple
potentials the behaviour is approximately independent of the shape.

\scatlen\born\gaussian

Table~\ref{table:as_aB} shows the scattering length $a_s$ for
different potential strengths $a\sub B$ for the Gaussian potential,
primarily for the cases studied in this work where $|a\sub B|/b$ is
close to unity.  The Born approximation equals the correct scattering
length only in the limit of weak attraction where the magnitude of the
scattering length $a_s$ is much smaller than the interaction range
$b$.

\begin{table}[htb]
  \centering
  \begin{tabular}{|c|c|c|}
    \hline
    $a\sub B/b$ & $a_s/b$ & $\mathcal N\sub B$ \\
    \hline\hline
    $+3.625$     & $+1.00$  & 0 \\
    $+1.00$    & $+0.565$  & 0 \\
    \hline
    $-0.3560$  & $-0.50$ & 0 \\
    $-0.500$  & $-0.84$ & 0 \\
    $-0.551$    & $-1.00$  & 0 \\
    $-1.00$  & $-5.98$ & 0 \\
    $-1.069$    & $-10.0$ & 0 \\
    $-1.110$    & $-15.7$ & 0 \\
    $-1.1761$   & $-100$  & 0 \\
    $-1.1860$  & $-401$  & 0 \\
    \hline
  \end{tabular}
  \hspace{.5cm}
  \begin{tabular}{|c|c|c|}
    \hline
    $a\sub B/b$ & $a_s/b$ & $\mathcal N\sub B$ \\
    \hline\hline
    $-1.18765$  & $-799$  & 0 \\
    $-1.1890$  & $-4212$ & 0 \\
    $-1.1893$  & $-85601$ & 0 \\
    \hline
    $-1.2028$  & $+100$  & 1 \\
    $-1.220$    & $+44.5$  & 1 \\
    $-1.3380$  & $+10.0$   & 1 \\
    $-1.35$    &  $+9.32$   & 1 \\
    $-1.50$     & $+5.31$ & 1 \\
    \hline
    $-6.868$    & $-1.00$  & 1 \\
    $-7.6612$   & $-10.0$ & 1 \\
    \hline
  \end{tabular}
  \caption [Scattering lengths for various Gaussian strengths]
  {The scattering length $a_s$ in units of $b$ for various
    strengths of a Gaussian potential measured as $a\sub B/b$.  
    The number $\mathcal N\sub B$ is the number of bound two-body
    states.}
  \label{table:as_aB}
\end{table}

\twobound\scatlen\gaussian

To exemplify, in experimental work $^{87}$Rb atoms have a scattering
length of $a_s\simeq100$ a.u.\footnote{The scattering lengths for
  relevant spin states of $^{87}$Rb atoms are according to Pethick and
  Smith \cite{pet01} all close to 100 a.u., where 1 a.u.
  $=0.529\cdot10^{-10}$ m.}  Assuming an interaction range around
$b=1$ nm we obtain $a_s/b=5.29$.  This can be modelled by a Gaussian
two-body interaction with $a\sub B/b\simeq-1.5$, where the lowest
solution corresponds to two-body bound states and the next accounts
for the properties of the dilute gas.  However, by applying an
external magnetic field it is possible to change the internal energy
levels in alkali atoms, e.g.~in $^{85}$Rb \cite{cor00}, and thereby
change the scattering length to almost any desired value.  This allows
experimental studies of a large range of scattering lengths.

\zerorangepot\scatlen

The short-range two-body interaction with $s$-wave scattering length
$a_s$ has in mean-field contexts \cite{dal99}, i.e.~with a Hartree
ansatz as in \smarteq{eq:hartree}, been modelled by the
three-dimensional zero-range potential
\begin{eqnarray}
  V_\delta(\boldsymbol r)
  =
  \frac{4\pi\hbar^2a_s}{m}\delta(\boldsymbol r)
  \label{eq:zerorange-potential}
  \;,
\end{eqnarray}
where $\delta$ is the Dirac delta function.  Only the scattering
length enters as the parameter characterizing the two-body
interaction.  This is usually assumed to be succesful when
$n|a_s|^3\ll1$, where $n$ is the density of the system.  For this
zero-range interaction \smarteq{eq:Born_appr} yields $a\sub B=a_s$,
which is rarely the case for finite-range interactions, as is obvious
for the cases illustrated in \reffig.~\ref{fig:scatlen}.

\gaussian

The finite-range Gaussian potential from
\refeq.~(\ref{eq:vpotGaussian}) is used in the following calculations.
In order to test the dependence on the short-range details of the
interaction, a linear combination of different Gaussians was also used
in some cases, although these results are not shown here.

\section{Analytical angular properties}

\label{sec:analyt-angul-prop}

Before solving numerically we investigate various limits analytically.
In the non- or weakly-interacting limit, the kinetic-energy
eigenfunctions are relevant for understanding the properties of the
many-body system.  When a two-body bound state is present, there is a
signature of it in the angular spectrum, which can also be studied
analytically.  A zero-range treatment incorporates the well-known
asymptotic two-body behaviour into the many-body wave function.  This
leads to an equation which has an analytic solution for a very dilute
system.  Finally, we average the interactions in a way that resembles
the mean field, i.e.~all correlations are neglected.  These different
analytic approaches provide a basis for understanding the numerical
solutions, which we turn to in section \ref{sec:numer-angul-solut}.

\subsection{Kinetic-energy eigenfunctions}

\label{sec:angul-kinet-energy}

First, non-interacting particles, that is $v=0$, are considered.  With
the transformation in \refeq.~(\ref{eq:transformed_alphak_kinetic}),
\refeq.~(\ref{eq:faddeev_eq}) becomes
\begin{eqnarray}
  \bigg[-\frac{d^2}{d\alpha^2}+
  \frac{(3N-7)(3N-9)}{4}\tan^2\alpha
  -\frac{9N-19}{2}
  -\lambda\bigg]
  \tilde\phi(\alpha)
  =
  0
  \label{eq:free_ang_eq}
  \;.
\end{eqnarray}
Here $\tilde\phi(\alpha)$ is a reduced angular wave function
\begin{eqnarray}
  \tilde\phi(\alpha)
  \equiv
  \sin\alpha\cos^{(3N-7)/2}\alpha
  \;\phi(\alpha)
  \label{eq:angular-equation-1}
  \;,
\end{eqnarray}
in analogy to the transformation from radial to reduced radial wave
function for the two-body problem.  Since $\phi$ for a physical state
cannot diverge at $\alpha=0$ or $\alpha=\pi/2$, the boundary condition
for the reduced angular wave function is
$\tilde\phi(0)=\tilde\phi(\pi/2)=0$.

\jacpol

Non-reduced solutions to \refeq.~(\ref{eq:free_ang_eq}) are given by
the Jacobi polynomials $\JacP$ \cite{abr} as
\begin{eqnarray}
  \phi_K(\alpha)
  =
  \JacP_\nu^{[1/2,(3N-8)/2]}(\cos2\alpha)
  \label{eq:free_ang_solution}
  \;.
\end{eqnarray} 
See further details in appendix~\ref{sec:prop-select-funct}.  The
hyperspherical quantum number $K$ is given by $K=2\nu=0,2,4,\ldots$
and denotes the angular kinetic-energy eigenfunction with $\nu$ nodes
in $\alpha$ space.  The corresponding angular eigenvalues are
$\lambda_K=K(K+3N-5)$.  This notation is consistent with the general
hyperspherical harmonics from \smarteqs{eq:ang_kin_eigenvalue_eq} and
(\ref{eq:hyperspherical_harmonics}).  The lowest eigenvalue is zero
corresponding to a constant eigenfunction $\JacP_0=1$.

\begin{figure}[htb]
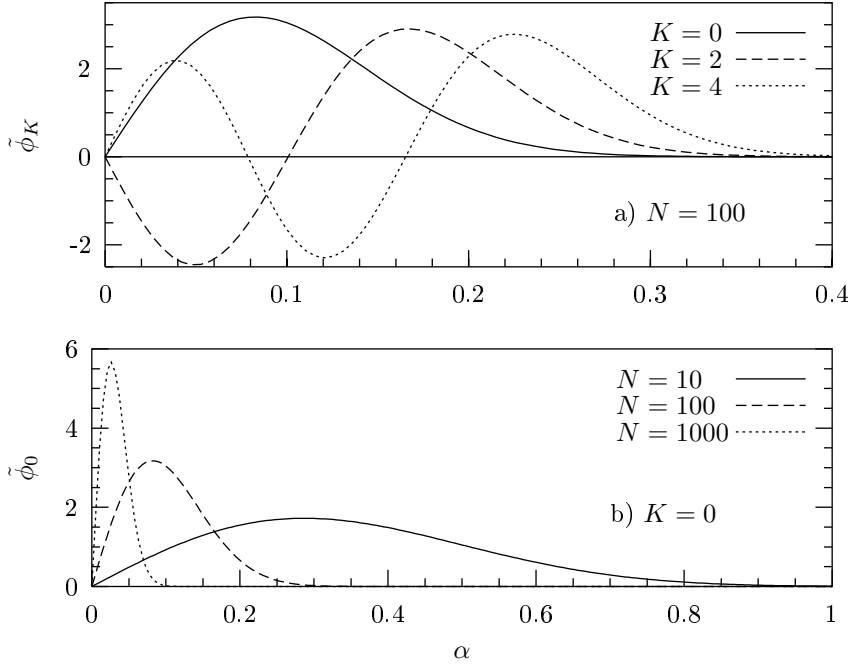

  \centering
  \input{jacobyp_N100}
  \input{jacobyp_N}
  \caption[Angular kinetic-energy eigenfunctions] 
  {The reduced angular wave function $\tilde\phi_K$, defined in
    \refeqs.~(\ref{eq:angular-equation-1}) and
    (\ref{eq:free_ang_solution}), for a) $N=100$ and $K=0,2,4$ and b)
    $K=0$ and $N=10,100,1000$.  The normalization is $\int_0^{\pi/2}
    d\alpha\;|\tilde\phi_K(\alpha)|^2 = 1$.}
  \label{fig13}
\end{figure}

\Reffig.~\ref{fig13}a shows the reduced angular kinetic-energy
eigenfunctions for $N=100$ and the lowest three eigenvalues.  The
constant wave function $\phi_{K=0}$ is in the figure represented by
$\tilde\phi_0(\alpha) = \sin\alpha\cos^{(3N-7)/2}\alpha$, where
$|\tilde\phi_0|^2$ then is the volume element in $\alpha$ space.  The
oscillations are located at relatively small $\alpha$ values.  As seen
in \reffig.~\ref{fig13}b, the location of the maximum changes as
$1/\sqrt N$ due to the centrifugal barrier proportional to
$\tan^2\alpha$ in \smarteq{eq:free_ang_eq}.  Thus, as $N$ increases,
the probability becomes increasingly concentrated in a smaller and
smaller region of $\alpha$ space around $\alpha=0$.

\fadeqN

Some solutions may be spurious, i.e.~each component $\phi$ is
non-vanishing, but the full wave function $\Phi$ in
\refeq.~(\ref{eq:faddeev_decomposition_swaves}) is identically zero:
\begin{eqnarray}
  \Phi
  =\sum_{i<j}^N \phi_{ij}
  =0
  \label{eq:spurious}
  \;.
\end{eqnarray}
\Refeq.~(\ref{eq:angular-equation1}) shows that such a component
$\phi$ with zero sum is an eigenfunction of the angular kinetic-energy
operator.  Here the $K=2$ eigenfunction from
\refeq.~(\ref{eq:free_ang_solution}) has a vanishing angular average, i.e.
\begin{eqnarray}
  \int d\tau\;\sum_{i<j}^N \phi_{K=2}(\alpha_{ij})=0
  \label{eq:K2spurious}
  \;,
\end{eqnarray}
see appendix \ref{sec:prop-select-funct}.  This criterion is not
identical to \refeq.~(\ref{eq:spurious}), but functions $\phi$ that
obey \refeq.~(\ref{eq:K2spurious}) are nevertheless inert to the
interaction potential as it occurs in the Faddeev-like
\smarteq{eq:angular-equation3} and in the angular variational
\smarteq{eq:simplified_var_eq}.  Solutions like $\phi_{K=2}$ obtained
by solving \smarteqs{eq:angular-equation3} or
(\ref{eq:simplified_var_eq}) are therefore independent of the
interactions and the eigenvalue is independent of $\rho$.  Since the
$K=2$ function is spurious in this sense, it must be avoided when
obtaining the solutions.

\angvar

\subsection{Asymptotic spectrum for two-body states}

\label{sec:lambda_largerho}

\fadeqN\twobound

For large values of $\rho$, the short-range two-body potential $v$
with range $b$ is non-vanishing only when $\alpha$ is smaller than a
few times $b/\rho$.  For larger values of $\alpha$, the
``potential-rotation'' terms $v\hatR\phi$ in the angular Faddeev-like
\refeq.~(\ref{eq:faddeev_eq}) can therefore be omitted.

We first assume that the rotation terms $\hatR\phi$ can be neglected
for smaller $\alpha$.  For $\alpha\ll1$ substitution of
$r\simeq\sqrt2\rho\alpha$ instead of $\alpha$ in
\refeq.~(\ref{eq:faddeev_eq}) then leads to the two-body equation with
energy $E^{(2)}$:
\begin{eqnarray}
  \bigg[
  -\frac{\hbar^2}{m}\frac{d^2}{dr^2}
  +V(r)
  -E^{(2)}
  \bigg]
  u(r) 
  =0
  \label{eq:twobody_eq}
  \;,
\end{eqnarray}
where $2m\rho^2E^{(2)}/\hbar^2 = \lambda+9N/2-9$ and
$u(\sqrt2\rho\sin\alpha) = \tilde\phi(\alpha)$.  A two-body bound
state with $E^{(2)}<0$ corresponds to an eigenvalue $\lambda$
diverging towards $-\infty$ as $-\rho^2$.  Moreover, the wave function
will be concentrated around $r\sim b$, which in terms of $\alpha$
means $\alpha\sim b/\rho\ll1$.  Such solutions do not produce
significant rotation terms, which is consistent with the omission in
the derivation.  The structure of the $N$-body system is given by the
fully symmetrized wave function for two particles in the bound state,
while all other particles are far away, thus producing the large
average distance.

A solution to \refeq.~(\ref{eq:faddeev_eq}) that does not correspond
to a two-body bound state has a wave function distributed over larger
regions of $\alpha$ space.  As the potentials then vanish for large
$\rho$, we are left with the free solutions, i.e.~the free spectrum of
non-negative $\lambda$ values is obtained in this limit of large
$\rho$.

\twothreshold

A two-body state with energy slightly below zero forces $\lambda$ to
diverge slowly as $-\rho^2$.  On the other hand, if the two-body
system is slightly unbound, $\lambda$ instead converges slowly to zero
which is the lowest eigenvalue of the free solutions.  Precisely at
the threshold, it seems that $\lambda$ should not be able to decide
and therefore must remain constant.  Thus, for infinitely large
two-body $s$-wave scattering length we are led to expect that one
angular eigenvalue approaches a negative constant for large $\rho$.
Similar predicted behaviours have been confirmed for three particles
\cite{nie01}.  In section~\ref{sec:angular-potentials} we turn to the
numerical verification for $N>3$.

\twobound

\subsection{Zero-range approximation}

\label{sec:zero-range-angular}

\zerorangeappr
\fadeqN

A zero-range treatment of three-body systems leads to an equation
which can be easily solved for the angular eigenvalue
\cite{nie01,fed01b}.  The basic assumption is that when the
hyperradius is large compared to the range of the interaction, two
interacting particles in the many-body system consider each other as
point particles.  Therefore, the details of the interaction potential
can be replaced by a boundary condition at zero separation.  Moreover,
it is seen from the Faddeev-like \smarteq{eq:faddeev_eq} that outside
the potential range the angular equation is just the kinetic-energy
eigenvalue equation with the solutions from section
\ref{sec:angul-kinet-energy}.  Combination of these observations leads
to an analytic solution as follows.

\twobound

The many-body wave function at small two-particle separation
approaches the two-body wave function.  The two-body wave function at
low energy according to \smarteq{eq:u_asymptote} then behaves as
\begin{eqnarray}
  \frac{1}{u(r)}\frac{du(r)}{dr}\bigg|_{r=0}=-\frac{1}{a_s}
  \label{eq:2bodlogderiv}
  \;.
\end{eqnarray}
In order to compare consistently, we use the many-body wave function
including the volume element in the angle $\alpha$ where $\alpha$ is
related to the two-body distance $r$ by $r=\sqrt2\rho\sin\alpha$,
i.e.~we assume that at small separations the two-body wave function
$u(r)$ is represented by $\alpha\Phi(\rho,\Omega)$.
\Smarteq{eq:2bodlogderiv} then becomes
\begin{eqnarray}
  \frac{\partial[\alpha\Phi(\rho,\Omega)]}
  {\partial\alpha}\bigg|_{\alpha=0}
  =
  -\frac{\sqrt{2}\rho}{a_s}
  \alpha\Phi(\rho,\Omega)\bigg|_{\alpha=0}
  \label{eq:zerorange1}
  \;.
\end{eqnarray}
By averaging over all angular coordinates except $\alpha$ we obtain
\begin{eqnarray}
  \Phi(\rho,\Omega)
  =
  \phi(\alpha)
  +2(N-2)\hatR_{13}^{(N-2)}\phi(\alpha)
  +\frac12(N-2)(N-3)\hatR_{34}^{(N-2)}\phi(\alpha)
  \;.\;
\end{eqnarray}

Outside the diagonal potential $v(\alpha)$ the solutions and
eigenvalues, with proper boundary condition at $\alpha=\pi/2$, are
\begin{eqnarray}
  &&
  \phi_\nu(\alpha) 
  =
  \tilde \JacP_\nu(-\cos2\alpha)
  \;,\qquad
  \tilde \JacP_\nu(x)
  \equiv
  \JacP_\nu^{[(3N-8)/2,1/2]}(x)
  \;,\\
  &&
  \lambda=2\nu(2\nu+3N-5)
  \label{eq:zerorange_freesol}
  \;.
\end{eqnarray}
Since we do not restrict $\phi_\nu$ for $\alpha\to0$, non-integer
values of $\nu$ are allowed.  For small $\alpha$ the solutions behave
as \cite{nie01}, see also appendix~\ref{sec:prop-select-funct},
\begin{eqnarray}
  &&
  \phi_\nu(\alpha)
  \simeq
  \frac{A}{\alpha}+B
  \;,\qquad
  A
  \equiv
  -\frac{\sin(\pi\nu)}{\sqrt\pi}
  \frac{\Gamma\big(\nu+\frac{3N-6}{2}\big)}{\Gamma\big(\nu+\frac{3N-5}2\big)}
  \;,\\
  &&
  B
  \equiv
  \cos(\pi\nu)\frac{2}{\sqrt\pi}\frac{\Gamma\big(\nu+\frac32\big)}
  {\Gamma(\nu+1)}
  \;.
\end{eqnarray}
Then at the edge of the zero-range potential we get
\begin{eqnarray}
  &&
  \alpha\Phi(\rho,\Omega)\Big|_{\alpha=0}
  =\alpha\phi_\nu(\alpha)\Big|_{\alpha=0}=A
  \;,\\
  &&
  \frac{\partial[\alpha\Phi(\rho,\Omega)]}{\partial\alpha}\bigg|_{\alpha=0}
  =B
  +2(N-2)\hatR_{13}^{(N-2)}\phi_\nu(0)
  \nonumber\\
  &&
  \qquad
  +\frac12(N-2)(N-3)\hatR_{34}^{(N-2)}\phi_\nu(0)
  \;,\\
  &&
  \hatR_{34}^{(N-2)}\phi_\nu(0)
  =\frac{2\gammafkt{3N-6}\Gamma\big(\nu+\frac32\big)}
  {\sqrt\pi\Gamma\big(\nu+\frac{3N-6}2\big)}
  \stackrel{\nu\to0}{\longrightarrow}
  1
  \;,\\
  &&
  \hatR_{13}^{(N-2)}\phi_\nu(0)
  =
  \frac{2\gammafkt{3N-6}}{\sqrt\pi\gammafkt{3N-9}}
  \bigg(\frac23\bigg)^{(3N-8)/2}
  \times
  \\ \nonumber
  &&
  \qquad\int_{-1}^{1/2} dx \sqrt{1+x}
  \Big(\frac12-x\Big)^{(3N-11)/2}\tilde \JacP_\nu(x)
  \stackrel{\nu\to0}{\longrightarrow}
  1
  \quad\textrm{for }N>3
  \;,
  \\
  &&
  \hatR_{13}^{(N-2)}\phi_\nu(0)
  =\frac{2\sin[(\nu+1)\pi/3]}{(\nu+1)\sqrt3}
  \stackrel{\nu\to0}{\longrightarrow}
  1
  \quad\textrm{for }N=3
  \;.
\end{eqnarray}
Combination of these results leads to $\rho/a_s$ as a function of
$\nu$:
\begin{eqnarray}
  &&
  \frac{\rho}{a_s}
  =
  \frac{\sqrt2\Gamma\big(\nu+\frac32\big)}{\sin(\pi\nu)}
  \frac{\gammafkt{3N-6}\Gamma\big(\nu+\frac{3N-5}2\big)}
  {\Gamma\big(\nu+\frac{3N-6}2\big)^2}
  \label{eq:zerorange5}
  \times
  \\
  &&
  \quad
  \bigg[
  \frac{\cos(\pi\nu)\Gamma\big(\nu+\frac{3N-6}2\big)}
  {\Gamma(\nu+1)\gammafkt{3N-6}}
  +2(N-2)
  \frac{\hatR_{13}^{(N-2)}\phi_\nu(0)}{\hatR_{34}^{(N-2)}\phi_\nu(0)}
  +\frac{(N-2)(N-3)}2
  \bigg]
  \nonumber
  \;.
\end{eqnarray}  
At small $|\nu|\ll1$, the square bracket yields $N(N-1)/2$, and then
$\nu$ becomes
\begin{eqnarray}
  \nu(\rho)
  \simeq\frac{N(N-1)}{2\sqrt{2\pi}}
  \frac{\gammafkt{3N-5}}{\gammafkt{3N-6}}
  \frac{a_s}{\rho}
  \;.
\end{eqnarray}
The angular eigenvalue $\lambda$ from \smarteq{eq:zerorange_freesol}
is then
\begin{eqnarray}
  \lambda(\rho)
  \simeq2\nu(3N-5)
  =
  \sqrt{\frac2\pi}
  N(N-1)\frac{\gammafkt{3N-3}}{\gammafkt{3N-6}}
  \frac{a_s}{\rho}
  \label{eq:zerorange10}
  \;.
\end{eqnarray}
This derivation is valid when $\nu\ll1$, or equivalently when $\rho\gg
N^{5/2}|a_s|$.

\fadeqN

As we shall see in the following section, the result in
\smarteq{eq:zerorange10} can be obtained otherwise.  However, when the
treatment of \smarteq{eq:zerorange5} is numerically extended to
smaller hyperradii, unmistakably wrong results are encountered.
Whether this is reminiscent of the initially expected deficiencies of
the Faddeev-like equation or it is a mistake in the treatment of
\smarteq{eq:zerorange5} is presently not sorted out.

\subsection{Average, non-correlated effects of interactions}

\label{sec:effect-short-range}

\lambdadelta

As discussed in section \ref{sec:mean-field-descr}, a mean-field wave
function corresponds to a constant angular wave function where no
correlations are included.  With a non-correlated, constant angular
wave function $\Phi_{K=0} = \sum_{i<j}^N\phi_{K=0}(\alpha_{ij})$, the
expectation value of the angular Hamiltonian $\hat h_\Omega$ becomes
\begin{eqnarray}
  \lambda_{K=0}
  =
  \langle\Phi_{K=0}|\hat h_\Omega|\Phi_{K=0}\rangle_\Omega
  =
  \Big\langle\Phi_{K=0}\Big|
  \sum_{k<l}^Nv_{kl}
  \Big|\Phi_{K=0}\Big\rangle_\Omega
  \label{eq:K0_expectation}
  \;,
\end{eqnarray}
without contribution from angular kinetic energy.  Proceeding in the
manner of the mean field we then have to assume the same ansatz for
the two-body interaction, i.e.~the $\delta$ function from
\smarteq{eq:zerorange-potential}.  With this zero-range interaction,
\smarteq{eq:K0_expectation} becomes
\begin{eqnarray}
  \lambda_\delta 
  \equiv
  \sqrt{\frac{2}{\pi}}
  N(N-1) 
  \gammafktB{3N-3}{3N-6}
  \frac{a_s}{\rho}
  \;
  \stackrel{N\gg1}{\longrightarrow}
  \;
  \frac32\sqrt{\frac3\pi}N^{7/2}\frac{a_s}{\rho}
  \label{eq:lambda_delta} 
  \;.
\end{eqnarray}
\etalcite{Bohn}{boh98} did a similar calculation, but since they did not
separate out the centre-of-mass motion, the present result for
$\lambda_\delta$ is effectively that of \cite{boh98} with $N$ replaced
by $N-1$ in the $\Gamma$ function.  For $N\gg1$ the results are
identical.

We note that the angular potential from \smarteq{eq:lambda_delta}
coincides with \smarteq{eq:zerorange10}, i.e.~the large-hyperradii
derivation from the zero-range model in section
\ref{sec:zero-range-angular}.  This indicates that the structure of
the two-body correlated ansatz for the many-body wave function catches
the essential information in agreement with the low-density result,
\smarteq{eq:lambda_delta}, which corresponds to the mean field.

Thus, the zero-range interaction from \smarteq{eq:zerorange-potential}
leads to reasonable energies in the dilute limit.  However, at larger
densities (smaller $\rho$) a negative scattering length $a_s$
potentially leads to unphysical behaviours.  We can understand this
problem by putting $\lambda_\delta\propto a_s/\rho$ into the radial
potential, \smarteq{eq:radial.potential}, which yields a term
$a_s/\rho^3$ that diverges faster than other terms as $\rho\to0$.  We
return to this problem in chapter \ref{kap:stability_validity}.

\twobound\lambdadelta

Thus, a zero-range two-body interaction in mean-field computations can
lead to a collapse.  This problem is not present for finite-range
interactions, and the present method allows the use of strongly
attractive potentials.  The $\delta$ interaction furthermore does not
allow a study of short-range properties such as bound two-body systems
and similar clusterizations.  Both problems are overcome by using a
finite-range potential in the present model.  When $\rho$ is much
larger than the potential range $b$, the expectation value of a
finite-range potential is of the same form as $\lambda_\delta$ in
\refeq.~(\ref{eq:lambda_delta})
\begin{eqnarray}
  \lambda_{K=0}\super{finite}
  \stackrel{\rho\gg b}\longrightarrow
  \sqrt{\frac{2}{\pi}}
  N(N-1)
  \gammafktB{3N-3}{3N-6}
  \frac{a\sub B}{\rho}
  \label{eq:lambda_shortrange}
  \;,
\end{eqnarray}
with the Born approximation $a\sub B$ from \smarteq{eq:Born_appr}
instead of the real scattering length $a_s$.

\gaussian

In the opposite limit, when $\rho\ll b$, the result is strongly
dependent on the shape of the potential.  For example, the Gaussian
potential from \smarteq{eq:vpotGaussian} yields
\begin{eqnarray}
  \lambda_{K=0}\super{finite}
  \stackrel{\rho\ll b}\longrightarrow
  \frac{4}{\sqrt\pi}N(N-1)\frac{a\sub B}{b}\Big(\frac{\rho}{b}\Big)^2
  \label{eq:lambda_smallrho}
  \;.
\end{eqnarray}
As seen from these two limits there are some scaling properties for
finite-range potentials.  The angular eigenvalue at a given $N$ value
depends only on $a\sub B/b$ and $\rho/b$.  For a Gaussian potential we
have
\begin{eqnarray}
  v_{kl}
  =
  \frac{2m\rho^2V_0}{\hbar^2}e^{-r_{kl}^2/b^2}
  =
  \frac{8a\sub B}{\sqrt\pi b}
  \Big(\frac{\rho}{b}\Big)^2
  e^{-2(\rho/b)^2\sin^2\alpha_{kl}}
  \;,
\end{eqnarray}
which implies that for a given value of $a\sub B/b$, the angular
eigenvalue $\lambda$ is only a function of $\rho/b$.  The radial
potential $U$ from \smarteq{eq:radial.potential}, which we return to
in chapter \ref{kap:radial}, can be scaled as
\begin{eqnarray}
  \frac{2mb^2U(\rho)}{\hbar^2}
  =
  \frac{\lambda}{(\rho/b)^2}+
  \frac{(3N-4)(3N-6)}{4(\rho/b)^2}+\frac{(\rho/b)^2}{(b\sub t/b)^4}
  \;,
\end{eqnarray}
where $b\sub t=\sqrt{\hbar/(m\omega)}$ is the characteristic length
for a harmonic trap of angular frequency $\omega$.  The scaled energy
$2mb^2E/\hbar^2$ is then for a given $N$ value only a function of
$a\sub B/b$ and $b\sub t/b$.  These scaling properties are useful in
model calculations.


\section{Numerical angular solutions}

\label{sec:numer-angul-solut}

In the previous section we discussed solutions to the angular equation
in the presence of no interactions, in the case of two-body bound
states, and in the zero-range limit.  However, solutions with general
two-body interactions have to be obtained numerically, which is the
quest of the present section.  We first comment on the numerical
procedure before discussing properties of the angular eigenvalues and
wave functions.

\subsection{Numerical method}

\label{sec:numerical-method}

\angvar

The angular eigenvalue equation was rewritten in
chapter~\ref{kap:hyperspherical_method} by a variational technique as
the second-order integro-differential
\smarteq{eq:simplified_var_eq} in the variable $\alpha$, where 
$r_{12}=\sqrt2\rho\sin\alpha$.  For neutral atoms in recent trapping
experiments the interaction range is very short compared to the
spatial extension of the $N$-body system.  Then this equation
simplifies to contain at most one-dimensional integrals.  The validity
of the approximations only relies on the small \emph{range} $b$ of the
potential, whereas the scattering length $a_s$ can be as large as
desired.

Even though the complexity of the angular equation does not increase
as the number of particles increases, the numerical solutions become
harder to handle for large $N$.  The origin of this problem is the
sharp peak in the angular volume element for large $N$, see section
\ref{sec:angul-kinet-energy}.

\subsubsection{Expansion on kinetic-energy eigenfunctions}

\label{sec:expans-kinet-eigenf}

\twobound

A usual method within the hyperspherical formalism is to expand the
angular wave function on kinetic-energy eigenfunctions
\cite{lin95,boh98}.  Such an expansion is successful when the physical
extension of the system is comparable to the interaction range.  The
hyperspherical harmonics contain oscillations at angles of the order
of magnitude $\alpha\sim \mathcal O(1/K)$, so for a given hyperradius
we need $K$ values of the order of $K\sub{max} \sim \mathcal
O(\rho/b)$ to describe potentials limited to $\alpha < b/\rho$.  Thus,
the angular kinetic-energy eigenfunctions constitute an ineffective
basis at large hyperradii since the diagonal potential in this case
will be sharply peaked around $\alpha=0$, and a huge number of terms
is necessary to account for the correct behaviour of the wave function
around $\alpha=0$.

For trapped particles the scale of the system is determined by the
trap length $b\sub t$ which for atomic gases usually is of order
$\mu$m.  Since the interaction range $b$ usually is in the nm region,
an expansion on kinetic-energy eigenfunctions converges slowly and is
not appropriate for the present treatment.

\subsubsection{Finite differences}

Instead of an expansion on hyperspherical harmonics we choose a basis
of discrete mesh points distributed in $\alpha$ space
$\phi(\alpha)\to\underline\phi \equiv [\phi(\alpha_1), \ldots,
\phi(\alpha_M)]$ to take into account the short range of the potential
and to keep sufficient information about small $\alpha$.  Derivatives
are then written as finite differences \cite{koo90} and integrations
like $\hatR\phi(\alpha)$ of \refeq.~(\ref{eq:Gkernel_0}) can be
expressed in matrix form, i.e.~$\hatR\phi(\alpha) \to
\underline{\underline R}\;\underline\phi$.

Numerical computation of the integrals becomes increasingly difficult
with decreasing interaction range.  This can be understood in terms of
the $\alpha$ coordinate, since the potential at a given $\rho$
and a given range $b$ of the interaction, is confined to an
$\alpha$ region of size $\Delta\alpha\sim b/\rho$, which for
Bose-Einstein condensates easily becomes very small and thus cannot be
handled directly numerically.

\fadeqN\angvar

Recently the method of finite elements was applied to the Faddeev-like
\smarteq{eq:faddeev_eq}.  With finite elements the basis functions are
smooth and yield more reliable matrix elements, especially those
involving derivatives due to the kinetic energy.  This proves easier
to handle, but is presently not implemented for the angular
variational \smarteq{eq:simplified_var_eq}.  For details about
finite-elements methods see references in \etalcite{Press}{pre89}.

Unless stated otherwise, the following numerical results are obtained
with the method of finite differences.

\subsection{Angular potentials}

\label{sec:angular-solutions}
\label{sec:angular-potentials}


\gaussian

\begin{figure}[htb]
  \centering
  \input{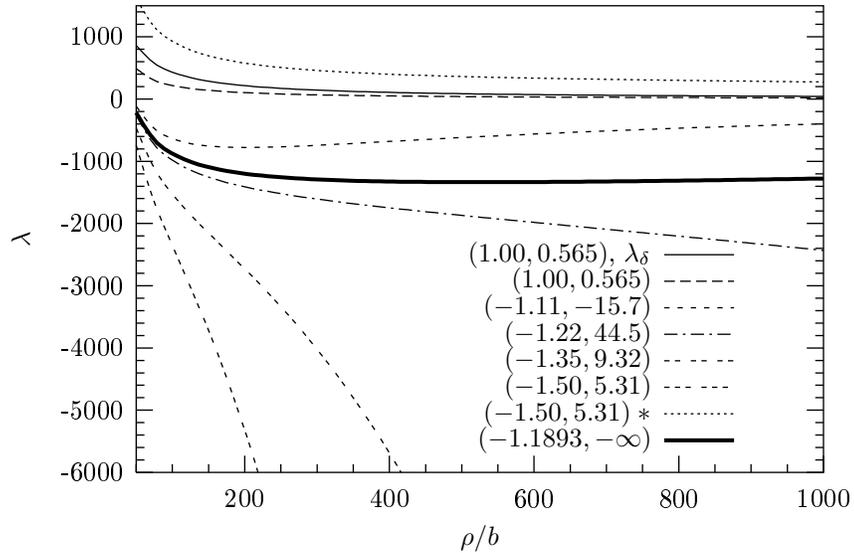}
  \caption [Angular eigenvalues for twenty particles]
  {Angular eigenvalues for $N=20$ and parameters $(a\sub B/b,a\sub
    s/b)$ as shown on the figure.  A star refers to the first excited
    state.  For $a\sub B/b=-1.1893$ we have $a_s/b=-85601$, see table
    \ref{table:as_aB}, which here is denoted by $-\infty$.}
  \label{fig:lambdavar}
\end{figure}

The angular eigenvalue depends on the number of particles, on the size
of the system through the hyperradius, and on the two-body potential.
\Reffig.~\ref{fig:lambdavar} shows the angular eigenvalue for the
 particle number $N=20$ and various Gaussian potential strengths.
Only the lowest $\lambda_0$ is shown unless otherwise indicated.


\twobound\lambdadelta

The long-dashed line shows the calculation for a purely repulsive
interaction with positive scattering length.  Here the angular
eigenvalue approaches zero at large hyperradii approximately as
$1/\rho$.  The thin, solid line shows $\lambda_\delta \propto
a_s/\rho$ from \refeq.~(\ref{eq:lambda_delta}) for the same scattering
length.  These two curves almost coincide at large hyperradii.  The
short-dashed curve shows the angular eigenvalue for a slightly
attractive two-body interaction without any two-body bound state and
with negative scattering length.  This angular potential approaches
zero from below as $1/\rho$, also in agreement with
\smarteq{eq:lambda_delta}.  For a larger attraction, when the
scattering length becomes very large, the angular eigenvalue (thick,
solid line) is almost constant for a large region of hyperradii.  This
agrees with the expectations in section \ref{sec:lambda_largerho}.
For a slightly larger attraction the scattering length turns positive
and a two-body bound state forms.  Then (dot-dashed line) the lowest
angular eigenvalue at some point diverges to minus infinity.  For even
larger attraction the binding energy of the bound state increases and
$\lambda$ diverges faster, see the sequence of the dot-dashed,
double-dashed, and triple-dashed lines.

The dotted line shows the angular eigenvalue for the next angular
solution for the strongest attraction.  This approaches zero from
above as $1/\rho$, which resembles the behaviour for a purely
repulsive interaction (long-dashed line).  This illustrates the use of
the terms ``effectively repulsive'' or ``effectively attractive'' in
the mean field, depending on the sign of $a_s$ even though the
interaction potential might be purely attractive.  See related
comments by Geltman and Bambini \cite{gel01}.


\twobound

\begin{figure}[htb]
  \centering
  \input{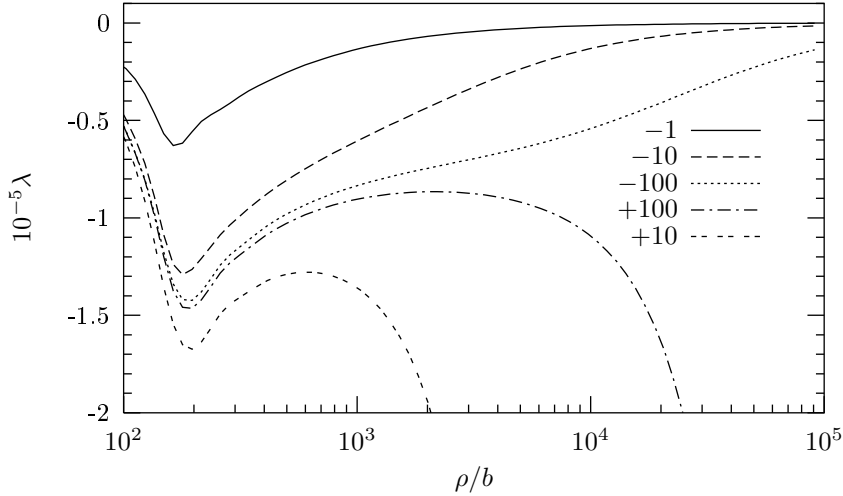}
  \caption[Angular eigenvalues with two-body bound states]
  {The lowest angular eigenvalues $\lambda$ for $N=100$ bosons
    interacting via a Gaussian two-body potential
    $V(r)=V_0\exp(-r^2/b^2)$ with zero or one bound two-body states.
    The scattering lengths $a_s /b$ are indicated on the figure.}
  \label{fig:lambda_numerical}
\end{figure}

For one hundred particles figure~\ref{fig:lambda_numerical} shows the
lowest angular potential for various attractive interactions.
Qualitatively the same behaviours as for $N=20$ are observed.  When
$a_s = -b$ (solid line) the system has no bound two-body states.  The
lowest angular eigenvalue is zero at $\rho=0$, decreases then through
a minimum as a function of $\rho$, and approaches zero at
large hyperradii as $a_s/\rho$.  A larger attraction (broken lines)
decreases all angular eigenvalues for all $\rho$ values.  The details
at smaller hyperradii hardly change with large variations of the
scattering length.  However, at larger distances the approach towards
zero is converted into a parabolic divergence as soon as the
scattering length jumps from negative (dotted line) to positive
(dot-dashed line) corresponding to the appearance of a bound two-body
state.  The faster divergence (double-dashed line) is again observed
for increasing binding energy.


The characteristic feature for both cases $N=20,100$ is the
large-distance asymptotic behaviours.  For repulsive potentials all
eigenvalues are positive and the lowest approaches zero from above.
The higher eigenvalues would then converge to $K(K+3N-5)$ as $1/\rho$,
where $K=4,6,8...$.  The solution for $K=2$ is not allowed,
corresponding to removal of the non-physical spurious solution, see
section~\ref{sec:angul-kinet-energy}.

\lambdadelta

For weak attractions the lowest $\lambda$ is negative and approaches
zero from below as $1/\rho$.  The higher angular eigenvalues approach,
again, $K(K+3N-5)$ corresponding to the spectrum for free particles.
The constant of proportionality to $\rho^{-1}$ for the lowest
eigenvalue is qualitatively recovered as the predicted dependence on
$a_s$.  Calculations with a two-body potential as a linear combination
of different Gaussians (not shown) confirm that the large-distance
angular potential only depends on the scattering length $a_s$ as in
$\lambda_\delta$.


\twobound

In the presence of a two-body bound state the divergence as $-\rho^2$
reflects the corresponding two-body binding energy, see
\refeq.~(\ref{eq:twobody_eq}).  Generally, an attractive finite-range
interaction can support a certain number $\mathcal N\sub B$ of
two-body bound states for both positive and negative scattering
lengths.  Then the lowest angular eigenvalues,
$\lambda_0,\lambda_1,\ldots,\lambda_{\mathcal N\subsub B-1}$, describe
these bound two-body states within the many-body system at large
hyperradii, i.e.~they diverge to $-\infty$ as seen in
figure~\ref{fig:lambda_numerical}.


\twobound

The next eigenvalue $\lambda_{\mathcal N\subsub B}$ converges to zero
at large distance and corresponds to the first ``two-body-unbound''
mode.  The higher eigenvalues would then, once more, converge to
$K(K+3N-5)$.  Increasing the attraction to allow another bound
two-body state would then shift the asymptotic spectrum such that one
more eigenvalue diverges while the non-negative energy spectrum
remains unchanged.  This yields qualitatively the same asymptotic
spectrum for the unbound modes irrespective of the number of bound
states below.  This invaluably eases the computations, i.e.~all the
bound states of the two-body system are not needed in order to
describe the unbound modes of the many-body system.  Therefore, the
two-body interaction does not have to be the real two-body
interaction, which allows all the two-body bound states that are known
to exist, but the interaction potential can be written in a way that
accounts for the investigated properties.  This is the case for the
potential from \etalcite{Geltman}{gel01}, \smarteq{eq:vpot_geltman},
and also for the Gaussian potential, \smarteq{eq:vpotGaussian},
applied in the present work.

\twobound\lambdadelta

\begin{figure}[htb]
  \centering
  \input{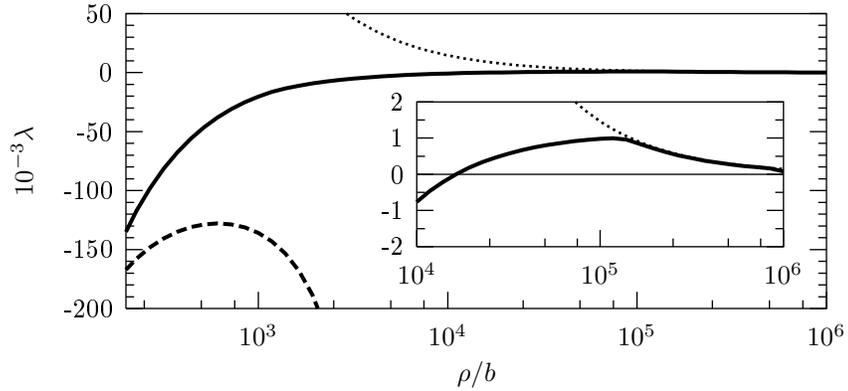}
  \caption
  [Two lowest angular eigenvalues for one bound two-body state] {The
    two lowest angular eigenvalues (dashed and solid curves) for
    $N=100$, $a_s/b=+10$, and one bound two-body state.  The dotted
    curve is $\lambda_\delta$ for the same scattering length.}
  \label{fig:trento2}
\end{figure}

These properties of the two lowest eigenvalues in the presence of one
two-body bound state are evident in \reffig.~\ref{fig:trento2}.  The
lowest eigenvalue (dashed curve) diverges to minus infinity
proportional to $\rho^2$.  This corresponds to the bound state.  The
second eigenvalue (solid curve) is negative at small hyperradii, but
turns positive at larger and approaches the asymptotic behaviour of
$\lambda_\delta\propto a_s/\rho$ (dotted curve, see details in the
inset).
Since the second eigenvalue at small and intermediate hyperradii is
negative, this might allow a self-bound system located at distances
far inside and independent of a confining external trap potential.
This feature is absent in a description with overall repulsive
potentials, corresponding to positive scattering lengths, for example
the zero-range interaction with $a_s>0$.  Then no attractive part is
possible.


\twothreshold

At each threshold for the appearance of a new bound two-body state, one
eigenvalue asymptotically approaches a negative constant as in
\smartfig{fig:lambdavar}.  This  eigenvalue is responsible for the
structure of the $N$-body system for very large scattering lengths.
This reflects the transition from unbound to bound two-body states,
that is the transition from convergence towards zero as $-1/\rho$ to
divergence as $-\rho^2$, see section
\ref{sec:lambda_largerho}.


\fadeqN\angvar

\begin{figure}[htb]
  \centering
  \input{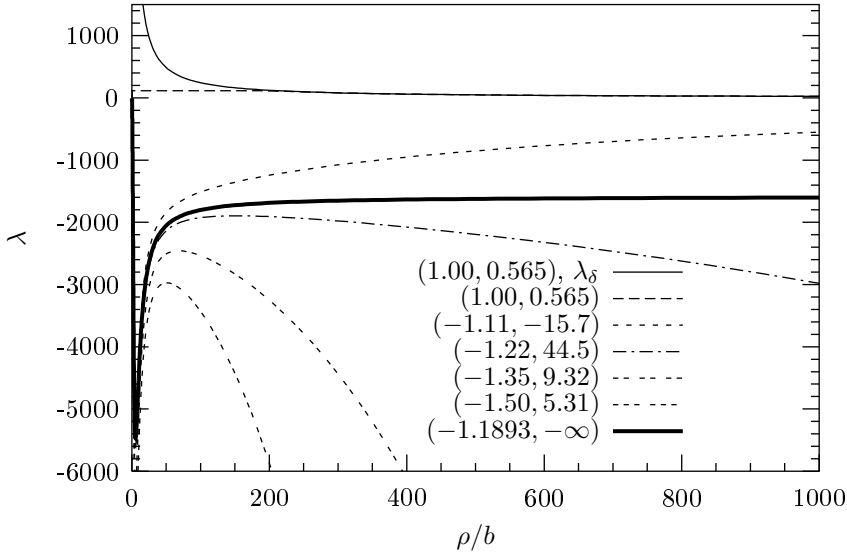}
  \caption [Angular eigenvalues for twenty particles
  (Faddeev calculation)] {Angular eigenvalues for $N=20$ and
    parameters $(a\sub B/b,a_s/b)$ as shown on the figure obtained
    from the Faddeev-like \smarteq{eq:faddeev_eq}.}
  \label{fig:lambdavarFEM}
\end{figure}

We finish the discussion of the angular eigenvalue from the
variational \smarteq{eq:simplified_var_eq} by comparing with the
results from a finite-element treatment of the Faddeev-like
\smarteq{eq:faddeev_eq}.
\Reffig.~\ref{fig:lambdavarFEM} shows the results from the
Faddeev-like equation for the same parameters as in
\reffig.~\ref{fig:lambdavar}.  At large hyperradius the results
agree, whereas they differ as $\rho\to0$.  This is probably due to the
short-range approximation of section \ref{sec:shortrangeappr},
although we recall the non-variational nature of the Faddeev-like
equation, as discussed in section \ref{sec:faddeev_eq}, as another
possible source.
However, due to the importance of correlations higher than two-body in
the denser regions, corrections to these short-distance results would
be in order even without the short-range approximation.  Moreover, for
a description of a dilute many-boson system we do not need the details
at such short distances, so they are not considered in the following.

\subsection{Angular wave function}


\twobound

\begin{figure}[htb]
  \centering
  \input{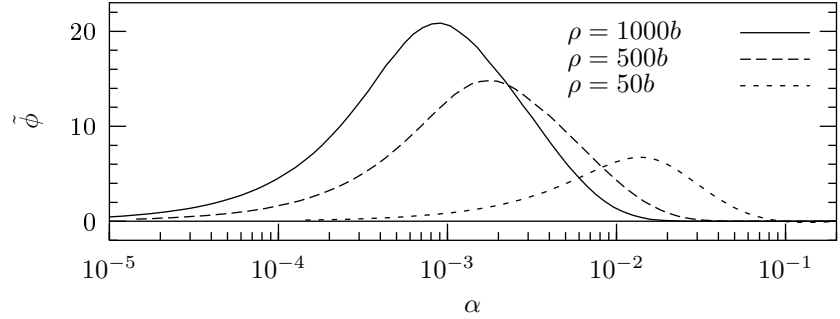}
  \caption [Angular wave functions for two-body bound states] 
  {The lowest reduced angular wave functions for $N=20$ and $a\sub
    B=-1.50b$, $a_s=5.31b$ for three values of the hyperradius.  This
    potential has one bound two-body state.}
  \label{fig:phibound}
\end{figure}

The total angular wave function is determined as the sum of two-body
components in \refeq.~(\ref{eq:faddeev_decomposition_swaves}).
\Reffig.~\ref{fig:phibound} shows the  lowest component wave
function, reduced as in \smarteq{eq:angular-equation-1}, for a
two-body potential with one bound two-body state.  With increasing
$\rho$ the amplitude concentrates at smaller and smaller values of
$\alpha$.  This reflects the convergence towards the two-body bound
state in agreement with the transformation $r_{12} = \sqrt{2}\rho \sin
\alpha$, see section~\ref{sec:lambda_largerho}.  The numerical
recovery of this behaviour is essential, since otherwise the
large-distance properties cannot be described.

\begin{figure}[htb]
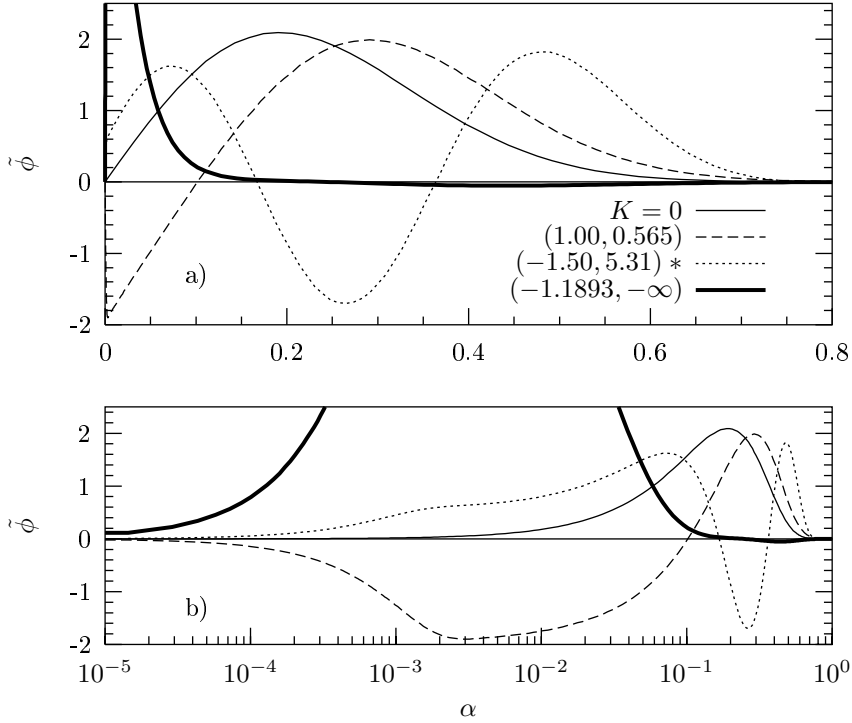

  \centering
  \input{phi.asvar}
  \input{phi.asvar.log}
  \caption [Angular wave functions for different interaction parameters]
  {a) Angular wave functions for $N=20$ and $\rho=500b$ for different
    interaction parameters $(a\sub B/b,a_s/b)$ as shown on the figure.
    The $K=0$ curve corresponds to a non-interacting system.  A star
    refers to the first excited state.  b) The same as a), but with
    logarithmic $\alpha$ axis.}
  \label{fig:phi.asvar}
\end{figure}

The angular eigenfunction varies with the strength of the interaction.
Examples of this variation are shown in
\reffig.~\ref{fig:phi.asvar}a.  The lowest non-interacting wave
function (thin, solid line) has only nodes at the endpoints.  The
repulsive case shows an oscillation (dashed line) which lowers the
angular energy due to the rotation terms.  The fast change at small
$\alpha$, which is emphasized in \reffig.~\ref{fig:phi.asvar}b, is
typical for interacting particles.  The wave function for the excited
state (dotted line) has an additional node.  The corresponding
lower-lying wave function was shown as the dashed line in
\reffig.~\ref{fig:phibound}.

\twothreshold

The wave function for infinite scattering length (thick, solid line in
\smartfig{fig:phi.asvar}) corresponds to an interaction where the
two-body bound state is at the threshold for occurrence.  This
eigenfunction resembles those where a bound two-body state is present,
compare with the results shown in \reffig.~\ref{fig:phibound}.
However, now (thick curve of \reffig.~\ref{fig:phi.asvar}b) the wave
function is located at larger $\alpha$ values.

\begin{figure}[htb]
  \centering
  \input{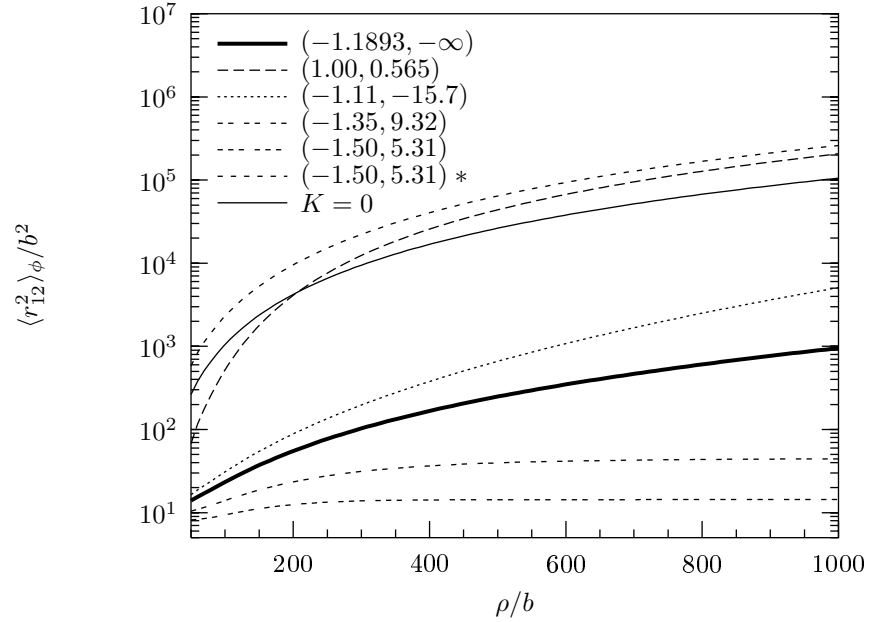}
  \caption [Second angular moment as a function of hyperradius]
  {The second moment $\langle r_{12}^2\rangle_\phi$ as a function of
    hyperradius for $N=20$ for solutions to the angular variational
    equation with different interaction parameters specified in the
    figure by $(a\sub B/b,a_s/b)$.  Also shown is the $K=0$ value.  A
    star refers to the first excited state.}
  \label{fig:moment}
\end{figure}

The properties of the component of the angular wave function is
further illustrated by the second moment defined by
\begin{eqnarray}
  \langle r_{12}^2
  \rangle_{\phi} \equiv 2\rho^2\langle \phi_{12} | \sin^2\alpha |\phi_{12}
  \rangle
  \;.
\end{eqnarray}
A number of these moments for different interactions are shown in
\reffig.~\ref{fig:moment} as functions of $\rho$.  For states obtained
from repulsive potentials, moderately attractive potentials without
bound two-body states, and for excited states of positive $\lambda$,
the moment $\langle r_{12}^2\rangle_\phi$ increases proportional to
$\rho^2$ for large $\rho$.  This resembles the behaviour of the
expectation value in the lowest angular state for a non-interacting
system, i.e.~$K=0$, where $\langle r_{12}^2\rangle_\phi =
2\rho^2/(N-1)$.  The qualitative explanation is that large $\rho$
implies the limit of a non-interacting spectrum with the corresponding
non-correlated wave functions.

\angvar

In contrast, a different behaviour is observed when the potential can
bind two particles, i.e.~$\langle r_{12}^2\rangle_\phi$ approaches a
constant at large $\rho$.  The angular equation in this limit
approaches the two-body \smarteq{eq:twobody_eq}.  The wave function in
the zero-range limit converges to $u(r)=\exp(-r/a_s)$.  The second
moment is then found as $\langle u|r^2|u\rangle = a_s^2/2$, which in
the limit of large $\rho$ reproduce the constant values for $\langle
r_{12}^2\rangle_\phi$ when $a_s/b=9.32$ and $a_s/b=5.31$, i.e.~the
double- and triple-dashed lines in \reffig.~\ref{fig:moment} approach
$9.32^2/2\simeq43$ and $5.31^2/2\simeq14$, respectively.

\twobound

Expressed differently, when a two-body bound state is present, the
angular wave function is at increasing $\rho$ squeezed inside the
potential since the range in $\alpha$ space decreases proportional to
$\rho^{-1}$.  This implies $\langle\phi_{12}| \sin^2\alpha
|\phi_{12}\rangle \propto 1/\rho^2$.  The distance between a pair of
particles is therefore independent of $\rho$ at large values of
$\rho$.  This means that pairwise the two-body bound state is
approached while all other particles are far away.  The symmetrization
does not affect this conclusion.  Thus, apart from this symmetrization
of the many-boson wave function, the attributes of the many-body
system in the presence of this two-body bound state show only small
deviations from the well-known properties of the isolated two-body
bound state.

\twothreshold

At the threshold for two-body binding, that is for infinite scattering
length, the intermediate behaviour once again emphasizes the
transition from bound to unbound, see the thick, solid line in
\reffig.~\ref{fig:moment}.

\section{Summary}

\label{sec:summ-angul-prop}

\twobound

Further numerical analysis allows us to construct a parametrization
for the behaviour of the lowest angular eigenvalue for attractive
two-body interactions in two different regimes: i) no bound two-body
states and $a_s<0$, and ii) $a_s>0$ and one bound two-body state of
energy $E^{(2)}$.  These details are previously published
\cite{sor03a} and collected in appendix \ref{sec:scal-with-scatt}.
Here we summarize the results, illustrate them, and then comment on
them in relation to the previous observations.

\lambdadelta\lambdaschematic\lambdainfty

\subsection{Parametrization}

For small hyperradii $\rho < \rho_0 \equiv
0.87N^{1/2}(b/|a_s|)^{1/3}b$ we use for all $a_s$ the perturbation
result obtained as the expectation value of the two-body interaction
$V(r)$ in a constant angular wave function, i.e.~for $N\gg1$
\begin{eqnarray}
  \lambda\sub{a} 
  =
  \frac{mV(0)N^2\rho^2}{\hbar^2}
  \quad
  {\rm for}\;\rho<\rho_0
  \label{eq:lambda0_smallrho}
  \;.
\end{eqnarray}
For hyperradii exceeding the lower limit $\rho_0$ the analytic
expressions from \refeqs.~(\ref{eq:scale_lambdaminus}),
(\ref{eq:scale_gminus}), (\ref{eq:scale_lambdaplus}), and
(\ref{eq:scale_gplus}) are expressed as
\begin{eqnarray}
  &&
  \lambda\sub{a}(N,\rho)
  =
  -|\lambda_\delta(N,\rho)|\bigg(1+\frac{0.92N^{7/6}b}{\rho}\bigg)
  \label{eq:schematic}
  \\ \nonumber
  &&
  \quad
  \times\left\{
    \begin{array}{ll}
      1-\exp\Big[-\frac{|\lambda_\infty(N)|}{|\lambda_\delta(N,\rho)|}\Big]
      & \textrm{when } a_s<0 \;, \\
      \frac{|\lambda_\infty(N)|}{|\lambda_\delta(N,\rho)|}+
      \frac{|\lambda^{(2)}(\rho)|}{|\lambda_\delta(N,\rho)|}
      & \textrm{when } a_s>0 \;,
    \end{array}
  \right.
  \quad
  \textrm{for }\rho>\rho_0
  \;,
\end{eqnarray}
with $\lambda_\delta$ from \smarteq{eq:lambda_delta} and
\begin{eqnarray}
  \lambda_\infty(N)
  &=&
  -1.59N^{7/3}
  \label{eq:lambda_infty}
  \label{eq:lambda_infty_copy}
  \;,
  \\
  \lambda^{(2)}(\rho)
  &=&
  \frac{2m\rho^2}{\hbar^2}
  E^{(2)}
  \;,\qquad
  E^{(2)}=-\frac{\hbar^2}{m|a_s|^2}c
  \;.
\end{eqnarray}
The number $c$ approaches unity when the scattering length becomes
very large.  The factor $(1+0.92N^{7/6}b/\rho)$ reflects dependence on
the finite range $b$ of the Gaussian two-body interaction.  At
$\rho\sim N^{7/6}|a_s|$ we find $\lambda_\delta \sim \lambda_\infty
\sim \lambda^{(2)}$.

\lambdainfty

The results of the parametrizations in
equations~({\ref{eq:lambda0_smallrho}) and (\ref{eq:schematic}) are
illustrated in figure~\ref{fig:schematic} for \mbox{$N=100$} and
various scattering lengths.
\begin{figure}[htb]
  \centering
  \input{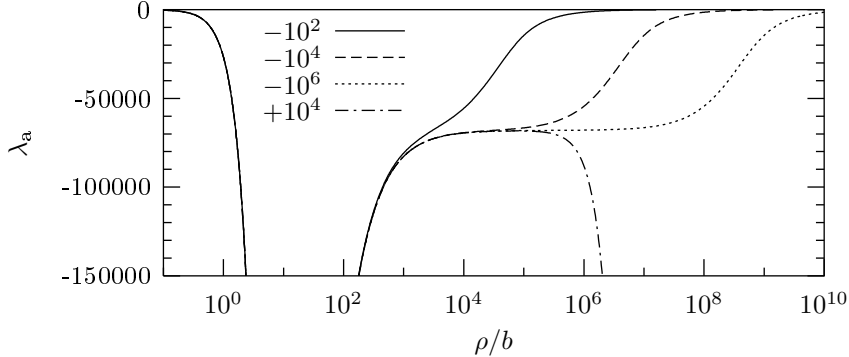}
  \caption
  [Angular eigenvalues from analytical approximation] {The angular
    eigenvalue $\lambda$, equations~(\ref{eq:schematic}) and
    (\ref{eq:lambda0_smallrho}), for $N=100$ as function of $\rho$ for
    the different scattering lengths given on the figure in units of
    the range $a_s/b$.}
  \label{fig:schematic}
\end{figure}
The pronounced deep minimum at $\rho\sim\rho_0$ is in the region
depending on the two-body potential and reflects the qualitative
behaviour of the lowest angular eigenvalue.  After this strongly
attractive region at small $\rho$ the eigenvalues approach zero.  As
the size of the scattering length increases, the eigenvalue develops a
plateau at a constant value $\lambda_\infty$ independent of $a_s$.
Eventually at large $\rho$ the eigenvalues vanish as $\lambda_\delta$
when $a_s<0$ and diverge to $-\infty$ when $a_s>0$.  This is
comparable with the sequence of the short-dashed, thick-solid, and
dot-dashed lines in \smartfig{fig:lambdavar}.

When $a_s<0$, the analytic and the correct eigenvalues both exceed the
asymptotic zero-range result, i.e.~$\lambda\sub{a} \ge
\lambda_\delta$ for all hyperradii.  This means that the ground-state
energy is higher than the energy obtained with the zero-range
interaction.  Thus, the ground state energy from the present model is
higher than the mean-field energy. The origin of this sequence of
energies is that the zero-range interaction inevitably leads to
diverging energies for smaller distances. The present model avoids
this non-physical short-range collapse.

When $a_s>0$ the interaction is effectively repulsive at large
hyperradii and an analytical expression in this case for the second
angular eigenvalue obeys $\lambda\sub{a} \le \lambda_\delta$ for all
hyperradii, due to the divergence of $\lambda_\delta\to+\infty$ as
$\rho\to0$.  Correspondingly, the energies are smaller than the
zero-range mean-field result in the positive-$a_s$ case.

\subsection{At the threshold}

\label{sec:at-threshold}


\lambdainfty

At intermediate hyperradii, that is when
\begin{eqnarray}
  b < \frac{\rho}{N^{7/6}} < |a_s|
  \;,
  \label{eq:intermediate_region}
\end{eqnarray}
the angular eigenvalue as obtained from \smarteq{eq:schematic} is
independent of both the short-range details of the two-body
interaction and the scattering length.  Then $\lambda$ approaches a
constant value given by \smarteq{eq:lambda_infty} as
$\lambda_\infty\simeq -1.59N^{7/3}$.  This plateau value can be
estimated by considering the angular eigenvalue for a two-body bound
state:
\begin{eqnarray}
  \lambda^{(2)}(\rho)\simeq-\frac{2\rho^2}{a_s^2}
  \label{eq:ld5}
  \;.
\end{eqnarray}
The plateau terminates at a hyperradius $\rho_a$ where this two-body
angular potential intersects with $\lambda_\delta$ from
\refeq.~(\ref{eq:lambda_delta}), i.e.
\begin{eqnarray}
  \lambda_\infty(N)
  = \lambda^{(2)}(\rho_a)
  =\lambda_\delta(N,\rho_a)
  \simeq
  \frac{3}{2}\sqrt{\frac{3}{\pi}}N^{7/2}\frac{a_s}{\rho_a}
  \;.
  \label{eq:ld3}
\end{eqnarray}
Combination of \refeqs.~(\ref{eq:ld5}) and (\ref{eq:ld3}) yields
\begin{eqnarray}
  \rho_a 
  &\simeq&
  \sqrt[3]{\frac34}N^{7/6}|a_s|
  \label{eq:rhoa_copy}
  \;,
  \\
  \lambda_\infty(N)
  &\simeq&
  -\sqrt[3]{\frac{9}{2}}N^{7/3}
  \simeq
  -1.65N^{7/3}
  \;,
\end{eqnarray}
which is in agreement with the numerical results in
\refeqs.~(\ref{eq:lambda_infty}) and (\ref{eq:intermediate_region}).
However, the $N$ dependence cannot be predicted from the angular
equation since the result is an interplay between the various terms in
\smarteq{eq:simplified_var_eq}.

The symbol $\lambda_\infty$ is chosen for this constant since the
relevant $\rho$ region extends to infinity in the limit of infinitely
large scattering length.  With no bound two-body states ($a_s<0$) the
lowest angular eigenvalue approaches zero at larger hyperradii,
whereas it diverges towards $-\infty$ as $\rho^2$ when a bound
two-body state is present ($a_s>0$).  On the threshold for a two-body
bound state $a_s= \pm \infty$ and the angular eigenvalue therefore
remains constant. For finite, but large, $a_s$ the eigenvalue lingers
and cannot decide which way to go until the hyperradius exceeds a size
$\rho_a$ proportional to the scattering length given by
\smarteq{eq:rhoa_copy}.

\subsection{Discussion}


\twobound

The two-body correlations built into the many-body wave function are
evident in the properties of the angular wave function.  The particles
feel pairwise repelled or attracted to each other, which is reflected
in the average two-body distance in the two-body amplitude.  The
presence of a two-body bound state is described by an angular
adiabatic potential proportional to the two-body binding energy.  The
angular wave function in this limit equals the wave function for the
two-body bound state.


The angular adiabatic potential reflects the effective interaction
between the bosons.  We recovered numerically the scattering length as
the determining parameter for a dilute system with large average
separation.  Deviations at larger densities resulted in a
parameter-free effective interaction $\lambda_\infty$ which is
interpreted in simple physical terms as the transition between the
shape-dependent and the scattering-length-dependent regions.  The
properties of the lowest angular eigenvalues are collected in
table~\ref{table:lambdabehaviour}.

\begin{table}[htb]
  \centering
  \begin{tabular}{c|c||c|c|c}
    $\mathcal N\sub B$ & $a_s$ & $\lambda_0$ & $\lambda_1$ & $\lambda_2$ \\
    \hline
    \hline
    0                & $>0$ & $\lambda_\delta$&$\lambda_{K=4}$  & $\lambda_{K=6}$ \\
    \hline
    0                & $<0$ & $\lambda_\delta$&$\lambda_{K=4}$  & $\lambda_{K=6}$ \\
    \hline
    threshold & $\mp\infty$ & $\lambda_\infty$ & constant & constant \\
    \hline
    1                & $>0$ & $2m\rho^2E_0^{(2)}/\hbar^2$     & $\lambda_\delta$ & $\lambda_{K=4}$ \\
    \hline
    1                & $<0$ & $2m\rho^2E_0^{(2)}/\hbar^2$     & $\lambda_\delta$ & $\lambda_{K=4}$ \\
    \hline
    threshold & $\mp\infty$ & $2m\rho^2E_0^{(2)}/\hbar^2$   &$\lambda_\infty$ & constant  \\
    \hline
    2                & $>0$ & $2m\rho^2E_0^{(2)}/\hbar^2$ &       $2m\rho^2E_1^{(2)}/\hbar^2$ & $\lambda_\delta$ \\
    \hline
    2                & $<0$ & $2m\rho^2E_0^{(2)}/\hbar^2$ & $2m\rho^2E_1^{(2)}/\hbar^2$ & $\lambda_\delta$ \\
    \hline
    threshold & $\mp\infty$ & $2m\rho^2E_0^{(2)}/\hbar^2$ & $2m\rho^2E_1^{(2)}/\hbar^2$ & $\lambda_\infty$  \\
    \hline
  \end{tabular}
  \caption[]
  {The behaviour of the lowest angular eigenvalues at large hyperradii
    as a function of the number $\mathcal N\sub{B}$ of bound two-body
    states and for different regions of the scattering length.  The
    attraction increases through the sequence.  $E_n^{(2)}$ is the
    energy of the $n$'th two-body state.}
  \label{table:lambdabehaviour}
\end{table}


Calculations with a correlated Jastrow wave function with the right
behaviour at small interparticle distances and with realistic
interactions \cite{cow01} and with various finite-range potentials
\cite{blu01} confirm that the ground-state energy of a dilute boson
system only depends on the scattering length and not on the details of
the potential.  The advantage of the present model is that it results
in a relatively simple one-dimensional differential equation which
provides analytical results in some limits, i.e.~the
scattering-length-only behaviour is recovered analytically in
section~\ref{sec:zero-range-angular}.  Furthermore, from the two-body
ansatz for the wave function no further assumptions are necessary in
order to obtain the large-distance scattering-length-only signatures.

\chapter{Hyperradial confinement and condensates}             
\label{kap:radial}
In chapter~\ref{kap:angular} we studied the attributes of the
solutions to the hyperangular equation for a fixed value of the
hyperradius.  This freezed the variation in the average distance
between the particles, but nevertheless showed a range of
characteristics depending on the nature of the two-body interaction.
In this chapter we complete the treatment of the degrees of freedom in
the centre-of-mass system by studying the radial equation and the
properties of its solutions.

\radpot

Besides the contributions from kinetic energy and interactions, the
radial equation contains a term due to an external field acting on the
particles.  As shown in chapter~\ref{kap:hyperspherical_method} this
separates for a harmonic field nicely into a centre-of-mass part and a
hyperradial part.  The inclusion of such a term is discussed in
section \ref{sec:trapped-bosons}.  Then the hyperangular contributions
due to interactions are included in section \ref{sec:prop-radi-potent}
in a study of the properties of the full radial potential and the
solutions to the radial equation.  Section \ref{sec:self-bound}
presents more details about negative-energy states, which include the
Efimov-like states that are described further in section
\ref{sec:efimov-like-states}.  In section \ref{sec:trap-states-or} we
discuss condensation before summing up in section
\ref{sec:summary-radial}.

\section{Trapped bosons}

\label{sec:trapped-bosons}

\traplength

In experiments neutral atoms, for instance evaporated sodium atoms
\cite{dav95}, are cooled and trapped by lasers, and then held and
further cooled in magnetic fields which interact with the magnetic
moments of the particles.  In a common set-up, the time-averaged
orbiting potential, a static magnetic field is combined with a
rotating magnetic field \cite{pet01}.  This effectively generates a
harmonic-oscillator field in which all particles move, e.g.~for
particle $i$ we have
\begin{eqnarray}
  V\sub{trap}(\boldsymbol r_i)
  =
  \frac12m\big(\omega_x^2x_i^2+\omega_y^2y_i^2+\omega_z^2z_i^2\big)
  \;,
\end{eqnarray}
where the position of particle $i$ is $\boldsymbol r_i =
(x_i,y_i,z_i)$, and the angular frequencies along the coordinate
directions $q=x,y,z$ are denoted by $\omega_q$.  These angular
frequencies $\omega_q$ depend on the magnetic moments of the atoms and
on the strengths of both the stationary field and the time-varying
magnetic field.  Experimentally it is possible to obtain the same
effective frequency along two axes, $x$ and $y$, and a different
frequency along the third axis, $z$.  E.g.~for $^{85}$Rb atoms the
effective frequencies $\nu_q=\omega_q/(2\pi)$ in recent experiments
are $\nu_x=\nu_y=17.5$ Hz, and $\nu_z\sim\nu_x/2$ \cite{don01}.  In
terms of the trap lengths $b_q=\sqrt{\hbar/(m\omega_q)}$ this is
$b_x=b_y=2591$ nm and $b_z\sim \sqrt{2}b_x$.

\traplength

We will address the general geometry in chapter~\ref{kap:deformed} and
here restrict ourselves to a spherically symmetric field,
$\omega=\omega_x=\omega_y=\omega_z$, which leaves a central potential
\begin{eqnarray}
  V\sub{trap}(r_i)=\frac12m\omega^2r_i^2
  \;.
\end{eqnarray}
Put differently, we treat the axial field as spherical with
$\omega=\sqrt[3]{\omega_x\omega_y\omega_z}$ as the geometric mean
angular frequency.  A set of parameters which we will use frequently
is for $^{87}$Rb-atoms with oscillator frequency
$\nu\sub{trap}=\omega/(2\pi)=200$ Hz \cite{boh98}, thus yielding
$b\sub t\equiv\sqrt{\hbar/(m\omega)}=763$ nm.  All lengths are then
scaled in units of the typical interaction range $b\simeq10$ a.u.,
which leads to $b\sub t/b\simeq 1442$.

\couplingterms\radpot

In the case of the free angular solutions from section
\ref{sec:angul-kinet-energy}, we have $\lambda_K=K(K+3N-5)$ with
$K=0,2,4,\ldots$.  In the general case, i.e.~when we include
dependences beyond the $s$-waves in one hyperangle, we can replace $K$
with $K_{N-1}$ from \smarteq{eq:angularKnumber}.  The radial solutions
are analytically obtained from \refeq.~(\ref{eq:noncoupled_radial_eq})
with the radial potential from \refeq.~(\ref{eq:radial.potential})
with zero coupling terms.  The radial wave functions are then given by
\begin{eqnarray}
  f_n(\rho)
  =
  e^{-\rho^2/(2b\sub t^2)}
  \rho^{l_{N,K}+1}
  \mathcal L_n^{(l_{N,K}+1/2)}\bigg(\frac{\rho^2}{b\sub t^2}\bigg)
  \;,\qquad
  l_{N,K}=\frac{3N-6}2+K
  \;,
\end{eqnarray}
where $\mathcal L_n$ for $n=0,1,\ldots$ is the generalized Laguerre
polynomial with $n$ radial nodes.  Here $l_{N,K}$ plays the role of a
generalized angular momentum due to the kinetic energy of the
many-body system.  Especially, $l_{2,0}=0$ reproduces the familiar
behaviour of the harmonic-oscillator solutions for the two-particle
system \cite{bra83}.  The energy is $E_n = \hbar\omega[3(N-1)/2+2n+K]$
with the subtraction of the centre-of-mass ground-state energy
$3\hbar\omega/2$.

The ground state with $n=0$ is usually associated with the mode of
Bose-Einstein condensation, see mean-field approaches
\cite{pet01,pit03} or related hyperspherical approaches
\cite{boh98,wat99}.  In terms of temperature the condition for onset
of Bose-Einstein condensation is that the thermal length scale $l\sub
T$ given by $k\sub BT\sim\hbar^2/(ml\sub T^2)$ is larger than the
average distance $\bar r$ between particles \cite{pet01}.  For bosons
in a trap $\bar r \sim \sqrt{k\sub BT/(m\omega^2)}/N^{1/3}$, which
yields that condensation occurs when
\begin{eqnarray}
  k\sub BT <
  N^{1/3}
  \hbar\omega
  \;.
\end{eqnarray}
When this equation is fulfilled, a large number of atoms prefer the
ground state which is a signature of the condensate.  The criterion is
equivalent to a sufficiently large level spacing of the modes of the
harmonic oscillator.  In the present thesis we assume no
finite-temperature effects, corresponding to a sufficiently large
level spacing.

In this non-interacting picture all quantum modes of the many-boson
system are represented by single-particle levels.  This means that the
Bose-Einstein distribution can be applied to the non-interacting
single-particle levels.  In general the interactions complicate
matters.  In the following sections we discuss the structure of the
radial solutions when interactions are included, and then in
section~\ref{sec:trap-states-or} return to this problem.

\section{Radial potential and solutions}

\label{sec:prop-radi-potent}

The effect of interactions for a fixed value of the hyperradius was
discussed in chapter~\ref{kap:angular} in terms of the angular
potentials.  These in themselves tell only a part of the story about
the many-boson system.  As indicated above for the non-interacting
case, the radial equation is the next step in obtaining knowledge
about the physical properties.  The angular potentials and angular
wave functions then enter the effective radial potential and transfer
information about the interactions to quantities like energy and size
of the system.

\couplingterms\radpot

For a dilute Bose gas the coupling terms of
\refeq.~(\ref{eq:coupling.terms}), which in the non-interacting case
are identically zero, contribute at most about 1 \% compared with
other terms of the full radial \refeq.~(\ref{eq:radial.equation}).  In
the following all coupling terms are therefore omitted and the
solutions to the uncoupled radial
\refeq.~(\ref{eq:noncoupled_radial_eq}) are considered.  This way only
the angular potential $\lambda$ itself plays a role and additional
information from the angular wave function $\Phi(\rho,\Omega)$ is
neglected.  We should however bear in mind that coupling terms might
play a role at larger densitites or scattering lengths.  The radial
potential then consists of three terms, where the repulsive
centrifugal barrier and the confining external field both are
positive.  The interaction term can be either repulsive or attractive.
The combination has structure depending on the strength of the
interaction.


The lowest potential for the non-interacting system, which was
discussed in section~\ref{sec:trapped-bosons}, is shown as the thick,
dashed line in \reffig.~\ref{fig:u_as_m0.5.N20}a for $N=20$ particles.
This has a global minimum at values of the hyperradius given by the
trap length, that is at $\rho\sim\rho\sub{trap} \equiv
\sqrt{3N/2}b\sub t$.  The schematic character of this non-interacting
potential is representative also for a very weak two-body attraction
and for a purely repulsive two-body potential.  Corresponding
solutions are confined to the region between the infinitely large
potential walls at small and large hyperradii.

\begin{figure}[htb]
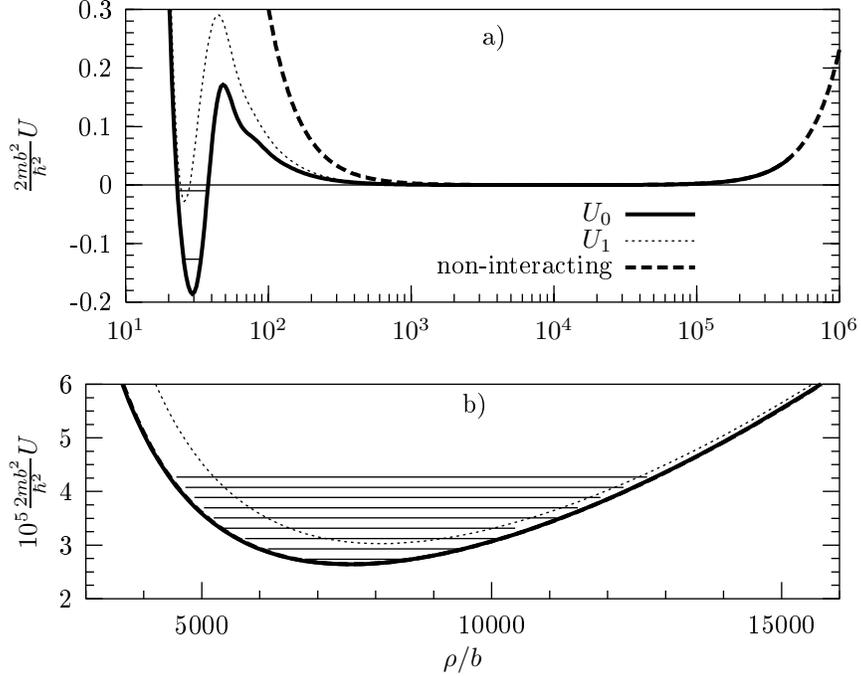

  \centering
  \input{u_as_m0.5.N20}
  \input{u_as_m0.5.N20.traponly}
  \caption
  [Radial potential for twenty particles and small scattering length]
  {a) Radial potentials $U_0$ (thick, solid line) and $U_1$ (dotted
    line) from \refeq.~(\ref{eq:radial.potential}) corresponding to
    the lowest two angular potentials for $N=20$, $a_s/b = -0.5$, and
    $b\sub t/b=1442$.  The thick, dashed line shows the lowest
    non-interacting potential, that is with $\lambda=0$.  Horizontal
    lines show the lowest two energy levels in $U_0$.  b) Details at
    larger hyperradii with the next nine energy levels in $U_0$.  Here
    $U_0$ and the curve for non-interacting particles are hardly
    distinguishable.}
  \label{fig:u_as_m0.5.N20}
\end{figure}


Also shown in \reffig.~\ref{fig:u_as_m0.5.N20}a are the lowest two
radial potentials for a Gaussian interaction with no bound two-body
states and a small, negative scattering length.  The deviations from
the weakly-interacting case are substantial.  For the lowest (thick,
solid line) a second minimum has developed due to the attraction
between the bosons.  This dominates at large densities, i.e.~at small
hyperradii.  A barrier separates this global minimum from another
minimum at large hyperradii, see details in
\reffig.~\ref{fig:u_as_m0.5.N20}b.  This second minimum almost
coincides with the minimum for the non-interacting case and these are
hardly distinguishable in the figure.


With this potential the diagonal radial equation is solved. The
solutions can be divided into groups related to either the first or
the second minimum.  The lowest two radial eigenstates in the lowest
potential have negative energies, indicated as horizontal lines in
\reffig.~\ref{fig:u_as_m0.5.N20}a, and the hyperradial wave function
is located in the global minimum at relatively low hyperradii.  They
are truly bound states as they cannot decay into continuum states at
large hyperradii.  Their properties are independent of the external
trap which only has an influence at much larger distances.  These
self-bound $N$-body states might decay into lower-lying states
consisting of various bound cluster states, e.g.~a number of diatomic
or triatomic clusters.  We discuss this further in chapter
\ref{kap:stability_validity}.  The possibility of self-bound many-body
systems even though the two- and three-body subsystems are unbound is
also discussed by Bulgac \cite{bul02}, who, however, considers a
three-body interaction strength as a determining parameter for the
properties of the self-bound many-boson system.

The group of states in the higher-lying minimum all have positive
energies.  These radial eigenstates are located in the trap minimum at
larger hyperradii, see \reffig.~\ref{fig:u_as_m0.5.N20}b, with
approximately equidistant spacing as for the non-interacting
oscillator.  The lowest of these ``trap states'' can be interpreted as
the state of the condensate.  Thus, the structure of the ``trap
states'' is similar for effectively attractive and repulsive
interactions, i.e.~for positive and negative scattering lengths.
However, an attraction produces a series of lower-lying states at
smaller hyperradii.


The radial potential $U_1$ corresponding to the second adiabatic
potential $\lambda_1$ is shown as the dotted line in
\reffig.~\ref{fig:u_as_m0.5.N20}.  This contains larger contributions
from hyperangular kinetic energy, but still has a second minimum at
small hyperradii.  Otherwise the structure is the same with a barrier
and a local minimum at larger hyperradii.


Increasing $N$ leaves quantitatively the same features for pure
repulsion.  Attraction leads to a decreasing barrier at intermediate
hyperradius and at some point this barrier vanishes altogether.  At
the same time the attractive minimum at smaller hyperradius becomes
deeper.  This leads to an increasing number of bound states in this
minimum as a function of $N$.  \Reffig.~\ref{fig:u_asneg} shows $U_0$
for a larger number of particles, $N=100$, and doubled scattering
length.  The increased effective attraction is pronounced at large
densities, that is at small $\rho$.  The barrier height is now small
compared to the potential depth at small hyperradii.  The rather deep
and narrow minimum occurs for $N=100$ about 150 times the range of the
interaction.  This corresponds to a root-mean-square two-body distance
of about 15 times the interaction range $b$.

\begin{figure}[htb]
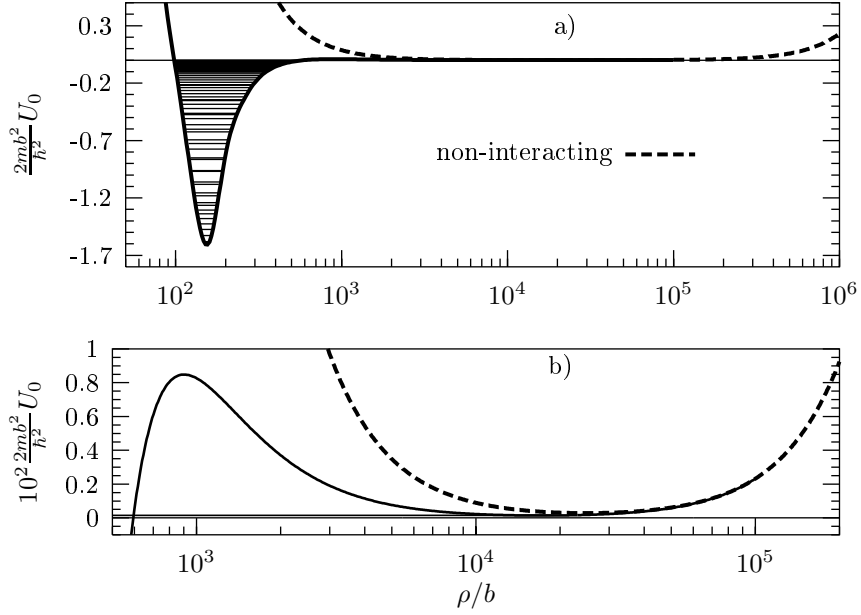

  \centering
  \input{u_asneg.N100}
  \input{u_asneg.largerho.N100}
  \caption[Radial potential for one hundred bosons]
  {a) Radial potential $U_0$ from \refeq.~(\ref{eq:radial.potential})
    corresponding to the lowest angular potential for $N=100$,
    $a_s/b=-1$, and $b\sub t/b=1442$.  Also shown as horizontal lines
    are the negative energies $E_{0,n}$, $n=0,\ldots,46$, in the
    lowest potential in the uncoupled radial
    \refeq.~(\ref{eq:radial.equation}).  b) Detail at larger
    hyperradii.  The energy of the first oscillator-like state is
    shown as a horizontal line close to zero.}
  \label{fig:u_asneg}
\end{figure}


As the scattering length increases, the barrier disappears and the
effective potential inside the trap approaches the $\rho^{-2}$
behaviour characteristic for Efimov states.  We return to a discussion
of such states in section~\ref{sec:efimov-like-states}.

We also study the hyperradial wave function $F$ which tells about the
root-mean-square displacement $\bar r_R$ from the centre of the system
defined by
\begin{eqnarray}
  \bar r_R^2
  \equiv
  \frac{1}{N} \sum_{i=1}^N(\boldsymbol r_i-\boldsymbol R)^2
  =
  \frac{\rho^2}{N}
  \;.
\end{eqnarray}
It is shown in \smartfig{fig:densprofile} for various scattering
lengths.
\begin{figure}[htb]
  \centering
  \input{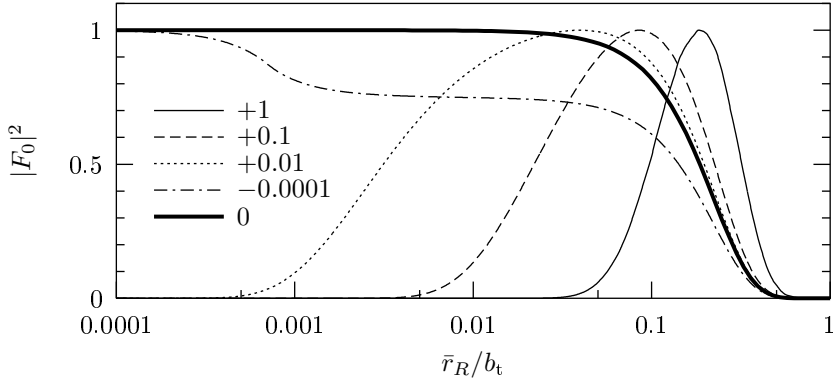}
  \caption[Density profile]
  {The probability distribution $|F_0|^2$ for the rms separation from
    the centre of the system $\bar r_R$ defined by $\bar r_R^2\equiv
    \sum_{i=1}^N(\boldsymbol r_i-\boldsymbol R)^2/N = \rho^2/N$ for
    $N=20$, $b\sub t/b=1442$, and scattering lengths indicated as
    $a_s/b$.  The angular potential was obtained from the
    parametrization in \refeq.~(\ref{eq:schematic}) for $a_s<0$ and
    from $\lambda_\delta$ for $a_s>0$.  The normalizations are
    different.}
  \label{fig:densprofile}
\end{figure}
The thick, solid curve shows the non-interacting result.  When
interactions are included, the expected result turns up, i.e.~that
repulsion forces the particles away from the centre whereas the
opposite holds for attraction.

\section{Self-bound many-body states}


\label{sec:self-bound}

Since the external field is negligible when $\rho\ll\sqrt{N} b\sub t$,
the radial potential is negative when $\lambda+(3N-4)(3N-6)/4<0$ and
$\rho$ is sufficiently small.  Then self-bound many-body states with
negative energies and finite extensions are possible.  The radial
equation corresponding to the relatively weak attraction between the
twenty bosons in the potential shown in
\reffig.~\ref{fig:u_as_m0.5.N20} has two negative-energy solutions
with the wave function located in the global negative minimum.


With the parametrization in \smarteq{eq:schematic}
\reffig.~\ref{fig:u_schematic} shows the analytical radial potential,
\smarteq{eq:radial.potential}, corresponding to one of the angular
eigenvalues from \reffig.~\ref{fig:schematic}.  The radial potential
is negative in a large range of hyperradii, which can be divided into
three different regions.  For small hyperradii, region 1, the radial
potential has a minimum.  For intermediate hyperradii, Efimov region,
the angular potential is constant and therefore the radial potential
behaves as $-1/\rho^2$.  This is from \reffig.~\ref{fig:schematic}
seen to appear for $\rho/b$ between $10^2$ and $10^4$.  For large
hyperradii, region 2, that is when $\rho\ge N^{7/6}|a_s|$, the angular
potential behaves as $-1/\rho$, so the radial potential vanishes as
$-1/\rho^3$.  Finally the trap $\propto\rho^2$ dominates with positive
contributions at large hyperradii $\rho\gg\sqrt Nb\sub t$.

\begin{figure}[thb]
  \centering
  \input{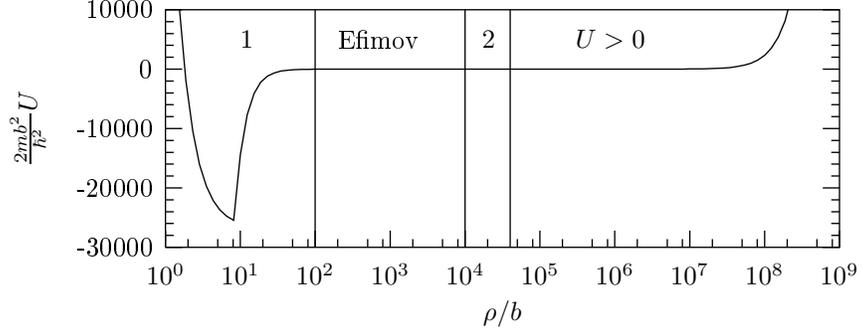}
  \caption  [Radial potentials from the parametrized angular eigenvalue]
  {Analytic radial potential obtained from
    \refeqs.~(\ref{eq:radial.potential}) and (\ref{eq:schematic}) for
    $N=100$, $a_s/b=-10^4$, and $b\sub t/b=1442$.}
  \label{fig:u_schematic}
\end{figure}

With the method described in \cite{khu02} it is possible to estimate
the number $\mathcal N$ of bound states in the different regions, i.e.
\begin{eqnarray} 
  \mathcal N
  \simeq
  \frac{\sqrt{2m}}{\pi\hbar}
  \int {\rm d} \rho  \sqrt{|U^{(-)}(\rho)|} 
  \label{eq:numberofstates}  
  \;,
\end{eqnarray}
where $U^{(-)}(\rho)$ denotes the negative part of the radial
potential $U(\rho)$.  The bound states in this potential can be
divided into groups according to their hyperradial extension. The
total number of such states is written as $\mathcal N = \mathcal N_1 +
\mathcal N_E+\mathcal N_2$ where $\mathcal N_1$, $\mathcal N\sub{E}$,
and $\mathcal N_2$ are the number of states located respectively in
the attractive pocket at small hyperradii, in the intermediate
$-1/\rho^2$ region, and at hyperradii large compared with the
scattering length.

The analytic expressions for the angular potential from
\smarteqs{eq:lambda0_smallrho} and (\ref{eq:schematic}) yield the
crude estimate that the number of self-bound states in the pocket is
$\mathcal N_1\simeq 1.3 N^{3/2}$.  The outer region supports bound
states when the trap length $b\sub{t}$ is sufficiently large, that is
$b\sub t \gg N|a_s|$, and analogously the number is estimated to be
$\mathcal N_2 \simeq 0.78 N^{7/6}$.  The intermediate region is
considered in the following section.

\section{Efimov-like many-body states}

\label{sec:efimov-like-states}

\efiN


When the scattering length is large, the three-body system exhibits
the so-called Efimov effect \cite{efi70} where many bound three-body
state turns up.  In the following we envistigate the properties of the
many-body system in this Efimov regime.  Much of the formulation is
quite similar to that in a recent description of three-body Efimov
states \cite{nie01}.

A large scattering length implies through the eigenvalue from
\refeq.~(\ref{eq:schematic}) an intermediate region in hyperradius
where the angular potential is almost constant, see section
\ref{sec:at-threshold}.  More specifically, when 
\begin{eqnarray}
  b < \frac{\rho}{N^{7/6}} < |a_s|
  \label{eq:intermediate_region_copy1}
  \;,
\end{eqnarray}
then \refeq.~(\ref{eq:lambda_infty}) yields $\lambda \simeq
\lambda_\infty=-1.59N^{7/3}$ and two of the terms in the radial
equation add to a negative value.  At the threshold for binding of the
two-body system, i.e.~$|a_s| = \infty$, the radial potential in
\refeq.~(\ref{eq:radial.potential}) then has the form
\begin{eqnarray}
  U(\rho) 
  &\simeq&
  \frac{\hbar^2}{2m}\bigg(\frac{-\xi^2-1/4}{\rho^2}
  +\frac{\rho^2}{b\sub t^4}\bigg)
  \label{eq:Efimov-potential}
  \;,\\
  \xi^2
  &\equiv&
  -\lambda_\infty-\frac{(3N-4)(3N-6)}{4}-\frac{1}{4}
  \;
  \stackrel{N\ggg1}{\longrightarrow}
  \;
  1.59N^{7/3}
  \label{eq:xi2}
  \;.
\end{eqnarray}
This implies that no repulsive barrier is present.  Then the effective
potential behaves as $-\rho^{-2}$ until the trap dominates.


\Reffig.~\ref{figs:asinfty} shows the radial potential for $N=20$ and
infinite scattering length corresponding to $\lambda_\infty\sim-1340$
or $\xi^2\sim 584$.  Deviations from the form in
\smarteq{eq:Efimov-potential} are only present at small hyperradii due
to the finite range of the interaction.

\begin{figure}[htb]
  \centering
  \input{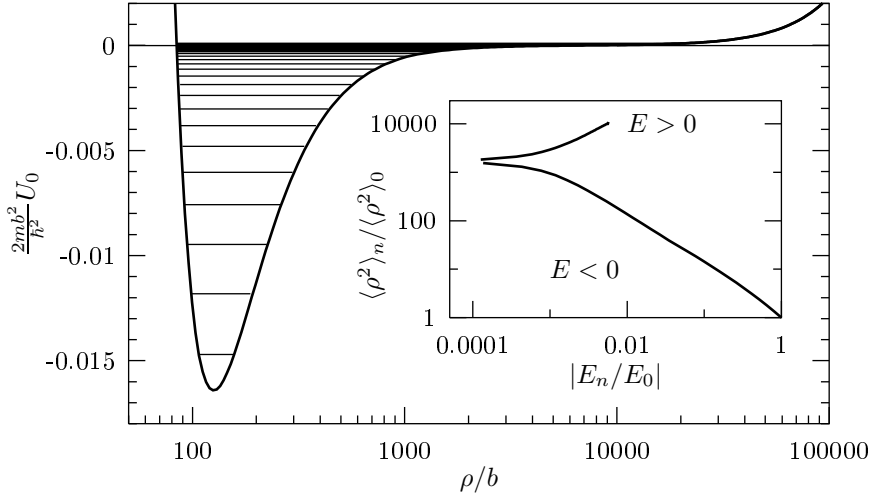}
  \caption
  [Radial potential and energy-length relations for infinite
  scattering length] {The lowest radial potential for $N=20$, $|a_s| =
    \infty$, and $b\sub t/b=1442$.  The horizontal lines indicate the
    $69$ lowest energy eigenvalues, with 30 below zero and 39
    \emph{very} close-lying above zero.  The inset relates their
    mean-square hyperradii with the absolute values of their energies.
    The lowest state has $2mb^2E_0/\hbar^2 \simeq -0.0147$ and
    $\sqrt{\langle\rho^2\rangle_0}/b \simeq 136$.}
  \label{figs:asinfty} 
\end{figure}


Without the external $\rho^2$ potential the $1/\rho^2$ potential in
\refeq.~(\ref{eq:Efimov-potential}) would produce infinitely many
radial solutions to the non-coupled radial
\refeq.~(\ref{eq:noncoupled_radial_eq}).  The radial wave function for
these states would behave like
\begin{eqnarray}
  f_\infty(\rho) =
  \sqrt\rho\sin\bigg[|\xi|\ln\bigg(\frac{\rho}{\rho\sub{sc}}\bigg)\bigg]
  \label{eq:frho_infty}
  \;,
\end{eqnarray}
with some hyperradius scale $\rho\sub{sc}$.  The energies and
mean-square hyperradii for such states are related by
\begin{eqnarray}
  E_n
  =
  -\frac{\hbar^2}{2m\langle\rho^2\rangle_n}\frac23(1+\xi^2)
  \;,\qquad
  E_n=E_0e^{-2\pi n/|\xi|}
  \label{eq:efimov_energy}
  \;,
\end{eqnarray}
where the exponential dependence on the strength $\xi$ of the
effective potential and the number $n$ of the excited state is
highlighted.  This relation can be written as
\begin{eqnarray}
  \frac{E_n}{E_{n+1}}
  =
  \frac{\langle\rho^2\rangle_{n+1}}{\langle\rho^2\rangle_n}
  =
  e^{2\pi/|\xi|}
  \label{eq:Efimov_E-rho-relations}
  \;.
\end{eqnarray}
With increasing quantum number these states become exponentially
larger with exponentially smaller energies approaching zero.

Around thirty states with this character are obtained for the
potential in \reffig.~\ref{figs:asinfty}.  The lower curve in the
inset of \reffig.~\ref{figs:asinfty} illustrates the relation in
\refeq.~(\ref{eq:Efimov_E-rho-relations}), and accordingly many states
are in this log-log plot represented by a point on the straight line
with slope $-1$.  The very lowest states deviate due to the attraction
at small $\rho$, and the states close to $E=0$ deviate due to the
external potential.  The denser positive energy spectrum in the upper
part of the inset approaches a straight line with slope $+1$ as
expected for a harmonic potential.  Using
\refeq.~(\ref{eq:hyperradius}) we get $2 \langle\rho^2\rangle = (N-1)
\langle r_{12}^2\rangle \simeq 2 (N-1) \langle r_1^2\rangle$.  Even
the most bound state with $\langle\rho^2\rangle^{1/2} \simeq 136 b$
then has a root-mean-square (rms) distance between two particles
$\langle r_{12}^2\rangle^{1/2} \simeq 44 b$, which is much larger than
the interaction range.  Also the rms distance from the centre of the
trap $\langle r_1^2\rangle^{1/2} \simeq 31 b$ is large.

The intermediate region responsible for the constant $\lambda$ is only
present when the scattering length is relatively large, i.e.~when
relation~(\ref{eq:intermediate_region_copy1}) is obeyed.  This
corresponds to hyperradiii larger than $\rho\sub{min} = N^{7/6}b$ and
smaller than $\rho\sub{max} = N^{7/6}|a_s|$.  The number of
Efimov-like states $\mathcal N\sub{E}$ located in this region is then
by \refeq.~(\ref{eq:numberofstates}) given as
\begin{eqnarray}
  \mathcal N\sub{E}
  \simeq
  \frac{|\xi|}{\pi}
  \ln\bigg(\frac{\rho\sub{max}}{\rho\sub{min}}\bigg)
  \simeq
  0.40N^{7/6}
  \ln\bigg(\frac{|a_s|}{b}\bigg)
  \label{eq:numberofEfimovstates}
  \;,
\end{eqnarray}
where \refeq.~(\ref{eq:xi2}) yielded the last estimate.  The number of
Efimov-like states $\mathcal N\sub E$ increases strongly with $N$.
This assumes that the external trap has no influence on the
hyperradial potential for $\rho <\rho\sub{max}$.  However, when the
trap length $b\sub{t}$ is sufficiently small, that is when
$\rho\sub{trap} = \sqrt{3N/2}b\sub{t} < N^{7/6}|a_s|$, the extension of
the plateau is truncated at large hyperradii.  The number of states is
then estimated by substituting $\rho\sub{max}$ with $\rho\sub{trap}$
in \refeq.~(\ref{eq:numberofEfimovstates}).  This yields
\begin{eqnarray}
  \mathcal N\sub{E}
  \simeq
  0.40N^{7/6}
  \ln\bigg(\frac{\sqrt{3/2}b\sub t}{N^{2/3}b}\bigg)
  \;.
\end{eqnarray}


When the trap length is large and does not terminate the plateau at
large distances, the mean-square hyperradii of the first and last
Efimov-like states are of the order $\rho\sub{min}^2 \sim N^{7/3}b^2$
and $\rho\sub{max}^2 \sim N^{7/3}|a_s|^2$, respectively.
\Refeq.~(\ref{eq:efimov_energy}) then yields the energies of the first
and last Efimov-like states
\begin{eqnarray}
  E\sub{first} \sim -\frac{\hbar^2}{2mb^2}
  \;,\qquad
  E\sub{last} \sim -\frac{\hbar^2}{2m|a_s|^2}
  \label{eq:e_first_last}
  \;.
\end{eqnarray}
These energies are independent of the particle number $N$ and remind
of the kinetic-energy scale of strongly bound two-body states and the
two-body binding energy, respectively.  However, the rms distances
$\bar r$ between two particles in these many-body states are
\emph{not} given by $b$ and $a_s$.  In fact, $\bar r$ contains an
additional $N$-dependent factor, i.e.~$\bar r \simeq N^{2/3}b,
N^{2/3}|a_s| $ for the two cases.  These constant energy limits imply
that the density of Efimov-like states increases with the particle
number.

These many-body states arise when the two-body scattering length is
large.  This is the condition for the occurrence of the three-body
Efimov states \cite{efi70,fed93}, that show characteristic properties
similar to \refeqs.~(\ref{eq:frho_infty}), (\ref{eq:efimov_energy}),
and (\ref{eq:Efimov_E-rho-relations}).  The author and co-workers
therefore proposed to call the many-body states with similar
attributes for many-body Efimov states \cite{sor02}.  A more correct
name is probably Efimov-\emph{like} many-body states since some
definitions of the term ``Efimov state'' read that infinitely many
$N$-body bound states occur when the $N-1$-body system is on the
threshold for binding.  According to \etalcite{Amado}{ama73} ``there
is no Efimov effect for four or more particles'' in the sense that
being on the threshold for binding in the $N-1$-body system does not
produce infinitely many $N$-body bound states when $N>4$.  This
statement is not in contradiction with the Efimov-like states
discussed here since the present Efimov-like $N$-body states occur
when the two-body system, and \emph{not} the $N-1$-body system, is on
the threshold for binding.  However, the quoted remark reminds us that
the three-, four-, \ldots, $N-1$-body systems are also bound and that
many of the $N$-body states might be resonances embedded in the
continua of dimer, trimer, and higher-order cluster states.  They
could be artifacts of the model where only special degrees of freedom
are treated, and where we recall the possible changes due to larger
coupling terms for large scattering lengths. \couplingterms However,
because the particles are far from each other and couplings to the
continuum states therefore could be weak, some of these states might
be distinguishable structures which could be relatively stable.  We
return to such considerations in section~\ref{sec:observ-efim-like}.

\efiN

\section{Trap states and ``the condensate''}

\label{sec:trap-states-or}

In the non-interacting case hyperradial many-body states are located
in the potential minimum created by a competition between the kinetic
energy and the external trap.  Similar behaviours were seen in the
cases of repulsion and attraction, see for instance in
\smartfig{fig:u_as_m0.5.N20} the excited states above the lowest two
located in the minimum at large hyperradii.  The corresponding density
profile of the lowest trap state is similar to that obtained in
experiments creating Bose-Einstein condensates \cite{dal99}.  We can
therefore call this trap state for ``the condensate'' or define a
condensate by the typical signatures of the lowest state located in
this minimum due to the external trap.

The attraction produces lower-lying many-body bound states with an
average distance between the particles much smaller than in the
condensate-like state.  The structure of these states could as well be
characterized as a condensate (condensed $N$-body state), but they are
much more unstable due to the much larger density and the larger
recombination probability.  These lower states have no parallels in
mean-field computations.

Through the derived adiabatic potential the two-body unbound mode is
responsible for the properties of atomic Bose-Einstein condensation
where no clusterization is allowed.  We focus on the state of the
condensate in the second minimum and in the present work only use the
lower-lying negative-energy states in connection with the possible
decay of the condensate.  However, first we discuss a definition of a
condensate state in the present context.

\subsection{A definition of ``condensate''}

\label{subsec:definition}

\bec

In mean-field treatments, with repulsive two-body potentials and
confining trap potentials, the condensate is uniquely defined as a
statistical mixture of single-particle states with the ground state
dominating \cite{pet01,pou02}.  This many-body state is mainly
determined by the properties of the trap.  It is at best only
approximately stationary due to the neglected degrees of freedom which
allow energetically favored di- and tri-atomic cluster states.  This
instability is also an experimental fact seen by permanent loss of
trapped atoms, e.g.~in recombination processes \cite{don01}.

Without any two-body interaction the properties of the many-body
system is determined by the thick, dashed potential curve in
\reffig.~\ref{fig:u_asneg}.  Then the condensate is a physical state
dominated by the ground-state component.  With attractive interactions
(full curve) the deep minimum at small hyperradius is produced.  Then
the ground state, located in this minimum, has nothing to do with a
condensate.  The density is so high that couplings to other degrees of
freedom would develop higher-order correlations and processes like
three-body recombinations would quickly destroy the single-atom nature
of the gas.  This ground state, before or after recombinations, does
not show the signature of a Bose-Einstein condensate where many
particles occupy one single-particle level.

The formulation in the present work does not use the concept of
single-particle levels.  Therefore we cannot talk about a statistical
distribution of particles with the majority in the lowest state.
However, we can talk about a many-particle system described as a
superposition of many-body eigenstates where the lowest states are
favored in thermal equilibrium.  To clarify, a quantum state is given
as the superposition of eigenstates $\Psi_n(\rho,\Omega)$ from
\refeq.~(\ref{eq:hyperspherical_wave_function}):
\begin{eqnarray}
  \Psi\sub{quantum state}(\rho,\Omega)
  =
  \sum_{n=0}^\infty c_n \Psi_n(\rho,\Omega)
  =
  \sum_{n=0}^\infty
  c_n
  \sum_{\nu=0}^\infty
  F_{\nu,n}(\rho)\Phi_\nu(\rho,\Omega)
  \;,\quad
\end{eqnarray}
with the normalization $\sum_{n=0}^\infty|c_n|^2=1$.  The spatial
extension of a condensate must be sufficiently large in order to
exceed a certain minimum interparticle distance $d\sub c$ below which
the atoms are too close and recombine very fast.  This distance
depends on the scattering length and on the number of particles.
Therefore, a state cannot be characterized as a condensate if
components with $\langle r_{12}^2\rangle \ll d\sub c^2$ are dominating
contributions in the wave function.

One of the stationary states in this model can be defined as the
``ideal condensate state'', i.e.~the state of lowest energy with one
component, labeled by the quantum numbers $\nu\sub c$ and $n\sub c$,
which has
\begin{eqnarray}
  \langle r_{12}^2\rangle_{\nu\sub c,n\sub c}
  \gtrsim
  d\sub c^2
  \;.
\end{eqnarray}
When no bound two-body states are included in the model, this ideal
state is determined by the adiabatic component in the lowest angular
potential, that is $\nu\sub c = 0$.  On the other hand, the states of
lowest energy with $\nu = 0$ might have an average particle distance
less than $d_c$.  The appropriate choice among these excited states
depends on the number of particles and on the scattering length.  The
ideal state is then characterized by one dominating component, that is
$\nu\sub{c} = 0$, $|c_{n\sub c}|\simeq1$, and $|c_{n\ne n\sub c}|\ll
1$.  If it is impossible to distinguish states with these features, it
probably makes little sense to define a condensate.  The possible
states would be too unstable.

If $d\sub c$ is significantly smaller than $b\sub t$, then the state
of lowest energy located in the second minimum can be identified as
the condensate.  This state is characterized by a radial wave function
$F(\rho)$ with the root-mean-square (rms) radius
$\langle\rho^2\rangle^{1/2}$ approximately equal to the hyperradius at
the second minimum of the adiabatic potential $U_0(\rho)$.

\Reffig.~\ref{fig:rho2_n.N100}a shows the rms interparticle distance
$\bar r_n$ given by $\bar r_n^2 \equiv \langle r_{12}^2\rangle_{n} =
2\langle\rho^2\rangle_{n}/(N-1)$ for the lowest excited states,
labeled by $n$, in the potential of \reffig.~\ref{fig:u_asneg}.  All
states with $n\le46$ have $20b \leq \bar r_n \leq 100b$, which implies
that the particles are well outside the range of the interaction with
each other.  Whether the average distance qualifies a state as a
condensate depends on the decay rate of this state.  From
\smartfig{fig:rho2_n.N100}b it is seen that the abrupt change in rms
distance does not influence the energy which changes smoothly for the
states in question.

\begin{figure}[htb]
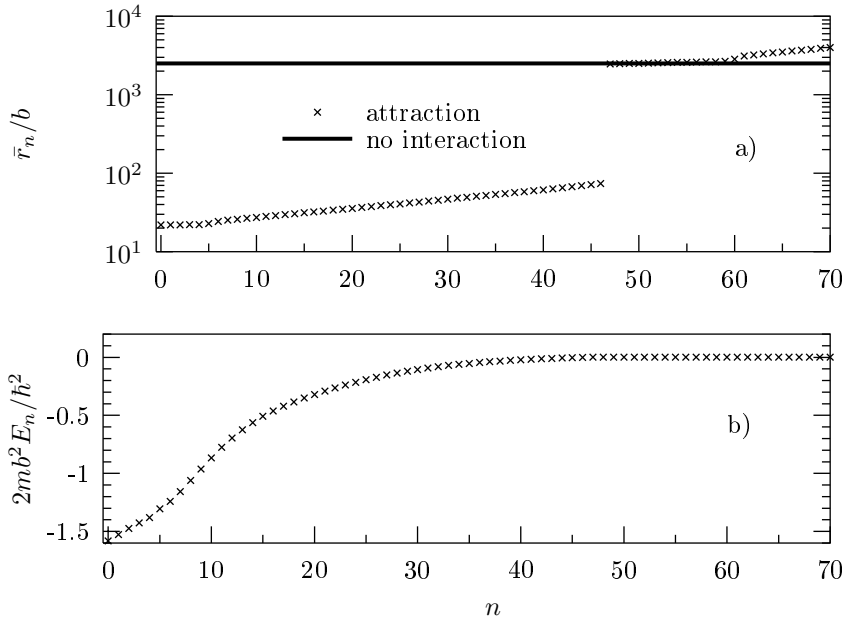

  \centering
  \input{rho2_n.N100}
  \input{E_n.N100}
  \caption [Interparticle distance and energy as a function of quantum
  number] {a) The root-mean-square distance $\bar r$ for $\nu=0$ as a
  function of the hyperradial quantum number $n$ for $N=100$, $a_s/b =
  -1$, and $b\sub t/b=1442$. b) The energy $E$ for the same case.}
  \label{fig:rho2_n.N100}
\end{figure}

For the positive-energy states ($n\ge 47$) the average particle
distance now exceeds $2000b$, that is $\bar r^2 \simeq 3b\sub t^2$
which approximately is obtained in the limit of a non-interacting gas.
This investigation repeated for the higher adiabatic potentials
$\nu\geq1$ results in the same pattern (not shown), although there are
fewer states with small interparticle distance.

In section~\ref{sec:decay} we return to a discussion of the
appropriate value for $d\sub{c}$, which then would characterize these
states as ideal for a condensate or not.

\bec

\subsection{Interaction energy}

The total energy of a state in the first minimum only depends on the
interaction since this state is bound even in the absence of the
external field.  Such a state has no analogue in mean-field
calculations.  Total energies of states in the second minimum are
dominated by the contribution from the confining field and therefore
are rather insensitive to anything else than this field.  It is then
more informative to study interaction energies where the large
background external-field contribution is removed.

\gpe

\Reffig.~\ref{fig:energy_N} shows the interaction energy per particle
as a function of the particle number for a relatively weak attraction
corresponding to the small scattering length $a_s/b=-0.84$.  The two
crosses for $N=20$ (on the left) show the results from the two-body
correlated model for the lowest adiabatic channel $\nu=0$ with quantum
numbers $n=7$ and $n=8$.
The interaction energy is negative for the lower state, whereas the
shown value for the upper state is positive due to the extra internal
kinetic energy.
This is repeated for the larger $N$ values $N=100$ and $N=900$,
i.e.~the lowest state characteristic for a condensate is shown along
with some of the higher-lying states.  For $N\gtrsim950$ there are no
trap states since the barrier has vanished.  However, the correlated
solutions are still stable due to the use of the finite-range
potential.

\begin{figure}[htb]
  \centering
  \input{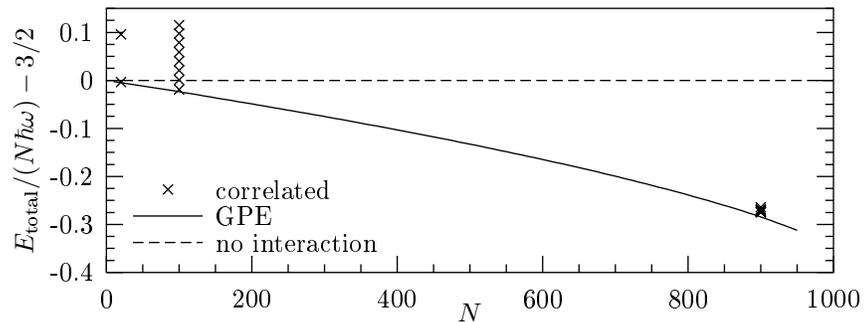}
  \caption[Interaction energies]
  {Interaction energy as a function of $N$ for $a_s/b=-0.84$ and
    $b\sub t/b=1442$.  The crosses are results of the present
    hyperspherical calculation for three numbers of particles.  The
    quantum numbers are $\nu=0$ and, for the lowest cross in each of
    the three cases, $n\sub c=7$ for $N=20$ ($N|a_s|/b\sub t=0.012$),
    $n\sub c=52$ for $N=100$ ($N|a_s|/b\sub t=0.058$), and $n\sub
    c=88$ for $N=900$ ($N|a_s|/b\sub t=0.52$).  The solid curve shows
    results of the GPE.}
  \label{fig:energy_N}
\end{figure}

We anticipate the mean-field discussion of chapter
\ref{kap:meanfield_validity}, i.e.~an ansatz for the many-body wave
function as a product of single-particle amplitudes and a zero-range
interaction potential lead to the mean-field Gross-Pitaevskii equation
(GPE).  Shown as the solid line in \smartfig{fig:energy_N} is the
interaction energy obtained from the GPE.
For small $N$ values the GPE solution is stable and the related
interaction energy is negative due to the attraction between the
particles.  A nearly linear behaviour is observed at small particle
numbers since each particle interacts with $N-1$ other particles.
We observe the similarity between the mean-field energy and the energy
for the lowest trap state in the hyperspherical model.
As $N$ increases, the mean-field attraction increases and a
non-physical collapse is inevitable.  This instability occurs for
$N|a_s|/b\sub t\simeq0.55$ which corresponds to $N\simeq950$ with the
present set of parameters.  There is no stable solution to the GPE for
$N|a_s|/b\sub t>0.55$.


Thus, using the correlated model we generally observe both
condensate-like and collapsed many-boson states.  In the present
example for relatively few particles $N$, it was easy to distinguish
due to the presence of the intermediate potential barrier.  In
chapter~\ref{kap:stability_validity} we discuss the case of small or
vanishing barrier and in chapter \ref{kap:meanfield_validity} compare
further to the mean field.

\section{Summary}

\label{sec:summary-radial}


The radial potential exhibits features from kinetic energy,
interactions, and external field, and thus combines the information
available within the hyperspherical model.  The structure of the
system depends mainly on the scattering length and the trap length.
Confined many-body states of negative energy may occur even without an
external confining potential.  This is possible when the effective
attraction between the bosons is sufficiently large.


Self-bound states with properties similar to the three-body Efimov
states occur when the scattering length is very large compared to the
range of the interaction.  These states may still have relatively low
density and thus avoid the instant collapse due to three-body
recombinations.  This will be further addressed in
chapter~\ref{kap:stability_validity}.


A description of the condensate as effectively a non-self-bound
many-body state confined by the external trap is possible within this
hyperspherical treatment.  Then the average properties are comparable
to those of the mean-field treatment, as we shall see in more detail
in the following chapter.  So, the main effect of correlations is to
allow the Efimov-like many-body states, and the finite interaction
range prevents an infinite collapse of the system as the density
increases.

\chapter{Mean field and validity}                             
\label{kap:meanfield_validity}
The mean field provides a study of the average properties of a
many-boson system.  Section \ref{sec:conn-mean-field} first presents
some features of the mean-field method, then discusses
density-dependent interactions, and finally compares mean-field
results with the results from the correlated model.  Section
\ref{sec:validity} contains a discussion of the ranges of validity for
both models.

\section{Comparison to mean field}

\label{sec:conn-mean-field}

\hartree\gpe

When studying a dilute system of particles, the first approach is
usually to apply a mean-field method where the many-body wave function
for identical bosons is factorized into one-particle amplitudes $\psi$
as described in section \ref{sec:mean-field-descr}.  This often leads
to the non-linear Gross-Pitaevskii equation (GPE) \cite{pit61simple},
where the single-particle function enters as a properly normalized
single-particle density $|\psi(\boldsymbol r_1)|^2$.  In the
stationary case the GPE is written as
\begin{eqnarray}
  \bigg[
  -\frac{\hbar^2}{2m}\frac{\partial^2}{\partial \boldsymbol r_1^2}
  +\frac12m\omega^2r_1^2
  +\frac{4\pi\hbar^2a_s}{m}(N-1)|\psi(\boldsymbol r_1)|^2
  -\mu
  \bigg]
  \psi(\boldsymbol r_1)
  =0
  \label{eq:gpe}
  \;,
\end{eqnarray}
where $m$ is the mass of the particles, $\omega$ is the angular
frequency of an external trapping potential, and $a_s$ is the two-body
$s$-wave scattering length.  The eigenvalue of this equation is the
chemical potential $\mu$ which is related to the total energy
$E\sub{total}$ by $\mu = \partial E\sub{total} / \partial N$.

The GPE was applied to a description of experiments with trapped
alkali atoms \cite{edw95,bay96} and is now routinely solved for the
density profile of a condensate \cite{dal99,pet01,pit03} and for the
dynamical evolution of a condensate with the time-dependent version of
\smarteq{eq:gpe} \cite{adh02d}.  \etalcite{Proukakis}{pro98} derived a
non-linear Schr\"odinger equation from a microscopic treatment of
binary interactions and in the low-density limit obtained the GPE.
The under-lying assumptions are valid when the scattering length $a_s$
is small compared to the inter-particle separations \cite{lie01},
i.e.~when $n|a_s|^3\ll1$ where $n$ is the density.\footnote{In
chapters \ref{kap:meanfield_validity} and \ref{kap:stability_validity}
$n$ denotes the density and \emph{not} the hyperradial quantum number
as in chapter \ref{kap:radial}.}  The mean-field validity condition is
then fulfilled, which means that the mean free path
$\textrm{\emph{æ}}$ given by $\textrm{\emph{æ}}=1/(na_s^2)$ is long
compared to the average distance which approximately is
$n^{-1/3}$.\footnote{The Danish letter ``\emph{æ}'' is pronounced
almost like the vowel in the English ``them'' or ``Thames''.}  The
low-energy scattering properties expressed by the scattering length
are then decisive.  In the following we first comment on the choice of
interaction, then discuss density-dependent interactions, and finally
collect some differences between the results from the mean-field
method and the hyperspherical adiabatic method.

\subsection{Two-body interactions}

\label{sec:mf_two-body-interaction}

\gpe

The origin of the interaction term in the Gross-Pitaevskii equation
(GPE) is the approximation
\begin{eqnarray}
  \int d\boldsymbol r_2\; V(\boldsymbol r_2-\boldsymbol r_1)|
  \psi(\boldsymbol r_2)|^2
  \simeq
  |\psi(\boldsymbol r_1)|^2
  \int d\boldsymbol r_2\; V(\boldsymbol r_1-\boldsymbol r_2)
\end{eqnarray}
for an interaction of much shorter range than the average distance.
This integral is given by the Born approximation to the scattering
length as we saw in section \ref{sec:model-an-interacting}
\begin{eqnarray}
  \int d\boldsymbol r_2\; V(\boldsymbol r_1-\boldsymbol r_2)
  =
  \frac{4\pi\hbar^2}{m}
  a\sub B
  \;,
\end{eqnarray}
where the GPE then occurs with $a\sub B$ replaced by $a_s$.  The
smallness of $a_s$ compared with the average distance between
particles is the criterion of validity \cite{pit61simple}.  This
corresponds to a scattering situation where the wave function hardly
changes due to the scattering, i.e.~the wave length is very large
corresponding to low energy and low density.

The mean-field treatment above corresponds to a zero-range interaction
with $a\sub B$ replaced by $a_s$
\begin{eqnarray}
  V_\delta(\boldsymbol r)
  =
  \frac{4\pi\hbar^2a_s}{m}
  \delta(\boldsymbol r)
  \;,\qquad
  \boldsymbol r=\boldsymbol r_2-\boldsymbol r_1
  \label{eq:meanfield_potential}
  \;,
\end{eqnarray}
see also section \ref{sec:model-an-interacting}.  This limit can be
obtained from a finite-range potential where the range approaches zero
and the strength is appropriately adjusted.  The finite-range Gaussian
interaction of \refeq.~(\ref{eq:vpotGaussian}) can be expressed as
\begin{eqnarray}
  &&
  V\sub G(\boldsymbol r)
  =\frac{4\pi\hbar^2a\sub B}{m}\delta\sub G(\boldsymbol r)
  \label{eq:Gauss-pot}
  \;,\\
  &&
  \delta\sub G(\boldsymbol r)
  \equiv\frac{1}{\pi^{3/2}b^3}e^{-r^2/b^2}
  \;,\qquad
  1=\int d\boldsymbol r
  \;\delta\sub G(\boldsymbol r)
  \;.
\end{eqnarray}
The Gaussian $\delta\sub G(\boldsymbol r)$ is in the limit when
$b\to0$ a representation of the Dirac $\delta$ function.  For
$a_s=a\sub B$ we then have
\begin{eqnarray}
  \lim_{b\to0}V\sub G(\boldsymbol r)=V_\delta(\boldsymbol r)
  \;.
\end{eqnarray}
However, $a_s = a\sub B$ is only valid when $|a\sub B|/b \to 0$, which
is rarely the case, see \reffig.~\ref{fig:scatlen}.

The aim of computing reliable energies in the mean-field approximation
can be achieved with \refeq.~(\ref{eq:meanfield_potential}) for dilute
systems \cite{esr99b}.  The interaction and the Hilbert space must be
consistent, that is to say that a renormalized interaction follows a
restricted space to produce the correct energy.  In this case the
Hilbert space is restricted to the mean-field product wave function.
Any inclusion of features outside this restricted space, for example
two-body cluster structures, would be disastrous \cite{fed01}.
Maintaining the finite-range interaction with the correct scattering
length then results in different properties of the interaction even
when the range approaches zero on any scale defined by the problem.
Thus, the mean-field product wave function with a realistic two-body
potential would also lead to wrong results, as we shall see later.

The full Hilbert space with the correct interaction must produce
correct answers to any proposed question. Whether a realistic
interaction combined with the present inclusion of two-body
correlations reproduces the main features is not obvious.  However,
the investigations in chapter~\ref{kap:radial} demonstrate that the
energy in the mean-field approximation for dilute systems is
reproduced.  This asymptotic behaviour is determined by the scattering
length which only implicitly is contained in a given combination of
range and strength of the Gaussian interaction.  This implies that the
Hilbert space of the model accounts properly for the correlations
crucial at large separations.

\subsection{Density-dependent interactions}

\label{sec:densdepint}

In section~\ref{sec:effect-short-range} we studied the angular
potential with a zero-range interaction as in
\smarteq{eq:meanfield_potential}.  This led to the angular eigenvalue
$\lambda_\delta$ from \refeq.~(\ref{eq:lambda_delta}), which in the
limit of large densities clearly is wrong.  A possible treatment at
large densities is to use a finite-range potential with the correct
scattering length as discussed in
section~\ref{sec:model-an-interacting}.  However, this immediately
requires a treatment beyond the mean field, e.g.~by the Jastrow or
Faddeev approaches.
Another common approach, especially in nuclear physics
\cite{sie87,coc03}, but also for atomic many-boson systems
\cite{lee57,lee57b,bra02b}, is to expand the interaction in
density-dependent terms.  In the present case the two-body interaction
can be written as
\begin{eqnarray}
  V(\boldsymbol r)=g_2(n)\delta(\boldsymbol r)
  \label{eq:densdeptwobod}
  \;,
\end{eqnarray}
where the low-density limit of the coupling strength $g_2(n)$ is
$g_2(0)=4\pi\hbar^2a_s/m$.

We relate the density $n$ to the root-mean-square (rms) hyperradius
$\bar\rho \equiv \sqrt{\langle\rho^2\rangle}$.  The density is related
to the rms separation $\bar r$, which is defined by $\bar
r^2\equiv\langle r_{12}^2\rangle$, by $1/n\simeq 4\pi\bar r^3/3$.  The
relation $\bar r^2= 2 \bar\rho^2 /(N-1)$, obtained from
\refeq.~(\ref{eq:hyperradius}), then yields
\begin{eqnarray}
  n\simeq
  \frac{3}{8\sqrt2\pi}
  \frac{N^{3/2}}{\bar\rho^3}
  \label{eq:dens_rho}
  \;.
\end{eqnarray}
We then replace $\bar\rho$ by $\rho$ and for a fixed hyperradius
calculate the angular potential as an expectation value of the
two-body interactions, \smarteq{eq:densdeptwobod}, in a constant
hyperangular function.  This yields
\begin{eqnarray}
  \lambda_{\delta n}(n)=\lambda_\delta\frac{g_2(n)}{g_2(0)}
  \;,\qquad
  \lambda_{\delta n}(0)=\lambda_\delta
  \;.
\end{eqnarray}
The reverse relation yields an expression for the density-dependent coupling strength, i.e.
\begin{eqnarray}
  g_2(n)
  =
  g_2(0)
  \frac{\lambda_{\delta n}(n)}{\lambda_\delta}
  \;.
\end{eqnarray}
We assume that the numerically obtained angular potentials $\lambda$,
as calculated in chapter~\ref{kap:angular} and parametrized in
\refeq.~(\ref{eq:schematic}), represent the density-dependent
potential rather well, so we identify $\lambda_{\delta n}$ with the
lowest angular potential for the case with no two-body bound states.
Here we use the above-mentioned translation between $n$ and $\rho$.
So \reffig.~\ref{fig:coupling_densdep}a shows for various $N$ values
$\lambda_{\delta n}(n)/\lambda_\delta=g_2(n)/g_2(0)$ as a function of
the density in the combination $N^2n|a_s|^3$.
\begin{figure}[htb]
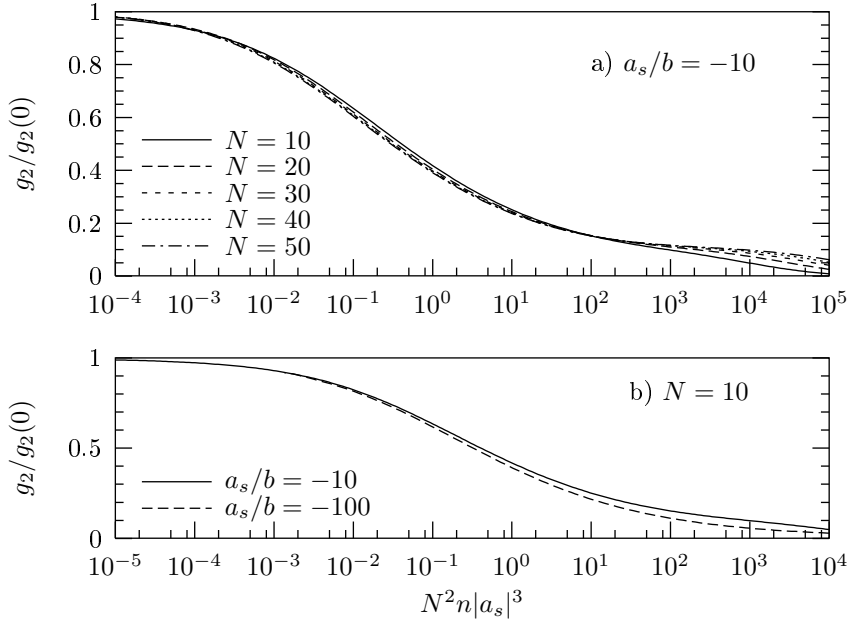

  \centering
  \input{coupling_densdep}
  \input{coupling_densdep2}
  \caption[Density-dependent coupling strengths]
  {a) The coupling strength in units of the zero-density value as a
    function of the density for $a_s/b=-10$ and various particle
    numbers.  b) The coupling strength in units of the zero-density
    value as a function of the density for $N=10$ and various
    scattering lengths.  The results are obtained by a finite-element
    treatment of the Faddeev-like \smarteq{eq:faddeev_eq}.}
  \label{fig:coupling_densdep}
\end{figure}

At low densities the ratio approaches unity which yields the correct
limit $g_2\simeq g_2(0)$.  At larger densities the deviations are
significant, and the coupling strength vanishes altogether as
$n|a_s|^3\to\infty$ since the finite-range interaction contains no
divergence.  \Reffig.~\ref{fig:coupling_densdep}b shows the same
quantities for different scattering lengths and the behaviours are
confirmed.

When the scattering length is negative and large, $n|a_s|^3\gg1/N^2$,
the angular potential assumes a constant value
$\lambda\simeq\lambda_\infty$, see \refeq.~(\ref{eq:lambda_infty}) in
section~\ref{sec:summ-angul-prop}.  This yields a coupling strength of
magnitude
\begin{eqnarray}
  \frac{g_2(n)}{g_2(0)}
  \simeq
  \frac{\lambda_\infty}{\lambda_\delta}
  \simeq
  \frac{0.48}{N^{2/3}n^{1/3}|a_s|}
  \simeq
  \frac{0.77\bar r}{N^{2/3}|a_s|}
  \;.
\end{eqnarray}
In this region the coupling strength decreases linearly with the rms
distance $\bar r$ between the bosons.  The interaction energy per
particle in this region is given by
\begin{eqnarray}
  \frac EN
  \simeq
  \frac12g_2(n)n
  \simeq
  -\frac{\pi\hbar^2n^{2/3}}{N^{2/3}m}
  \;.
\end{eqnarray}
This is independent of the scattering length.  A Jastrow calculation
in the denser region for \emph{positive} scattering lengths by
\etalcite{Cowell}{cow01} yielded $E/N\simeq13\hbar^2n^{2/3}/m$, which
reminds of the present result for negative scattering length.  Besides
the sign change and the different factor, the present result contains
an additional dependence on the number of particles.

This expansion in $(n^{1/3}|a_s|)^{-1}$ is in clear contrast to
density-expansions where the energy functional is written as
expansions in $n^{1/3}|a_s|$ \cite{lee57,bra02b}.  Such a low-density
expansion clearly diverges at large densities where it is not intended
to work.  The present results also need corrections at large densities
due to higher-order correlations, but they might provide a modified
zero-range interaction $V(\boldsymbol r)=g_2(n)\delta(\boldsymbol r)$
which could possibly be implemented in a GPE-like treatment of systems
denser than usually within reach.

In order to avoid collapse due to an attractive two-body $\delta$
interaction, some methods apply a repulsive three-body contact
interaction.  This can be written as
\begin{eqnarray}
  V_3(\boldsymbol r_{12},\boldsymbol r_{13})
  =
  g_3(n)
  \delta(\boldsymbol r_{12})
  \delta(\boldsymbol r_{13})
  \;,
\end{eqnarray}
where we allow a density-dependent coupling strength $g_3$.  The
expectation value of the three-body interactions in a constant angular
wave function yields the angular potential
\begin{eqnarray}
  \lambda\sub{3-body}(n)
  =
  \frac{\sqrt3}{8\pi^2}N(N-1)(N-2)(N-3)
  \bigg(N-\frac53\bigg)\bigg(N-\frac73\bigg)
  \frac{g_3(n)m}{\hbar^2\rho^4}
  \label{eq:lambda_3body}
  \;.
\end{eqnarray}
Inspired by \etalcite{Gammal}{gam99} we parametrize the coupling
strength as
\begin{eqnarray}
  g_3(n)=\frac{g_2^2(0)k_3(n)}{\hbar\omega}
  \;,
\end{eqnarray}
which for $N\gg1$ yields, in units of $\lambda_\delta$,
\begin{eqnarray}
  \frac{\lambda\sub{3-body}(n)}{\lambda_\delta}
  =
  \frac{4\sqrt\pi}{3[\rho/(\sqrt Nb\sub t)]^3}
  \frac{Na_s}{b\sub t}
  k_3(n)
  \;.
\end{eqnarray}
\etalcite{Gammal}{gam99} use values $0\le k_3\le0.03$ which yields
$\lambda\sub{3-body}/\lambda_\delta\ll1$ for a system with
$\rho\sim\sqrt Nb\sub t$.  However, at larger densities we have
$\lambda\sub{3-body}>\lambda_\delta$.  Even though this three-body
contact interaction can not account for the details when three
particles approach each other, it might provide a step towards an
explicit inclusion of three-body correlations.\footnote{On this train
of thought the work by \etalcite{Bohn}{boh98} was an initiation for
the present study of two-body correlations.}

\fermion

A treatment of fermion antisymmetry in the hyperangular equations
probably becomes too complicated when many particles are involved.
However, the effect of two-body correlations for fermions might be
included by a modified zero-range coupling strength as described for
bosons above.  The density dependence could possibly be extracted for
a few fermions and then applied for a large number of particles.

\subsection{Properties of the wave functions}

\hartree

In the dilute limit the Hartree wave function is closely related to
the hyperradial function and the Jastrow correlated wave function is
closely related to the Faddeev-like decomposition of the wave
function, see section~\ref{sec:wave-function}.  A direct comparison of
the wave functions is in general not possible as this requires an
expansion on a complete set of basis functions in one of the
coordinate systems.  The necessary calculations involve non-reducible
high-dimensional integrals.

Instead we use the indirect relations provided in section
\ref{sec:mean-field-descr} where energy and average distance between
particles are characteristic features of the solutions.  For a given
scattering length the energy $E$ is numerically obtained for identical
bosons as a function of the particle number.  The interaction energy
is next calculated as $E-E_0$ where $E_0 = 3N\hbar\omega/2$ is the
energy of the non-interacting trapped gas.  The results for
$a_s/b=-0.84$ are shown in figure~\ref{fig:energy_N_copy1}a.  The
discussion of stability which follows in section
\ref{sec:stability-criterion} shows that in terms of variational
average distance the GPE energy for attractive potentials has a local
minimum at large average distances and much lower energies at small
average distances. The physically meaningful mean-field solution is
located in the minimum at large average distance.  This minimum
becomes unstable for sufficiently large particle numbers.  In the
example of figure~\ref{fig:energy_N_copy1}a no stable mean-field
solution (solid, thin line) exists for $N=1000$.  This is consistent
with the experimentally established stability criterion
$N|a_s|/b\sub{t}< 0.55$ \cite{cla03} as seen from the upper
$N|a_s|/b\sub t$-axis.

\begin{figure}[htb]
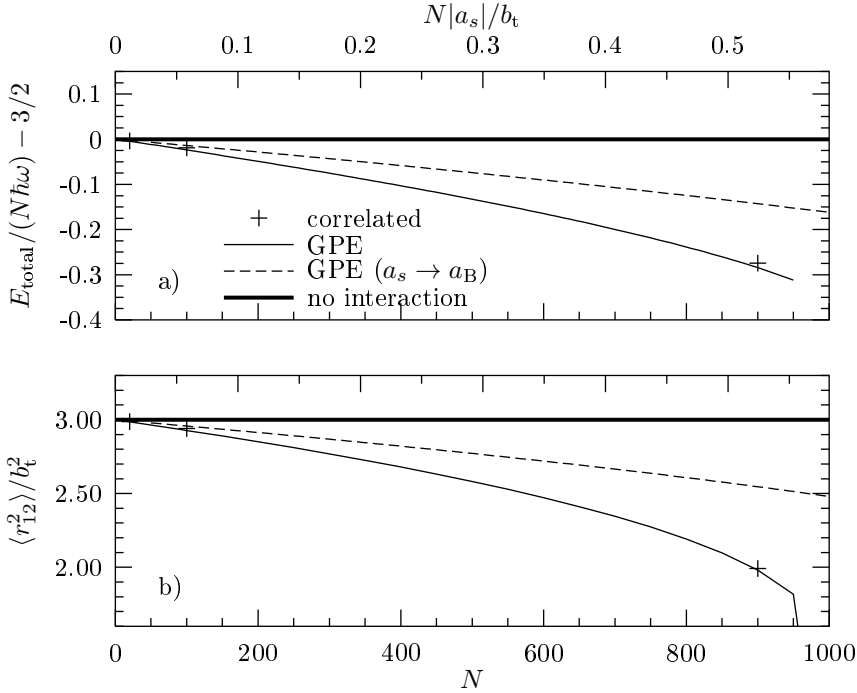

  \centering
  \input{energy_N_B}
  \input{rrms_N}
  \caption[Energy and distance from Gross-Pitaevskii equation and two-body
  correlated model] {a) Interaction energy as a function of $N$ for
    $a_s/b=-0.84$ ($a\sub B/b=-0.5$) and $b\sub t/b=1442$.  The thin
    solid line shows GPE results and the plusses are obtained from the
    two-body correlated model; see also \smartfig{fig:energy_N}.  The
    dashed line shows the GPE results for $a_s/b=-0.5$.  The upper
    $N|a_s|/b\sub t$-axis applies for $a_s/b=-0.84$.  b) Mean-square
    distance between the particles for the same cases.}
    \label{fig:energy_N_copy1}
\end{figure}

The same figure shows results obtained with the present two-body
correlated method for three different particle numbers (plusses).  The
correlated and mean-field interaction energies are remarkably similar.
It may at first appear odd that the mean-field interaction energy is
marginally lower than by use of the correlated wave function which
includes an extra degree of freedom.  The reason is that the
mean-field result is obtained with an effective interaction which only
in the Born approximation has the correct scattering length, while the
correlated solution is obtained for an interaction with the correct
scattering length.  The mean-field interaction is effectively more
attractive as discussed in section \ref{sec:summ-angul-prop}.

The proper comparison is then a GPE calculation with $a_s/b=-0.5$,
corresponding to the Born approximation for a Gaussian of $a\sub
B/b=-0.5$ with true scattering length $a_s/b=-0.84$.  As seen in
\smartfig{fig:energy_N_copy1}a (dashed curve), now the mean-field
interaction energies are numerically smaller.  This comparison does
not include the negative-energy states supported by the attractive
pocket at short distance, see e.g.~\smartfig{fig:u_as_m0.5.N20}.  They
would appear far below the ``condensate-like'' state shown in
figure~\ref{fig:energy_N_copy1}a.

\gpe

Using equations~(\ref{eq:hyperradius}) and (\ref{e21}), we compare in
figure~\ref{fig:energy_N_copy1}b $\langle r_{12}^2\rangle$ for the
solutions to the mean-field approximation and the hyperspherical
methods.  Also this quantity is very similar for the two methods,
whereas we again observe the discrepancy when we for the GPE method
replace $a_s$ by $a\sub B$.  The mean-square distance decreases with
increasing particle number for calculations with an attractive
potential.  As $N$ approaches $1000$, the Gross-Pitaevskii mean-field
radius approaches zero due to the unavoidable collapse.  The same
behaviour is seen for radii and interaction energies, i.e.~the average
distance between particles decreases until the condensate collapses
and the size vanishes in the mean field, while many-body bound states
with smaller extension play a role in the present hyperspherical
description.  Then also higher-order correlations can be expected to
be essential and result in recombination processes as will be
discussed in section \ref{sec:decay}.

In conclusion, for weak interactions or very small scattering lengths
a stationary many-body state can be approximated by a product of
single-particle amplitudes. However, stronger attraction between
particles must invoke other degrees of freedom like clusterization.
Then a single-particle description is not valid.  This is in agreement
with general expectations, and thus confirmed with the present point
of departure.

\section{Validity conditions}

\label{sec:validity}

We conclude the chapter by estimating validity criteria for the
models.  Of special importance in relation to trapped atomic gases is
a radial wave function with rms hyperradius $\rho \sim \sqrt{N}b\sub
t$.  Accurate angular eigenvalues in this region are therefore crucial
for a proper description.  If these hyperradii are sufficiently large,
that is $\rho \sim \sqrt{N}b\sub t > N^{7/6}|a_s|$, the angular
eigenvalue has reached its asymptotic value $\lambda
\simeq\lambda_\delta$.  This condition is equivalent to $N|a_s|/b\sub
t<N^{1/3}$, which is obeyed by stable systems with $N|a_s|/b\sub t <
0.55 <N^{1/3}$.

The different models are valid if appropriately designed, i.e.~the
present two-body correlated model reproduces the correct effective
interaction for the correct scattering length for any short-range
interaction, whereas the Gross-Pitaevskii equation (GPE) reproduces
this same correct effective interaction by using the Born
approximation.  Interaction energies and sizes would be very similar
for the states corresponding to the condensate.

From \smarteq{eq:dens_rho} a given rms hyperradius $\bar\rho$ is
related to the density $n$ of the system by $n \sim N^{3/2} /
\bar\rho^3$.  The zero-range mean-field method (GPE) is usually valid
for condensates when $n|a_s|^3\ll1$, see \cite{cow01,pit03}.  Then the
number of particles within a scattering volume $4 \pi |a_s|^3/3$ is on
average much smaller than one.  From the present model, in the
zero-range asymptotic region of $\bar\rho>N^{7/6}|a_s|$, we find that
$n \bar\rho^3>N^{7/2}n |a_s|^3$ or $n|a_s|^3<1/N^2\ll1$, which means
that the system is very dilute and both the GPE method and the
correlated method are valid.

For $\bar\rho<N^{7/6}|a_s|$ the large-distance asymptotic
$\lambda\simeq\lambda_\delta$ is not valid.  This corresponds to that
we cannot use a non-correlated model with the zero-range interaction.
This is here interpreted as an indication of the inadequacies of the
GPE in the region where $\bar\rho<N^{7/6}|a_s|$ or equivalently where
$n|a_s|^3>1/N^2$.  This appears different than the usual criterion of
validity $n|a_s|^3\ll1$, but could potentially specify what is meant
by ``much smaller than unity''.

The present adiabatic hyperspherical method with two-body correlations
explicitly allowed in the form of the wave function is valid in the
region $\bar\rho>N^{7/6}|a_s|$ where correlations are expected to be
insignificant.  The inclusion of two-body correlations is expected to
allow smaller hyperradii $\bar\rho<N^{7/6}|a_s|$.  When higher-order
clusterizations occur, any method without correlations higher than
two-body breaks down.  The absolute lower criterion must be that the
distance between two particles on average exceeds the interaction
range $b$, i.e.~$\bar\rho > N^{1/2} b$.  We quote this as the
criterion even though explicit calculations might prove that
higher-order correlations alter the lower limit.

In conclusion, the validity regions for the two-body correlated method
and the GPE version of the mean field are estimated to be
\begin{eqnarray}
  \bar\rho > \sqrt N b 
  &\quad \textrm{for two-body correlated method}
  \;,
  \\
  \bar\rho > N^{7/6}|a_s| 
  &\quad \textrm{for GPE (the present result)}
  \;.
\end{eqnarray}
These relations can with equation~(\ref{eq:dens_rho}) be expressed via
the density as
\begin{eqnarray}
  nb^3 < 1
  &\quad \textrm{for two-body correlated method}
  \;,
  \\
  n|a_s|^3 < \frac{1}{N^2}
  &\quad \textrm{for GPE (the present result)}
  \;,
  \\
  n|a_s|^3 \ll 1
  &\quad \textrm{for GPE (usual expectation)}
  \;,
\end{eqnarray}
where we also collected the usual criterion for validity of the GPE.
When the density is low, both descriptions are valid and the energies
are similar.  For larger densities the importance of correlations
increases and the mean-field approximation breaks down.  At even
higher density also two-body correlations are insufficient and the
particles may clusterize further.

\begin{figure}[htb]
  \centering
  \input{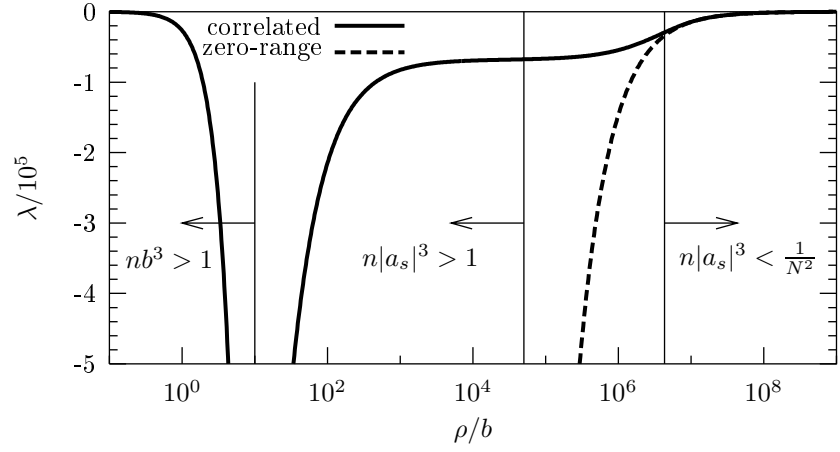}
  \caption[Angular eigenvalue with outlined validity regions]
  {The lowest angular eigenvalue for $a_s/b=-10^4$, $N=100$, and no
    bound two-body states (solid line).  The dashed curve is
    $\lambda_\delta$ for the same scattering length.  The vertical
    lines indicate regions of different density.}
  \label{fig:trento3}
\end{figure}

These conclusions are illustrated in \reffig.~\ref{fig:trento3} where
the lowest angular eigenvalue for a case with no two-body bound state
and negative scattering length is compared to the zero-range angular
potential for the same scattering length.  For low density
$n|a_s|^3<1/N^2$ the effective energy of the two methods coincide.
For larger densities the GPE energy diverges, while the energy from
the finite-range model remains finite.  Moreover, it deviates in a
region where the density is still relatively low $n|a_s|^3<1$ so
higher-order correlations, especially three-body, do not play a role
yet.  As the average distance becomes smaller, we expect corrections
due to higher-order correlations.

\chapter{Macroscopic stability and decay}                     
\label{kap:stability_validity}
Chapter~\ref{kap:radial} opened a discussion of the macroscopic boson
systems known as condensates, and a definition of the condensate was
given in the present hyperspherical context.  In this chapter we
discuss stability and consider possible dynamics, primarily related to
this lowest trap state or ``condensate''.  In terms of the degrees of
freedom explicitly included in the two-body correlated model, section
\ref{sec:stability-criterion} presents a criterion for macroscopic
stability.  This is similar to the discussion presented by
\etalcite{Bohn}{boh98}.  Section \ref{sec:decay} ventures into a
discussion of degrees of freedom that in the strictest sense are
beyond the present model.  This especially involves three-body
recombination events that potentially ignite dynamics of the system as
one macroscopic whole.

\section{Stability criterion}

\label{sec:stability-criterion}

\macstab\bec

Macroscopic instability of systems of bosons has been investigated
thoroughly the last eight years.  Macroscopic stability for a
Bose-Einstein condensate (BEC) means that the BEC state is well
defined and has a sufficiently long lifetime when considering possible
decay modes.  In this sense it was originally expected that
Bose-Einstein condensation could not be realized for atoms with
effectively attractive interactions.  However, this was achieved in
1995 for $^7$Li atoms \cite{bra95}.  Since then numerous experiments,
e.g.~\cite{rob01b,cla03}, have tested the critical region.  Also the
collapse process itself has been studied \cite{sac99,don01,rob01}.

The criterion for stability of a system with negative scattering
length $a_s$ can be expressed as a critical combination of the number
of particles, the scattering length, and the trap length $b\sub
t\equiv\sqrt{\hbar/(m\omega)}$, where $\omega$ is a geometric mean,
see section \ref{sec:trapped-bosons}.  Recently it was measured that
when $N|a_s|/b\sub t>0.55$, a condensate of $^{85}$Rb-atoms is
unstable \cite{cla03}.  This can be understood as a competition
between the kinetic energy which is effectively repulsive and the
two-body attractive interaction.  When the kinetic energy subdues the
net attraction, a meta-stable system with signatures of a condensate
exists.  The number of kinetic-energy terms equals the number of
particles $N$ and the number of interactions equals the number of
particle pairs $N^2/2$.  Therefore, as $N$ or the scattering length
$a_s$ increases, the interactions at some point win and for attraction
lead to collapse.  When the trap length $b\sub t$ is small, the system
is compressed too much and is according to the criterion unstable.
This means that the interactions win at larger densities.

In the mean field this can be formulated variationally with a
Gaussian-Hartree amplitude with a variational width $w$,
i.e.~$\psi(r_1)=\exp[-r_1^2/(2w^2)]$, see also Pethick and Smith
\cite{pet01}.  The kinetic energy per particle is proportional to
$1/w^2$, the external field energy $w^2/b\sub t^4$, and the
interaction energy $Na_s/w^3$.  This leads to the variational total
energy
\begin{eqnarray}
  \frac{2m}{\hbar^2}E\sub{total}(w)
  =
  \frac{3N}{2w^2}
  +\frac{3Nw^2}{2b\sub t^4}
  +\sqrt{\frac2\pi}
  N^2
  \frac{a_s}{w^3}
  \label{eq:mf_varenergy}
  \;.
\end{eqnarray}
This yields energy curves in the length scale $w$, analogous to the
hyperradial potential curves in chapter~\ref{kap:radial}, which has
stable points for a sufficiently weak attraction.  The critical value
is then found to be about $0.67$ \cite{dal99,pet01}.  More detailed
analysis of the Gross-Pitaevskii \smarteq{eq:gpe}, incorporating time
dependence, yields a value of 0.55 \cite{gam01} in agreement with the
experimentally measured value \cite{cla03}.

By analogy with the mean-field discussion we here take a closer look
at the derivation of the stability criterion in the hyperspherical
frame.  This is also equivalent to the derivation performed by
\etalcite{Bohn}{boh98}.  The criterion is obtained by estimating when
the radial barrier disappears.  The effective hyperradial potential
$U(\rho)$ from \refeq.~(\ref{eq:radial.potential}) can in the
asymptotic region when $\rho > N^{7/6}|a_s|$, that is when
$\lambda\simeq\lambda_\delta$ from \refeq.~(\ref{eq:lambda_delta}),
for $N\gg1$ be written as
\begin{eqnarray}
  \frac{2mU(\rho)}{\hbar^2}
  =
  \frac{9N^2}{4\rho^2}
  +\frac{\rho^2}{b\sub t^4}
  +\sqrt{\frac{2}{\pi}}
  N^2
  \bigg(\frac{3N}{2}\bigg)^{3/2}
  \frac{a_s}{\rho^3}
  \label{eq:u_delta}
  \;.
\end{eqnarray}
If we rescale \smarteq{eq:u_delta} with $\rho=\sqrt{3N/2}w$, this
hyperradial potential is identical to the mean-field variational
energy from \smarteq{eq:mf_varenergy}, i.e.~$U(w)=E\sub{total}(w)$.
For a large and negative $a_s$ or a large value of $N$ there is no
stable region in such a potential.  In general, a potential of the
form
\begin{eqnarray}
  u(x)=\frac{A}{x^2}+Bx^2-\frac{C}{x^p}
  \;,\qquad
  \{A,B,C\}>0
  \label{eq:uABC_x}
  \;,
\end{eqnarray}
diverges to $+\infty$ as $x\to\infty$ and if $p>2$ to $-\infty$ as
$x\to0$.  For a sufficiently small constant $C$ there will always be a
local minimum.  There will be no local minimum when $p\ge2$ and
\begin{eqnarray}
  C 
  >
  \frac{8}{p(p+2)}
  \frac{A^{(p+2)/4}}{B^{(p-2)/4}}
  \bigg(\frac{p-2}{p+2}\bigg)^{(p-2)/4}
  \label{eq:uABC_x_stabilitycrit}
  \;.
\end{eqnarray}
For the present case the power $p$ is given by $p=3$ and $A$, $B$, and
$C$ are given by comparing \refeqs.~(\ref{eq:u_delta}) and
(\ref{eq:uABC_x}).  Then the radial potential for negative scattering
length has a local minimum only when
\begin{eqnarray}
  \frac{N|a_s|}{b\sub t} < \frac{2\sqrt{2\pi}}{5^{5/4}} \simeq 0.67
  \label{eq:stabilitycrit_localmin}
  \;.
\end{eqnarray}
This is identical to the value obtained variationally from the mean
field since the $a_s$-only dependence at large hyperradii corresponds
to the mean field with a zero-range interaction, as discussed in
chapter~\ref{kap:angular} and directly evident by comparing
\smarteqs{eq:mf_varenergy} and (\ref{eq:u_delta}).  This gives the
right order of magnitude of the critical combination of particle
number, scattering length, and trap length.  The discrepancy from the
experimental value can be accounted for by the deformation of the
external field and the zero-point energy from motion in the trap.
This will be discussed in chapter~\ref{kap:deformed}.


The criterion is in \reffig.~\ref{fig:simul} illustrated by the radial
potential, with the angular eigenvalue obtained from
\refeq.~(\ref{eq:schematic}), as a function of the hyperradius for a
series of different particle numbers and scattering lengths.  The
strongly-varying short-distance dependence is omitted to allow focus
on intermediate and large hyperradii.  When an intermediate barrier is
present, the condensate is described as the state of lowest energy
located in the minimum at large hyperradius.  This minimium exists for
$a_s<0$ when $N|a_s|/b\sub t < 0.67$ as established above.

In \smartfig{fig:simul}a-\ref{fig:simul}d the particle number is fixed
at $N=6000$ while only the scattering length $a_s$ varies.  In
\smartfig{fig:simul}d-\ref{fig:simul}f the scattering length is fixed
at $a_s/b = -0.35$ and $N$ is varied.
\begin{figure}[htb]
  \centering
  \input{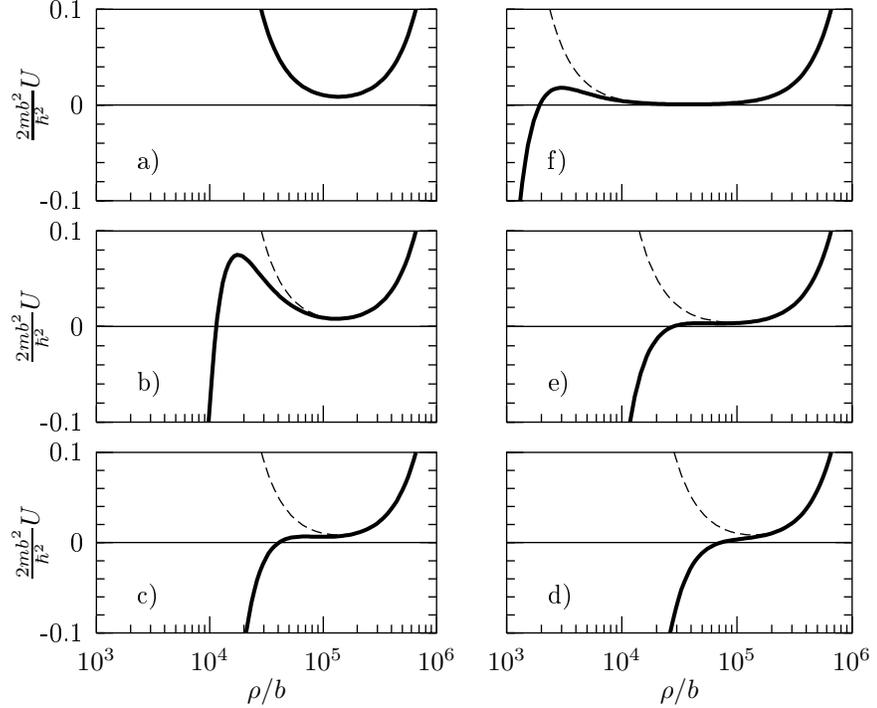}
  \caption
  [Radial potentials as a function of particle number and scattering
  length] {Radial potentials with $b\sub{t}/b=1442$ and a) $N=6000$,
    $a_s=0$; b) $N=6000$, $a_s/b=-0.05$; c) $N=6000$, $a_s/b=-0.18$;
    d) $N=6000$, $a_s/b=-0.35$; e) $N=3000$, $a_s/b=-0.35$; f)
    $N=500$, $a_s/b=-0.35$.  The dashed lines are obtained with
    $a_s=0$.}
  \label{fig:simul}
\end{figure}
In \smartfig{fig:simul}a the two-body interaction is zero, that is
$a_s=0$, which leads to a vanishing lowest angular eigenvalue
$\lambda=0$.  The effective radial potential then consists only of the
centrifugal barrier and the external field with one minimum.  In
\smartfig{fig:simul}b an attractive potential with $a_s=-0.05b$ is
sufficiently strong to overcompensate for the centrifugal repulsion
and create a second minimum in the radial potential at smaller
hyperradius; the final divergence $U(\rho)\to+\infty$ when $\rho\to0$
is not included in the scale of the figure.  An intermediate barrier
is left between the two minima at small and large hyperradii.  A
further increase of the attraction in \smartfig{fig:simul}c removes
the barrier while leaving a smaller flat region.  The
negative-potential region around the minimum at small hyperradius is
now even more pronounced.  This tendency is continued in
\smartfig{fig:simul}d with a stronger attraction.  With the scattering
length from \smartfig{fig:simul}d, i.e.~$a_s=-0.35b$, and a decreasing
number of particles the intermediate barrier is slowly restored.  In
\smartfig{fig:simul}e for $N=3000$ a barrier is about to occur, and in
\smartfig{fig:simul}f for $N=500$ an intermediate barrier is again
present between a minimum at small and large hyperradii.

The discussion in this section involved the static properties of
states located in the local minimum at large hyperradii.  Other
factors are important when we in the following section to some extent
include time dependence.

\macstab

\section{Decay}

\label{sec:decay}

\recthree\maccol\bec

The Bose-Einstein condensate (BEC) is intrinsically unstable and
decays spontaneously, e.g.~into lower-lying dimer states.
Recombination of two particles into a lower-lying state is possible by
emission of a photon, but the rate is enhanced when a third particle
is involved instead of the photon.  This three-body recombination
process inevitable occurs in a system of bosons when at least one
two-body bound state exists.  This has been suggested to be important
for a BEC \cite{adh01b,ued03}.  The related change of the surrounding
medium could lead to an instability which involves many particles, and
thus result in much faster decays which could be described as a
collapse \cite{adh02b}.

\Reffig.~\ref{fig:cartoon} illustrates the different behaviours by
using the angular eigenvalues parametrized through
\refeqs.~(\ref{eq:lambda0_smallrho}) and (\ref{eq:schematic}).
In \reffig.~\ref{fig:cartoon}a the scattering length is relatively
small and a large barrier separates the outer minimum from the inner
region.  When the scattering length increases, the barrier decreases
first into a relatively flat region as in \reffig.~\ref{fig:cartoon}b
and then disappears completely as in \reffig.~\ref{fig:cartoon}c when
the trap length is exceeded.

\begin{figure}[htb]
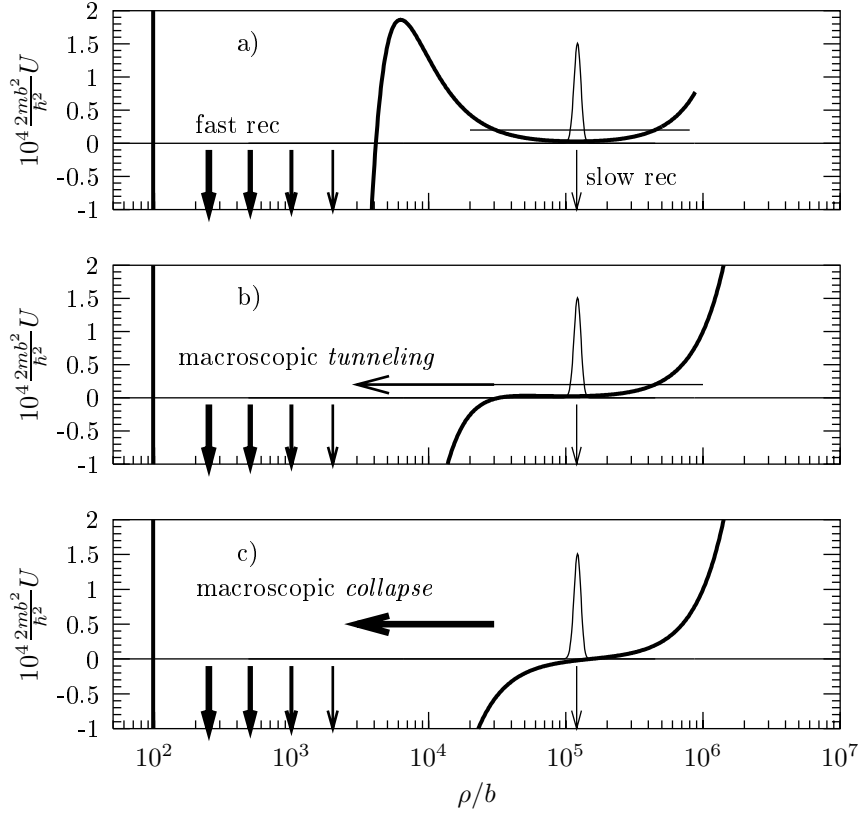

  \centering
  \input{cartoon1}
  \input{cartoon2}
  \input{cartoon4}
  \caption[Sequence of radial potentials as the barrier vanishes]
  {The radial potential from the parametrization for $N=100$, $b\sub
    t/b=10^4$, and a) $a_s/b=-6$, b) $a_s/b=-50$, and c)
    $a_s/b\to-\infty$.  The shown wave function is the lowest radial
    solution in the non-interacting case.  The horizontal lines in
    parts a) and b) indicate an energy level (not to scale).}
  \label{fig:cartoon}
\end{figure}

The discussion of macroscopic dynamics in this picture involves
various isolated ideas which lead to simple decay rates.  These are
then incorporated in a description of the experimental collapse
situation.

\subsection{Three-body recombination}

\label{sec:decay-rates}

\recthree

The condensate state is unstable due to the couplings into degrees of
freedom different than the coherent many-body mode.  The formation of
bound-state dimers is possible by a three-body process where the third
particle ensures conservation of energy and momentum.  The number of
these three-body recombination (rec) events per unit volume and time
can be estimated by the upper limit given in \cite{nie99,bed00}:
\begin{eqnarray}
  \nu\sub{rec}
  \simeq
  68\frac{\hbar |a_s|^4n^3}{m}
  \label{eq:recrate}
  \;,
\end{eqnarray}
where $n$ is the density.  This expression can be converted into an
estimate of the recombination rate for a given root-mean-square (rms)
hyperradius $\bar\rho \equiv \sqrt{\langle\rho^2\rangle}$ by using the
relation between density and rms hyperradius from
\refeq.~(\ref{eq:dens_rho}),
i.e.~$n\simeq3N^{3/2}/(8\sqrt2\pi\bar\rho^3)$.  With the volume
$\mathcal{V} = N/n$ the total recombination rate then is
\begin{eqnarray}
  \frac{\Gamma\sub{rec}}{\hbar} 
  = \nu\sub{rec} \mathcal{V}
  \simeq 0.5 \frac{N^4\hbar |a_s|^4}{m\bar\rho^6}
  \label{eq:Gammarec}
  \;.
\end{eqnarray}
The recombination rate increases rapidly with decreasing $\bar\rho$,
as indicated by the vertical arrows in \reffig.~\ref{fig:cartoon}
where we interchange $\rho$ and $\bar\rho$ to illustrate this effect.

\bec

If $N(t)$ is the number of particles present in the coherent many-body
state as time $t$ goes by, the recombination time scale
$\tau\sub{rec}$ can be defined by $N(t) = N(0)\exp(-t/\tau\sub{rec})$.
This leads to the rate $\Gamma\sub{rec}/\hbar = -dN/dt =
N/\tau\sub{rec}$, so
\begin{eqnarray}
  \tau\sub{rec}
  =
  \frac{N\hbar}{\Gamma\sub{rec}}
  \simeq
  \frac{2m\bar\rho^6}{N^3\hbar |a_s|^4}
  \simeq
  \frac{m\bar r^6}{4\hbar |a_s|^4}
  \label{eq:threebodyrectime}
  \;.
\end{eqnarray}
Since the condensate forms in the external trap, the system must be
stable versus recombination events on a time scale $\tau\sub{trap}$
which is given by the oscillator time scale $2\pi/\omega$, that is
$\tau\sub{rec} > \tau\sub{trap} \equiv 2\pi/\omega$.  With the
relation $1/\omega=mb\sub t^2/\hbar$ the criterion for a stable
condensate becomes
\begin{eqnarray}
  \bar r > \sqrt[6]{8\pi}|a_s|^{2/3}b\sub t^{1/3} = d\sub{c}
  \;.
\end{eqnarray}
Here an expression for the minimal separation $d\sub{c}$, as
introduced in section~\ref{subsec:definition}, is obtained.  In units
of $b\sub t$ we have
\begin{eqnarray}
  \frac{d\sub{c}}{b\sub t}
  =
  \sqrt[6]{8\pi}\bigg(\frac{|a_s|}{b\sub t}\bigg)^{2/3}
  \;,
\end{eqnarray}
where the determining combination is $|a_s|/b\sub t$.

Thus, for $|a_s|/b\sub t\ll1$ also $d\sub c/b\sub t\ll1$.  The rms
distance $\bar r$ for a state located in the second minimum is of the
order $b\sub t$ and therefore $\bar r > d\sub c$, i.e.~for this state
$\bar r$ is larger than the critical stability length $d\sub c$.  This
state then qualifies as a condensate.  For $^{87}$Rb atoms with $a_s
\simeq 100$ a.u.~and trapped in a field with $\nu\sub{trap}
\simeq 100$ Hz, we obtain $\tau\sub{rec}\sim7$ days.

\bec

\subsection{Macroscopic tunneling and recombination}

\mactun

The second decay process is macroscopic tunneling through the small
barrier as indicated in \reffig.~\ref{fig:cartoon}b.  The model
provides stationary eigenstates which by definition are time
independent.  Thus, strictly the states do not tunnel through the
barrier.  However, an exponentially small tail extends to small
hyperradii or large density.  All particles thus approaching each
other would recombine into molecular clusters because the density is
very large in the inner region.  The rate of this two-step decay with
tunneling through the barrier and subsequent recombination is
determined by the bottleneck.  The rate of recombination due to
macroscopic tunneling can be estimated by \cite{boh98}
\begin{eqnarray}
  &&
  \frac{\Gamma\sub{tunnel}}{\hbar} 
  \simeq
  \frac{N\nu\sub{tunnel}} {1 + e^{2\sigma}}
  \label{eq:gammatunnel}
  \;,\qquad
  \nu\sub{tunnel} = \frac{1}{2\pi}
  \sqrt{\frac{1}{m}\frac{d^2U(\rho)}{d\rho^2}\bigg|_{\rho\subsub{min}}}
  \label{eq:gammanu}
  \;,\\
  &&
  \sigma = \int_{\rho\subsub{in}}^{\rho\subsub{out}}
  d\rho\sqrt{\frac{2m}{\hbar^2} \Big[U(\rho)-E\Big]}
  \label{eq:gammasigma}
  \;,
\end{eqnarray}
where the multiplication by the factor $N$ gives the total number of
recombined particles.  Here $\rho\sub{min}$ is the position of the
second minimum of $U$, and $\rho\sub{in}$ and $\rho\sub{out}$ are the
points where the barrier height equals the energy $E$.

When $N|a_s|/b\sub t\ll 1$, the barrier is large and the very small
rate can be estimated through \refeqs.~(\ref{eq:gammatunnel}) and
(\ref{eq:gammasigma}).  The action integral is then large and given by
\begin{eqnarray}
  \sigma
  \simeq
  \frac{3}{2} N \ln\bigg(\frac{b\sub t}{N|a_s|} \bigg)
  \;.
\end{eqnarray}
Classical turning points of the potential are present when
$N|a_s|/b\sub t \leq 0.53$.  Close to this threshold, i.e.~when the
barrier is small, the exponent is
\begin{eqnarray}
  \sigma
  \simeq
  1.7 N \bigg(
  1-\frac{N|a_s|}{0.53b\sub t}\bigg)
  \;,
\end{eqnarray}
which is valid when $N|a_s|/(0.53b\sub t)$ is close to unity.

At the threshold for macroscopic stability then $N|a_s|/b\sub t \sim
0.5$, which due to the factor of $N$ implies that $|a_s|/b\sub t\ll
1$.  Close to this threshold we have $\bar r \sim b\sub t\gg d\sub c$,
which means that the average distance between the bosons is so large
that the three-body recombination is slow compared to the typical time
scale for oscillation in the harmonic-oscillator trap.  Therefore, the
three-body recombination does not limit the macroscopic stability of a
condensate.  In the limit $\sigma\ll1$ we get explicitly
\begin{eqnarray}
  \frac{\Gamma\sub{rec}}{\Gamma\sub{tunnel}}
  \simeq
  \frac{1}{7.0 N^4}
  \ll
  1
  \label{eq:gammarectun}
  \;,
\end{eqnarray}
implying that the macroscopic tunneling process dominates.  With
$\sigma\ll1$ then $\Gamma\sub{tunnel}/\hbar\simeq 0.5N\nu\sub{tunnel}$
and $\nu\sub{tunnel}\simeq\nu\sub{trap}$, which yields a tunneling
time of about $1/\nu\sub{trap}$.  For the case with
$\nu\sub{trap}\simeq 100$ Hz the macroscopic tunneling time scale is
$10$ ms.  This is much smaller than the three-body recombination time
scale which close to stability is given by the reciprocal of
\smarteq{eq:gammarectun}, that is
\begin{eqnarray}
  \frac{\tau\sub{rec}}{\tau\sub{tunnel}}
  =
  \frac{\Gamma\sub{tunnel}}{\Gamma\sub{rec}}
  \simeq
  7.0N^4\gg1
  \;.
\end{eqnarray}

The three-body recombination rate is in \reffig.~\ref{fig:lifetimes}
shown as a function of the hyperradius (solid curve) and compared with
the macroscopic tunneling rate (dashed curve) where all particles in
the condensate simultaneously disappear during contraction.  At small
hyperradii the three-body recombination rate is much larger than the
macroscopic tunneling rate, whereas the opposite holds for large
hyperradii. For the chosen set of parameters the two time scales are
roughly equal around the second minimum where the condensate is
located.  However, the tunneling rate depends strongly on the barrier
through the combination $N|a_s|/b\sub t$.
\begin{figure}[htb]
  \centering
  \input{lifetimes}
  \caption [Time scales involved in the macroscopic decay]
  {The three-body recombination rate from \refeq.~(\ref{eq:Gammarec})
    in units of the oscillator frequency
    $\nu\sub{trap}=\omega/(2\pi)$, which typically is 10-100 Hz, as a
    function of hyperradius for $N=100$, $a_s/b=-50$, and $b\sub
    t/b=10^4$.  Shown as the horizontal, dashed line is the
    macroscopic tunneling rate from \refeq.~(\ref{eq:gammatunnel}).
    Shown as the horizontal, dotted line is the macroscopic collapse
    rate from \refeq.~(\ref{eq:collapserate}) when the scattering
    length is much larger than the trap length.}
  \label{fig:lifetimes}
\end{figure}
Variation of either of the three quantities then moves the tunneling
rate up or down in \reffig.~\ref{fig:lifetimes}.  For a larger barrier
the condensate would only decay slowly by recombination.  For a
smaller barrier macroscopic tunneling would dominate and all the
bosons in the condensate would participate in a ``collective
recombination'' in a short time interval.

When a few particles recombine into dimers and leave the condensate,
the system is no longer in an eigenstate of the corresponding new
Hamiltonian.  An adiabatic adjustment of Hamiltonian and wave function
could then take place.  Since fewer particles and unchanged $a_s$ and
$b\sub t$ means a larger barrier, the macroscopic stability of the new
system is therefore increased.

\recthree\mactun

\subsection{Macroscopic collapse}

\maccol

Scenarios where the boson system develops with time as one unified
body are open for investigations in experiments where the effective
interaction almost instantaneously is changed by tuning close to a
resonance \cite{ino98,don01,rob01}.  An initially small magnitude of
the scattering length, corresponding to a stable condensate state in
the second minimum of the hyperradial potential, can be changed to a
value where the barrier is removed.

In one experiment \cite{don01} a condensate is first created with
effectively zero interaction, i.e.~zero scattering length as in
\reffig.~\ref{fig:simul}a.  The radial wave function is then located
at relatively large distances in the minimum created by the compromise
between centrifugal barrier and external field.  The attractive pocket
at small distances is not present and the condensate forms as the
ground state in this potential.  \Smartfig{fig:colsequence} shows both
the radial potential (thick, dashed line) and the wave function (thin,
dashed line) for schematic model parameters.

In the experiment the effective interaction was then suddenly changed
by tuning a Feshbach resonance \cite{cor00} to obtain a large and
negative scattering length \cite{don01}.  The measurement showed a
burst and a remnant of coherent atoms.  This was interpreted and
explained as formation of dimers via the two-body resonance, a burst
of dissociating dimers, and a remnant of an oscillating mixture of
coherent atoms and coherent molecules \cite{kok02b,mac02,koh03b}.

In the present formulation the effective potential is suddenly altered
by a change of the underlying two-body interaction.  The corresponding
new radial potential, shown as the thick solid line in
\reffig.~\ref{fig:colsequence}, has a pronounced attractive region
which is able to support a number of self-bound many-body states.

\begin{figure}[thb]
  \centering
  \input{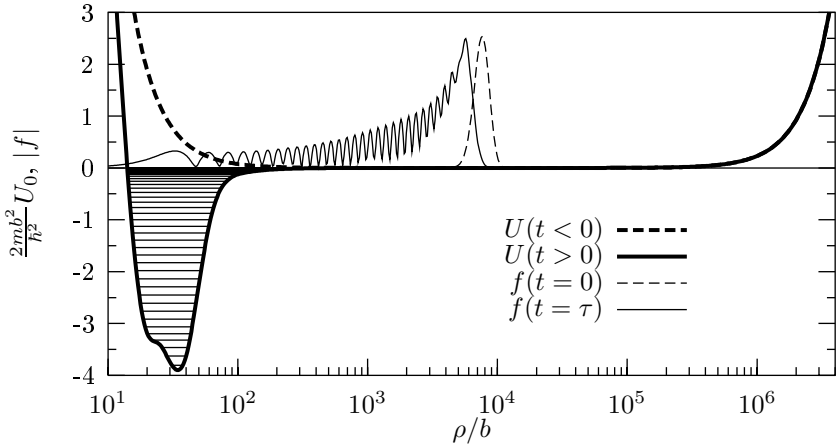}
  \caption
  [Collapsing many-body wave function in the potential for twenty
  bosons] {Wave functions $f$ and effective hyperradial potentials $U$
    in dimensionless units as a function of hyperradius for $N=20$ and
    $b\sub t/b=1442$.  The scattering length is zero up to the time
    $t=0$ and then suddenly changes to be large and negative at later
    times $t>0$.  Potentials and the corresponding wave functions are
    sketched for $t=0$ and at a time $\tau\sim0.1$ ms after half a
    period.  The horizontal lines show the stationary negative-energy
    states for $t>0$.}
  \label{fig:colsequence}
\end{figure}

Since the initial wave function is not a stationary state in the new
potential, a motion is started towards smaller hyperradii where it
would be reflected off the ``wall'' of the centrifugal term.  This
macroscopic contraction or collapse is indicated by the large arrow in
\reffig.~\ref{fig:cartoon}c.  If no degrees of freedom beyond the
model assumptions are involved, the system would then oscillate
between the centrifugal repulsion and the wall of the external field.
This corresponds to an oscillation in density.  However, during the
macroscopic contraction the rate for dimer production via the
three-body process increases and dimers are produced and subsequently
ejected from the trap.  Since the rate explodes as the contraction
culminates, all particles should recombine instantly.  Thus, the
relevant time scale, i.e.~the bottleneck, is the time scale for
macroscopic contraction.

If the only excitations are the degrees of freedom contained in the
lowest new hyperspherical potential with $s$-waves, we can get
quantitative information by expanding on the new eigenfunctions.  The
dominating states in this expansion are on the transition point
between the lowest-lying positive-energy states with energies
comparable to the initial condensate and the highest-lying Efimov-like
many-body states, now present because of the large scattering length,
see section~\ref{sec:efimov-like-states}.  These states have a spatial
extension as large as that of the initial state.  The time scale for
evolution of the initial state in the new potential is then determined
by the energy differences between such levels.  The states of positive
energy and large spatial extension confined by the trap are roughly
separated by the oscillator quantum of energy $\hbar\omega$.  The
corresponding rate for populating smaller distances with the
consequence of immediate recombination is then crudely estimated to be
\begin{eqnarray}
  \frac{\Gamma\sub{collapse}}{\hbar}
  \sim
  \frac{1}{\tau\sub{trap}}
  =
  \frac{\omega}{2\pi}
  \label{eq:collapserate}
  \;.
\end{eqnarray}
The resulting non-stationary wave function provides a specific
oscillation time.  After half a period the extension of the system has
reached its minimum.  The wave function at this time $\tau\sim
\tau\sub{trap}/2\sim0.1$ ms is also shown in
\reffig.~\ref{fig:colsequence}.  Experimentally \cite{don01} the
macroscopic-collapse time is verified to be of the order $\sim
1/\omega$.  These time scales agree on the order-of-magnitude level.

The rate of macroscopic collapse is also shown in
\reffig.~\ref{fig:lifetimes}.  This is larger than the tunneling rate.
The motion in the potential is slow compared to the recombination time
for distances in the minimum at small $\rho$, whereas the opposite
holds for distances in the minimum at large $\rho$.  The time
evolution after the sudden removal of the barrier could then be as
follows.  A macroscopic collapse towards smaller hyperradii sets in.
This is followed by emission of dimers which lowers the number of
remaining particles and results in a reappearing barrier.  The part of
the wave function trapped at large distances in the second minimum can
then stabilize into a condensate with fewer particles.  The time scale
for these processes should then be between the macroscopic-collapse
time and the recombination time at the second minimum.

This makes the assumption that no other degrees of freedom are
exploited, for example the angular dependence of the wave function or
molecular bound states described by other adiabatic potentials.
Direct population of two-body bound states requires inclusion of the
adiabatic potential asymptotically describing these states.  This is
possible within the model, but constitutes a major numerical
investigation of coherent atoms and molecules, oscillations between
them, and three-body recombinations within the same framework.  Other
time scales due to these neglected degrees of freedom could possibly
turn up in such a complete study of the dynamics of a many-boson
system.

\maccol

\subsection{Observation of Efimov-like states}

\label{sec:observ-efim-like}

\efiN

The recombination probability increases with decreasing hyperradius
due to the higher density, i.e.~several particles are close in space
and therefore much more likely recombine into molecular states.  The
time scale $\tau\sub{rec}$ for three-body recombination is given by
$N(t)=N(0)\exp(-t/\tau\sub{rec})$ where $N$ is the number of atoms in
the condensate.  This is as a function of the root-mean-square
hyperradius $\bar\rho$ estimated by
\refeq.~(\ref{eq:threebodyrectime}).  This recombination time for the
highest-lying Efimov-like states with $\bar\rho \sim N^{7/6} |a_s|$,
see \smarteq{eq:intermediate_region}, can then be compared to the time
scale for motion in the condensate which is given by $\tau\sub{trap} =
2\pi/\omega$.  With $\bar\rho\simeq N^{7/6}|a_s|$ we obtain
\begin{eqnarray}  
  \frac{\tau\sub{rec}}{\tau\sub{trap}}
  \simeq
  \frac{N^2}{\pi}
  \bigg(\frac{N|a_s|}{b\sub t}\bigg)^2
  \label{e47}
  \;.
\end{eqnarray}
Thus, close to the limit of stability, i.e.~$N|a_s|/b\sub t \sim 0.5$,
we have $\tau\sub{rec} \gg \tau\sub{trap}$ for $N \gg 1$, so the
recombination process is rather slow for these highest-lying
Efimov-like states.  Even though the lifetime is shorter than for the
initially created condensate, it might be long enough for an
observation of these states.

If the Efimov-like states are populated in experiments where the
potential suddenly is changed from \reffig.~\ref{fig:simul}a to
\reffig.~\ref{fig:simul}d, they could possibly be indirectly observed.
A signature of this many-body Efimov effect would be observation of
the diatomic molecules formed in the recombination process and with
the estimated time scale from \refeq.~(\ref{e47}). The rate should
then be inversely proportional to the square of the scattering length
reached after changing the potential.  The dimers can probably not be
distinguished from this and other processes, but the measured rate can
possibly be separated into different characteristic components.  Since
their lifetime due to recombination processes can be very large
compared to the time scale defined by the external field, these
negative-energy self-bound many-body states should essentially
maintain their spatial extension after the external field is switched
off.  This is in contrast to positive-energy states where only the
trap prevents expansion.  Thus, a relatively slow time evolution of
the density distribution without external field should be
characteristic for these many-body Efimov-like states.  A later
measurement of a system denser than expected for a positive-energy
system could then be a signature of the self-bound many-body state.

\efiN

\section{Summary}

The stability criterion for the many-boson system is verified within
the hypersperical framework.  The avoided collapse at large densities,
compared to collapse of the Gross-Pitaevskii description, allows a
more detailed study of what happens during the macroscopic-collapse
process.  Time scales are estimated on the basis of three-body
recombination at different densities.  Macroscopic-tunneling times and
oscillation times in a ``free-fall'' collapse allow the existence of
the self-bound states without recombination to other
cluster-structures.  However, since the critical region occurs when a
zero-range interaction in a non-correlated model describes the system
rather well, the inclusion of two-body correlations does not alter the
criterion for macroscopic stability.  This is seen by considering a
trapped gas with $n\sim1/b\sub t^3$.  Close to threshold $N|a_s|/b\sub
t\sim1$ implies $n|a_s|^3\sim1/N^3$.  This means that
$n|a_s|^3<1/N^2$, which is the asymptotic region where
$\lambda\simeq\lambda_\delta$ and the GPE method is in agreement with
the hyperspherical correlated method, see section \ref{sec:validity}.
In this sense the correlations do not modify the expectations obtained
from a mean-field consideration.  However, the possibility for a study
of couplings between the coherent many-body system and the bound
two-body channels is clearly beyond the mean field.  This is a goal
for future investigations of decays.

\chapter{Deformed boson system}                               
\label{kap:deformed}
The atom traps in experiments are of cylindrical geometry, as
described in section \ref{sec:trapped-bosons}.  For $N$ attractive
atoms the stability criterion, as described in section
\ref{sec:stability-criterion}, is experimentally established to be $N
|a_s| / b\sub t < 0.55$ \cite{cla03}, where $a_s$ is the scattering
length and $b\sub t\equiv \sqrt{\hbar/(m \omega)}$ is the relevant
length scale of the harmonic trap of geometric average frequency
$\omega \equiv\sqrt[3]{\omega_x\omega_y\omega_z}$.  A reduction from
three dimensions to effectively one or two dimensions was observed
experimentally \cite{gor01b} in the limit when the interaction energy
is small compared to the level spacing in the tightly-confining
dimension.  Experiments with continuous variation of the trap geometry
from three to either one or two effective dimensions \cite{gor01b},
with a two-dimensional structure \cite{gre01,ryc03}, and an effective
one-dimensional geometry \cite{tol03} request a corresponding
theoretical description.

Theoretical interpretations and the underlying analyses are frequently
based on model assumptions of spherical symmetry \cite{boh98,adh02b},
as discussed in section~\ref{sec:trapped-bosons}.  Confinement to
lower dimensions can also be studied directly without the
three-dimensional starting point.  This has been done with a
variational calculation in Gross-Pitaevskii equation (GPE)
\cite{bay96} and more recently in the GPE with variational dimensionality
\cite{mck02b}.  Also effects on stability of deformed
external fields have been investigated by use of the GPE formulation
\cite{bay96,gam01,adh01c}.  Extreme deformations could result in effective
one-dimensional or two-dimensional systems which can be described by
effective interactions of corresponding discrete lower dimensions
\cite{ols98,pet00,pet01b,lee02}.

In the present chapter, which corrects and extends the discussions in
reference \cite{sor03d}, we rewrite the hyperspherical formulation
from chapter \ref{kap:hyperspherical_method} to account for a general
deformation of the external field.  Since two-body correlations are
not yet included in the wave function, this hyperspherical approach
reminds of a mean-field treatment.  We investigate the stability
criterion in section \ref{sec:stabilitydeformed}.  Section
\ref{sec:effdim} contains an approach to an effective dimension which
depends on the deformation of the external field.  Since the
interactions are presently not included in this effective dimension,
we therefore in section \ref{sec:effint} introduce them on top of the
derived $d$-dimensional Hamiltonian.  Although the choice of
interactions is not unique, we can with some guess obtain an
alternative stability criterion and subsequently interpret the results
in terms of a deformation-dependent coupling strength, which is
finally compared with known results.

\section{Hyperspherical description of deformation}

As described in section \ref{sec:trapped-bosons} a combination of
magnetic fields results in an effective trapping potential, which can
be described as the deformed harmonic oscillator potential
$V\sub{trap}$ acting on all the identical particles of mass $m$
\begin{eqnarray}
  V\sub{trap}(\boldsymbol r_i)
  =
  \frac{1}{2}m
  \big(\omega_x^2 x_{i}^2
  +\omega_y^2y_{i}^2
  +\omega_z^2z_{i}^2\big)
  \;.
\end{eqnarray}
The hyperradius $\rho$ is the principal coordinate, which is separated
into the components $\rho_x$, $\rho_y$, and $\rho_z$ along the
different axes, i.e.
\begin{eqnarray}
  \rho^2
  =
  \frac{1}{N}
  \sum_{i<j}^Nr_{ij}^2 = \rho_x^2+\rho_y^2+\rho_z^2
  = \rho_\perp^2 + \rho_z^2
  \;,\qquad
  \rho_\perp^2\equiv \rho_x^2+\rho_y^2
  \;,
\end{eqnarray}
where $\boldsymbol r_{ij} = \boldsymbol r_j - \boldsymbol r_i$.  In
the centre-of-mass system the remaining coordinates are given as
angles collectively denoted by $\Omega$, see analogies in chapter
\ref{kap:hyperspherical_method} and more details in
appendix~\ref{sec:hypercyl-coord}.

An application here of the method presented in
chapter~\ref{kap:hyperspherical_method} is to assume a relative wave
function as a sum of two-particle components.  In the case of a
spherical trapping field each two-body component only needs dependence
on $\rho$ and the two-body distance $r_{ij}=\sqrt2\rho\sin\alpha_{ij}$
through an angle $\alpha_{ij}$.  For a deformed external field it also
needs dependence on the angle $\vartheta_{ij}$ between the interatomic
vector $\boldsymbol r_{ij}$ and the axis of the external field.  The
two-body component should in the cylindrical case then be on the form
$\phi(\rho,\alpha_{ij},\vartheta_{ij})$, which would lead to an
angular equation in the two variables $\alpha_{12}$ and
$\vartheta_{12}$ with more complicated integrals than those appearing
in \smarteqs{eq:faddeev_eq} and (\ref{eq:simplified_var_eq}).  We will
not investigate this, but here restrict ourselves to no dependence on
hyperangles.  This is expected to dominate for dilute systems where
the large distances average out directional dependence.

Thus, we neglect correlations in analogy to a mean-field treatment, so
in the dilute limit the hyperangular average of the relative
Hamiltonian is
\begin{eqnarray}
  &&
  \langle \hat H\rangle_\Omega
  \;\to\;
  \hat H
  =
  \hat H_{x} + \hat H_{y} + \hat H_{z}
  +  \hat V
  \;,\qquad
  \hat V = \sum_{i<j}^N
  \langle V_{ij}  \rangle_\Omega
  \;,
  \\
  &&
  \frac{2m\hat H_q}{\hbar^2}
  =
  -\frac{1}{\rho_q^{d(N-1)-1}}
  \frac{\partial}{\partial\rho_q}
  \rho_q^{d(N-1)-1}
  \frac{\partial}{\partial\rho_q}
  +\frac{\rho_q^2}{b_q^4}
  \label{eq:gend1}
  \;,
\end{eqnarray}
where $d=1$ and $b_q^2 \equiv \hbar/(m\omega_q)$ for $q=x,y,z$. The
interactions $V_{ij}$ are averaged over all angles $\Omega$, which for
the zero-range interaction $4\pi\hbar^2a_s\delta(\boldsymbol
r_{ij})/m$ from \smarteq{eq:zerorange-potential} for $N\gg1$ yields
\begin{eqnarray}
  \hat V
  = \frac{4\pi\hbar^2 a_s}{m} \sum_{i<j}^N  \langle
  \delta(\boldsymbol r_{ij})   \rangle_\Omega  = 
  \frac{\hbar^2}{2m}
  \frac{1}{2\sqrt{\pi}}
  N^{7/2}
  \frac{a_s}{\rho_x \rho_y \rho_z}  
  \;.
\end{eqnarray}
If we replace as $\rho_x=\rho_y=\rho_z=\rho/\sqrt3$, this is identical
to $\hbar^2\lambda_\delta/(2m\rho^2)$ with $\lambda_\delta$ from
\smarteq{eq:lambda_delta}.

We define the following dimensionless coordinates and parameters:
\begin{eqnarray}
  &&
  x\equiv \frac{\rho_x}{b_x} \sqrt{\frac{2}{N}}
  \;, \qquad
  y\equiv \frac{\rho_y}{b_y} \sqrt{\frac{2}{N}}
  \;, \qquad
  z\equiv \frac{\rho_z}{b_z} \sqrt{\frac{2}{N}}
  \;,
  \\
  &&
  \beta \equiv \frac{b_x^2 + b_y^2}{2 b_z^2}
  \;, \qquad
  \gamma \equiv \frac{b_x^2 - b_y^2}{2 b_z^2}
  \;, \qquad
  s\equiv \frac{Na_s}{b\sub t}
  \;, \qquad
  b\sub t^3\equiv b_x b_y b_z
  \;.\;
\end{eqnarray}
The deformation along the different axes is then described by $\beta$
and $\gamma$, and $s$ is the effective interaction strength.  The
Schr\"odinger equation $\hat H F(\rho_x,\rho_y,\rho_z) =
EF(\rho_x,\rho_y,\rho_z)$ is rewritten with the transformation
\begin{eqnarray}
  f(x,y,z)
  \propto
  (xyz)^{(N-2)/2} F(\rho_x,\rho_y,\rho_z)
  \label{eq:fFtrans}
\end{eqnarray}
in order to avoid first derivatives.  We then obtain
\begin{eqnarray}
  &&
  \bigg[
  -\frac{1}{\beta+\gamma}\frac{\partial^2}{\partial x^2}
  -\frac{1}{\beta-\gamma}\frac{\partial^2}{\partial y^2}
  -\frac{\partial^2}{\partial z^2}
  \nonumber\\
  &&
  \qquad
  +\frac{N^2u(x,y,z)-\varepsilon}{2\sqrt[3]{\beta^2-\gamma^2}}
  \bigg]
  f(x,y,z)=0
  \label{eq:gend16}
  \;,\\
  &&
  u(x,y,z)
  =
  \frac12\sqrt[3]{\beta^2-\gamma^2}\bigg[
  \frac{1}{\beta+\gamma}\Big(x^2+\frac{1}{x^2}\Big)
  \nonumber\\
  &&
  \qquad
  +\frac{1}{\beta-\gamma}\Big(y^2+\frac{1}{y^2}\Big)
  +z^2+\frac{1}{z^2}\bigg]
  +\sqrt{\frac{2}{\pi}}
  \frac{s}{xyz}
  \label{eq:gend17}
  \;,\quad
\end{eqnarray}
where $\varepsilon \equiv 2NE/(\hbar\omega)$.  Without interaction,
i.e.~$a_s=0$, the ground-state solution is
\begin{eqnarray}  \label{eq:gend27}
  f(x,y,z) = (xyz)^{(N-2)/2} \exp[-N(x^2 + y^2 + z^2)/4] 
  \;,\quad
\end{eqnarray}
which for $N\gg1$ is peaked at $(x,y,z)=(1,1,1)$.

\begin{figure}[htb]
  \centering
  \psfig{figure=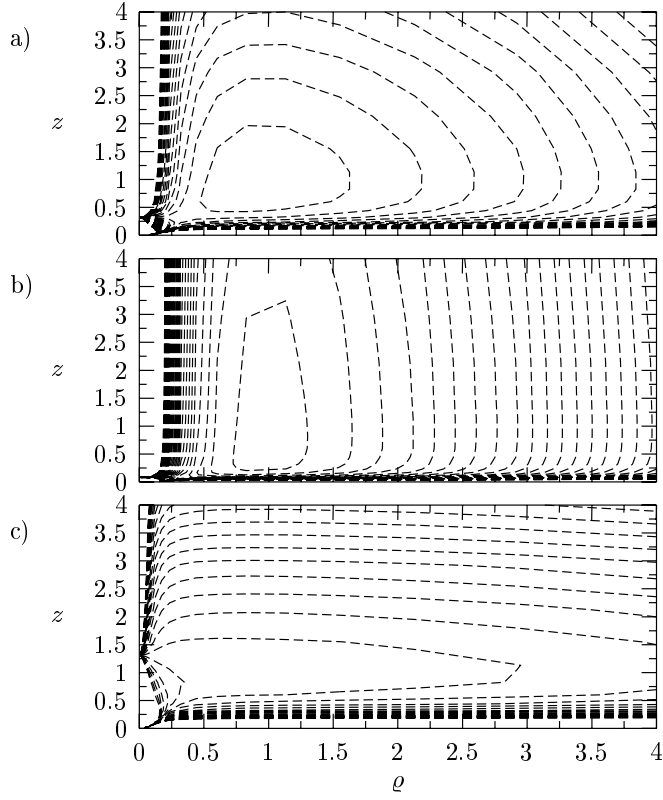,%
    bbllx=3.7cm,bblly=9cm,bburx=13.8cm,bbury=20.5cm,angle=0,width=9cm}
  \caption[Contour plots of effective potential for deformed boson system]
  {Contour plots of $u(x,y,z)$, \refeq.~(\ref{eq:gend17}), with
    $x=y=\varrho$ as a function of $(\varrho,z)$ for
    $s=-0.4\beta^{1/6}$ corresponding to $Na_s/b_\perp=-0.4$ for three
    deformations. The values for the contours change by 2, 2, and 5,
    respectively for a) $\beta=1$ (spherical), b) $\beta=1/16$
    (cigar-shaped or prolate), and c) $\beta=16$ (pancake-shape or
    oblate).}
  \label{olel3fig1}
\end{figure}

Here we do not solve this equation, but instead investigate the
character of the effective potential $u$.  For axial symmetry around
the $z$ axis the $x$ and $y$ directions cannot be distinguished, that
is when $\gamma=0$ and $\beta=b_\perp^2 / b_z^2$ with $b_\perp^2\equiv
b_xb_y$.  This symmetry amounts to replacing $\rho_x^2$ and $\rho_y^2
$ by $\rho_{\perp}^2/2$ in the equations.  A convenient definition for
this case is $2\varrho^2 \equiv x^2 + y^2$.  Equipotential contours of
$u$ in the $(\varrho,z)$ plane for $\varrho=x=y$ are shown in
\reffig.~\ref{olel3fig1} for attractive interactions.  For $a_s<0$
($s<0$) there is always a divergence towards $ - \infty$ when
$(\varrho,z) \rightarrow (0,0)$, see \refeq.~(\ref{eq:gend17}).
However, a stationary minimum is seen in both
\reffigs.~\ref{olel3fig1}a (spherical symmetry) and \ref{olel3fig1}b
(prolate) close to $(\varrho,z)=(1,1)$, whereas this minimum has
disappeared for the oblate system in \reffig.~\ref{olel3fig1}c.  For
very weak attraction a stationary minimum is present for all
deformations.

\Reffig.~\ref{olel3fig2} shows cuts of the potential $u(\varrho,z)$ along
paths close to the bottom of the valleys (see inset).  The spherical
minimum (full line) is shielded by a relatively small barrier from the
divergence for $\varrho\to0$.  The minimum for the prolate deformation
(dashed curve) is extremely stable although the divergence for
$\varrho\to0$ still exists.  For the oblate deformation (dot-dashed
line) the local minimum has vanished for this attraction strength.
\begin{figure}[htb]
  \centering
  \input{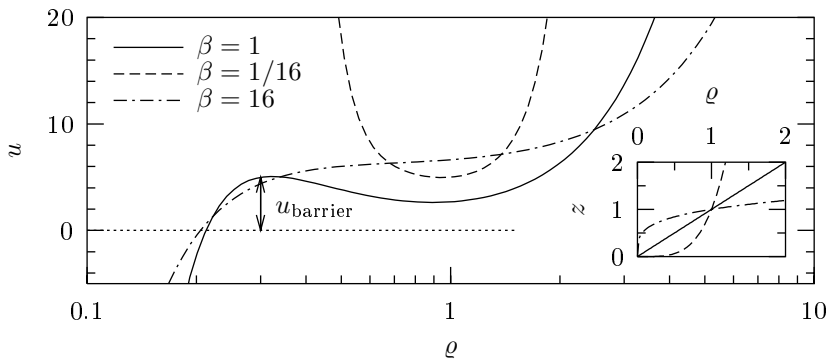}
  \caption[Effective potential along cuts for different deformations]
  {The potential $u(x,y,z)$ for $s=-0.4\beta^{1/6}$ as a function of
    $\varrho=x=y$ along cuts of the $(\varrho,z)$ plane where
    $z=\varrho^{1/\sqrt{\beta}}$.  The height $u\sub{barrier}$ of the
    local maximum at top of the barrier is indicated for the spherical
    case $\beta=1$.  The inset shows corresponding trajectories in the
    $(\varrho,z)$ plane, compare with \smartfig{olel3fig1}, for the
    three deformations.}
  \label{olel3fig2}
\end{figure}

\section{Stability criterion for bosons in a deformed trap}

\label{sec:stabilitydeformed}

The barrier height depends on the deformation of the external field,
see \reffigs.~\ref{olel3fig1} and \ref{olel3fig2}.  Extrema
$(x_0,y_0,z_0)$ of $u$ in \refeq.~(\ref{eq:gend17}) obey the three
equations obtained from
\begin{eqnarray} 
  \frac{b\sub t^2}{b_x^2}(x_0^4-1)
  =
  \sqrt{\frac{2}{\pi}}\frac{sx_0}{y_0z_0}
  \label{eq:1}
\end{eqnarray}
and symmetric permutations of $x$, $y$, and $z$.  This can be used to
determine the critical strength $s$ when a local minimum disappears.
The results for axial symmetry are shown as the thin solid line in
\reffig.~\ref{olel3fig3}.  In these units the critical strength $s$ is
largest for a geometry very close to spherical.  Since $s=Na_s/b\sub
t$ and $b\sub t^3=b_xb_yb_z$, this means that at fixed $b\sub t^3$, or
fixed volume, the scattering length can assume the largest negative
value for the spherical trap.  \etalcite{Gammal}{gam01} performed a
time-dependent study with the Gross-Pitaevskii equation (GPE) which
resulted in the critical strengths here shown as the dotted line.
This is in large regions lower than the present result, which might be
due to our neglect of quantum effects and time dependence.  The energy
gain due to correlations is not included in the simple expectation
value with the assumed angle-independent wave function.  The recent
value for the experimental stability region
\cite{cla03} is shown as the plus and agrees with the mean-field
model.  We perform two alternative derivations withing the
hyperspherical model in order to see if the assumptions need
modifications.

\begin{figure}[htb]
  \centering
  \input{threshold_gammal}
  \caption[Critical strength as a function of deformation] {The
    critical strength $|s|=N|a_s|/b\sub t$ as a function of the
    deformation $\beta = b_\perp^2/b_z^2$ from the potential in
    \refeq.~(\ref{eq:gend17}) (thin solid line), from
    \refeq.~(\ref{eq:gend18}) (dashed line), and from a mean-field
    Gross-Pitaevskii computation by \etalcite{Gammal}{gam01} (dotted
    line).  The thick solid line is \refeq.~(\ref{eq:crit_zeropoint})
    obtained by considering the zero-point energy.  The plus is the
    experimentally measured value \cite{cla03}.  Regions below curves
    are considered stable in the separate treatments.  The double- and
    triple-dashed lines indicate the effective cross-overs to two
    (2D*) and one (1D*) dimensions \cite{gor01b}.}
  \label{olel3fig3}
\end{figure}

These results can be compared to an analytical ``spherical''
approximation where the radial motion is described by only $\rho$
while the deformed external field remains the same.  The effective
radial potential $U$ is then obtained by adding centrifugal barrier
and the contributions from zero-range interaction and the external
field, see \refeq.~(\ref{eq:radial.potential}).  The angular average
replaces each of the three components $\rho_q^2$ and $R_q^2$ by
$\rho^2/3$ and $R^2/3$, where $\boldsymbol R$ is the centre-of-mass
coordinates, i.e.
\begin{eqnarray}
  &&
  \sum_{i=1}^N \langle V\sub{trap}(\boldsymbol r_i) \rangle_{\Omega}
  =
  \frac12m \frac{\omega_x^2+\omega_y^2+\omega_z^2}{3} (\rho^2 + N R^2)
  \;,
  \\
  &&
  \frac{2m \hat V}{\hbar^2}
  =
  8\pi a_s
  \sum_{i<j}^N \langle
  \delta(\boldsymbol r_{ij})   \rangle_\Omega 
  = 
  \frac{3}{2}
  \sqrt{\frac{3}{\pi}}
  N^{7/2}
  \frac{a_s}{\rho^3}
  \label{eq:vdelta_copy1}
  \;,
  \\
  &&
  \frac{2mU(\rho)}{\hbar^2}  
  =
  \frac{3}{2}
  \sqrt{\frac{3}{\pi}}
  N^{7/2}
  \frac{a_s}{\rho^3} + 
  \frac{9N^2}{4\rho^2}+
  \frac{\rho^2}{l_2^4} 
  \;,
  \label{eq:ueff_deformed}
\end{eqnarray}
where $3l_2^{-4} \equiv b_x^{-4}+b_y^{-4}+b_z^{-4}$.  By comparison
with \refeqs.~(\ref{eq:u_delta}), (\ref{eq:uABC_x}), and
(\ref{eq:uABC_x_stabilitycrit}) the stability condition becomes
\begin{eqnarray}
  &&
  \frac{N|a_s|}{b\sub t} <  k(\beta,\gamma)
  \;,\qquad
  k(1,0)=\frac{2\sqrt{2\pi}}{5^{5/4}} \simeq 0.67   
  \label{eq:gend18}
  \;,
  \\
  &&
  k(\beta,\gamma)
  =
  k(1,0)
  \sqrt[4]{\frac{3(\beta^2-\gamma^2)^{4/3}}
    {2\beta^2+2\gamma^2+(\beta^2-\gamma^2)^2}}
  \label{eq:gend29}
  \;.
\end{eqnarray}

The spherical limit corresponds to $\gamma=0$ and $\beta=1$ where the
barrier is present when $|s|= N |a_s| / b\sub t < 0.67$.  The result
for a cylinder (only $\gamma=0$) is shown as the dashed line in figure
\ref{olel3fig3} and is noticably different from, but numerically
almost coincides with, the ``deformed'' treatment, thin solid line.
An extreme oblate deformation corresponds to the two-dimensional limit
where $b_z\ll b_\perp$ and $\beta\to\infty$.  Here
\refeq.~(\ref{eq:gend29}) yields the critical strength $k \simeq
0.4\sqrt[4]{0.6}\sqrt{\pi/2}\beta^{-1/3}$.  As seen from the contour
plot in \reffig.~\ref{olel3fig1}c, the motion is now almost confined
at $z=1$.  From $x=y=\varrho$ we see that $u(\varrho,\varrho,1)$ only
has a local minimum when $|s|<\sqrt{\pi/2}\beta^{-1/3}$, which is
larger than the value where the $z$ motion is not fixed.  This is
reasonable since more degrees of freedom in the model lowers the
energy.  Baym and Pethick \cite{bay96} obtained with a variational
study of the GPE the criterion $|s|<\sqrt{\pi/2}\beta^{-1/3}$ provided
that the variational width in the axial direction does not change due
to the interactions.  This is identical to the criterion from studying
the potential $u(\varrho,\varrho,1)$, i.e.~consistent with the fixed
value $z=1$.  This again emphasizes the equivalence between the
present hyperspherical non-correlated model and the mean-field GPE.

Analogously, in the extreme prolate limit (one-dimensional) where
$\beta\to0$, \refeq.~(\ref{eq:gend29}) yields the critical strength
$k\simeq0.25\sqrt[4]{1.25}\sqrt{\pi}\beta^{1/6}$.  However, fixing
$x=y=1$ in \smarteq{eq:gend17} yields no critical strength since
$u(1,1,z)$ always has a global minimum.  Therefore, the other degrees
of freedom are essential in this prolate limit.

A modified stability criterion can be obtained by considering the
ground-state energy $E_0$ of the boson system, which in the
non-interacting case is $E_0=\hbar(\omega_x+\omega_y+\omega_z)(N-1)/2$
where the centre-of-mass energy is subtracted.  The system is unstable
when this energy is larger than the barrier height $U\sub{barrier}$,
see the indication in figure \ref{olel3fig2} of the corresponding
height $u\sub{barrier}$ for the reduced potential.  With this
condition the criterion of stability is
\begin{eqnarray}
  \frac{N|a_s|}{b\sub t}
  <
  \frac12
  \sqrt{\frac\pi3}\frac{l_1}{b\sub t}
  \sqrt{1+\frac{1}{12}\frac{l_1^4}{l_2^4}}
  \;,
  \label{eq:crit_zeropoint}
\end{eqnarray}
where $3l_1^{-2} \equiv b_x^{-2}+b_y^{-2}+b_z^{-2}$.  This is seen in
figure~\ref{olel3fig3} (thick solid line) to be below the GPE
calculations \cite{gam01}.  The improvement is here substantial
compared to when the zero-point energy is neglected.  In particular,
for the spherical case we get $N|a_s|/b\sub t\simeq0.53$ instead of
$N|a_s|/b\sub t\simeq0.67$.

The estimate of \smarteq{eq:crit_zeropoint} describes the stability
problem better since it includes the quantum effect due to the
zero-point energy.  The GPE calculation is time dependent and thus
describes the dynamics even better and is also closest to the
experimental value.  Since we expect the present non-correlated
hyperspherical treatment to be in agreement with the mean field, we
expect that a time-dependent treatment in this frame would yield the
same result as obtained by \etalcite{Gammal}{gam01}.

A recent variational Monte Carlo investigation of the stability
criterion in elongated, almost one-dimensional, traps yielded the
stability criterion $n\sub{1D}a\sub{1D} \lesssim 0.35$ \cite{ast03},
where $n\sub{1D}\sim N/b_z$ is the density in one dimension and
$a\sub{1D} = -b_\perp(b_\perp/a_s-1.0326)$.
\Refeq.~(\ref{eq:crit_zeropoint}) can in the one-dimensional limit be
written as $N|a_s|/b_\perp \lesssim 0.66$.  The deviation between the
two results might be due to our use of a three-dimensional zero-range
interaction in this non-correlated model, whereas
\etalcite{Astrakharchik}{ast03} used a one-dimensional model with a
zero-range interaction with coupling strength proportional to
$1/a\sub{1D}$ as well as a full 3D correlated model with hard-sphere
or finite-range potentials.  An effective potential analogous to
$\delta(x)/a\sub{1D}$ in the general case with intermediate
deformations would be a rewarding goal.

According to \etalcite{G\"orlitz}{gor01b} the interaction energy is
smaller than the energy in the tightly-confining dimension when
$|s|\le\sqrt{32/225}\beta^{-5/6}$ for the 1D limit and when $|s|\le
\sqrt{32/225}\beta^{5/3}$ for the 2D limit.  These cross-overs are 
indicated by double-dashed (two-dimensional) and triple-dashed
(one-dimensional) lines in \smartfig{olel3fig3}.  Since the critical
region in each limit is below the relevant cross-over, stable and
strongly deformed systems can be regarded as effectively one- or
two-dimensional in the sense of these energy relations.

\section{Effective dimension}

\label{sec:effdim}

The deformation of the external field effectively changes the
dimension $d$ of the space where the particles move.  The field
changes continuously and $d$ could correspondingly vary from three to
either two or one.  In order to arrive at such a description, we aim
at an effective $d$-dimensional Hamiltonian analogous to
\refeq.~(\ref{eq:gend1}) with only one radial variable $\rho$, a
deformation-dependent dimension $d$, and an effective trap length
$b_d$, i.e.
\begin{eqnarray}
  \frac{2m\hat H_d}{\hbar^2}
  =
  -\frac{1}{\rho^{d(N-1)-1}}
  \frac{\partial}{\partial\rho}
  \rho^{d(N-1)-1}
  \frac{\partial}{\partial\rho}
  +\frac{\rho^2}{b_d^4} + \frac{2m\hat V}{\hbar^2}
  \label{eq:gend6}
  \;,
\end{eqnarray}
where $\hat V$ represents all particle interactions in $d$ dimensions.
The requirement is that the Schr\"odinger equation $\hat H_d G_d = E_d
G_d$ with $d$-dimensional eigenfunction $G_d$ and eigenvalue $E_d$ is
obeyed, at least on average, i.e.
\begin{eqnarray}
  \int d\rho\;\rho^{d(N-1)-1}
  G_d^*(\rho)(\hat H_d-E_d)G_d(\rho)
  =
  0
  \;.
  \label{eq:gend7}
\end{eqnarray}

The lowest free solution, that is with $\hat V=0$, is given by
\smarteq{eq:gend27}.  In the cylindrical case we can relate the
$d$-dimensional function $G_d$ to this by performing the average with
respect to the angle $\theta$ in the parametrization $(\rho_\perp,
\rho_z)=\rho( \sin \theta, \cos \theta)$.  With inclusion of the
corresponding volume elements, see appendix \ref{sec:hypercyl-coord},
this leads to
\begin{eqnarray}
  \rho^{d(N-1)-1}|G_d(\rho)|^2 
  =
  \rho^{3(N-1)-1}
  \int_0^\pi d\theta \cos^{N-2}\theta\sin^{2N-3}\theta 
  |F(\rho,\theta)|^2
  \label{eq:gend40}
  \;,
\end{eqnarray}
where $F(\rho,\theta)$ can be obtained by rewriting
\smarteqs{eq:gend27} and (\ref{eq:fFtrans}).

The characteristic energy and length can be defined by
\begin{eqnarray}
  E_d =
  \frac{d\hbar^2}{2mb_d^2}(N-1)
  \;,\qquad
  db_d^2=2b_\perp^2+b_z^2
  \;,
  \label{eq:gend3}
\end{eqnarray}
which clearly is correct in the three limits, i.e.~spherical: $d=3$
and $b_d = b_z = b_\perp$, two-dimensional: $d=2$ and $b_\perp\gg
b_z$, and one-dimensional: $d=1$ and $b_z\gg b_\perp$.

In general it is not possible to find one $\rho$-independent set of
constants ($E_d,b_d,d$) such that $\hat H_d G_d = E_d G_d$.  Instead
we insist on the average condition in \refeq.~(\ref{eq:gend7}) with
$G_d$ and $E_d$ from \refeqs.~(\ref{eq:gend40}) and (\ref{eq:gend3}).
The result for axial geometry is a second-degree equation in $d$ with
one physically meaningful root, see details in
appendix~\ref{sec:deriv-eff-dim}.  The results for various $N$ values
are shown in \reffig.~\ref{olel3fig4}.
\begin{figure}[htb]
  \centering
  \input{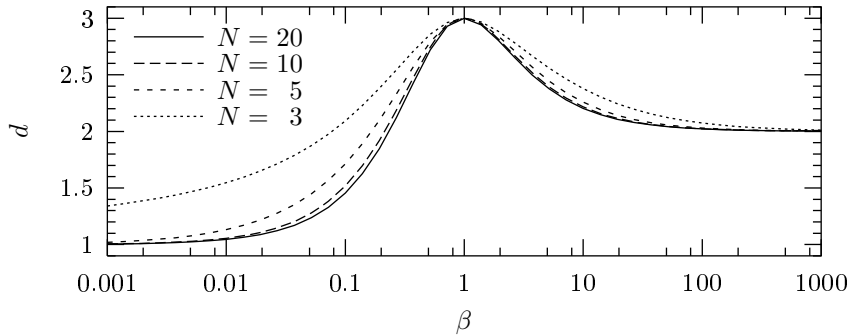}
  \caption[Effective dimension as a function of deformation] 
  {The effective dimension $d$ obtained as a function of the
    deformation parameter $\beta=b_\perp^2/b_z^2$.  Curves for larger
    $N$ are very close to that for $N=20$.}
  \label{olel3fig4}
\end{figure}
The effective dimension depends on $N$ for relatively small particle
numbers.  When $N>20$, the curve is essentially fixed.  Furthermore,
the asymptotic values of both $d=1$ (small $\beta$) and $d=2$ (large
$\beta$) are reached faster for larger $N$ since many particles feel
the geometric confinement stronger than few particles.\footnote{It may
be amusing to speculate on the meaning of $d=2$ for elongated
cigar-shaped confinement ($0.1\lesssim\beta\lesssim0.2$).}  Since
these effective dimensions are obtained as average values over $\rho$,
the system might look spherical at large distances and strongly
deformed at small distances, on average resulting in the curves in
\reffig.~\ref{olel3fig4}.

\section{Deformation-dependent interactions}

\label{sec:effint}

The effective dimension for the non-interacting system possibly
changes when interactions are included.  The steps of the previous
section should in principle be repeated with the interactions.
However, this would be complicated and miss the goal which is a simple
effective Hamiltonian with a renormalized interaction in lower
dimension, see analogies in the references \cite{ols98,pet00,lee02}.

We therefore start out with a two-body contact interaction with a
coupling strength which is modified due to the deformation.  This is
in line with the renormalization in section \ref{sec:densdepint} due
to the inclusion of two-body correlations.  So we write a
$d$-dimensional zero-range interaction with a dimension-dependent
coupling strength $g(d)$ as
\begin{eqnarray}
  V_d(r_{ij})
  =
  g(d) \delta^{(d)}(r_{ij})
  \;,\qquad
  g(3)=\frac{\hbar^2a_s}{m}
  \label{eq:gend36}
  \;,
\end{eqnarray}
where this ``$d$-dimensional $\delta$ function'' is defined by
$\delta^{(d)}(r)=0$ for $r\ne0$ and $\int_0^\infty
dr\;r^{d-1}\delta^{(d)}(r) = 1$.  The distance between two particles,
e.g.~particle 1 and 2, is in hyperspherical coordinates defined by
$r_{12}=\sqrt{2} \rho \sin \alpha$, where the angle $\alpha$ enters
the angular volume element as
\begin{eqnarray}
  d\Omega_\alpha
  = d\alpha\sin^{d-1}\alpha \cos^{d(N-2)-1}\alpha
  \;.
\end{eqnarray}
This is valid at least for $d=1,2,3$, see
appendix~\ref{sec:n-particles-d}.  The effective interaction $\hat V$
in \refeq.~(\ref{eq:gend6}) is for $N\gg1$ then given by the average
over all coordinates except $\rho$:
\begin{eqnarray}
  \hat V
  =
  \frac{N^2}{2}
  \frac{\int_0^{\pi/2}d\Omega_\alpha \;
    V_d(\sqrt2\rho\sin\alpha)}
  {\int_0^{\pi/2}d\Omega_\alpha}
  =
  \frac{\hbar^2}{2m}
  \frac{2N^2(Nd/4)^{d/2}}{\Gamma(d/2)}
  \frac{a_s}{\rho^d}
  \frac{g(d)}{g(3)}
  \label{eq:hatvd}
  \;.
\end{eqnarray}
However, this does not yield instability for $d<2$ since the power $d$
in $\rho^{-d}$ is smaller than two, see section
\ref{sec:stability-criterion}.

We therefore pursue another approach.  Inspired by the forms of
\refeqs.~(\ref{eq:vdelta_copy1}) and (\ref{eq:hatvd}), we write $\hat
V$ as
\begin{eqnarray}
  \hat V
  =
  \frac{\hbar^2}{2m}
  \frac{2N^2(Nd/4)^{p/2}}{\Gamma(d/2)}
  \frac{a_d}{\rho^p}
  \;,\qquad
  a_3=a_s
  \;,
\end{eqnarray}
which with $a_3=a_s$ coincides with the result for $d=3$ if we choose
$p=3$.  The effective potential $U_{d}$ in the $d$-dimensional
Schr\"odinger equation corresponding to \refeq.~(\ref{eq:gend6}) is
then
\begin{eqnarray}
  \frac{2mU_{d}(\rho)}{\hbar^2}  =
  \frac{2N^2(Nd/4)^{p/2}}{\Gamma(d/2)}
  \frac{a_d}{\rho^p}
  + 
  \frac{d^2N^2}{4\rho^2}+
  \frac{\rho^2}{b_d^4} 
  \;.
\end{eqnarray}
For $p<2$ this potential always has a global minimum and thus no
collapse is present.  For $p>2$ there is always divergence to
$-\infty$ when $\rho\to0$.  For weak attraction, i.e.~small $|a_d|$,
there is a local minimum.  This disappears at larger $|a_d|$ when
\begin{eqnarray}
  \frac{N|a_d|}{b\sub t}
  >
  \frac{b_d^{p-2}}{b\sub t}
  \frac{2^{1+p/2}d(p-2)^{(p-2)/4}\Gamma(d/2)}{p(p+2)^{(p+2)/4}}
  \;.
  \label{eq:stability_effective_interaction}
\end{eqnarray}
The criterion in \refeq.~(\ref{eq:gend18}) was also obtained by
estimating when the critical point vanished.  \Smarteq{eq:gend18} is
valid for all deformations, i.e.~any $d$.  In order to be able to
compare \smarteqs{eq:gend18} and
(\ref{eq:stability_effective_interaction}), we therefore choose $p>2$
such that \smarteq{eq:stability_effective_interaction} always is
applicable.  When \refeqs.~(\ref{eq:stability_effective_interaction})
and (\ref{eq:gend18}) agree, the effective interaction strength $a_d$
is given by
\begin{eqnarray}
  \frac{a_d}{a_s}
  =
  \frac{b_d^{p-2}}{b\sub t}
  \frac{2^{(p-1)/2}\Gamma(\frac{d}{2})5^{5/4}(p-2)^{(p-2)/4}}
  {\sqrt{\pi}(p+2)^{(p+2)/4}\beta^{1/6}p/d}
  \sqrt[4]{\frac{2+\beta^2}{3}}
  \label{eq:def_intstr}
  \;.
\end{eqnarray}
This effective interaction strength is in
\reffig.~\ref{fig:def_intstr} shown as a function of the deformation
for various choices of the power $p$.  The solid line shows the result
for $p=3$, which is known to be correct for $\beta=1$ ($d=3$).
Similarly the dashed line shows the result with $p=d$, which does not
work for $d<2$ ($\beta\lesssim0.2$).
\begin{figure}[htb]
  \centering
  \input{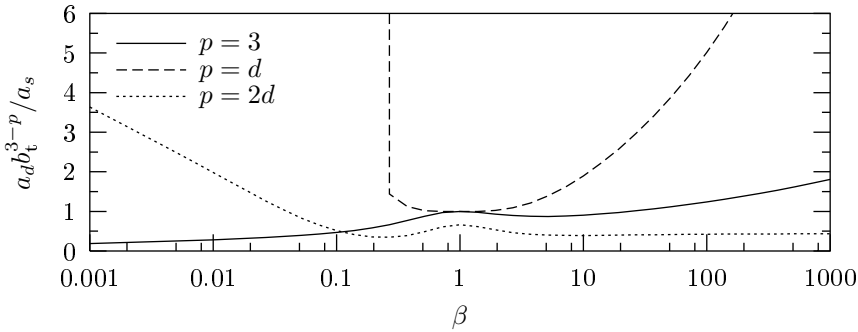}
  \caption
  [Effective interaction strength as a function of deformation] {The
    effective interaction strength $a_d$ from
    \refeq.~(\ref{eq:def_intstr}) obtained as a function of the
    deformation parameter $\beta=b_\perp^2/b_z^2$ in the large-$N$
    limit, i.e.~the connection between deformation and effective
    dimension obtained from the calculation for $N=20$ is used for
    this illustration.  The vertical divergence of the dashed line
    indicates the inadequacy of the corresponding method when $d<2$.}
  \label{fig:def_intstr}
\end{figure}
Since the effective coupling strength depends strongly on the power
$p$, we need further information about how the interactions enter the
effective potential.

An extreme deformation might lead to effectively one-dimensional or
two-dimensional properties.  Pitaevskii and Stringari \cite{pit03}
collected results for the effective coupling strength in two
dimensions that yields $g(2)=\sqrt{8\pi}\hbar^2a_s/(mb_z)$, whereas
the result from \smarteq{eq:def_intstr} in that limit is larger by the
factor $5^{5/4}/4\simeq 1.9$.  Even though the results differ by a
factor close to two, the right combination of lengths shows that we
have incorporated the degrees of freedom in the correct manner.  This
was also the case in the previous comparison of the stability
criterion with the one obtained by Baym and Pethick \cite{bay96}.
However, as was also mentioned by Pitaevskii and Stringari
\cite{pit03}, in the low-density limit in two dimensions the coupling
constant becomes density-dependent, which is beyond the present model
where correlations are neglected.

Since $p=d$ for $d=1$ does not yield a meaningful interpretation of
the stability criterion, a one-dimensional system needs a different
treatment.

\section{Discussion}

In conclusion, the hyperspherical method with a non-correlated
approach yielded stability criteria as a function of the deformation
of the external field.  For constant volume the highest stability was
found for spherical traps.  Effective dimensions $d$ continuously
varying between 1 and 3 were calculated as a function of the
deformation.  The system can possibly be described by a
$d$-dimensional effective radial potential with a $d$-dimensional
effective interaction.  However, this does not have an unambigious
form.  Applications to restricted geometries become simpler, where the
obtained two-dimensional coupling strength compares reasonably with a
coupling strength obtained by an axial average of a three-dimensional
contact interaction.  For the one-dimensional case an effective
coupling strength was not obtained.

A previous approach to a $d=1$ treatment by \etalcite{Gammal}{gam00}
shows that a three-body contact interaction is necessary for the GPE
to produce collapse in one spatial dimension.  In the present
framework a three-body contact interaction for a constant angular wave
function produces a hyperradial potential proportional to
$\rho^{-2d}$, compare with \refeq.~(\ref{eq:lambda_3body}), which for
any $d\ge1$ leads to instability if the three-body coupling strength
is sufficiently negative.  The dotted line in
\smartfig{fig:def_intstr} shows the effective coupling strength for
$p=2d$, corresponding to this three-body zero-range interaction.

\threecor

According to \etalcite{Astrakharchik}{ast03,ast03b} a Jastrow ansatz
for a correlated wave function and inclusion of two-body interactions
lead to collapse in one spatial dimension.  According to preliminary
Faddeev calculations with the two-body correlated model presented in
chapter~\ref{kap:hyperspherical_method}, a two-body interaction and
inclusion of only two-body correlations in one spatial dimension do
not lead to collapse.  It seems that at least three-body correlations
or three-body interactions are necessary in order to achieve a
realistic description of collapse in one dimension.

\chapter{Conclusions and perspectives}                        
\label{kap:conclusion}

The present thesis studied few-body correlations within a many-body
system, especially effects beyond the commonly applied mean field by a
method that is usually applied to clusterized systems.

Chapter~\ref{kap:hyperspherical_method} presented a hyperspherical
framework for including two-body correlations explicitly in the wave
function.  For bosons this was done as a sum of two-body amplitudes,
the Faddeev decomposition of the wave function.  One advantage of this
wave function is that it contains a signature of the average distance
between all particles.  In this respect it reminds of the mean field,
and it is indeed possible to relate the wave functions when the
interactions are sufficiently weak.  On the other hand, in the dilute
limit where three-body encounters are rare, this wave function reminds
of the Jastrow factorization into two-body amplitudes.  Thus, it
catches the information from the encounter of two particles at the
same time as remembering the background cloud of other particles.  In
the shape of a variational equation, the Schr\"odinger equation was
then reduced to a one-dimensional differential equation in a
hyperangle, plus a simple one-dimensional equation in the hyperradius.
A similar result was previously obtained by \etalcite{de la
Ripelle}{rip88} in terms of Faddeev-like equations.  However, the
present equation is variational and the complications compared to the
simpler Faddeev-like equations are not severe.

In chapter~\ref{kap:angular} we discussed analytical estimates of the
angular eigenvalues.  These provide results in the dilute limit in
agreement with expectations based on mean-field-like assumptions.
Then the numerical solutions confirm these results in the dilute limit
and furthermore provide significant deviations at larger densities.
In the regime of a very large two-body $s$-wave scattering length the
angular potential approaches a constant value which only depends on
the number of particles.  Also the signatures of a two-body bound
state are recognized, providing a possible link between the scattering
channels and the two-body bound channels within the many-boson system.

The macroscopic properties of a trapped system of bosons were
investigated in chapter~\ref{kap:radial}, where the radial equation
was solved for the size scale and total energy of the system.  Some of
the stationary solutions in the hyperradial potential turn out to have
a much smaller spatial extension than the typical size scale of a
Bose-Einstein condensate.  Furthermore, the total energy of many
states might be negative due to a large average attraction at large
densities, i.e.~when the bosons are close to each other.  The
Bose-Einstein condensate is usually unstable if the external
confinement is removed, whereas these negative-energy modes are
self-bound, and thus confined even if the trap is turned off.

\fermion

Chapter~\ref{kap:meanfield_validity} presented the basic assumptions
in the mean-field Gross-Pitaevskii (GP) treatment of a dilute boson
system.  We related the obtained angular potentials to a possible
density-dependent zero-range interaction.  This allows a possible
renormalization of the mean-field interaction for calculations with
the GP equation for denser systems.  The energies and average
distances for a dilute boson system obtained from the mean field are
very close to those obtained from the two-body correlated model.
However, the states with larger densities and negative energies have
no parallels in the mean-field model.  In this region we obtained
interaction energies independent of the scattering length.  Concerning
validity, chapter \ref{kap:meanfield_validity} gave estimates of the
validity ranges, extending the validity region for the two-body
correlated model to larger densities with deviations from the mean
field.

The stability criterion for a Bose-Einstein condensate is in the
present hyperspherical treatment of pairwise correlations obtained in
a way which is similar to the derivation from the mean field.  This
was evident in chapter~\ref{kap:stability_validity} where the
criterion was derived in terms of the hyperradial potential.
Three-body recombination and macroscopic tunneling were discussed
qualitatively, estimating the effects of degrees of freedom that are
not explicitly included in the two-body correlated model.  This is
related to discussions by \etalcite{Bohn}{boh98}.  The discussion of
macroscopic collapse when suddenly changing the underlying two-body
interactions is possible within the model.  This provides an estimate
of the collapse time which agrees on the order-of-magnitude level with
the measured time scale.

A possible improvement is the explicit inclusion of three-body
correlations in the ansatz for the wave function in order to study
three-body recombinations in the many-boson system.  This can also
indicate the validity of the present model, and tell if the self-bound
many-body states described in chapter \ref{kap:radial} have physical
relevance or if they are artificial products of the present ansatz
with two-body correlations.  A quantitative study of collapse dynamics
can be performed by studying the time-dependent problem with inclusion
of couplings between the different channels in the adiabatic expansion
of the wave function.

The investigation showed that the lifetime of some of the self-bound
many-body states might be so large that they can be observed in
experiments.  This might be done by turning off the trap, waiting for
some time, and then altering the two-body interaction such that the
particles repel each other and the system expands.  By extrapolating
back to the density profile before expansion, it might be concluded
that the system did not expand in the time period between the external
trap was turned off and the two-body interactions were made
effectively repulsive.

The effects of deformation were discussed in
chapter~\ref{kap:deformed}, which described the trapped boson system
in a non-correlated hyperspherical frame.  The external confining
field alters the stability criterion in agreement with the
experimentally measured criterion.  For a fixed volume any deviation
from the spherical geometry decreases the stability.  This was also
concluded by \etalcite{Gammal}{gam01} from a study of the
time-dependent GP equation.  The present treatment provides an
analytical stability criterion as a function of deformation, which
agrees well, but not perfectly, with the experimentally measured
value.  The deviations are probably due to the crudeness of the
approximations made for the analytical estimates, but not due to the
lack of correlations since these are also absent in a mean-field
treatment.  We furthermore seeked to use only one length scale to
describe the deformation.  For the non-interacting case this results
in an effective dimension, which then enters an effective Hamiltonian
in a single length scale.  However, the problem is the inclusion of
interactions, which does not appear trivial.  A proposed effective
potential provides a stability criterion, but does so in a
non-transparent way where the coupling to the two-body interactions
has vanished.  It seems that higher-order correlations or higher-order
interactions are crucial for a full understanding of stability
phenomena in effectively lower dimensions.


An immediate extension of the present work is to complete the
treatment of two-body correlations in lower dimensions and in the
general deformed system.  This could provide the wanted effective
interaction in lower dimensions and possibly confirm the results of
other approaches, e.g.~the references \cite{pet00,lee02} for the
two-dimensional cases.

The treatments in this thesis are performed at zero temperature.  The
effects of a finite temperature can possibly be included as a
statistical distribution of many-body states, where couplings between
the states then will play a larger role.  This might yield information
about the effect of pairwise correlations on the condensate fraction
and the transition temperature.

\fermion

Experiments with trapping of fermionic gases raise many questions
about the modification of correlations for fermions.  Especially the
problem of binary correlations between identical fermions is a great
challenge, but can potentially be built on top of a hyperspherical
frame.  Another approach is an extraction of a density-dependent
coupling strength for a fermion system.  This might also provide
answers to questions in other fields of physics, e.g.~in molecular
physics and nuclear physics, where the mean field is inadequate for
studies of exotic problems, for example nuclei close to a drip line.

In conclusion, the present study of two-body correlations yielded
insight into mechanisms that in the dilute limit can be accounted for
by a mean field, and yielded deviations especially in the presence of
a two-body bound state or a resonance.  The inclusion of three-body
correlations can be the crucial next step which provides answers to
questions about three-body recombination and the structure of
lower-dimensional systems.

\chapter{Sammendrag på dansk (Summary in Danish)}             
\label{kap:dansk}
\emph{Resum\'e: Afhandlingen ``Korte tilfældige sammenstød. Parvise
korrelationer blandt boson\-er'' omhandler teoretiske modeller for
parvise påvirkninger mellem atomer, der befinder sig i en gas af ens
partikler ved meget lav temperatur.  Teorien bag en model for opdeling
af et mangepartikelsystem i små grupper præsenteres, og denne model
anvendes i tilfældet hvor kun to partikler skiller sig ud.  Specielt
undersøges betydningen af store tætheder og kraftig vekselvirkning
mellem atomerne.  Dette viser afvigelser i forhold til en
middelfeltsmodel, hvor atomerne ikke har mulighed for at indrette sig
efter hinanden.  Studiet af cigarformede eller pandekageformede
systemer indikerer, at man må inkludere påvirkningen mellem tre atomer
for at forstå disse systemer til bunds.}

\vspace{.4cm}

Denne afhandling beskræftiger sig med korrelationer i bosonsystemer,
især relateret til de mange eksperimenter udført med meget kolde
alkaligasser (lithium, natrium, rubidium og cæsium) i de seneste 10
år.  Specielt studeres afvigelser fra middelfeltet.  Efter det
introducerende kapitel \ref{kap:introduction} præsenterede vi i
kapitel \ref{kap:hyperspherical_method} en hypersfærisk beskrivelse af
fåpartikelkorrelationer i et mangepartikelsystem.  Til laveste orden
inkluderede vi topartikelkorrelationer i form af en sum af
topartikelamplituder.  Når to partikler kommer tæt på hinanden, ændrer
mangepartikelbølgefunktionen sig fra den sædvanlige
enkeltpartikelstruktur, som kendes fra en middelfeltsbeskrivelse.  På
denne måde minder den hypersfæriske bølgefunktion for et tyndt system
om en kombination af en Jastrow-beskrivelse og en Hartree-beskrivelse.
Schr\"odinger-ligningen omskrives med sådanne topartikelamplituder til
en Faddeev-agtig ligning i en hypervinkel, der relateres til
topartikelafstanden, samt en simpel ligning i hyperradius, der
beskriver den samlede udstrækning af mangepartikelsystemet.  I den ene
hypervinkel udledte vi også en mere kompliceret variationsligning, der
ses som et alternativ til den Faddeev-agtige ligning.  I grænsen, hvor
middelafstanden mellem partiklerne er meget større end den typiske
vekselvirkingsrækkevidde, er det muligt at reducere komplikationerne
ved denne variationsligning, så et anvendeligt redskab fremkommer.

I kapitel 3 diskuteredes analytiske egenskaber af vinkelligningen.
For et fortyndet system giver dette resultater i overensstemmelse med
en ukorreleret antagelse for bølgefunktionen sammen med en
vekselvirkning, der normalt anvendes i en middelfeltsbeskrivelse.  De
numeriske løsninger bekræfter denne grænse, men viser også afvigelser
ved større tætheder, bl.a.~er tilstedeværelsen af en bundet tilstand
mellem to partikler bestemmende for en af egenværdierne, hvilket ikke
forekommer i middelfeltet.  Dette kan muligvis danne grundlaget for en
beskrivelse af koblinger mellem kondensatfasen og de bundne tilstande.

I kapitel 4 sammenføjede vi effekten fra vekselvirkningerne med
signaturerne fra en ydre fælde.  Dette resulterer i forskellige typer
løsninger til det fulde radiale problem.  Specielt forekommer
mangepartikeltilstande, der er bundet selv uden den ydre fældes
indflydelse.

I kapitel 5 beskrev vi middelfeltsantagelserne, og løsningerne fra den
korrelerede metode sammenlignedes med middelfeltsløsninger.  Det er
muligt at udlede en tæthedsafhængig vekselvirkning fra de korrelerede
beregninger.  Slutteligt vistes det, at gyldigheden af den korrelerede
metode strækker sig til områder med større tætheder og store
afvigelser fra middelfeltet.

Stabilitetskriterier og tidsskalaer for forskellige henfaldsmuligheder
diskuteredes i kapitel 6, dog uden at vi studerede de tidsafhængige
ligninger, og uden at vi inkluderede koblinger imellem de forskellige
faser eksplicit.  Dette er en mulig udvidelse af metoden.  Specielt
inklusionen af trepartikelkorrelationer ses som en nærliggende
fremtidig undersøgelse.  Tilstande med negativ energi kan muligvis
observeres i eksperimenter, da deres levetid er tilstrækkeligt stor,
og da deres rumlige udstrækning vil udvikle sig anderledes i tiden end
for systemer med positiv energi.

Deformationens indvirking på et bosonsystems egenskaber studeredes i
kapitel 7.  En ukorreleret fremgangsmåde giver analytiske
stabilitetskriterier, hvor det bedste er i nogenlunde overensstemmelse
med et tidsafhængigt middelfeltsstudie og med den eksperimentelt målte
værdi.  Vi udledte en effektiv dimension, der kan anvendes i studiet
af effektive vekselvirkninger for deformerede systemer eller i en
ekstrem grænse med diskret, lavere dimension.  Formen for den
effektive vekselvirkning er dog uklar og må i et nøjere studie af
korrelationer udledes fra de effektive potentialer.  Foreløbige
undersøgelser af et \'endimensionalt system viser, at
topartikelkorrelationer er utilstrækkelige for en beskrivelse af
kendte strukturer.  Sandsynligvis bør man inkludere
trepartikelkorrelationer for at opnå en tilfredsstillende beskrivelse.

Problematikken omkring fermiongasser kan måske udredes med en form for
topartikelkorrelationer indarbejdet i en hypersfærisk beskrivelse.
Dette kan også vise sig frugtbart inden for andre områder af fysikken,
f.eks.~i studiet af dripliniekerner eller af molekylære klynger.

Umiddelbare udvidelser af de beskrevne metoder er færdiggørelse af
studiet af korrelationer i deformerede systemer og inklusion af
eksplicit tidsafhængighed.  Desuden forventes en model indeholdende
trepartikelkorrelationer at kunne besvare mange spørgsmål, da
korrelationer af højere orden sandsynligvis er ubetydelige selv ved
forholdsvist store tætheder.  Samtidig kan dette teste gyldigheden af
resultaterne opnået vha.~antagelsen om topartikelkorrelationer.

\appendix                                                     
\chapter{Coordinate transformations}

\label{sec:app.coord-transf}

From Schaum's \cite{sch68} p.~124-125 we generalize transformations
between large sets of coordinates.  We start with a set of $M$
coordinates denoted by $\boldsymbol s=(s_1,s_2,\ldots,s_M)$ and
replace these by another set of $M$ coordinates $\boldsymbol
q=(q_1,q_2,\\ \ldots,q_M)$, i.e.~$s_i=s_i(\boldsymbol q)$.  We obtain
the relation between volume elements and Laplacians as follows.  First
define
\begin{eqnarray}
  h_j\equiv
  \bigg|{\partial\boldsymbol s\over\partial q_j}\bigg|
  \;,\qquad 
  \mathcal{H}\equiv\prod_{j=1}^M h_j
  \label{eq:37}
  \;.
\end{eqnarray}
The Laplacian operators and the volume elements are then connected by
\begin{eqnarray}
  &&
  \sum_{i=1}^M\frac{\partial^2}{\partial s_i^2}
  =
  \sum_{j=1}^M
  \hat\Delta_j
  \;,\qquad
  \hat\Delta_j
  \equiv
  \frac{\partial}{\partial q_j}
  \frac{\mathcal{H}}{h_j^2}\frac{\partial}{\partial q_j}
  \;,\\
  &&
  \prod_{i=1}^M
  ds_i
  =
  \mathcal H\prod_{j=1}^M dq_j
  \;.
  \label{eq:36}
\end{eqnarray}

\section{Jacobi coordinates for $N$ identical particles}

We start with the coordinate vectors $\boldsymbol r_i$ for $N$
identical particles in $d$ spatial dimensions.  These are then
transformed to the centre-of-mass coordinates
\begin{eqnarray}
  \boldsymbol R={1\over N}\sum_{i=1}^N\boldsymbol r_i
\end{eqnarray}
and $N-1$ relative Jacobi coordinates $\boldsymbol\eta_k$ for
$k=1,2,\ldots,N-1$:
\begin{eqnarray}
  &&
  \boldsymbol\eta_{N-1}=
  {1\over\sqrt2}(\boldsymbol r_2-\boldsymbol r_1)
  \;,\qquad\ldots\;,
  \nonumber\\
  &&
  \boldsymbol\eta_k=
  \sqrt{N-k\over N-k+1}\Big(\boldsymbol r_{N-k+1}-
  {1\over N-k}\sum_{i=1}^{N-k}\boldsymbol r_i\Big)
  \;.
\end{eqnarray}
The inverse relations are 
\begin{eqnarray}
  &&
  \boldsymbol r_i
  =
  \boldsymbol R-\sum_{k=1}^{N-i}{1\over\sqrt{(N-k)(N-k+1)}}\boldsymbol\eta_k+
  \sqrt{i-1\over i}\boldsymbol\eta_{N-i+1}
  \;,
  \nonumber\\
  &&
  \ldots\;,\qquad
  \boldsymbol r_N
  =
  \boldsymbol R-\sqrt{\frac{N-1}N}\boldsymbol\eta_1
  \;.
\end{eqnarray}
The notation in relation to \refeq.~(\ref{eq:37}) is
\begin{eqnarray}
  \boldsymbol s=
  \left\{
    \begin{array}{ll}
      (r_{1x},r_{1y},r_{1z},r_{2x},\ldots,r_{Nz}) & \textrm{for }d=3\;,\\
      (r_{1x},r_{1y},r_{2x},\ldots,r_{Ny}) & \textrm{for }d=2\;,\\
      (r_{1x},r_{2x},\ldots,r_{Nx}) & \textrm{for }d=1\;,
    \end{array}
  \right.
\end{eqnarray}
and $\boldsymbol q=(\boldsymbol R, \boldsymbol \eta_{N-1}, \ldots,
\boldsymbol \eta_1)$.  The $i$'th component of the centre-of-mass
coordinates obeys $h_{R_i}=N^{1/2}$ and for the relative components
$h_{\eta_{k,i}}=1$.  Then the volume element is
\begin{eqnarray}
  \prod_{i=1}^Nd\boldsymbol r_i
  =
  N^{d/2}d\boldsymbol R\prod_{k=1}^{N-1}d\boldsymbol\eta_k
  \;,
\end{eqnarray}
where each vector denotes $d$ degrees of freedom.  The total Laplacian
is
\begin{eqnarray}
  \hat\Delta\sub{total}
  \equiv
  \sum_{i=1}^N{\partial^2\over\partial\boldsymbol r_i^2}
  = \hat\Delta_R
  + \sum_{k=1}^{N-1}{\partial^2\over\partial\boldsymbol\eta_k^2}
  \;,\qquad
  \hat\Delta_R
  \equiv
  {1\over N}{\partial^2\over\partial\boldsymbol R^2}
  \;.
\end{eqnarray}

\section{Hyperspherical coordinates}

\subsection{Three particles in three dimensions}

For simplicity we first study the case of hyperspherical coordinates
for $d=3$ spatial dimensions and $N=3$ particles.  The Jacobi vectors
are
\begin{eqnarray}
  \boldsymbol \eta_2=\rho\sin\alpha\left(
    \begin{array}{c}
      \sin\vartheta_2\cos\varphi_2\\
      \sin\vartheta_2\sin\varphi_2\\
      \cos\vartheta_2
    \end{array}
  \right)
  \;,\quad
  \boldsymbol \eta_1=\rho\cos\alpha\left(
    \begin{array}{c}
      \sin\vartheta_1\cos\varphi_1\\
      \sin\vartheta_1\sin\varphi_1\\
      \cos\vartheta_1
    \end{array}
  \right)
  \;.
\end{eqnarray}
So, we start with six relative coordinates $\boldsymbol
s=(\eta_{1x},\ldots,\eta_{2z})$ and wish to obtain the volume element
and Laplacian operator in the (new) set of hyperspherical coordinates
$\boldsymbol q =
(\rho,\alpha,\vartheta_2,\varphi_2,\vartheta_1,\varphi_1)$.  In this
case $h_\rho=1$, $h_\alpha=\rho$, $h_{\vartheta_2}=\rho\sin\alpha$,
$h_{\varphi_2}=\rho\sin\alpha\sin\vartheta_2$,
$h_{\vartheta_1}=\rho\cos\alpha$, and
$h_{\varphi_1}=\rho\cos\alpha\sin\vartheta_1$, which yields
\begin{eqnarray}
  \mathcal{H}=\rho^5\sin^2\alpha\cos^2\alpha\sin\vartheta_2\sin\vartheta_1
  \;.
\end{eqnarray}
The terms in the Laplacian then become
\begin{eqnarray}
  &&
  \hat\Delta_\rho
  =
  {1\over\mathcal{H}}
  {\partial\over\partial\rho}
  {\mathcal{H}\over1^2}{\partial\over\partial\rho}=
  {1\over\rho^5}{\partial\over\partial\rho}
  \rho^5{\partial\over\partial\rho}
  \;,\\
  &&
  \hat\Delta_\alpha
  =
  {1\over\mathcal H}{\partial\over\partial\alpha}
  {\mathcal{H}\over\rho^2}{\partial\over\partial\alpha}=
  {1\over\rho^2\sin^2\alpha\cos^2\alpha}
  {\partial\over\partial\alpha}
  \sin^2\alpha\cos^2\alpha{\partial\over\partial\alpha}
  \;,\\
  &&
  \hat\Delta_{\vartheta_2}+\hat\Delta_{\varphi_2}
  =
  -{\hat{\boldsymbol l}_2^2\over\rho^2\sin^2\alpha}
  \;,
  \\
  &&
  \hat{\boldsymbol l}_k^2
  =-
  {1\over\sin\vartheta_k}{\partial\over\partial\vartheta_k}
  \sin\vartheta_k{\partial\over\partial\vartheta_k}
  -{1\over\sin^2\vartheta_k}{\partial^2\over\partial\varphi_k^2}
  \;,\\
  &&
  \hat\Delta_{\vartheta_1}+\hat\Delta_{\varphi_1}
  =
  -{\hat{\boldsymbol l}_1^2\over\rho^2\cos^2\alpha}
  \;,
  \\
  &&
  \hat\Delta
  \equiv
  \hat\Delta\sub{total}-\hat\Delta_{R}
  =
  {1\over\rho^5}{\partial\over\partial\rho}
  \rho^5{\partial\over\partial\rho}-
  {\hat\Lambda_2^2\over\rho^2}
  \;,\\
  &&
  \hat\Lambda_2^2
  \equiv
  -{1\over\sin^2\alpha\cos^2\alpha}{\partial\over\partial\alpha}
  \sin^2\alpha\cos^2\alpha{\partial\over\partial\alpha}+
  {\hat{\boldsymbol l}_2^2\over\sin^2\alpha}+
  {\hat{\boldsymbol l}_1^2\over\cos^2\alpha}  
  \;.
\end{eqnarray}
In this notation $\hbar\hat{\boldsymbol l}_k$ is the angular momentum operator
associated with $\boldsymbol \eta_k$.

\subsection{$N$ particles in three dimensions}

The hyperspherical coordinates are related to the Jacobi coordinates
by
\begin{eqnarray}
  &&
  \boldsymbol\eta_k=\rho_k\sin\alpha_k\left(
    \begin{array}{c}
      \sin\vartheta_k\cos\varphi_k\\
      \sin\vartheta_k\sin\varphi_k\\
      \cos\vartheta_k
    \end{array}
  \right)
  \;,\qquad
  k=1,2,\ldots,N-1\;,
  \\
  &&
  \rho_k=\rho_{k+1}\cos\alpha_{k+1}
  =\rho\prod_{j=k+1}^{N-1}\cos\alpha_j
  \;,\qquad
  \rho\equiv\rho_{N-1}
  \;.
\end{eqnarray}
The Jacobi coordinates are $\boldsymbol s =
(\eta_{1x},\ldots,\eta_{N-1,z})$.  With $\alpha_1=\pi/2$ the new set
of coordinates is $\boldsymbol q =
(\rho,\alpha_{N-1},\alpha_{N-2},\ldots,\alpha_2,\vartheta_k,\varphi_k)$.
This yields $h_\rho=1$, $h_{\alpha_k}=\rho_k$,
$h_{\vartheta_k}=\rho_k\sin\alpha_k$,
$h_{\varphi_k}=\rho_k\sin\alpha_k\sin\vartheta_k$, and the volume
element is
\begin{eqnarray}
  \mathcal H
  =
  \rho^{3N-4}
  \cdot
  \Bigg(\prod_{k=1}^{N-1}\sin\vartheta_k\Bigg)
  \cdot
  \Bigg(\prod_{k=2}^{N-1}\sin^2\alpha_k\cos^{3k-4}\alpha_k\Bigg)
  \label{eq:29}
  \;.
\end{eqnarray}
  
Each degree of freedom contributes to the Laplacian as follows:
\begin{eqnarray}
  &&
  \hat\Delta_\rho
  =
  {1\over\mathcal H}{\partial\over\partial\rho}
  {\mathcal H\over h_\rho^2}{\partial\over\partial\rho}
  =
  {1\over\rho^{3N-4}}{\partial\over\partial\rho}
  \rho^{3N-4}{\partial\over\partial\rho}
  \;,\\
  &&
  \hat\Delta_{\alpha_k}
  =
  {1\over\rho^2\prod_{j=k+1}^{N-1}\cos^2\alpha_j}
  {\cos^{4-3k}\alpha_k\over\sin^2\alpha_k}{\partial\over\partial\alpha_k}
  {\sin^2\alpha_k\cos^{3k-4}\alpha_k}{\partial\over\partial\alpha_k}
  \;,\qquad\\
  &&
  \hat\Delta_{\vartheta_k}
  =
  {1\over\rho^2\cdot(\prod_{j=k+1}^{N-1}\cos^2\alpha_j)}
  {1\over\sin^2\alpha_k}
  {1\over\sin\vartheta_k}{\partial\over\partial\vartheta_k}
  \sin\vartheta_k{\partial\over\partial\vartheta_k}
  \;,\\
  &&
  \hat\Delta_{\varphi_k}
  =
  {1\over\rho^2\cdot(\prod_{j=k+1}^{N-1}\cos^2\alpha_j)}
  {1\over\sin^2\alpha_k}
  {1\over\sin^2\vartheta_k}{\partial^2\over\partial\varphi_k^2}
  \;.
\end{eqnarray}    
We then note
\begin{eqnarray}
  \hat\Delta_{\varphi_k}+\hat\Delta_{\vartheta_k}=
  {1\over\rho^2\cdot(\prod_{j=k+1}^{N-1}\cos^2\alpha_j)}
  {-\hat{\boldsymbol l}_k^2\over\sin^2\alpha_k}
  \;.
\end{eqnarray}
All the terms can be collected in
\begin{eqnarray}
  &&
  \hat\Delta
  = \hat\Delta_\rho+
  \sum_{i=2}^{N-1}\hat\Delta_{\alpha_k}+
  \sum_{i=1}^{N-1}(\hat\Delta_{\varphi_k}+\hat\Delta_{\vartheta_k})
  = \hat\Delta_\rho-{\hat\Lambda^2_{N-1}\over\rho^2}
  \;,\\
  &&
  \hat\Lambda_k^2=\hat\Pi_k^2+{\hat\Lambda_{k-1}^2\over\cos^2\alpha_k}
  +{\hat{\boldsymbol l}_k^2\over\sin^2\alpha_k}
  \;,\qquad
  \hat\Lambda_1^2=\hat{\boldsymbol l}_1^2
  \;,\\
  &&
  \hat\Pi_k^2=
  -{1\over\sin^2\alpha_k\cos^{3k-4}\alpha_k}
  {\partial\over\partial\alpha_k}
  {\sin^2\alpha_k\cos^{3k-4}\alpha_k}
  {\partial\over\partial\alpha_k}
  \;.
\end{eqnarray}
A convenient transformation is
\begin{eqnarray}
  \hat\Pi_k^2
  &=&
  {1\over\sin\alpha_k\cos^{(3k-4)/2}\alpha_k}
  \bigg[
  -
  {\partial^2\over\partial\alpha_k^2}
  -{9k-10\over2}
  \nonumber\\
  &&
  +{(3k-4)(3k-6)\over4}\tan^2\alpha_k
  \bigg]
  \sin\alpha_k\cos^{(3k-4)/2}\alpha_k
  \;.
\end{eqnarray}

\subsection{$N$ particles in $d$ dimensions}

\label{sec:n-particles-d}

Without repeating the steps in the derivation we collect here the
results in $d$ spatial dimensions.  In the general dimension the
hyperspherical coordinates can be defined in the same way, when
applicable, as for three dimensions.  This means that for the integer
dimensions we have the set of coordinates
\begin{eqnarray}
  \boldsymbol q=
  \left\{
    \begin{array}{ll}
      (\rho, \alpha_{N-1}, \ldots, \alpha_2, \varphi_{N-1},
      \ldots, \varphi_1, \vartheta_{N-1}, \ldots, \vartheta_1)
      &
      \textrm{ for }d=3
      \;,\\
      (\rho, \alpha_{N-1}, \ldots, \alpha_2,
      \varphi_{N-1}, \ldots, \varphi_1)
      &\textrm{ for }d=2
      \;,\\
      (\rho,\alpha_{N-1}, \ldots, \alpha_2)
      &\textrm{ for }d=1
      \;.
    \end{array}
  \right.
\end{eqnarray}
While the coordinates $\alpha_k\in[0,\pi/2]$ for $d=3$ and $d=2$, it
is for $d=1$ convenient to include the sign of a Jacobi coordinate in
the definition of the corresponding hyperangle, and thus the
appropriate range for $d=1$ is $\alpha_k \in [-\pi/2,\pi/2]$.  The
volume element is given by
\begin{eqnarray}
  &&
  \prod_{k=1}^{N-1}d\boldsymbol\eta_k
  =
  d\rho\rho^{d(N-1)-1}d\Omega_{N-1}
  \;,\qquad
  d\Omega_k=d\Omega_\alpha^{(k)}d\Omega_\eta^{(k)}d\Omega_{k-1}
  \;,\quad\\
  &&
  d\Omega_1=d\Omega_\eta^{(1)}
  \;,\qquad
  d\Omega_\alpha^{(k)}=d\alpha_k\sin^{d-1}\alpha_k\cos^{d(k-1)-1}\alpha_k
  \;,\\
  &&
  d\Omega_\eta^{(k)}=\left\{\begin{array}{ll}
      d\varphi_kd\vartheta_k\sin\vartheta_k &  \textrm{ for } d=3
      \;,\\
      d\varphi_k &  \textrm{ for } d=2
      \;,\\
      1 & \textrm{ for } d=1
      \;.\\
    \end{array}\right.
  \;
\end{eqnarray}
The relative Laplacian becomes
\begin{eqnarray}
  &&
  \hat\Delta
  =\hat\Delta_\rho-\frac{\hat\Lambda^2_{N-1}}{\rho^2}
  \;,\qquad
  \hat\Delta_\rho=\frac{1}{\rho^{d(N-1)-1}}
  \frac{\partial}{\partial\rho}
  \rho^{d(N-1)-1}
  \frac{\partial}{\partial\rho}
  \;,\\
  &&
  \hat\Lambda_k^2=\hat\Pi_k^2
  +\frac{\hat\Lambda_{k-1}^2}{\cos^2\alpha_k}
  +\frac{\hat{\boldsymbol l}_k^2}{\sin^2\alpha_k}
  \;,\qquad
  \hat\Lambda_1^2=\hat{\boldsymbol l}_1^2
  \label{eq:hatlambdak2}
  \;,\\
  &&
  \hat\Pi_k^2=
  -\frac{1}
  {\sin^{d-1}\alpha_k\cos^{d(k-1)-1}\alpha_k}
  \frac{\partial}{\partial\alpha_k}
  {\sin^{d-1}\alpha_k\cos^{d(k-1)-1}\alpha_k}
  \frac{\partial}{\partial\alpha_k}
  \label{eq:pik2gend}
  \;,\qquad\\
  &&
  \hat{\boldsymbol l}_k^2
  =\left\{\begin{array}{ll}
      -
      {1\over\sin\vartheta_k}{\partial\over\partial\vartheta_k}
      \sin\vartheta_k{\partial\over\partial\vartheta_k}
      -{1\over\sin^2\vartheta_k}{\partial^2\over\partial\varphi_k^2}
      & \textrm{ for }d=3
      \;,\\
      -\frac{\partial^2}{\partial\varphi_k^2} & \textrm{ for }d=2
      \;,\\
      0 & \textrm{ for }d=1
      \;.\\
    \end{array}\right.
\end{eqnarray}
Useful transformations of the operators are
\begin{eqnarray}
  &&
  \hat\Pi_k^2=
  \frac{1}
  {\sin^{(d-1)/2}\alpha_k\cos^{l_{d,k}+1}\alpha_k}
  \bigg[
  -\frac{\partial^2}{\partial\alpha_k^2}
  +l_{d,k}(l_{d,k}+1)\tan^2\alpha_k
  \\ \nonumber
  &&
  \qquad
  +\frac{(d-1)(d-3)}{4}\cot^2\alpha_k
  +\frac{1-d^2(k-1)}{2}
  \bigg]
  \sin^{(d-1)/2}\alpha_k\cos^{l_{d,k}+1}\alpha_k
  \;,\\
  &&
  \hat\Delta_\rho=\frac{1}
  {\rho^{l_{d,N}+1}}
  \bigg[\frac{\partial^2}{\partial\rho^2}
  -\frac{l_{d,N}(l_{d,N}+1)}{\rho^2}\bigg]
  {\rho^{l_{d,N}+1}}
  \;,\quad
  l_{d,k} \equiv \frac{d(k-1)-3}{2}
  \;.\qquad\quad
\end{eqnarray}  
with non-negative integers $\nu_k=0,1,2,\ldots$.

\section{``Hypercylindrical'' coordinates}

\label{sec:hypercyl-coord}

Apart from using the same coordinates as in the spherical case, there
are, at least, two alternative methods for describing a system with
cylindrical symmetry or different geometries along all three
coordinate axes.

\subsection*{Combination of one and two dimensions}

\label{sec:combination-one-two}

The relative coordinates, which are important when describing
correlations, can be described by the usual $N-1$ Jacobi vectors,
which are now related to two hyperradii and corresponding hyperangles
by
\begin{eqnarray}
  \boldsymbol\eta_k = \left(\begin{array}{l}
      \rho_{\perp,k}\sin\alpha_k \; \cos\varphi_k \\
      \rho_{\perp,k}\sin\alpha_k \; \sin\varphi_k \\
      \rho_{z,k}\sin\beta_k
    \end{array}\right)
  \;,\qquad
  k=1,2,\ldots,N-1
  \;,
  \label{eq:gend13}
\end{eqnarray}
where $\rho_{\perp,N-1}=\rho_\perp$ and $\rho_{\perp,k} =
\rho_{\perp}\cos\alpha_{N-1}\cdots\cos\alpha_{k+1}$ for
$k=1,2,\ldots,N-2$.  Analogue relations hold for $\rho_{z,k}$,
especially $\rho_{z,N-1} = \rho_z$.  The recursions stop at
$\beta_1=\alpha_1=\pi/2$.  We collectively denote the angles by
$\Omega$.\footnote{The description can be extended to describe
deformations along all axes.  Then the $\rho_\perp$-part is separated
into $\rho_x$- and $\rho_y$-parts, that are similar to the
$\rho_z$-part in the present description.}  The volume element becomes
\begin{eqnarray}
  &&
  \prod_{k=1}^{N-1} d\boldsymbol\eta_k
  =
  d\rho_\perp \rho_\perp^{2(N-1)-1} \;
  d\rho_z \rho_z^{(N-1)-1} \;
  d\varphi_1
  \times
  \\
  &&
  \qquad
  \prod_{k=2}^{N-1}
  \Big[d\alpha_k\sin\alpha_k\cos^{2(k-1)-1}\alpha_k \;
  d\beta_k\cos^{(k-1)-1}\beta_k \;
  d\varphi_k\Big]
  \;.\nonumber
\end{eqnarray}
The relative Laplacian operator becomes
\begin{eqnarray}
  &&
  \hat\Delta
  =
  \hat \Delta_{\rho_z}
  + \hat \Delta_{\rho_\perp}
  - \frac{\hat\Lambda^2_{2,N-1}}{\rho_\perp^2}
  - \frac{\hat\Lambda^2_{1,N-1}}{\rho_z^2}
  \;,
  \\
  &&
  \hat\Delta_{\rho_q}
  =
  \frac{1}{\rho_q^{d(N-1)-1}}
  \frac{\partial}{\partial\rho_q}
  \rho_q^{d(N-1)-1}
  \frac{\partial}{\partial\rho_q}
  \;,
\end{eqnarray}
where $d=1$ for $q=z$, $d=2$ for $q=\perp$, and $\hat\Lambda_{d,N-1}$
is the operator in $d$ spatial dimensions given previously by
\refeq.~(\ref{eq:hatlambdak2}).  The angles to enter
\refeq.~(\ref{eq:hatlambdak2}) are for $d=1$ the $\beta_k$'s, and for
$d=2$ the $\alpha_k$'s and $\varphi_k$'s .

\subsection*{Parametrization of the hyperradius}

\label{sec:hypercyl_parametric}

Alternatively, one common hyperradius can be used along with an angle
$\theta$ which parametrizes the axial and plane contributions as
follows:
\begin{eqnarray}
  \boldsymbol\eta_k = \rho \left(\begin{array}{l}
      \sin \theta \; \cos\alpha_{N-1}\cdots\cos\alpha_{k+1}\sin\alpha_k
      \; \cos\varphi_k \\
      \sin \theta \; \cos\alpha_{N-1}\cdots\cos\alpha_{k+1}\sin\alpha_k 
      \; \sin\varphi_k \\
      \cos \theta \; \cos\beta_{N-1}\cdots\cos\beta_{k+1}\sin\beta_k
    \end{array}\right)
  \;,
\end{eqnarray}
with $\beta_1=\alpha_1=\pi/2$.  The volume element and the relative
Laplacian becomes
\begin{eqnarray}
  &&
  \prod_{k=1}^{N-1} d\boldsymbol\eta_k
  =
  d\rho \rho^{3(N-1)-1}
  d\theta \cos^{(N-1)-1}\theta\sin^{2(N-1)-1}\theta
  \times
  \\ \nonumber
  &&
  \qquad
  d\varphi_1
  \prod_{k=2}^{N-1}
  \Big[d\alpha_k\sin\alpha_k\cos^{2(k-1)-1}\alpha_k \;
  d\beta_k\cos^{(k-1)-1}\beta_k \;
  d\varphi_k\Big]
  \;,\\
  &&
  \hat \Delta
  =
  \hat\Delta_\rho
  + \hat\Delta_\theta
  - \frac{\hat\Lambda^2_{2,N-1}}{\rho^2\sin^2\theta}
  - \frac{\hat\Lambda^2_{1,N-1}}{\rho^2\cos^2\theta}
  \;,
  \\
  &&
  \hat\Delta_\rho
  =
  \frac{1}{\rho^{3N-4}}
  \frac{\partial}{\partial\rho}
  \rho^{3N-4}
  \frac{\partial}{\partial\rho}
  \;,
  \\
  &&
  \hat\Delta_\theta
  =
  \frac{1}{\rho^2}
  \frac{1}{\cos^{N-2}\theta\;\sin^{2N-3}\theta}
  \frac{\partial}{\partial \theta}
  \cos^{N-2}\theta\;\sin^{2N-3}\theta
  \frac{\partial}{\partial \theta}
  \;,
\end{eqnarray}
where $\hat\Lambda_{d,N-1}$ is the operator in $d$ spatial dimensions
as before.

\chapter{Hyperangular matrix elements}

\label{sec:hyper-matr-elem}

\section{Alternative Jacobi trees}

For use in the calculation of matrix elements different Jacobi trees
have to be chosen \cite{smi77simple}.  The relevant ones in the context of
the Faddeev- and angular variational equations are shown in
\reffig.~\ref{fig:jacobi.trees}.

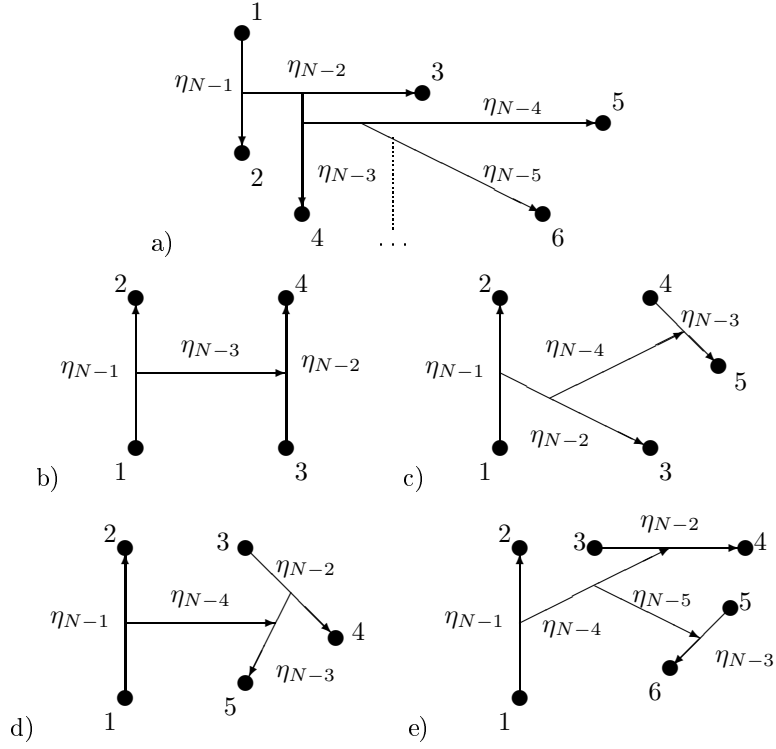
\begin{figure}[htb]
  \setlength{\unitlength}{2mm}
  \centering
  a)\begin{picture}(30,16.5)(-4.5,-2.5)
\linethickness{1pt}
\thinlines
\put(0,12){\vector(0,-1){7.6}}
\put(0,8){\vector(1,0){11.6}}
\put(4,8){\vector(0,-1){7.6}}
\put(4,6){\vector(1,0){19.6}}
\put(8,6){\vector(2,-1){11.8}}
\put(1,14){\makebox(0,0)[t]{$1$}}
\put(1,3){\makebox(0,0)[t]{$2$}}
\put(13,10){\makebox(0,0)[t]{$3$}}
\put(5,-1){\makebox(0,0)[t]{$4$}}
\put(25,8){\makebox(0,0)[t]{$5$}}
\put(21,-1){\makebox(0,0)[t]{$6$}}
\put(0,4){\circle*{1}}
\put(0,12){\circle*{1}}
\put(12,8){\circle*{1}}
\put(4,0){\circle*{1}}
\put(24,6){\circle*{1}}
\put(20,0){\circle*{1}}
\put(-2.5,8){\makebox(0,0)[b]{$\eta_{N-1}$}}
\put(5,9){\makebox(0,0)[b]{$\eta_{N-2}$}}
\put(7,2){\makebox(0,0)[b]{$\eta_{N-3}$}}
\put(18,6.5){\makebox(0,0)[b]{$\eta_{N-4}$}}
\put(18,2){\makebox(0,0)[b]{$\eta_{N-5}$}}
\put(10.2,-2){\makebox(0,0)[t]{\ldots}}
\qbezier[20](10,5)(10,2)(10,-1)
\end{picture}
  \vspace{.2cm}\\
  b)\begin{picture}(19,14)(-5,-2.5)
\linethickness{1pt}
\thinlines
\put(0,0){\vector(0,1){9.6}}
\put(0,5){\vector(1,0){10}}
\put(10,0){\vector(0,1){9.6}}
\put(-1,-1){\makebox(0,0)[t]{$1$}}
\put(-1,11.5){\makebox(0,0)[t]{$2$}}
\put(11,-1){\makebox(0,0)[t]{$3$}}
\put(11,11.5){\makebox(0,0)[t]{$4$}}
\put(0,0){\circle*{1}}
\put(10,0){\circle*{1}}
\put(0,10){\circle*{1}}
\put(10,10){\circle*{1}}
\put(-3,4.6){\makebox(0,0)[b]{$\eta_{N-1}$}}
\put(13,5){\makebox(0,0)[b]{$\eta_{N-2}$}}
\put(5,6){\makebox(0,0)[b]{$\eta_{N-3}$}}
\end{picture}
  \hspace{.4cm}
  c)\begin{picture}(21,14)(-5,-2.5)
\linethickness{1pt}
\thinlines
\put(0,0){\vector(0,1){9.6}}
\put(0,5){\vector(2,-1){9.6}}
\put(3.33,3.33){\vector(2,1){8.9}}
\put(10,10){\vector(1,-1){4.3}}
\put(-1,-1){\makebox(0,0)[t]{$1$}}
\put(-1,11.5){\makebox(0,0)[t]{$2$}}
\put(11,-1){\makebox(0,0)[t]{$3$}}
\put(11,11.5){\makebox(0,0)[t]{$4$}}
\put(16,5){\makebox(0,0)[t]{$5$}}
\put(0,0){\circle*{1}}
\put(10,0){\circle*{1}}
\put(0,10){\circle*{1}}
\put(10,10){\circle*{1}}
\put(14.5,5.5){\circle*{1}}
\put(-3,4.6){\makebox(0,0)[b]{$\eta_{N-1}$}}
\put(4,0){\makebox(0,0)[b]{$\eta_{N-2}$}}
\put(14,8){\makebox(0,0)[b]{$\eta_{N-3}$}}
\put(5,6){\makebox(0,0)[b]{$\eta_{N-4}$}}
\end{picture}
  \vspace{.2cm}\\
  d)\begin{picture}(21,14)(-6,-2.5)
\linethickness{1pt}
\thinlines
\put(0,0){\vector(0,1){9.6}}
\put(0,5){\vector(1,0){10}}
\put(8,10){\vector(1,-1){5.7}}
\put(11,7){\vector(-1,-2){2.8}}
\put(-1,-1){\makebox(0,0)[t]{$1$}}
\put(-1,11.5){\makebox(0,0)[t]{$2$}}
\put(7,0){\makebox(0,0)[t]{$5$}}
\put(6.5,11){\makebox(0,0)[t]{$3$}}
\put(15.5,5){\makebox(0,0)[t]{$4$}}
\put(0,0){\circle*{1}}
\put(0,10){\circle*{1}}
\put(8,1){\circle*{1}}
\put(8,10){\circle*{1}}
\put(14,4){\circle*{1}}
\put(-3,4.6){\makebox(0,0)[b]{$\eta_{N-1}$}}
\put(12,8){\makebox(0,0)[b]{$\eta_{N-2}$}}
\put(12,1){\makebox(0,0)[b]{$\eta_{N-3}$}}
\put(5,6){\makebox(0,0)[b]{$\eta_{N-4}$}}
\end{picture}
  \hspace{.4cm}
  e)\begin{picture}(23,15)(-6,-2.5)
\linethickness{1pt}
\thinlines
\put(0,0){\vector(0,1){9.6}}
\put(0,5){\vector(2,1){10}}
\put(5,10){\vector(1,0){9.6}}
\put(5,7.5){\vector(2,-1){7}}
\put(14,6){\vector(-1,-1){3.8}}
\put(-1,-1){\makebox(0,0)[t]{$1$}}
\put(-1,11.5){\makebox(0,0)[t]{$2$}}
\put(4,11){\makebox(0,0)[t]{$3$}}
\put(16,11){\makebox(0,0)[t]{$4$}}
\put(15,7){\makebox(0,0)[t]{$5$}}
\put(9,1){\makebox(0,0)[t]{$6$}}
\put(0,0){\circle*{1}}
\put(0,10){\circle*{1}}
\put(5,10){\circle*{1}}
\put(15,10){\circle*{1}}
\put(14,6){\circle*{1}}
\put(10,2){\circle*{1}}
\put(-3,4.6){\makebox(0,0)[b]{$\eta_{N-1}$}}
\put(10,11){\makebox(0,0)[b]{$\eta_{N-2}$}}
\put(15,2){\makebox(0,0)[b]{$\eta_{N-3}$}}
\put(3.5,4){\makebox(0,0)[b]{$\eta_{N-4}$}}
\put(9.5,6){\makebox(0,0)[b]{$\eta_{N-5}$}}
\end{picture}
  \caption[Various Jacobi trees]
  {Jacobi trees: a) standard, b) (12)(34), c) (123)(45), d) (12)(345),
    and e) (12)(34)(56).}
  \label{fig:jacobi.trees}
\end{figure}

The coordinates of the standard tree of \reffig.~\ref{fig:jacobi.trees}a
are defined by
\begin{eqnarray}
  \label{eq:standard}
  \boldsymbol\eta_{N-1}&=&
  \frac{1}{\sqrt2}(\boldsymbol r_2-\boldsymbol r_1)
  \;,\\
  \boldsymbol\eta_{N-2}&=&
  \sqrt{\frac{2}{3}}\bigg[\boldsymbol r_3-\frac12(\boldsymbol r_2+\boldsymbol r_1)\bigg]
  \;,\qquad\ldots
  \;,\\
  \boldsymbol\eta_1&=&
  \sqrt{\frac{N-1}{N}}\bigg[\boldsymbol r_N-
  \frac{1}{N-1}(\boldsymbol r_{N-1}+\ldots+\boldsymbol r_1)\bigg]
  \;.\qquad
\end{eqnarray}

In the (12)(34)-tree of \reffig.~\ref{fig:jacobi.trees}b two of the
vectors are different from the standard tree:
\begin{eqnarray}
  \label{eq:12.34}
  \boldsymbol{\eta}_{N-2}&=&\frac1{\sqrt2}(\boldsymbol r_4-\boldsymbol r_3)
  \;,
  \\
  \boldsymbol{\eta}_{N-3}&=&\frac12(\boldsymbol r_4+\boldsymbol r_3-\boldsymbol r_2-\boldsymbol r_1)
  \;.
\end{eqnarray}

In the (123)(45)-tree of \reffig.~\ref{fig:jacobi.trees}c two of the
vectors differ from the standard tree:
\begin{eqnarray}
  \label{eq:123.45}
  \boldsymbol{\eta}_{N-3}&=&\frac1{\sqrt2}(\boldsymbol r_5-\boldsymbol r_4)
  \;,
  \\
  \boldsymbol{\eta}_{N-4}&=&\sqrt{\frac65}\bigg[\frac12(\boldsymbol r_5+\boldsymbol r_4)
  -\frac13(\boldsymbol r_3+\boldsymbol r_2+\boldsymbol r_1)\bigg]
  \;.\;
\end{eqnarray}

In the (12)(345)-tree of \reffig.~\ref{fig:jacobi.trees}d three of the
vectors deviate from the standard tree:
\begin{eqnarray}
  \label{eq:12.345}
  \boldsymbol{\eta}_{N-2}&=&\frac1{\sqrt2}(\boldsymbol r_4-\boldsymbol r_3)
  \;,
  \\
  \boldsymbol{\eta}_{N-3}&=&\sqrt{\frac23}\bigg[\boldsymbol r_5-\frac12(\boldsymbol r_4+\boldsymbol r_3)\bigg]
  \;,
  \\
  \boldsymbol{\eta}_{N-4}&=&\sqrt{\frac65}\bigg[\frac13(\boldsymbol r_5+\boldsymbol r_4+\boldsymbol r_3)
  -\frac12(\boldsymbol r_2+\boldsymbol r_1)\bigg]
  \;.\;
\end{eqnarray}

In the (12)(34)(56)-tree of \reffig.~\ref{fig:jacobi.trees}e four
vectors are different:
\begin{eqnarray}
  \label{eq:12.34.56}
  \boldsymbol{\eta}_{N-2}&=&\frac1{\sqrt2}(\boldsymbol r_4-\boldsymbol r_3)
  \;,
  \qquad
  \boldsymbol{\eta}_{N-3}=\frac1{\sqrt2}(\boldsymbol r_6-\boldsymbol r_5)
  \;,\qquad
  \\
  \boldsymbol{\eta}_{N-4}&=&\frac12(\boldsymbol r_4+\boldsymbol r_3-\boldsymbol r_2-\boldsymbol r_2)
  \;,
  \\
  \boldsymbol{\eta}_{N-5}
  &=&
  \sqrt{\frac43}
  \bigg[
  \frac12(\boldsymbol r_6+\boldsymbol r_5)-
  \frac14(\boldsymbol r_4+\boldsymbol r_3+\boldsymbol r_2+\boldsymbol r_1)
  \bigg]
  \;.\;
\end{eqnarray}

Since only inter-relations between $\boldsymbol\eta_{N-1}$,
$\boldsymbol\eta_{N-2}$, and $\boldsymbol\eta_{N-3}$ are needed in
evaluating the matrix elements, we use the common notation:
\begin{eqnarray}
  &&
  \eta_{N-1}=\rho\sin\alpha
  \;,\qquad
  \eta_{N-2}=\rho\cos\alpha\sin\beta
  \\
  &&
  \eta_{N-3}=\rho\cos\alpha\cos\beta\sin\gamma
  \;,\qquad
  \boldsymbol\eta_k\cdot\boldsymbol\eta_l=\eta_k\eta_l\cos\vartheta_{k,l}
  \;.
\end{eqnarray}
Here $\vartheta_{k,l}$ is the angle between the $k$'th and $l$'th Jacobi
vectors.  We abbreviate $\vartheta_{N-1,N-2}\to\vartheta_x$,
$\vartheta_{N-1,N-3}\to\vartheta_y$, and $\vartheta_{N-2,N-3}\to\vartheta_z$.  An
azimuthal angle $\varphi$ determining the projection of
$\boldsymbol\eta_{N-3}$ onto the plane of $\boldsymbol\eta_{N-1}$ and
$\boldsymbol\eta_{N-2}$ is defined in the usual way such that
\begin{eqnarray}
  \cos\vartheta_z
  =
  \sin\vartheta_x\sin\vartheta_y\cos\varphi+
  \cos\vartheta_x\cos\vartheta_y
  \;.
\end{eqnarray}
With $\tau=\{\beta,\gamma,\vartheta_x,\vartheta_y,\varphi\}$ a matrix
element of an arbitrary function $f$ of all the variables $\alpha$ and
$\tau$ then becomes
\begin{eqnarray}
  &&
  \int d\tau\; f(\alpha,\tau)
  =
  \frac{\int d\tilde\tau \; f(\alpha,\tau)}{\int d\tilde\tau}
  \label{eq:6}
  \;,\\
  &&
  \int d\tilde\tau \; g(\alpha,\tau)
  =
  \angleint{\beta}{3N-10}\angleint{\gamma}{3N-13}
  \nonumber\\
  &&
  \qquad
  \times
  \int_0^\pi d\vartheta_x\sin\vartheta_x
  \int_0^\pi d\vartheta_y\sin\vartheta_y
  \int_0^{2\pi}d\varphi \;g(\alpha,\tau)
  \;.
\end{eqnarray}
The normalization is explicitly $\int d\tau=1$.  In the following
matrix elements we need relations for interparticle distances and
therefore define $\boldsymbol\eta_{ij} \equiv (\boldsymbol
r_j-\boldsymbol r_i)/\sqrt2$ and the angle $\alpha_{ij}$ related to
$\eta_{ij} = \rho\sin\alpha_{ij}=r_{ij}/\sqrt2$.

\section{Matrix elements: Faddeev}

\label{sec:app.fadd-like-equat}

\Refeqs.~(\ref{eq:intf34}) and (\ref{eq:intf13}) are evaluated as follows.

In the integral $\int d\tau\; \phi(\alpha_{34})$ a convenient choice
of coordinates is the alternative Jacobi (12)(34)-tree of
\reffig.~\ref{fig:jacobi.trees}b.  The angle $\alpha_{34}$ is
associated with the distance $r_{34}=\sqrt2\eta_{34}$ by the relation
\begin{eqnarray}
  \eta_{34}=
  \eta_{N-2}=
  \rho\cos\alpha\sin\beta=
  \rho\sin\alpha_{34}
  \;\Longleftrightarrow\;
  \sin\alpha_{34}=\cos\alpha\sin\beta
  \;.
  \label{eA1}
\end{eqnarray}
The integrand $\phi(\alpha_{34})$ only depends on $\alpha_{34}$, which
is a function of $\alpha$ and $\beta$.  Therefore at fixed $\alpha$
\refeq.~(\ref{eq:6}) reduces to
\begin{eqnarray}
  &&
  \int d\tau\; \phi(\alpha_{34})
  =
  \frac{\angleint{\beta}{3N-10}\;\phi(\alpha_{34})}
  {\int_0^{\pi/2}d\beta\;\sin^2\beta\cos^{3N-10}\beta}
  =
  \\ \nonumber
  &&
  \qquad
  \frac{4}{\sqrt\pi}\gammafktB{3N-6}{3N-9}
  \int_0^{\pi/2}d\beta\;\sin^2\beta\cos^{3N-10}\beta\;\phi(\alpha_{34})
  \equiv
  \hatR_{34}^{(N-2)}\phi(\alpha)
  \;.
  \label{eq:7}
\end{eqnarray}

To describe three particles in $\int d\tau\;\phi(\alpha_{13})$
simultaneously, Jacobi vectors of the standard tree are needed.  The
distance between particles $1$ and $3$ is related to the corresponding
Jacobi vector
\begin{eqnarray}
  \boldsymbol\eta_{13}=\frac{1}{\sqrt2}(\boldsymbol r_3-\boldsymbol r_1)=
  \frac12\boldsymbol\eta_{N-1}+\frac{\sqrt3}{2}\boldsymbol\eta_{N-2}
  \;,
\end{eqnarray}
The hyperangle $\alpha_{13}$, associated with the distance between
particles $1$ and $3$ through
$\eta_{13}=r_{13}/\sqrt2=\rho\sin\alpha_{13}$, is then
\begin{eqnarray}
  \label{eq:5}
  \sin^2\alpha_{13}
  =
  \frac{1}{4}\sin^2\alpha+
  \frac{3}{4}\cos^2\alpha\sin^2\beta+
  \frac{\sqrt3}{2}\sin\alpha\cos\alpha\sin\beta\cos\vartheta_x
  \;,
\end{eqnarray}
where $\vartheta_x$ is the angle between the Jacobi vectors
$\boldsymbol\eta_{N-1}$ and $\boldsymbol\eta_{N-2}$.  Note that $\phi(\alpha_{13})$,
through $\alpha_{13}$, for fixed $\alpha$ depends on $\beta$ and
$\vartheta_x$, which leaves a two-dimensional integral. Therefore
\refeq.~(\ref{eq:6}) becomes
\begin{eqnarray}
  &&
  \int d\tau\; \phi(\alpha_{13})
  =
  \frac{\int_0^{\pi/2}d\beta\;\sin^2\beta\cos^{3N-10}\beta
    \int_0^\pi d\vartheta_x\;\sin\vartheta_x\;\phi(\alpha_{13})}
  {\int_0^{\pi/2}d\beta\;\sin^2\beta\cos^{3N-10}\beta
    \int_0^\pi d\vartheta_x\;\sin\vartheta_x}
  \\ \nonumber
  &&\qquad
  =
  \frac{2}{\sqrt\pi}\gammafktB{3N-6}{3N-9}
  \int_0^{\pi/2}d\beta\;\sin^2\beta\cos^{3N-10}\beta
  \int_0^\pi d\vartheta_x\;\sin\vartheta_x\;\phi(\alpha_{13})
  \;.
\end{eqnarray}
This integral can be reduced to one dimension by a partial
integration.  The final one-dimensional integral becomes
\begin{eqnarray}
  &&
  \int d\tau\;\phi(\alpha_{13})
  =
  \frac{4}{\sqrt{3\pi}}\gammafktB{3N-6}{3N-7}
  \sin^{-1}\alpha\cos^{8-3N}\alpha
  \times
  \nonumber\\
  &&
  \Bigg[\int_{(\alpha-\pi/3)\Theta(\alpha>\pi/3)}^{\pi/2-|\pi/6-\alpha|}
  d\alpha_{13}\;\cos^{3N-9}\gamma^+\sin\alpha_{13}\cos\alpha_{13}
  \phi(\alpha_{13})-
  \label{eq:40}
  \\ \nonumber
  &&
  \int_0^{(\pi/3-\alpha)\Theta(\pi/3>\alpha)}
  d\alpha_{13}\;\cos^{3N-9}\gamma^-\sin\alpha_{13}\cos\alpha_{13}
  \phi(\alpha_{13})\Bigg]
  \equiv\hatR_{13}^{(N-2)}\phi(\alpha)
  \;,
\end{eqnarray}
where $\sin^2\gamma^{\pm} = 4(\sin^2\alpha +\sin^2\alpha_{13}
\mp\sin\alpha\sin\alpha_{13})/3$, and $\Theta$ is the truth function.


\section{Matrix elements: variational}

\label{sec:int.var.eqn}

We first divide the integrals of \refeq.~(\ref{eq:angvar2}) into similar terms,
then compute them in general, and finally in the short-range limit.

\subsection{Numbers of different terms}

\label{sec:counting-terms}

We have to evaluate the double sums of \refeq.~(\ref{eq:angvar2})
including the potential:
\begin{eqnarray}
  \sum_{k<l}^Nv_{kl}
  \sum_{i<j}^N\phi_{ij}
  \;.
\end{eqnarray}
Three types of terms occur, due to the fact that we vary the wave
function component $\phi_{12}^*$ in \refeq.~(\ref{eq:angvar1}): the
potential concerning particles $1$ and $2$, the potential concerning
one of the particles $1$ or $2$ and a third particle and the potential
concerning neither particle $1$ nor $2$, but a third and a fourth
particle.  We obtain
\begin{eqnarray}
  &&
  \sum_{k<l}^Nv_{kl}
  =
  v_{12}
  +\sum_{l=3}^Nv_{1l}
  +\sum_{l=3}^Nv_{2l}
  +\sum_{3\le k<l}^Nv_{kl}
  \nonumber\\
  &&
  \qquad
  \to
  v_{12}+2(N-2)v_{13}+\frac12(N-2)(N-3)v_{34}
  \;,
\end{eqnarray}
where the arrow indicates the identity of the terms after integration
over all angles except $\alpha_{12}$, i.e.~analogously to the steps leading
up to \refeq.~(\ref{eq:angular-equation3}).  Treating each of these in
the quadruple sum, where the repeated use of arrows ($\to$) has the
meaning given just above:

Fixing $\phi_{12}^*$ and $v_{12}$ yields three different terms:
\begin{eqnarray}
  &&
  v_{12}
  \sum_{i<j}^N\phi_{ij}
  =
  v_{12}\bigg(
  \phi_{12}
  +\sum_{j=3}^N\phi_{1j}
  +\sum_{j=3}^N\phi_{2j}
  +\sum_{3\le i<j}^N\phi_{ij}
  \bigg)
  \nonumber\\
  &&
  \qquad
  \to
  v_{12}
  \bigg[
  \phi_{12}+2(N-2)\phi_{13}+\frac12(N-2)(N-3)\phi_{34}
  \bigg]
  \;,
\end{eqnarray}
as shown in \reffig.~\ref{fig:phi12v12}.
\begin{figure}[htbp]
  \setlength{\unitlength}{2mm}
  \centering
  a)\begin{picture}(10,15)(-7,-3)
\linethickness{1pt}
\thinlines
\put(0,0){\line(0,1){10}}
\put(-1,-1){\makebox(0,0)[t]{$1$}}
\put(-1,11.5){\makebox(0,0)[t]{$2$}}
\put(0,0){\circle*{1}}
\put(0,10){\circle*{1}}
\put(-3.5,4.6){\makebox(0,0)[b]{$\phi^*,v,\phi$}}
\end{picture}
  \hspace{.2cm}
  b)\begin{picture}(16,15)(-4,-3)
\linethickness{1pt}
\thinlines
\put(0,0){\line(1,0){10}}
\put(0,0){\line(0,1){10}}
\put(-1,-1){\makebox(0,0)[t]{$1$}}
\put(-1,11.5){\makebox(0,0)[t]{$2$}}
\put(11,-1){\makebox(0,0)[t]{$3$}}
\put(0,0){\circle*{1}}
\put(0,10){\circle*{1}}
\put(10,0){\circle*{1}}
\put(-2.5,4.6){\makebox(0,0)[b]{$\phi^*,v$}}
\put(5,-2){\makebox(0,0)[b]{$\phi$}}
\end{picture}
  \hspace{.2cm}
  c)\begin{picture}(15,15)(-3.3,-3)
\linethickness{1pt}
\thinlines
\put(0,0){\line(0,1){10}}
\put(10,0){\line(0,1){10}}
\put(-1,-1){\makebox(0,0)[t]{$1$}}
\put(-1,11.5){\makebox(0,0)[t]{$2$}}
\put(11,-1){\makebox(0,0)[t]{$3$}}
\put(11,11.5){\makebox(0,0)[t]{$4$}}
\put(0,0){\circle*{1}}
\put(10,0){\circle*{1}}
\put(10,10){\circle*{1}}
\put(0,10){\circle*{1}}
\put(-2.5,4.6){\makebox(0,0)[b]{$\phi^*,v$}}
\put(8.7,4.6){\makebox(0,0)[b]{$\phi$}}
\end{picture}
  \caption[Diagonal potential terms]
  {Illustration of $\phi_{12}^*v_{12}$-terms.}
  \label{fig:phi12v12}
\end{figure}

Fixing $\phi_{12}^*$ and $v_{13}$ yields seven different terms.  These
can be identified in two steps, the first of which separates into four
different sums:
\begin{eqnarray}
  &&v_{13}
  \sum_{i<j}^N\phi_{ij}
  =
  v_{13}
  \bigg(
  \sum_{j=2}^N\phi_{1j}+
  \sum_{j=3}^N\phi_{2j}+
  \sum_{j=4}^N\phi_{3j}+
  \sum_{4\le i<j}^N\phi_{ij}
  \bigg)
  \;.
\end{eqnarray}
Each of these four terms are then identified as:
\begin{eqnarray}
  &&
  v_{13}
  \sum_{j=2}^N\phi_{1j}
  =
  v_{13}\bigg(\phi_{12}+\phi_{13}+\sum_{j=4}^N\phi_{1j}\bigg)
  \nonumber\\
  &&
  \qquad
  \to
  v_{13}\big[\phi_{12}+\phi_{13}+(N-3)\phi_{14}\big]
  \;,\\
  &&
  v_{13}
  \sum_{j=3}^N\phi_{2j}
  =
  v_{13}\bigg(\phi_{23}+\sum_{j=4}^N\phi_{2j}\bigg)
  \to
  v_{13}\big[\phi_{23}+(N-3)\phi_{24}\big]
  \;,
  \\
  &&
  v_{13}
  \sum_{j=4}^N\phi_{3j}
  \to
  v_{13}(N-3)\phi_{34}
  \;,
  \\
  &&
  v_{13}
  \sum_{4\le i<j}^N\phi_{ij}  
  \to
  v_{13}\frac12(N-3)(N-4)\phi_{45}
  \;.
\end{eqnarray}
The resulting seven types are shown in \reffig.~\ref{fig:phi12v13}.
\begin{figure}[htbp]
  \setlength{\unitlength}{2mm}
  \centering
  a)\begin{picture}(16,15)(-4,-3)
\linethickness{1pt}
\thinlines
\put(0,0){\line(1,0){10}}
\put(0,0){\line(0,1){10}}
\put(-1,-1){\makebox(0,0)[t]{$1$}}
\put(-1,11.5){\makebox(0,0)[t]{$2$}}
\put(11,-1){\makebox(0,0)[t]{$3$}}
\put(0,0){\circle*{1}}
\put(0,10){\circle*{1}}
\put(10,0){\circle*{1}}
\put(-2.5,4.6){\makebox(0,0)[b]{$\phi^*,\phi$}}
\put(5,-1.6){\makebox(0,0)[b]{$v$}}
\end{picture}
  \hspace{.2cm}
  b)\begin{picture}(14,15)(-2,-3)
\linethickness{1pt}
\thinlines
\put(0,0){\line(1,0){10}}
\put(0,0){\line(0,1){10}}
\put(-1,-1){\makebox(0,0)[t]{$1$}}
\put(-1,11.5){\makebox(0,0)[t]{$2$}}
\put(11,-1){\makebox(0,0)[t]{$3$}}
\put(0,0){\circle*{1}}
\put(0,10){\circle*{1}}
\put(10,0){\circle*{1}}
\put(-1.3,4.6){\makebox(0,0)[b]{$\phi^*$}}
\put(5,-2){\makebox(0,0)[b]{$v,\phi$}}
\end{picture}
  \\
  c)\begin{picture}(14,15)(-2,-3)
\linethickness{1pt}
\thinlines
\put(0,0){\line(1,0){10}}
\put(0,0){\line(0,1){10}}
\put(0,10){\line(1,-1){10}}
\put(-1,-1){\makebox(0,0)[t]{$1$}}
\put(-1,11.5){\makebox(0,0)[t]{$2$}}
\put(11,-1){\makebox(0,0)[t]{$3$}}
\put(0,0){\circle*{1}}
\put(0,10){\circle*{1}}
\put(10,0){\circle*{1}}
\put(-1.3,4.6){\makebox(0,0)[b]{$\phi^*$}}
\put(5,-1.5){\makebox(0,0)[b]{$v$}}
\put(5.4,5.4){\makebox(0,0)[b]{$\phi$}}
\end{picture}
  \hspace{.2cm}
  d)\begin{picture}(14,15)(-2,-3)
\linethickness{1pt}
\thinlines
\put(0,0){\line(1,0){10}}
\put(0,0){\line(0,1){10}}
\put(0,0){\line(1,1){10}}
\put(-1,-1){\makebox(0,0)[t]{$1$}}
\put(-1,11.5){\makebox(0,0)[t]{$2$}}
\put(11,-1){\makebox(0,0)[t]{$3$}}
\put(11,11.5){\makebox(0,0)[t]{$4$}}
\put(0,0){\circle*{1}}
\put(10,0){\circle*{1}}
\put(10,10){\circle*{1}}
\put(0,10){\circle*{1}}
\put(-1.3,4.6){\makebox(0,0)[b]{$\phi^*$}}
\put(5,-1.5){\makebox(0,0)[b]{$v$}}
\put(4.7,6){\makebox(0,0)[b]{$\phi$}}
\end{picture}
  \hspace{.2cm}
  e)\begin{picture}(14,15)(-2,-3)
\linethickness{1pt}
\thinlines
\put(0,0){\line(1,0){10}}
\put(0,0){\line(0,1){10}}
\put(0,10){\line(1,0){10}}
\put(-1,-1){\makebox(0,0)[t]{$1$}}
\put(-1,11.5){\makebox(0,0)[t]{$2$}}
\put(11,-1){\makebox(0,0)[t]{$3$}}
\put(11,11.5){\makebox(0,0)[t]{$4$}}
\put(0,0){\circle*{1}}
\put(10,0){\circle*{1}}
\put(10,10){\circle*{1}}
\put(0,10){\circle*{1}}
\put(-1.3,4.6){\makebox(0,0)[b]{$\phi^*$}}
\put(5,-1.5){\makebox(0,0)[b]{$v$}}
\put(5,8){\makebox(0,0)[b]{$\phi$}}
\end{picture}
  \\
  f)\begin{picture}(14,15)(-2,-3)
\linethickness{1pt}
\thinlines
\put(0,0){\line(1,0){10}}
\put(0,0){\line(0,1){10}}
\put(10,0){\line(0,1){10}}
\put(-1,-1){\makebox(0,0)[t]{$1$}}
\put(-1,11.5){\makebox(0,0)[t]{$2$}}
\put(11,-1){\makebox(0,0)[t]{$3$}}
\put(11,11.5){\makebox(0,0)[t]{$4$}}
\put(0,0){\circle*{1}}
\put(10,0){\circle*{1}}
\put(10,10){\circle*{1}}
\put(0,10){\circle*{1}}
\put(-1.3,4.6){\makebox(0,0)[b]{$\phi^*$}}
\put(5,-1.5){\makebox(0,0)[b]{$v$}}
\put(8.7,4.6){\makebox(0,0)[b]{$\phi$}}
\end{picture}
  \hspace{.2cm}
  g)\begin{picture}(24,15)(-2,-3)
\linethickness{1pt}
\thinlines
\put(0,0){\line(1,0){10}}
\put(0,0){\line(0,1){10}}
\put(10,10){\line(1,0){10}}
\put(-1,-1){\makebox(0,0)[t]{$1$}}
\put(-1,11.5){\makebox(0,0)[t]{$2$}}
\put(11,-1){\makebox(0,0)[t]{$3$}}
\put(9,9){\makebox(0,0)[t]{$4$}}
\put(21,11.5){\makebox(0,0)[t]{$5$}}
\put(0,0){\circle*{1}}
\put(10,0){\circle*{1}}
\put(10,10){\circle*{1}}
\put(0,10){\circle*{1}}
\put(20,10){\circle*{1}}
\put(-1.3,4.6){\makebox(0,0)[b]{$\phi^*$}}
\put(5,-1.5){\makebox(0,0)[b]{$v$}}
\put(15,8){\makebox(0,0)[b]{$\phi$}}
\end{picture}
  \caption[Semi-diagonal potential terms]
  {Illustration of $\phi_{12}^*v_{13}$-terms.}
  \label{fig:phi12v13}
\end{figure}

Fixing $\phi_{12}^*$ and $v_{34}$ yields six different terms,
identified as follows.  The first step is:
\begin{eqnarray}
  &&v_{34}
  \sum_{i<j}^N\phi_{ij}
  =
  v_{34}
  \bigg(
  \sum_{j=2}^N\phi_{1j}+
  \\
  &&
  \qquad
  \sum_{j=3}^N\phi_{2j}+
  \sum_{j=4}^N\phi_{3j}+
  \sum_{j=5}^N\phi_{4j}+
  \sum_{5\le i<j}^N\phi_{ij}
  \bigg)
  \nonumber
  \;.
\end{eqnarray}
In the next step the sums are treated:
\begin{eqnarray}
  &&
  v_{34}
  \sum_{j=2}^N\phi_{1j}
  =
  v_{34}\bigg(
   \phi_{12}+\phi_{13}+\phi_{14}+\sum_{j=5}^N\phi_{1j}
  \bigg)
  \nonumber
  \\
  &&
  \qquad
  \to
  v_{34}\big[\phi_{12}+2\phi_{13}+(N-4)\phi_{15}\big]
  \;,
  \\
  &&
  v_{34}
  \sum_{j=3}^N\phi_{2j}
  =
  v_{34}\bigg(
   \phi_{23}+\phi_{24}+\sum_{j=5}^N\phi_{2j}
  \bigg)
  \to
  v_{34}\big[2\phi_{13}+(N-4)\phi_{15}\big]
  \;,\qquad
  \\
  &&
  v_{34}
  \sum_{j=4}^N\phi_{3j}
  =
  v_{34}\bigg(
   \phi_{34}+\sum_{j=5}^N\phi_{3j}
  \bigg)
  \to
  v_{34}\big[\phi_{34}+(N-4)\phi_{35}\big]
  \;,
  \\
  &&
  v_{34}
  \sum_{j=5}^N\phi_{4j}
  \to
  v_{34}
  (N-4)\phi_{35}
  \;,
  \\
  &&
  v_{34}
  \sum_{5\le i<j}^N\phi_{ij}
  \to
  v_{34}
  \frac12(N-4)(N-5)\phi_{56}
  \;.\quad
\end{eqnarray}
See the six types in \reffig.~\ref{fig:phi12v34}.
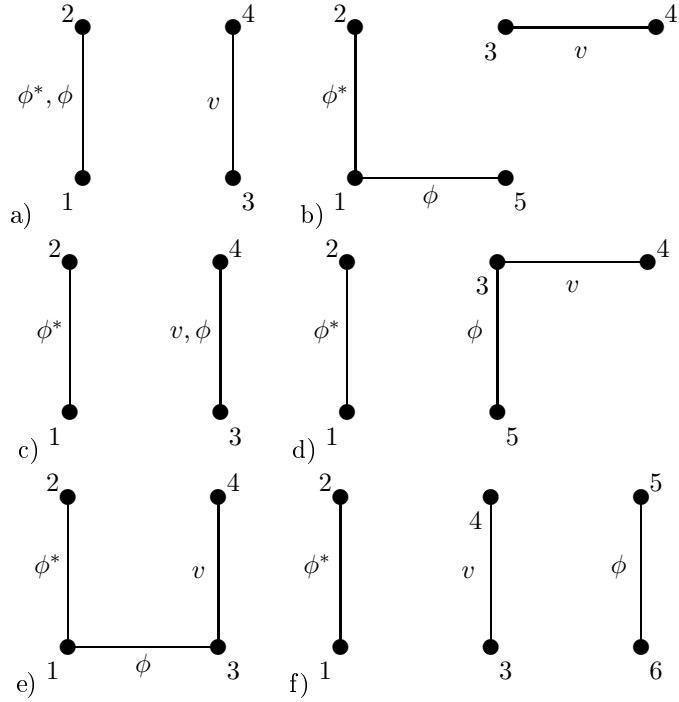
\begin{figure}[htbp]
  \setlength{\unitlength}{2mm}
  \centering
  a)\begin{picture}(15,15)(-3.3,-3)
\linethickness{1pt}
\thinlines
\put(0,0){\line(0,1){10}}
\put(10,0){\line(0,1){10}}
\put(-1,-1){\makebox(0,0)[t]{$1$}}
\put(-1,11.5){\makebox(0,0)[t]{$2$}}
\put(11,-1){\makebox(0,0)[t]{$3$}}
\put(11,11.5){\makebox(0,0)[t]{$4$}}
\put(0,0){\circle*{1}}
\put(10,0){\circle*{1}}
\put(10,10){\circle*{1}}
\put(0,10){\circle*{1}}
\put(-2.5,4.6){\makebox(0,0)[b]{$\phi^*,\phi$}}
\put(8.7,4.6){\makebox(0,0)[b]{$v$}}
\end{picture}
  \hspace{.2cm}
  b)\begin{picture}(24,15)(-2,-3)
\linethickness{1pt}
\thinlines
\put(0,0){\line(1,0){10}}
\put(0,0){\line(0,1){10}}
\put(10,10){\line(1,0){10}}
\put(-1,-1){\makebox(0,0)[t]{$1$}}
\put(-1,11.5){\makebox(0,0)[t]{$2$}}
\put(11,-1){\makebox(0,0)[t]{$5$}}
\put(9,9){\makebox(0,0)[t]{$3$}}
\put(21,11.5){\makebox(0,0)[t]{$4$}}
\put(0,0){\circle*{1}}
\put(10,0){\circle*{1}}
\put(10,10){\circle*{1}}
\put(0,10){\circle*{1}}
\put(20,10){\circle*{1}}
\put(-1.3,4.6){\makebox(0,0)[b]{$\phi^*$}}
\put(5,-2){\makebox(0,0)[b]{$\phi$}}
\put(15,8){\makebox(0,0)[b]{$v$}}
\end{picture}
  \\
  c)\begin{picture}(14,15)(-2,-3)
\linethickness{1pt}
\thinlines
\put(0,0){\line(0,1){10}}
\put(10,0){\line(0,1){10}}
\put(-1,-1){\makebox(0,0)[t]{$1$}}
\put(-1,11.5){\makebox(0,0)[t]{$2$}}
\put(11,-1){\makebox(0,0)[t]{$3$}}
\put(11,11.5){\makebox(0,0)[t]{$4$}}
\put(0,0){\circle*{1}}
\put(10,0){\circle*{1}}
\put(10,10){\circle*{1}}
\put(0,10){\circle*{1}}
\put(-1.3,4.6){\makebox(0,0)[b]{$\phi^*$}}
\put(8,4.6){\makebox(0,0)[b]{$v,\phi$}}
\end{picture}
  \hspace{.2cm}
  d)\begin{picture}(24,15)(-2,-3)
\linethickness{1pt}
\thinlines
\put(0,0){\line(0,1){10}}
\put(10,0){\line(0,1){10}}
\put(10,10){\line(1,0){10}}
\put(-1,-1){\makebox(0,0)[t]{$1$}}
\put(-1,11.5){\makebox(0,0)[t]{$2$}}
\put(11,-1){\makebox(0,0)[t]{$5$}}
\put(9,9){\makebox(0,0)[t]{$3$}}
\put(21,11.5){\makebox(0,0)[t]{$4$}}
\put(0,0){\circle*{1}}
\put(10,0){\circle*{1}}
\put(10,10){\circle*{1}}
\put(0,10){\circle*{1}}
\put(20,10){\circle*{1}}
\put(-1.3,4.6){\makebox(0,0)[b]{$\phi^*$}}
\put(8.5,4.6){\makebox(0,0)[b]{$\phi$}}
\put(15,8){\makebox(0,0)[b]{$v$}}
\end{picture}
  \\
  e)\begin{picture}(14,15)(-2,-3)
\linethickness{1pt}
\thinlines
\put(0,0){\line(1,0){10}}
\put(0,0){\line(0,1){10}}
\put(10,0){\line(0,1){10}}
\put(-1,-1){\makebox(0,0)[t]{$1$}}
\put(-1,11.5){\makebox(0,0)[t]{$2$}}
\put(11,-1){\makebox(0,0)[t]{$3$}}
\put(11,11.5){\makebox(0,0)[t]{$4$}}
\put(0,0){\circle*{1}}
\put(10,0){\circle*{1}}
\put(10,10){\circle*{1}}
\put(0,10){\circle*{1}}
\put(-1.3,4.6){\makebox(0,0)[b]{$\phi^*$}}
\put(5,-2){\makebox(0,0)[b]{$\phi$}}
\put(8.7,4.6){\makebox(0,0)[b]{$v$}}
\end{picture}
  \hspace{.2cm}
  f)\begin{picture}(24,15)(-2,-3)
\linethickness{1pt}
\thinlines
\put(0,0){\line(0,1){10}}
\put(10,0){\line(0,1){10}}
\put(20,0){\line(0,1){10}}
\put(-1,-1){\makebox(0,0)[t]{$1$}}
\put(-1,11.5){\makebox(0,0)[t]{$2$}}
\put(11,-1){\makebox(0,0)[t]{$3$}}
\put(9,9){\makebox(0,0)[t]{$4$}}
\put(21,11.5){\makebox(0,0)[t]{$5$}}
\put(21,-1){\makebox(0,0)[t]{$6$}}
\put(0,0){\circle*{1}}
\put(0,10){\circle*{1}}
\put(10,0){\circle*{1}}
\put(10,10){\circle*{1}}
\put(20,10){\circle*{1}}
\put(20,0){\circle*{1}}
\put(-1.5,4.6){\makebox(0,0)[b]{$\phi^*$}}
\put(8.5,4.6){\makebox(0,0)[b]{$v$}}
\put(18.5,4.6){\makebox(0,0)[b]{$\phi$}}
\end{picture}
  \caption[Non-diagonal potential terms]
  {Illustration of $\phi_{12}^*v_{34}$-terms.}
  \label{fig:phi12v34}
\end{figure}

\subsection{Evaluation of terms}

\label{sec:evaluation-terms}

The term of \reffig.~\ref{fig:phi12v12}a is trivial since the integrand is
independent of $\tau$.  The terms of \reffigs.~\ref{fig:phi12v12}b
\ref{fig:phi12v12}c, \ref{fig:phi12v13}a, \ref{fig:phi12v13}b,
\ref{fig:phi12v34}a, and \ref{fig:phi12v34}c can be evaluated by
\refeqs.~(\ref{eq:7}) and (\ref{eq:40}).

The term of \reffig.~\ref{fig:phi12v13}c becomes with the use of the
standard Jacobi tree of \reffig.~\ref{fig:jacobi.trees}a

\begin{eqnarray}
  &&
  \int d\tau\; f(\alpha_{13})\;g(\alpha_{23})
  =
  \frac{2}{\sqrt\pi}\gammafktB{3N-6}{3N-9}
  \times
  \nonumber\\
  &&
  \qquad
  \int_0^{\pi/2} d\beta\;\sin^2\beta\cos^{3N-10}\beta
  \int_0^\pi d\vartheta\;\sin\vartheta\;f(\alpha_{13})\;g(\alpha_{23})
  \;,
  \\
  &&
  \sin^2\alpha_{13,23}=\frac{3}{4}\cos^2\alpha\sin^2\beta+
  \frac{1}{4}\sin^2\alpha\pm
  \frac{\sqrt3}{2}\cos\alpha\sin\alpha\sin\beta\cos\vartheta
  \;.\qquad
\end{eqnarray}

The term of \reffig.~\ref{fig:phi12v34}f becomes with the use of the
alternative (12)(34)(56)-tree of \reffig.~\ref{fig:jacobi.trees}e
\begin{eqnarray}
  &&
  \int d\tau\; v(\alpha_{34})\;\phi(\alpha_{56})
  =
  \frac{2A_N}{\pi}
  \times
  \\ \nonumber
  &&
  \qquad
  \int d\beta\;\sin^2\beta\cos^{3N-10}\beta
  \int d\gamma\;\sin^2\gamma\cos^{3N-13}\gamma\;
  v(\alpha_{34})\;\phi(\alpha_{56})
  \;,\qquad
  \\
  &&
  \sin\alpha_{34}=\cos\alpha\sin\beta
  \;,\qquad
  \sin\alpha_{56}=\cos\alpha\cos\beta\sin\gamma
  \;,\\
  &&
  A_N\equiv (3N-8)(3N-10)(3N-12)
  \;.
\end{eqnarray}

The terms of \reffigs.~\ref{fig:phi12v13}g, \ref{fig:phi12v34}b, and
\ref{fig:phi12v34}d are evaluated using the (123)(45)- and
(12)(345)-trees of \reffigs.~\ref{fig:jacobi.trees}c and
\ref{fig:jacobi.trees}d, so
\begin{eqnarray}
  &&
  \int d\tau\; I_5(\alpha,\tau)
  =
  \frac{A_N}{\pi}
  \int_0^{\pi/2}d\beta\;\sin^2\beta\cos^{3N-10}\beta
  \times
  \\ \nonumber
  &&
  \qquad
  \int_0^{\pi/2}d\gamma\;\sin^2\gamma\cos^{3N-13}\gamma
  \int_0^\pi d\vartheta_{x,z}\;\sin\vartheta_{x,z}\;
  I_5(\alpha,\tau)
  \;,
\end{eqnarray}
where $I_5(\alpha,\tau)$ can be either
$v(\alpha_{34})\phi(\alpha_{35})$ or $f(\alpha_{13})g(\alpha_{45})$.
The relevant angles are 
\begin{eqnarray}
  &&
  \sin\alpha_{34}
  =
  \cos\alpha\sin\beta
  \;,\qquad
  \sin\alpha_{45}=\cos\alpha\cos\beta\sin\gamma
  \;,\\
  &&
  \sin^2\alpha_{35}
  =
  \frac{\cos^2\alpha}4\Big(3\cos^2\beta\sin^2\gamma
  +\sin^2\beta
  \nonumber\\
  &&\qquad
  +2\sqrt3\cos\beta\sin\beta\sin\gamma\cos\vartheta_z\Big)
  \;,\qquad\quad
\end{eqnarray}
and $\alpha_{13}$ given by \refeq.~(\ref{eq:5}).  Note the identity
$\int d\tau\; v(\alpha_{34})\phi(\alpha_{15})= \int d\tau\;
\phi(\alpha_{13})\;v(\alpha_{45})$.

The terms of \reffigs.~\ref{fig:phi12v13}d, \ref{fig:phi12v13}e,
\ref{fig:phi12v13}f, and \ref{fig:phi12v34}e are evaluated using the
standard Jacobi tree.  Then \refeq.~(\ref{eq:6}) reduces to, with
$i=1,2,3$,
\begin{eqnarray}
  &&
  \int d\tau\; f(\alpha_{13})\;g(\alpha_{i4})
  =
  \frac{A_N}{4\pi^2}
  \int_0^{\pi/2}d\beta\;\sin^2\beta\cos^{3N-10}\beta
  \int_0^\pi d\vartheta_x\;\sin\vartheta_x
  \times
  \nonumber\\
  &&
  \int_0^{\pi/2}d\gamma\;\sin^2\gamma\cos^{3N-13}\gamma
  \int_0^\pi d\vartheta_y\;\sin\vartheta_y
  \int_0^{2\pi} d\varphi\;
  f(\alpha_{13})\;g(\alpha_{i4})
  \;.
\end{eqnarray}
The angles $\alpha_{ij}$ can be determined by $\rho\sin\alpha_{ij} =
\eta_{ij}$ through the relations
\begin{eqnarray}
  &&\boldsymbol\eta_{13}
  =
  \frac{\sqrt3}{2}\boldsymbol\eta_{N-2}+\frac12\boldsymbol\eta_{N-1}
  \;,\\
  &&
  \boldsymbol\eta_{14}=\sqrt{\frac{2}{3}}\boldsymbol\eta_{N-3}+
  \frac{1}{2\sqrt3}\boldsymbol\eta_{N-2}+\frac12\boldsymbol\eta_{N-1}
  \;,
  \\
  &&\boldsymbol\eta_{24}=\sqrt{\frac{2}{3}}\boldsymbol\eta_{N-3}+
  \frac{1}{2\sqrt3}\boldsymbol\eta_{N-2}-\frac12\boldsymbol\eta_{N-1}
  \;,
  \\
  &&
  \boldsymbol\eta_{34}=\sqrt{\frac{2}{3}}\boldsymbol\eta_{N-3}-
  \frac{1}{\sqrt3}\boldsymbol\eta_{N-2}
  \;.\qquad\quad
\end{eqnarray}

\subsection{Results in the short-range limit}

\label{sec:results-delta-limit}

The integrals in the short-range limit, when the range $b$ of
$V(r_{ij})$ is much smaller than the size scale $\rho$, are:
\begin{eqnarray}
  &&
  \int d\tau\; v(\alpha_{34})\phi(\alpha_{13})\;\simeq\;
  v_1(\alpha)\hatR_{3413}\super{(2)}\phi(\alpha)
  \;,
  \\
  &&
  \int d\tau\; v(\alpha_{34})\phi(\alpha_{15})\;\simeq\;
  v_1(\alpha)\hatR_{13}^{(N-3)}\phi(\alpha)
  \;,
  \\
  &&
  \int d\tau\; v(\alpha_{34})\phi(\alpha_{34})=
  \hatR_{34}^{(N-2)}v\phi(\alpha)
  \simeq v_1(\alpha)\phi(0)
  \;,\;\qquad
  \\
  &&
  \int d\tau\; v(\alpha_{34})\phi(\alpha_{35})\;\simeq\;
  v_1(\alpha)\hatR_{3435}\super{(1)}\phi(\alpha)
  \;,
  \\
  &&
  \int d\tau\; v(\alpha_{34})\phi(\alpha_{56})\;\simeq\;
  v_1(\alpha)\hatR_{34}^{(N-3)}\phi(\alpha)
  \;,
  \\
  &&
  \int d\tau\; v(\alpha_{13})\phi(\alpha_{13})=
  \hatR_{13}^{(N-2)}v\phi(\alpha)
  \simeq v_2(\alpha)\phi(0)
  \;,\qquad
  \\
  &&
  \int d\tau\; v(\alpha_{13})\phi(\alpha_{i4})\simeq
  v_2(\alpha)\hatR_{1314}\super{(2)}\phi(\alpha)
  \;;\; i=1,3
  \;,\quad
  \\
  &&
  \int d\tau\; v(\alpha_{13})\phi(\alpha_{23})\simeq
  v_2(\alpha)\phi(\alpha)
  \;,
  \\
  &&
  \int d\tau\; v(\alpha_{13})\phi(\alpha_{24})\simeq
  v_2(\alpha)\hatR_{1324}\super{(2)}\phi(\alpha)
  \;,
  \\
  &&
  \int d\tau\; v(\alpha_{13})\phi(\alpha_{45})\simeq
  v_2(\alpha)\hatR_{1345}\super{(1)}\phi(\alpha)
  \;.
\end{eqnarray}
The integrals are given by
\begin{eqnarray}
  \hatR_{ijkl}\super{(1)}\phi(\alpha)
  \equiv
  \frac4{\sqrt\pi}\gammafktB{3N-9}{3N-12}\angleint{\gamma}{3N-13}
  \;\phi(\alpha_{kl}^0)
  \;,\qquad
\end{eqnarray}
where $\sin\alpha_{35}^0\equiv\sqrt3\cos\alpha\sin\gamma/2$,
$\sin\alpha_{45}^0\equiv\cos\alpha\cos\beta_0\sin\gamma$,
$\sin\beta_0\equiv\tan\alpha/\sqrt3$, and
\begin{eqnarray}
  &&
  \hatR_{ijkl}\super{(2)}\phi(\alpha)
  \equiv
  \frac2{\sqrt\pi}\gammafktB{3N-9}{3N-12}\angleint{\gamma}{3N-13}
  \nonumber\\
  &&\qquad
  \times
  \int_0^\pi d\vartheta_x\sin\vartheta_x\;\phi(\alpha_{kl}^0)
  \;,
  \\
  &&
  \sin^2\alpha_{14}^0
  \equiv
  \frac19\sin^2\alpha+
  \frac23\cos^2\alpha\cos^2\beta_0\sin^2\gamma
  \nonumber\\
  &&\qquad
  +\frac{2\sqrt2}{3\sqrt3}\sin\alpha\cos\alpha\cos\beta_0\sin\gamma\cos\vartheta_x
  \;,
  \\
  &&
  \sin^2\alpha_{24}^0
  \equiv
  \frac49\sin^2\alpha+
  \frac23\cos^2\alpha\cos^2\beta_0\sin^2\gamma
  \nonumber\\
  &&\qquad
  +\frac{4\sqrt2}{3\sqrt3}\sin\alpha\cos\alpha\cos\beta_0\sin\gamma\cos\vartheta_x
  \;,
  \\
  &&
  \sin^2\alpha_{13}^0
  \equiv
  \frac14\sin^2\alpha+
  \frac12\cos^2\alpha\sin^2\gamma
  \nonumber\\
  &&\qquad
  +\frac1{\sqrt2}\sin\alpha\cos\alpha\sin\gamma\cos\vartheta_x
  \;.
\end{eqnarray}
The two-dimensional integral $\hatR_{ijkl}\super{(2)}\phi(\alpha)$
can be reduced to a one-dimensional integral, analogously to
\refeq.~(\ref{eq:40}), by a transformation of the general form
\begin{eqnarray}
  &&
  I=
  \int_0^{\pi/2}d\gamma\sin^2\gamma\cos^p\gamma
  \int_0^\pi d\vartheta\sin\vartheta \;\phi(\alpha')
  \;,\\
  &&
  \sin^2\alpha'
  =f^2(\alpha)
  +g^2(\alpha)\sin^2\gamma
  +2f(\alpha)g(\alpha)\sin\gamma\cos\vartheta
  \;,
\end{eqnarray}
to the one-dimensional integral
\begin{eqnarray}
  I
  &=&
  \frac{1}{2f(\alpha)g(\alpha)(p+1)}
  \Bigg[
  \int
  _{\textrm{MAX}\{0,\alpha'^-\}}
  ^{\alpha'^+}
  d\alpha'\sin2\alpha'\cos^{p+1}\gamma^+ \;\phi(\alpha')
  \nonumber\\
  &&
  -
  \int
  _{0}
  ^{\textrm{MAX}\{0,-\alpha'^-\}}
  d\alpha'\sin2\alpha'\cos^{p+1}\gamma^- \;\phi(\alpha')
  \Bigg]
  \;,\\
  \sin\gamma^\pm
  &\equiv&
  \frac{\pm\sin\alpha' - f(\alpha)}{g(\alpha)}
  \;,\quad
  \sin\alpha'^\pm
  =f(\alpha)\pm g(\alpha)
  \;.
\end{eqnarray}
The function MAX$\{x,y\}$ outputs the largest of the two numbers $x$
and $y$.

\chapter{Properties of Jacobi functions}

\label{sec:prop-select-funct}

The Jacobi function $\JacP_\nu^{(a,b)}(x)$ is related to the hyperangular
kinetic-energy eigenfunctions.  More specifically, an eigenfunction to
the operator $\hat\Pi_k^2$ from \smarteq{eq:pik2gend} is
$\JacP_\nu^{(a,b)}(x)$ with $x=\cos2\alpha_k$, $a=(d-2)/2$, and
$b=d(k-1)/2-1$.  Some relevant properties of these functions in the
relation to the hyperspherical treatments are given by
\etalcite{Nielsen}{nie01}.  The important properties in this context
are the following \cite{abr,nie01}.

The Jacobi function $\JacP_\nu^{(a,b)}(x)$ is a solution to the
differential equation
\begin{eqnarray}
  (1+x^2)y''(x)+[a+b+(a+b+1)x]y'(x)+\nu(\nu+a+b+1)y(x)=0
  \;.
\end{eqnarray}
A second solution is $\JacP_\nu^{(b,a)}(-x)$, which for integer $\nu$ is
identical to $\JacP_\nu^{(a,b)}(x)$.  For non-integer $\nu$ the Jacobi
function $\JacP_\nu(x)$ is regular at $x=1$ and irregular at $x=-1$.  We
will for integer $\nu$ not consider the irregular solution, which
diverges at both $x=\pm1$.

Important relations for the Jacobi function are
\begin{eqnarray}
  &&
  \JacP_\nu^{(a,b)}(x)
  =
  \frac{\Gamma(\nu+a+1)}{\Gamma(\nu+1)\Gamma(a+1)}
  \hypgeoF\bigg(-\nu,\nu+a+b+1;a+1;\frac{1-x}2\bigg)
  \;,\\
  &&
  \JacP_\nu^{(b,a)}(x)
  =
  \frac{\Gamma(-a)}{\Gamma(\nu+1)\Gamma(-\nu-a)}
  \hypgeoF\bigg(-\nu,\nu+a+b+1;1+a;\frac{1+x}2\bigg)
  \qquad
  \nonumber\\
  &&
  \qquad
  +\frac{\Gamma(a)\Gamma(\nu+b+1)}{\Gamma(\nu+1)\Gamma(-\nu)\Gamma(nu+a+b+1)}
  \bigg(\frac{1+x}2\bigg)^{-a}
  \nonumber\\
  &&
  \qquad
  \times
  \hypgeoF\bigg(\nu+b+1,-\nu-a;1-a;\frac{1+x}2\bigg)
  \;,
\end{eqnarray}
where $\hypgeoF$ is the hypergeometric function $_2\hypgeoF_1$ \cite{abr}.

For small values of the argument $\alpha_k$ they can be rewritten via
\begin{eqnarray}
  \JacP_\nu^{(a,b)}(x)
  &=&
  \JacP_\nu^{(b,a)}(-x)\cos\pi\nu
  -\JacQ_\nu^{(b,a)}(-x)\sin\pi\nu
  \;,\\
  \JacP_\nu^{(a,b)}(\cos2\alpha)
  &=&
  \frac{\Gamma(\nu+a+1)}{\Gamma(\nu+1)\Gamma(a+1)}
  \;,\\
  \JacQ_\nu^{(a,b)}(\cos2\alpha)
  &=&
  \frac{\Gamma(a)\Gamma(\nu+b+1)}{\pi\Gamma(\nu+a+b+1)}\alpha^{-2a}
  \qquad\textrm{for }a>0
  \;.
\end{eqnarray}
Some properties of the gamma function $\Gamma(x)$ are
$\Gamma(x+1)=x\Gamma(x)$, \\ $\Gamma(x)\Gamma(1-x)=\pi/\sin(\pi x)$,
and $\Gamma(A+x)/\Gamma(A)\to A^x$ for $A\gg|x|$.

\section*{Rotation properties}

For integer values of the quantum number $\nu$ the Jacobi function for
$d=3$ and $k=N-1$, i.e.~$a=1/2$ and $b=3N/2-4$, can be written as the
polynomial (omitting upper indices)
\begin{eqnarray}
  \JacP_{\nu}(\cos2\alpha)
  &=&
  \sum_{n=0}^{\nu}c_{\nu,n}\sin^{2n}\alpha
  \;,
  \\
  c_{\nu,0}
  &=&1
  \;,
  \\
  c_{\nu,n+1}
  &=&
  -c_{\nu,n}{(3N-5)(\nu-n)+2(\nu^2-n^2)\over(n+1)(2n+3)}
  \;.
\end{eqnarray}
Some rotation properties of these kinetic-energy eigenfunctions are
\begin{eqnarray}
  \hatR_{13}^{(N-2)} \JacP_0
  &=&
  \JacP_0
  \;,\qquad
  \hatR_{34}^{(N-2)} \JacP_0
  =
  \JacP_0
  \;,
  \\
  \hatR_{13}^{(N-2)} \JacP_1
  &=&
  {N-5\over4(N-2)}\JacP_1
  \;,
  \\
  \hatR_{34}^{(N-2)} \JacP_1
  &=&
  -{1\over N-2}\JacP_1
  \;.
\end{eqnarray}
Related properties are
\begin{eqnarray}
  \hatR_{13}^{(N-2)} \sin^2\alpha 
  &=&
  {1\over4(N-2)}\big[3+(N-5)\sin^2\alpha\big]
  \;,
  \\
  \hatR_{34}^{(N-2)} \sin^{2n}\alpha 
  &=&
  \frac{\Gamma\big({3N-6\over2}\big)} {\Gamma\big({3N-6\over2}+n\big)}
  \frac{\Gamma\big({3\over2}+n\big)} {\Gamma\big({3\over2}\big)}
  \cos^{2n}\alpha
  \;.
\end{eqnarray}

\chapter{Numerical scalings of angular potential}

\label{sec:scal-with-scatt}

Chapter \ref{kap:angular} contained an account of the properties of
the numerically obtained angular eigenvalues.  We collect here some of
the details behind the parametrization of the angular eigenvalue in
\smarteqs{eq:lambda0_smallrho} and (\ref{eq:schematic}).  They were
published as a part of a larger article \cite{sor03a} and are kept
here for completeness.

\section{Effective dependence on the scattering length}

The angular eigenvalue spectrum coincides with the free spectrum at
both small and large hyperradii; at $\rho = 0$ because all
interactions are multiplied by $\rho^2$ and at $\rho = \infty$ because
the short-range interaction has no effect at infinitely large
distances.  Thus, perturbation theory for a Gaussian potential shows
that for small $\rho$ the eigenvalues all change from their
hyperspherical values $\lambda_\nu(0) = 2\nu(2\nu+3N-5)$ with
$\nu=0,1,\ldots$ as
\begin{eqnarray}
  \lambda_\nu(\rho)-\lambda_\nu(0)
  =
  \frac{m V(0)}{\hbar^2} N(N-1) \rho^2
  \;.
\end{eqnarray}

If the two-body potential is attractive, but too weak to support a
bound state, the eigenvalues reach a minimum and then return to one of
the finite hyperspherical values.  For a two-body bound state of
energy $E^{(2)}$ one eigenvalue diverges as $\lambda = 2mE^{(2)}\rho^2
/\hbar^2 $.  The corresponding structure describes, appropriately
symmetrized, one pair of particles in that bound state and all others
far apart from the pair and from each other.  In addition to this
finite number of such eigenvalues the hyperspherical spectrum emerges
at large distances.

To illustrate we show in \reffig.~\ref{fig:as_var} a number of angular
eigenvalues $\lambda$ as functions of hyperradius for different
potentials.  The entirely positive (solid line) corresponds to a
repulsive Gaussian.  The curves diverging at large hyperradii (dotted
and thick dot-dashed lines) correspond to potentials with one bound
two-body state.  

\begin{figure}[htb]
  \centering
  \input{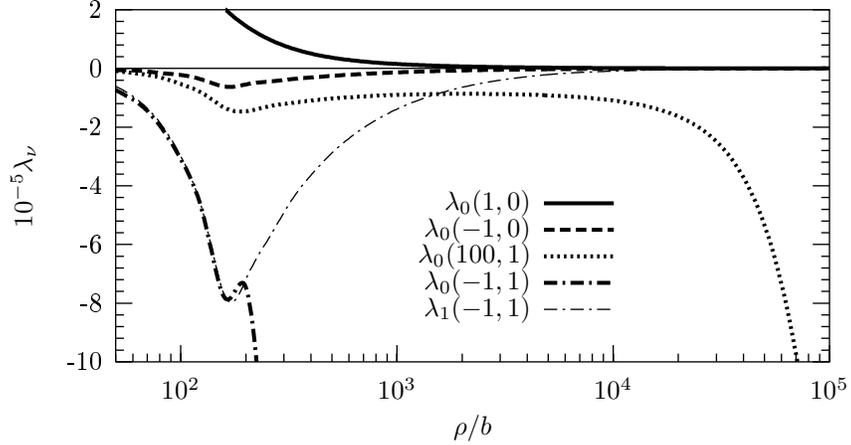}
  \caption [Angular eigenvalues for different scattering lengths]
  {Angular eigenvalues $\lambda_\nu$ (divided by $10^5$) as functions
    of hyperradius divided by interaction range, $\rho /b$, for
    $N=100$, for scattering lengths $a_s/b$ and numbers of bound
    two-body states $\mathcal N\sub B$ indicated as $\lambda_\nu(a\sub
    s/b,\mathcal N\sub B)$ on the figure.}
  \label{fig:as_var}
\end{figure}

The convergence of $\lambda$ as $\rho \rightarrow 0$ is due to the
finite range of the potential and depends on the interaction range
$b$.  The deep minima at small to intermediate distances depend
strongly on both the number of particles and the strength of the
attraction.  Increasing the strength of the attraction leads to larger
negative values of $\lambda$. This trend continues by increasing the
attraction even further until the same scattering length is reached
but now with one bound two-body state.

The asymptotic behaviour of $\lambda$ is compared to the zero-range
result $\lambda_\delta$ in \reffig.~\ref{fig:as_var.transformed}.  The
convergence to the limiting value is fastest for the smallest value of
$a_s$ (dashed curve) already reflecting that the correlations arising
for large scattering lengths (dotted line) cannot be accounted for by
the simple zero-range result.  This is well understood for three
particles where the Efimov effect (very large $a_s$) extends
correlations in hyperradius to distances around four times the average
scattering length \cite{fed93,jen97}.  These effects are not present
in the mean-field type of zero-range expectation value contained in
$\lambda_\delta$.  When $\rho$ exceeds $a_s$ by a sufficiently large
amount $\lambda_\delta$ is approached.

\begin{figure}[htb]
  \centering
  \input{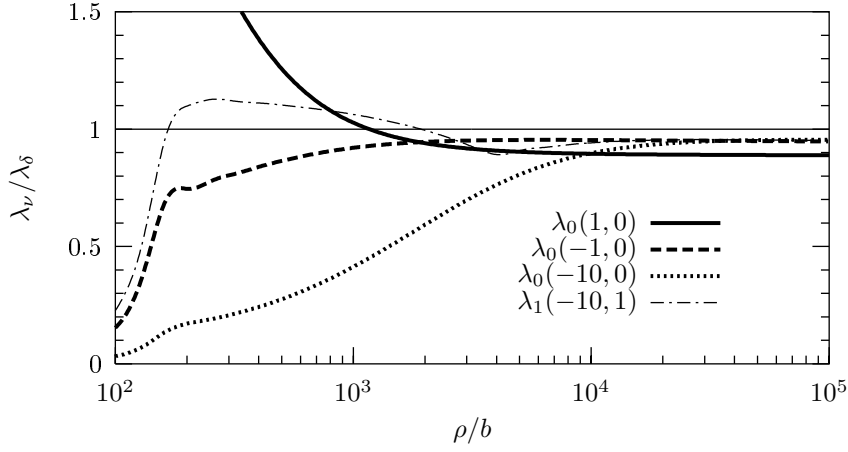}
  \caption
  [Angular eigenvalues in units of zero-range result] {Same as figure
    \ref{fig:as_var}, but the angular potential is shown in units of
    the zero-range result in \refeq.~(\ref{eq:lambda_delta}).}
  \label{fig:as_var.transformed}
\end{figure}

The positive scattering length also leads to an eigenvalue approaching
$\lambda_\delta$ at large distance with a similar convergence rate
(solid curve).  A stronger attraction corresponding to one bound
two-body state produces one diverging eigenvalue while the second
eigenvalue converges towards the lowest hyperspherical value
(dot-dashed curve).  It almost coincides with the lowest eigenvalue
for the same scattering length but for a potential without bound
two-body states (dotted curve).

The deviations from $\lambda_\delta$ at large distance is in all cases
less than 10\%.  The asymptotic behaviour is very smooth but still
originating in systematic numerical inaccuracies which can be cured by
increasing the number of integration points in the finite-difference
scheme.

\section{Dependence on the number of particles}

The angular eigenvalues increase rapidly with $N$ as seen from the
$N^{7/2}$-de\-pen\-dence in $\lambda_\delta$,
\smarteq{eq:lambda_delta}.  The major variation in magnitude is then
accounted for by using this large-distance zero-range result as the
scaling unit.  \Reffig.~\ref{fig:lambda.401} shows a series of
calculations for the same two-body interaction for different numbers
of atoms.  All curves are similar, i.e.~there is a systematic increase
in the characteristic hyperradius, where the curves bend over and
approach the zero-range result.  The large-distance asymptote is
determined by the scattering length.  A characteristic length $\rho_a$
is conveniently defined by
\begin{eqnarray}
  \rho_a(N)
  \equiv
  N^{7/6}
  |a_s|
  \label{eq:rhoa}
  \;,
\end{eqnarray}
where the power is obtained numerically to be very close to the
indicated value $7/6$.

\begin{figure}[htb]
  \centering
  \input{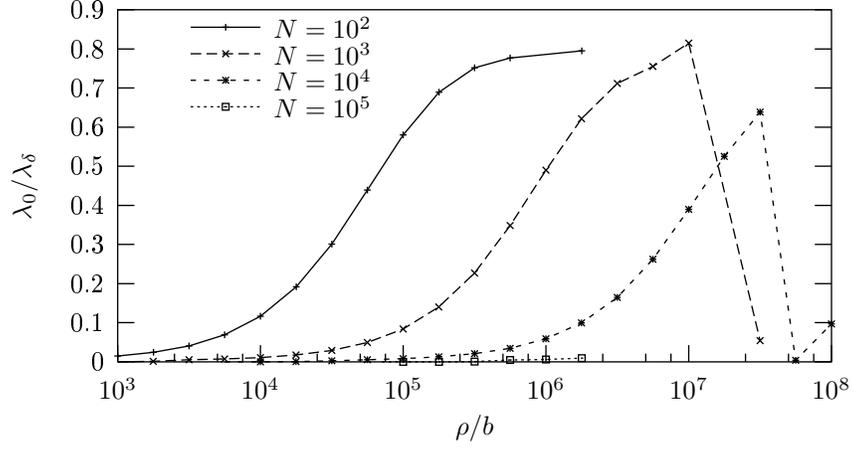}
  \caption[Angular eigenvalues for different particle numbers]
  {The lowest angular eigenvalue as a function of hyperradius for
    $a_s/b=-401$ for four different numbers of particles
    $N=10^2,10^3,10^4,10^5$.  The angular potentials are in units of
    $\lambda_\delta$.}
  \label{fig:lambda.401}
\end{figure}

The quality of this scaling is seen in
\reffig.~\ref{fig:lambda.401.transformed} where all curves
essentially coincide for distances smaller than $\rho_a$.  At larger
hyperradii the zero-range result of $+1$ should be obtained. However,
here numerical inaccuracies produce systematic deviations from a
common curve, i.e.~the deviations increase with $N$.  

\begin{figure}[htb]
  \centering
  \input{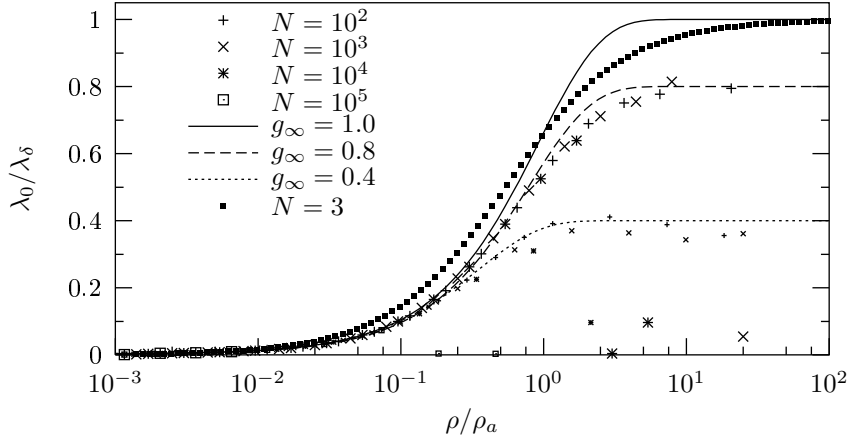}
  \caption[Angular eigenvalues as a function of a scaled hyperradius] 
  {The same as \reffig.~\ref{fig:lambda.401}, but with $\rho$ in units
    of $\rho_a$.  The larger points following the intermediate curve
    ($g_\infty=-0.8$) are calulated with the highest numerical
    accuracy and the smaller points along the upper curve
    ($g_\infty=-0.4$) are obtained with lower accuracy. The curve for
    $g_\infty=-1.0$ is the expected correct asymptotic behaviour.}
  \label{fig:lambda.401.transformed}
\end{figure}

The numerical curves can be rather well reproduced by the function
\begin{eqnarray}
  \lambda^{(-)}(N,\rho)
  =
  |\lambda_\delta(N,\rho)|\cdot g^{(-)}(\rho/\rho_a)
  \label{eq:scale_lambdaminus}
  \;,\\
  g^{(-)}(x)
  =
  g_{\infty} \big(1-e^{-x/x_a}\big)\Big(1+\frac{x_b}{x}\Big)
  \label{eq:scale_gminus}
  \;,
\end{eqnarray}
where $g_{\infty}$ has the value $-1$ in accurate calculations,
because $a_s<0$.  The exponential term reproduces the rather steep
approach to the asymptotic value as seen in
\reffig.~\ref{fig:lambda.401.transformed}.  The behaviour at smaller
distance, depending on the range of the interaction, is here simulated
by the $x_b$-term.  The extreme limit of $\rho \rightarrow 0$ is not
computed and not included in the approximate function in
\refeq.~(\ref{eq:scale_gminus}).

The two groups of computations in
\reffig.~\ref{fig:lambda.401.transformed} are reasonably reproduced by
the parameter sets $x_a \simeq 0.74$, $x_b \simeq 2.3\cdot10^{-3}$,
and $ g_{\infty} \simeq -0.8$ or $g_{\infty} \simeq -0.4$ for the high
and low accuracy, respectively.  These parameters may also depend on
the scattering length.  Table~\ref{table:parameters} gives the best
choice of parameters for different $a_s$.

\begin{table}[htb]
  \centering
  \begin{tabular}{|c||c|c|c|c|}
    \hline    
    $a_s/b$ & $-5.98$ & $-401$ & $-799$ & $-4212$ \\
    \hline
    \hline
    $-g_{\infty}$ & 0.99 & 0.80 & 0.65 & 0.30  \\
    \hline
    $x_a$ & 1.06 & 0.74 & 0.59 & 0.28  \\
    \hline
    $-g_{\infty}/x_a$ & 0.93 & 1.081 & 1.099 & 1.077  \\
    \hline
    $x_b$ & $0.15$ & $2.3\cdot10^{-3}$ & $1.15\cdot10^{-3}$ & 
    $2.2\cdot10^{-4}$  \\
    \hline
    $x_b/(b/|a_s|)$ & 0.92 & 0.922 & 0.919 & 0.927  \\
    \hline
  \end{tabular}
  \caption[Parameters for the scaling of angular potential]
  {Numerical values of $g_{\infty}$, $x_a$, and $x_b$ for four
    scattering lengths.}  
  \label{table:parameters}
\end{table}

It should be noticed that $-g_{\infty}$ and $x_a$ both are of order
unity, and that the fraction $g_{\infty}/x_a$ is almost constant,
except for the smallest scattering length.  The parameter $x_b$,
introduced to account for the finite interaction range, is close to
$b/|a\sub s|$.

At large hyperradii, where $x_a\ll \rho/\rho_a$ or equivalently $\rho
\gg N^{7/6}|a_s|$, $\lambda^{(-)}$ approaches $g_{\infty}
|\lambda_\delta|$.  The expected large-distance asymptotic behaviour
is $\lambda^{(-)} \to \lambda_\delta$ and $g_{\infty}$ should
therefore approach $-1$ in increasingly accurate calculations.  The
results for $N=100$ and different scattering lengths, see
\reffig.~\ref{fig:as_var.transformed}, confirm this conclusion by
deviating less than $10\%$ from $\lambda_\delta$ at large hyperradii.

A well-established result for $N=3$ identical bosons is the
large-distance behaviour \cite{jen97}
\begin{eqnarray}
  \lambda_\delta(N=3,\rho) =
  \frac{48a_s}{\sqrt2\pi\rho}
  \;,
\end{eqnarray}
which is in agreement with $\lambda_\delta$ obtained from
\smarteq{eq:lambda_delta} for $N=3$.  Then the universal function
$g^{(-)}$ asymptotically approaches $g_{\infty}=-1$ for all scattering
lengths.

The function $g^{(-)}$ is almost independent of $N$.  This combined
with the conclusion for $N=3$ implies that $g_{\infty}=-1$ is valid
for all scattering lengths and particle numbers.

The angular eigenvalue is given by $g^{(-)}(x) \simeq g_\infty x/x_a$
for $x_b\ll \rho/\rho_a \ll x_a$.  Numerical calculations in this
intermediate region of hyperradii therefore rather accurately
determines the fraction $g_{\infty}/x_a \simeq - 1.08 $ as given in
table~\ref{table:parameters}.  With $g_{\infty}=-1$ this implies that
$x_a\simeq1/1.08\simeq0.92$.  The parameters of $g^{(-)}(x)$ in
\refeq.~(\ref{eq:scale_gminus}) can now be collected to be
\begin{eqnarray}
  g_{\infty}=-1
  \;,\qquad
  x_a \simeq 0.92
  \;,\qquad
  x_b \simeq 0.92 \frac{b}{|a_s|}
  \label{eq:scale_parameters}
  \;.
\end{eqnarray}

The accuracy of the parametrization is seen in
\reffigs.~\ref{fig:lambda_analytic}a-d, where the angular eigenvalues
are shown in units of $\lambda^{(-)}$ with the individual set of
parameters from table~\ref{table:parameters}.  A fairly good agreement
is found for $\rho/\rho_a>x_b$.  The remaining deviations occur at
small hyperradii, which is not included in the
$g^{(-)}$-parametrization, and at large hyperradii where the numerical
inaccuracy increases with increasing scattering lengths.  On the other
hand the large-distance behaviour is known from analytic
considerations, which renders numerical computations at these
distances superfluous.

\begin{figure}[htb]
  \centering
  \psfig{figure=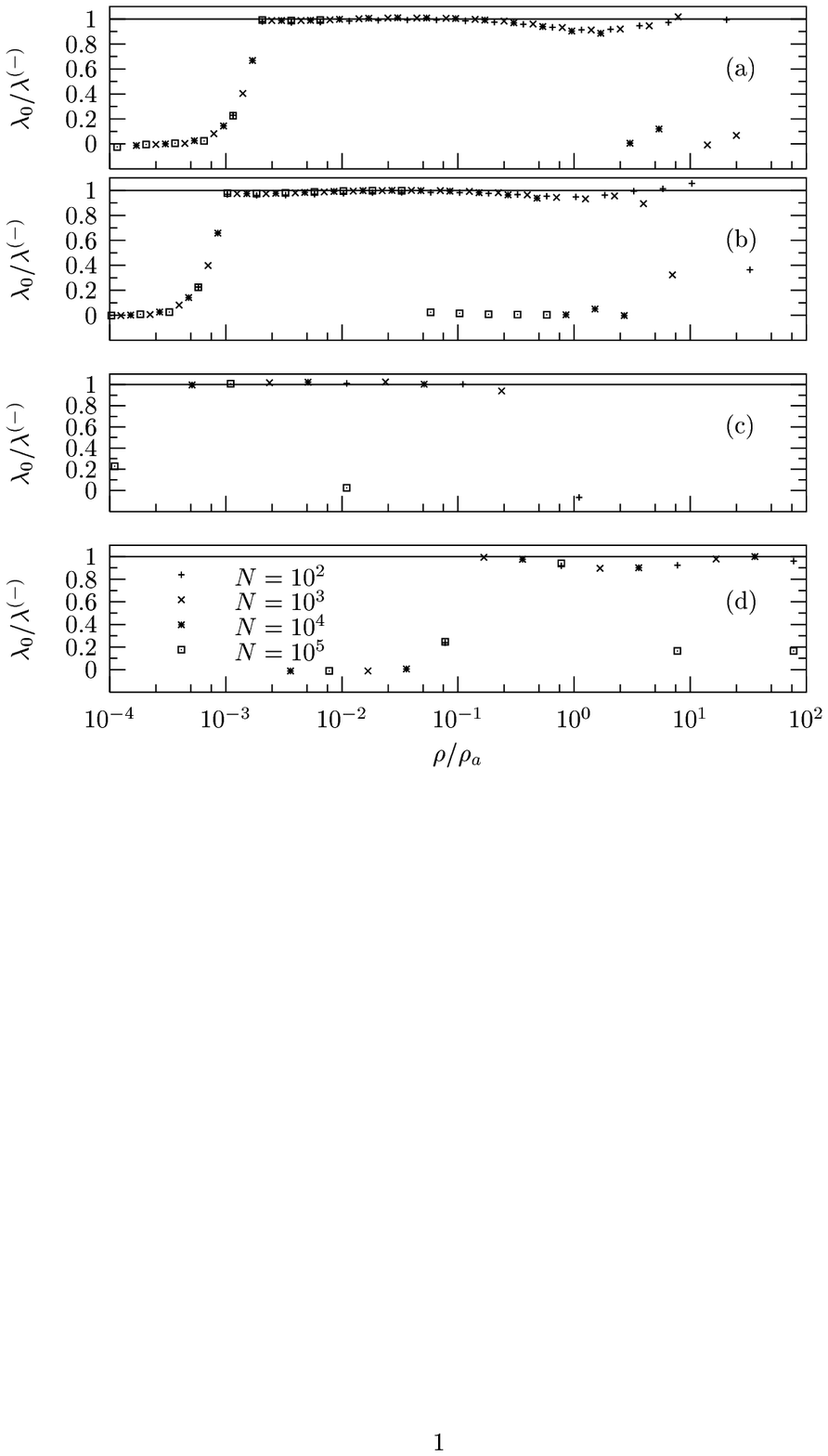,%
    bbllx=4.5cm,bblly=14.8cm,bburx=16.3cm,bbury=25.6cm,angle=0,width=12cm}
  \caption[Angular eigenvalues in units of analytical approximation] 
  {The lowest angular eigenvalue $\lambda$ in units of
    $\lambda^{(-)}$, \refeqs.~(\ref{eq:scale_lambdaminus}) and
    (\ref{eq:scale_gminus}) and table~\ref{table:parameters}, as
    functions of the hyperradius in units of $\rho_a$,
    \refeq.~(\ref{eq:rhoa}).  The scattering lengths are given by a)
    $a_s/b=-401$, b) $a_s/b=-799$, c) $a_s/b=-4212$, and d)
    $a_s/b=-5.98$.  The different $N$-values are as indicated.}
  \label{fig:lambda_analytic}
\end{figure}

\section{Bound two-body state}

In the presence of a bound two-body state one angular eigenvalue
eventually diverges at large hyperradii as
\begin{eqnarray} 
  \lambda^{(2)}(\rho)
  =\frac{2m\rho^2}{\hbar^2}E^{(2)}
  \;,\qquad
  E^{(2)}<0
  \label{eq:lambda_2bound}
  \;,
\end{eqnarray}
where $E^{(2)}$ is the energy of the two-body bound state.  In the
limit of weak binding, or for numerically large scattering lengths,
the energy of the two-body bound state is given by
\begin{eqnarray} 
  E^{(2)}=-\frac{\hbar^2}{m a_s^2}c
  \label{eq:twobodyenergy}
  \;,
\end{eqnarray}
where $c$ approaches unity for large scattering lengths.

The angular eigenvalue corresponding to a two-body bound state is
parametrized by an expression similar to
\refeqs.~(\ref{eq:scale_lambdaminus}) and (\ref{eq:scale_gminus}).
The effect of the bound two-body state only shows up at large
distances where the behaviour corresponds to
\refeq.~(\ref{eq:twobodyenergy}).  The small and intermediate
distances resemble the behaviour when no bound state is present.
Therefore the angular eigenvalue is given by the parametrization
\begin{eqnarray} 
  &&
  \lambda^{(+)}(N,\rho)
  =
  |\lambda_\delta(N,\rho)| \; g^{(+)}(\rho/\rho_a)
  \label{eq:scale_lambdaplus}
  \;,\\
  &&
  g^{(+)}(x)
  =
  x
  \Big(1+\frac{x_b}{x}\Big)
  \bigg(\frac{g_{\infty}}{x_a} - c\frac{4}{3}\sqrt{\frac{\pi}{3}}x^2\bigg)
  \label{eq:scale_gplus}
  \;,
\end{eqnarray}
with the notation and estimates from
\refeq.~(\ref{eq:scale_parameters}), i.e.~$x_b\simeq0.92b/|a_s|$ and
$g_{\infty}/x_a\simeq -1.08$.  The terms in the second bracket of this
expression only aim at the correct behaviour in the limits of small to
intermediate and large hyperradii.  The exact transition between these
regions is not reproduced.

\Reffig.~\ref{fig:lambda_analytic.plus100} shows a comparison of the
parametrization in \refeqs.~(\ref{eq:scale_lambdaplus}) and
(\ref{eq:scale_gplus}) with the computed angular eigenvalues for a
potential with one bound two-body state.  For the large scattering
length in \reffig.~\ref{fig:lambda_analytic.plus100}a one smooth curve
applies for all the particle numbers; numerical inaccuracies set in at
larger hyperradii, which is most obvious for the largest particle
numbers.  This smooth curve is in a large interval of hyperradii at
most deviating by 20\% from the parametrized form, and even less than
10\% at large hyperradii before numerical instabilities set in.

The shape at intermediate distances could be improved for example by
inclusion of a linear term in \refeq.~(\ref{eq:scale_gplus}).  The
smooth curve at small hyperradii is outside the range of validity of
the parametrization, i.e.~this is within the range of the two-body
potential and therefore depends on details of the interaction.

\begin{figure}[htb]
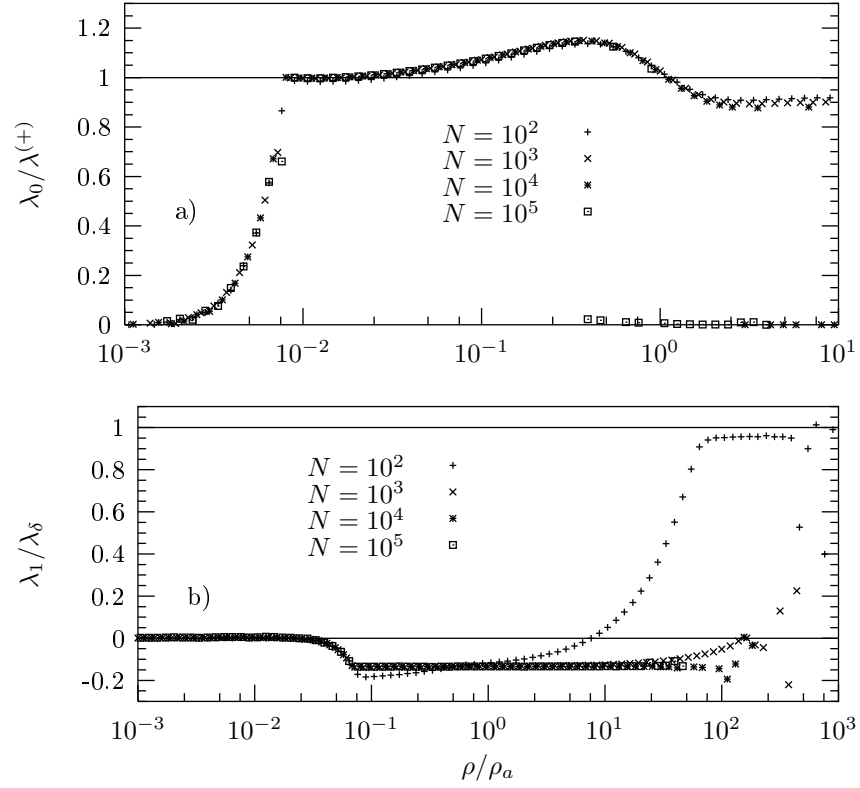

  \centering
  \input{lambda_analytic.plus100}
  \input{lambda_excited_analytic.plus10}
  \caption
  [Angular eigenvalues for two-body states in units of approximation]
  {a) The lowest angular eigenvalue $\lambda_0$ in units of
    $\lambda^{(+)}$, \refeqs.~(\ref{eq:scale_lambdaplus}) and
    (\ref{eq:scale_gplus}), for $a_s/b=+100$ and $c=1.02$, when the
    potential holds one bound two-body state.  The number of particles
    is indicated on the figure.  The parameters are
    $g_{\infty}/x_a=-1.09$ and $x_b=9.2\cdot10^{-3}$.  b) The first
    excited angular eigenvalue $\lambda_1$ in units of
    $\lambda_\delta$ for $a_s/b = +10$.}
  \label{fig:lambda_analytic.plus100}
\end{figure}

The lowest eigenvalue diverges at large hyperradius as described in
connection with \reffig.~\ref{fig:lambda_analytic.plus100}a and
\refeq.~(\ref{eq:scale_lambdaplus}).  If the two-body potential only
has one bound state the second eigenvalue is expected to approach zero
at large distances as $\lambda_\delta$.  This pattern should be
repeated with more than one bound two-body state, i.e.~the first
non-divergent angular eigenvalue should behave as $\lambda_\delta$ for
large $\rho$.  \Reffig.~\ref{fig:lambda_analytic.plus100}b therefore
compares the computed first excited angular eigenvalue with
$\lambda_\delta$ for different $N$.  As in
\reffig.~\ref{fig:lambda.401.transformed} smooth and almost universal
curves are obtained at small $\rho$, where the approach to unity sets
in exponentially fast depending on $N$, but now for $\rho$ one or two
orders of magnitude larger than $\rho_a$.  A parametrization would
also here be possible.  The large-distance asymptotic behaviour of the
first excited state corresponds to a repulsive potential as seen by
the approach to $+1$.  However, at small and intermediate hyperradii
the potential is attractive ($\lambda_1<0$).

\chapter{Derivation of effective dimension}

\label{sec:deriv-eff-dim}

For the case of $N$ non-interacting identical bosons trapped in a
cylindrically deformed harmonic field the Hamiltonian can be written
with the choice of coordinates in
appendix~\ref{sec:hypercyl_parametric} as
\begin{eqnarray}
  \hat H
  &=&
  \hat H_\rho
  +\hat H_\theta
  +\hat T_\Omega
  \;,\\
  \frac{2m\hat H_\rho}{\hbar^2}
  &=&
  -\frac{1}{\rho^{3N-4}}
  \frac{\partial}{\partial\rho}
  \rho^{3N-4}
  \frac{\partial}{\partial\rho}
  +\frac{\rho^2}{b_\perp^4}
  \;,
  \\
  \frac{2m\rho^2\hat H_\theta}{\hbar^2}
  &=&
  -\frac{1}{\cos^{N-2}\theta\sin^{2N-3}\theta}
  \frac{\partial}{\partial \theta}
  {\cos^{N-2}\theta\sin^{2N-3}\theta}
  \frac{\partial}{\partial \theta}
  \nonumber\\
  &&
  +\rho^4\bigg(\frac{1}{b_z^4}-\frac{1}{b_\perp^4}\bigg)\cos^2\theta
  \;.
\end{eqnarray}
Here $\hat T_\Omega$ is the angular kinetic energy operator which is
neglected later when there is no dependency on the internal angles.
The ground state total wave function is known to be
\begin{eqnarray}
  \Psi\sub{total}
  =
  \prod_{i=1}^N
  \exp \bigg( - \frac{r_{i,z}^2}{2b_z^2}
  - \frac{r_{i,\perp}^2}{2b_\perp^2} \bigg)
  \;,
\end{eqnarray}
with $b_q \equiv \sqrt{\hbar/(m\omega_q)}$.  
We change to the coordinates $\{\rho,\theta\}$ given by
$\rho_z=\rho\cos \theta$ and $\rho_\perp=\rho\sin \theta$, see
appendix~\ref{sec:hypercyl_parametric}, and obtain the wave function
\begin{eqnarray}
  \Psi\sub{total}
  &=&
  \exp\bigg(
  -\frac{NR_z^2}{2b_z^2}
  -\frac{NR_\perp^2}{2b_\perp^2}
  \bigg)
  \;
  F(\rho,\theta)
  \;,\\
  F(\rho,\theta)
  &=&
  \exp\bigg(
  -\frac{\rho^2}{2b_\perp^2}
  \bigg)
  \exp\bigg[
  -\rho^2\cos^2\theta\bigg(\frac{1}{2b_z^2}-\frac{1}{2b_\perp^2}\bigg)
  \bigg]
  \;.
  \label{eq:gend_Frel}
\end{eqnarray}
We write the Schr\"odinger equation and integrate over the coordinate
$\theta$ as follows:
\begin{eqnarray}
  0
  &=&
  \int d\theta
  \;
  \Omega_\theta(\theta)
  F^*(\rho,\theta)\Big(\hat H_\rho+\hat H_\theta-E\Big)F(\rho,\theta)
  \;,\\
  \Omega_\theta(\theta)
  &\equiv&
  \cos^{N-2}\sin^{2N-3}\theta
  \;.\;
  \label{eq:gend_Sch_thetaaveraged}
\end{eqnarray}
Doing this we end up with terms only depending on $\rho$.  Including
formally the integration over hyperradius, but without completing it,
we get
\begin{eqnarray}
  0&=&
  \int d\rho\;\rho^{3(N-1)-1}
  \int d\theta \;\Omega_\theta(\theta)
  F^*(\rho,\theta)\Big(\hat H_\rho+\hat H_\theta-E\Big)F(\rho,\theta)
  \nonumber\\
  &=&
  \int d\rho\;\rho^{3(N-1)-1}
  B(\rho)\Big[-E+\frac{\hbar^2}{2m}(N-1)(2C_\perp+C_z)\Big]
  \label{eq:gend4}
  \;,\\
  B(\rho)
  &\equiv&
  \frac{\Gamma(a)\Gamma(2a)}{2\Gamma(3a)}
  e^{-C_\perp\rho^2}
  \mathcal M\Big(a,3a,-\Delta_1\rho^2\Big)
  \;,\qquad
  a\equiv\frac{N-1}{2}
  \;,
\end{eqnarray}
where $\mathcal M(a,b,z)$ is the Kummer function, i.e.~identical to
the confluent hypergeometric function $_1\mathcal F_1(a,b,z)$
\cite{abr}, $C_q \equiv 1/b_q^2$, and $\Delta_1 \equiv C_z-C_\perp$.
Furthermore, it is clear that
\begin{eqnarray}
  E
  =
  \frac{\hbar^2}{2m}(N-1)(2C_\perp+C_z)
  =\hbar\omega_\perp(N-1)+\frac{1}{2}\hbar\omega_z(N-1)
  \;.
\end{eqnarray}

We desire to write an effective $d$-dimensional Hamiltonian as
\begin{eqnarray}
  \hat H_d=\frac{\hbar^2}{2m}
  \Big[
  -\frac{1}{\rho^{d(N-1)-1}}
  \frac{\partial}{\partial\rho}
  \rho^{d(N-1)-1}
  \frac{\partial}{\partial\rho}
  +\frac{\rho^2}{b_d^4}
  \Big]
  \label{eq:gend6_copy1}
\end{eqnarray}
with an effective dimension $d$ and a general length scale $b_d$.  The
normalization of the corresponding $d$-dimensional Schr\"odinger
equation, with eigenvalue $E_d$, is
\begin{eqnarray}
  \int d\rho\;\rho^{d(N-1)-1}
  G_d^*(\rho)\Big(\hat H_d-E_d\Big)G_d(\rho)
  =
  0
  \;,
  \label{eq:gend7_copy1}
\end{eqnarray}
where $G_d$ is a $d$-dimensional wave function.  We want to
approximate the correct \refeq.~(\ref{eq:gend4}) by this equation.
From the two normalizations we identify the $d$-dimensional wave
function $G_d$ by the relation 
\begin{eqnarray}
  \rho^{3(N-1)-1}B(\rho) =
  \rho^{d(N-1)-1}|G_d(\rho)|^2
  \;.
\end{eqnarray}
Using this in \refeq.~(\ref{eq:gend7_copy1}) we obtain
\begin{eqnarray}
  0&=&
  \int d\rho\;\rho^{d(N-1)-1}
  G_d^*(\rho)\Big(\hat H_d-E_d\Big)G_d(\rho)
  \label{eq:gend8}
  \\ \nonumber
  &=&
  \int d\rho\;\rho^{3(N-1)-1}
  B(\rho)
  \bigg\{-E_d+\frac{\hbar^2}{2m}\Big[(N-1)(2C_\perp+C_z)+v(\rho^2)\Big]\bigg\}
  \;,\\
  v(x)
  &\equiv&
  -x(C_z^2-C_d^2)
  -\frac{1}{x}(3-d)a\big[a(3+d)-2\big]
  \nonumber\\
  &&
  +\frac{4}{3}C_\perp\Delta_1x\mu(-\Delta_1x)
  +\frac{4}{9}\Delta_1^2x\mu^2(-\Delta_1x)
  \label{eq:gend9}
  \;,\\
  \mu(z)
  &\equiv&
  \frac{\mathcal M(a,3a+1,z)}{\mathcal M(a,3a,z)}
  \label{eq:gend10}
  \;.
\end{eqnarray}
If we subtract \refeq.~(\ref{eq:gend8}) from \refeq.~(\ref{eq:gend4}) we get
\begin{eqnarray}
  \int d\rho\;\rho^{3(N-1)-1}
  B(\rho)
  \Big[E-E_d+\frac{\hbar^2}{2m}v(\rho^2)\Big]
  =
  0
  \label{eq:gend11}
  \;.
\end{eqnarray}  
We are now in a position to study the three limits i) spherical
($b_z=b_\perp$), ii) two-dimensional ($b_\perp\gg b_z$), iii) one-dimensional
($b_z\gg b_\perp$).

When the external trap is spherically symmetric, i.e.~$b_z=b_\perp$,
we have $C_z=C_\perp$, $\Delta_1=0$, and $\mu(0)=1$.  This yields
\begin{eqnarray}
  v(\rho^2)=-\rho^2(C_z^2-C_d^2)-\frac{a}{\rho^2}(3-d)[a(3+d)-2]
  \label{eq:gend12}
  \;.
\end{eqnarray}
The bracket of \refeq.~(\ref{eq:gend11}) is zero for all hyperradii when
three conditions are true:
\begin{eqnarray}
  E_d 
  &=& E 
  = \frac{\hbar^2}{2m}(N-1)(2C_\perp+C_z) 
  = \frac{3}{2}\hbar\omega(N-1)
  \;,\\
  C_d
  &=& C_z \;\Longleftrightarrow \;
  b_d=b_z=b_\perp
  \;,\qquad
  d=3
  \;.
\end{eqnarray}
$E$ is the ground state energy minus the centre-of-mass energy for $N$
identical non-interacting particles of mass $m$ in a three-dimensional
oscillator of frequency $\omega$.

The two-dimensional geometry occurs when the external trap is squeezed
along the $r$-plane such that $b_\perp\gg b_z$.  This leads to $C_z\gg
C_\perp$ and $\Delta_1=C_z-C_\perp\simeq C_z>0$.  In this case the
hyperradius is determined by the radial trap length, i.e.~$\rho\sim
\sqrt{N}b_\perp$, which implies that typically $-\Delta_1\rho^2\sim
-NC_z b_\perp^2 = -NC_z/C_\perp \to -\infty$.  We therefore need the
limit of $\mu(z)$ when $z\to-\infty$:
\begin{eqnarray}
  \mu(z)
  \simeq
  \frac{3}{2}\Big(1+\frac{a}{z}+\frac{2a^2-a}{z^2}\Big)
  \;.
\end{eqnarray}
This leads to 
\begin{eqnarray}
  v(\rho^2)
  \simeq
  -2aC_z
  -\rho^2(C_\perp^2-C_d^2)
  -\frac{a}{\rho^2}(2-d)[a(2+d)-2]
  \;.
\end{eqnarray}
The bracket of \refeq.~(\ref{eq:gend11}) becomes to order $\rho^{-2}$
\begin{eqnarray}
  E-E_d+\frac{\hbar^2}{2m}
  \Big\{
  -2aC_z
  -\rho^2(C_\perp^2-C_d^2)
  -\frac{a}{\rho^2}(2-d)[a(2+d)-2]
  \Big\}
  \;.
\end{eqnarray}
This is zero for all hyperradii when
\begin{eqnarray}
  E_d
  &=& E-\frac{\hbar^2}{2m}2aC_z
  = \frac{\hbar^2}{2m}(N-1)2C_\perp
  = \hbar\omega_\perp(N-1)
  \;,\\
  C_d&=&C_\perp
  \;\Longleftrightarrow\;
  b_d=b_\perp
  \;,\qquad
  d=2
  \;.
\end{eqnarray}

The system is effectively one-dimensional when the external trap is
squeezed along the $z$-axis such that $b_z\gg b_\perp$.  This leads to
$C_\perp\gg C_z$ and $\Delta_1=C_z-C_\perp\simeq -C_\perp<0$.  In this
case the hyperradius is determined by the axial trap length,
i.e.~$\rho \sim \sqrt{N}b_z$, which implies that typically
$-\Delta_1\rho^2\sim NC_\perp b_z^2 = NC_\perp/C_z \to +\infty$.  We
therefore need the limit of $\mu(z)$ when $z\to+\infty$:
\begin{eqnarray}
  \mu(z)
  \simeq
  \frac{3a}{z}\Big(1-\frac{a-1}{z}\Big)
  \;.
\end{eqnarray}
This leads to
\begin{eqnarray}
  v(\rho^2)
  \simeq
  -4aC_\perp
  -\rho^2(C_z^2-C_d^2)
  -\frac{a}{\rho^2}(1-d)[a(1+d)-2]
  \;.
\end{eqnarray}
The bracket of \refeq.~(\ref{eq:gend11}) becomes to order $\rho^{-2}$
\begin{eqnarray}
  E-E_d+\frac{\hbar^2}{2m}
  \Big\{
  -4aC_\perp
  -\rho^2(C_z^2-C_d^2)
  -\frac{a}{\rho^2}(1-d)[a(1+d)-2]
  \Big\}
  \;.
\end{eqnarray}
This is zero for all hyperradii when
\begin{eqnarray}
  E_d
  &=& E-\frac{\hbar^2}{2m}4aC_\perp
  = \frac{\hbar^2}{2m}(N-1)C_z
  = \frac{1}{2}\hbar\omega_z(N-1)
  \;,\\
  C_d&=&C_z
  \;\Longleftrightarrow\;  b_d=b_z
  \;,\qquad
  d=1
  \;.
\end{eqnarray}

The results are collected in table~\ref{table1}.
\begin{table}[htb]
  \centering
  \begin{tabular}{c||c|c|c}
    limit & spherical & oblate & prolate  \\
    \hline\hline
    condition & $b_z\simeq b_\perp$ & $b_\perp\gg b_z$ & $b_z\gg b_\perp$ \\
    \hline
    $z=-\Delta_1\rho^2$ & 0 & $-\infty$ & $+\infty$ \\
    \hline
    $\mu(z)$ & 1 & $\frac32(1+\frac az+\frac{2a^2-a}{z^2})$ & $\frac{3a}{z}(1-\frac{a-1}{z})$ \\
    \hline
    $E_d/(N-1)$ & $\frac32\hbar\omega$ & $\hbar\omega_\perp$ & $\frac12\hbar\omega_z$ \\
    \hline
    $b_d$ & $b_z\simeq b_\perp$ & $b_\perp$ & $b_z$ \\
    \hline
    $d$ & $3$ & $2$ & $1$ \\
    \hline
  \end{tabular}
  \caption[]
  {The (typical) values of $z$, $\mu(z)$, $E_d$, $b_d$, and $d$
    in the three limits: spherical, oblate, and prolate.}
  \label{table1}
\end{table}
In all three cases the energy is given by
\begin{eqnarray}
  E_d =
  \frac{d}{2}\hbar\omega_d(N-1)
  \;,\qquad
  \omega_d \equiv \frac{\hbar}{mb_d^2}
  \;.
\end{eqnarray}

In the general case, when we cannot assume $b_z = b_\perp$ or $b_z \gg
b_\perp$ or $b_z \ll b_\perp$, it is not possible to obtain a simple
expansion in $\rho$ of the bracket in \refeq.~(\ref{eq:gend11}).
Instead we study the integrated equation.

We rescale the equation in a convenient length scale $b_0$ given by
\begin{eqnarray}
  b_0^2 \equiv 2b_\perp^2+b_z^2
  \;.
\end{eqnarray}
In the three special limits we found that $E_d = \hbar^2ad/(mb_d^2)$
and $b_d^2 = b_0^2/d$.  We therefore introduce the parameters $e_d$
and $\beta_d$ given by 
\begin{eqnarray}
  E_d \equiv 
  \frac{\hbar^2ae_d}{mb_0^2}
  \;,\qquad
  \beta_d \equiv \frac{b_0^2}{b_d^2}
  \;.
\end{eqnarray}
This leads from \refeq.~(\ref{eq:gend11}) to the integrated equation
\begin{eqnarray}
  &&
  0
  =
  e_d-\frac12\beta_d^2+g(d)f_1(\beta)-f_2(\beta)
  \;,\qquad
  f_1(\beta)
  \equiv
  a^2I[x^{-2}]
  \;,\\
  &&
  f_2(\beta)
  \equiv
  2\beta_r
  +\beta_z
  -\frac12\beta_z^2
  +\frac43\beta_r(\beta_z-\beta_r)I\{x^2\mu[x^2(\beta_r-\beta_z)]\}
  \nonumber\\
  &&
  \qquad
  +\frac49(\beta_z-\beta_r)^2I\{x^2\mu^2[x^2(\beta_r-\beta_z)]\}
  \;,\\
  &&
  g(d)
  \equiv
  (3-d)(3+d-2/a)
  \;,\\
  &&
  \beta\equiv
  \frac{b_\perp^2}{b_z^2}
  \;,\qquad
  \beta_r
  \equiv
  \frac{b_0^2}{b_\perp^2} = 2+\frac{1}{\beta}
  \;,\qquad
  \beta_z
  \equiv
  \frac{b_0^2}{b_z^2} = \frac{2}{\beta}+1
  \;,\\
  &&
  I[f(x)]
  \equiv
  \frac{\beta_z^a\beta_r^{2a}}{a\Gamma(3a)}
  \int_0^\infty dx \; x^{6a-1}\exp(-\beta_rx^2)
  \nonumber\\
  &&
  \qquad
  \times
  \mathcal M\Big[a,3a,x^2(\beta_r-\beta_z)\Big]f(x)
  \;.
\end{eqnarray}

If we use the expectations that
\begin{eqnarray}
  e_d=d^2(1+\varepsilon)
  \;,\qquad
  \beta_d=d(\beta+\delta)^{1/2}
  \;,\qquad
  \varepsilon\ll 1
  \;,\qquad
  \delta\ll 1
  \label{eq:gend15}
\end{eqnarray}
we get that
\begin{eqnarray}
  &&\mathcal Ad^2+\mathcal Bd+\mathcal C=0
  \;,\qquad
  \mathcal A\equiv 1/2+\varepsilon-\delta/2-f_1(\beta)
  \;,\\
  &&
  \mathcal B\equiv 2f_1(\beta)/a
  \;,\qquad
  \mathcal C\equiv (9-6/a)f_1(\beta)-f_2(\beta)
  \;.\qquad
\end{eqnarray}
If we demand only one solution, i.e.~$\mathcal B^2-4\mathcal A\mathcal
C=0$, we get
\begin{eqnarray}
  d=-\frac{\mathcal B}{2\mathcal A}=-\frac{2\mathcal C}{\mathcal B}
  \;.
\end{eqnarray}
The results for various $N$-values are shown in
\reffig.~\ref{olel3fig4}.

Furthermore, we check that $\varepsilon-\delta/2$ is small
($<10^{-2}$) so the approximations in \refeq.~(\ref{eq:gend15}) are
valid.  The conclusion is that generally we can use
$E_d=\hbar^2ad/(mb_d^2)$ and $b_d^2=b_0^2/d$ as the relevant energy
and length scales, respectively, with $d$ given by
\reffig.~\ref{olel3fig4}.

\chapter{List of notations}

\setlongtables

\begin{longtable}{l|l|l}
  Notation & Description & Chapter\\
  \hline\hline
  \endhead
  \hline\hline
  \endfoot
  $a_s$ & two-body $s$-wave scattering length & 3-7\\
  $a\sub B$ & Born approximation to $a_s$ & 2,3,5\\
  $a_d$ & $d$-dimensional interaction parameter & 7\\
  $a\sub{1D}$ & one-dimensional scattering length & 7\\
  $b$ & interaction range & 2-6\\
  $b\sub t$ & trap length & 2-7\\
  $b_d,b_q,b_x,b_y,b_z,b_\perp$ & trap lengths & 4,7\\
  $d$ & spatial dimension & 7\\
  $d\sub c$ & condensate length & 4,6\\
  $E,E_n$ & total relative energy & 2,4-7\\
  $E\sub{total}$ & total energy & 5,6\\
  $E^{(2)}$ & two-body energy & 3\\
  $E_d$ & energy in $d$ dimensions & 7\\
  $f,f_\nu$ & reduced hyperradial wave function & 2,4,7\\
  $f_\infty$ & Efimov hyperradial wave function & 4\\
  $F,F_\nu$ & hyperradial wave function & 2,4,7\\
  $g,g_2,g_3$ & coupling strengths & 5,7\\
  $G$ & kernel for variational angular equation & 2\\
  $G_d$ & $d$-dimensional wave function & 7\\
  $\hat h_\Omega$ & reduced angular Hamiltonian operator & 2,3\\
  $\hbar$ & Planck's constant & 2-7\\
  $\hat H,\hat H_q$ & Hamiltonian operator & 2,7\\
  $\hat H_d$ & Hamiltonian operator in $d$ dimensions & 7\\
  $\cmplxI$ & complex number $\cmplxI=\sqrt{-1}$ & 2\\
  $i,j,k,l$ & indices & 2-7\\
  $k\sub B$ & Boltzmann's constant & 4\\
  $K=K_{N-1}$ & grand hyperangular momentum & 2-4\\
  $K_k$ & hyperangular momentum & 2-3\\
  $l_{N,K}$ & generalized angular momentum & 4\\
  $l_k$ & angular momentum & 2\\
  $l_T$ & thermal length & 4\\
  $l_1,l_2$ & trap lengths & 7\\
  $L_k$ & collection of angular momenta & 2\\
  $\tilde L=L_{N-1}$ & total relative angular momentum & 2\\
  $\mathcal L$ & Laguerre polynomial & 4\\
  $m$ & particle mass & 2-7\\
  $m_k$ & projection of angular momentum & 2\\
  $M$ & total mass & 2\\
  $M_k$ & projection of $L_k$ & 2\\
  $\tilde M=M_{N-1}$ & projection of $\tilde L$ & 2\\
  $n$ & single-particle density & 2\\
  $n$ & density & 3,5,6\\
  $n$ & hyperradial quantum number & 4\\
  $N$ & number of particles & 2-7\\
  $\mathcal N$ & number of bound states & 4\\
  $\mathcal N\sub B$ & number of bound two-body states & 3\\
  $\mathcal N\sub E$ & number of Efimov-like states & 4\\
  $\hat O$ & operator & 2\\
  $p_i$ & momentum & 2\\
  $\hat\permO$ & permutation operator & 2\\
  $P_R$ & total momentum & 2\\
  $\JacP,\tilde\JacP$ & Jacobi function & 2,3\\
  $q$ & degree of freedom (e.g.~$x,y,z$) & 4,7\\
  $Q_{\nu\nu'}$ & coupling term & 2\\
  $r=r_{12},r_{ij}$ & interparticle coordinates & 2-7\\
  $r_i$ & particle coordinates & 2-7\\
  $\bar r,\bar r_n$ & root-mean-square (rms) distance & 4-6\\
  $\bar r_R$ & rms separation from mass centre & 4\\
  $R,R_q$ & centre-of-mass coordinates & 2,7\\
  $\hatR_{ij},\hatR_{ijkl}$ & rotation operators & 2,3\\
  $R\sub{eff}$ & effective range & 3\\
  $s$ & reduced interaction strength & 7\\
  $t$ & time & 6\\
  $T$ & temperature & 4\\
  $\hat T$ & kinetic-energy operator & 2\\
  $\hat T_\rho$ & hyperradial kinetic-energy operator & 2\\
  $u$ & reduced two-body wave function & 3\\
  $u$ & reduced hyperradial potential & 6,7\\
  $U,U_\nu$ & hyperradial potential & 2-4,6,7\\
  $v,v_{ij}$ & reduced two-body potential & 2,3\\
  $v_1,v_2$ & rotations of potential & 2\\
  $V,V_{ij}$ & two-body potential & 2-3,5,7\\
  $\hat V$ & total interaction potential & 7\\
  $V_d$ & two-body potential in $d$ dimensions & 7\\
  $V\sub{trap}$ & external trapping potential & 2,4,7\\
  $V\sub G$ & Gaussian two-body potential & 3,5\\
  $V_\delta$ & zero-range interaction potential & 3,5,7\\
  $V_3$ & three-body interaction potential & 5\\
  $\mathcal V$ & volume & 6\\
  $w$ & variational width & 6\\
  $x,y,z$ & cartesian axes & 2-7\\
  $Y$ & spherical harmonics & 2\\
  $\mathcal Y,\tilde{\mathcal Y}$ & hyperspherical harmonics & 2\\
  \emph{æ} & mean-free path & 5\\
  \hline
  $\alpha=\alpha_{N-1}=\alpha_{12},\alpha'$ & primary hyperangle & 2,3,7\\
  $\alpha_{ij}$ & hyperangle related to $r_{ij}$ & 2,3,7\\
  $\alpha_k$ & hyperangle related to $\eta_k$ & 2,3\\
  $\beta$ &  deformation parameter & 7\\
  $\gamma$ &  deformation parameter & 7\\
  $\Gamma$ &  Euler's gamma function & 2,3,7\\
  $\Gamma\sub{rec},\Gamma\sub{tunnel},\Gamma\sub{collapse}$ & decay widths & 6\\
  $\delta$ &  Dirac delta function & 3,5,7\\
  $\delta^{(d)}$ &  $d$-dimensional delta function & 7\\
  $\delta\sub G$ & Gaussian representation of $\delta$ & 5\\
  $\hat\Delta$ & Laplacian operator & A\\
  $\varepsilon$ & reduced energy & 7\\
  $\eta_k$ & Jacobi coordinates & 2\\
  $\theta$ & polar angle for hyperradius & 7\\
  $\Theta$ & step or truth function & 2,3\\
  $\vartheta,\vartheta_k,\vartheta_{ij,k},\vartheta_{ij}$ & polar angle & 2,7\\
  $\kappa$ & wave vector or wave number & 2,3\\
  $\lambda,\lambda_\nu$ & hyperangular eigenvalue & 2-6\\
  $\lambda_K$ & kinetic-energy hyperangular eigenvalue & 2-4\\
  $\lambda_\delta$ & $\lambda$ with expectation value of $V_\delta$ & 3-7\\
  $\lambda_\infty$ & plateau value for angular eigenvalue & 3-6\\
  $\lambda^{(2)}$ & angular potential for two-body state & 3\\
  $\lambda\sub{3-body}$ & $\lambda$ from three-body interaction & 5\\
  $\hat\Lambda_{N-1}$ & grand hyperangular momentum operator & 2\\
  $\hat\Lambda_k$ & hyperangular momentum operator & 2\\
  $\mu$ & chemical potential & 5\\
  $\nu=\nu_{N-1},\nu_k$ & hyperangular quantum number & 2-4\\
  $\nu\sub{rec}$ & three-body recombination rate & 6\\
  $\nu\sub{tunnel}$ & macroscopic tunneling rate & 6\\
  $\nu\sub{trap},\nu_q,\nu_x,\nu_y,\nu_z$ & trap frequencies & 3-7\\
  $\xi$ & Efimov scale & 4\\
  $\hat\Pi,\hat\Pi_k,\hat\Pi_{ij}$ & partial hyperangular momenta & 2\\
  $\rho$ & hyperradius & 2-7\\
  $\rho_q,\rho_\perp,\rho_z$ & components of hyperradius & 7\\
  $\bar\rho$ & root-mean-square hyperradius & 3-6\\
  $\varrho$ & planar projection of coordinate & 7\\
  $\sigma$ & action integral & 6\\
  $d\tau$ & reduced hyperangular volume element & 2,3\\
  $\tau,\tau\sub{rec},\tau\sub{tunnel},\tau\sub{trap}$ & time scales & 6\\
  $\Upsilon$ & centre-of-mass wave function & 2\\
  $\phi,\phi_\nu$ & Faddeev component & 2,3,7\\
  $\phi_K$ &  Faddeev kinetic-energy eigenfunction & 3\\
  $\tilde\phi$ & reduced Faddeev component & 3\\
  $\Phi,\Phi_\nu$ & angular wave function & 2-4\\
  $\Phi_K$ & angular kinetic-energy eigenfunction & 3\\
  $\varphi,\varphi_k$ & azimuthal angle & 2\\
  $\chi$ & phase shift of two-body wave function & 3\\
  $\psi$ & wave function component & 2,5,6\\
  $\Psi$ & total wave function & 2,4\\
  $\omega,\omega_x,\omega_y,\omega_z,\omega_q$ & angular trap frequencies & 2-7\\
  $\Omega,\Omega_{N-1}$ & all hyperangles & 2-4,7\\
  $\Omega_k$ & part of hyperangles & 2\\
  $d\Omega_\alpha^{(k)}$ & volume element for $\alpha_k$ & 2,7\\
  $\Omega_\eta^{(k)}$ & angles for direction of $\eta_k$ & 2\\
\end{longtable}

\cleardoublepage                                              
\addcontentsline{toc}{chapter}{List of figures}
\listoffigures

\cleardoublepage
\addcontentsline{toc}{chapter}{\bibname}                      
\bibliographystyle{alpha}
\bibliography{/usr/users/oles/Few-body/Skriblerier/bibliografi}



\end{document}